\pdfoutput=1 
\documentclass{cau} 

\usepackage{tocloft,calc}

\AtBeginDocument{\addtolength\cftchapnumwidth{\widthof{\bfseries Chapter }}}

\usepackage[paper=b5j,layout=b5j,includefoot,top=30mm,bottom=20mm,left=25mm,right=25mm,headheight=10mm,headsep=10mm,footskip=15mm,hoffset=0mm, ,nomarginpar]{geometry}
\usepackage[utf8]{inputenc}
\usepackage{amsfonts}
\usepackage{amsmath}
\usepackage{amssymb}
\usepackage{bm}
\usepackage{derivative}
\usepackage{enumitem}
\usepackage{hyperref}
\usepackage{mathtools}
\usepackage{physics} 
\usepackage{titlesec}
\usepackage{xcolor}

\usepackage{graphicx}
\graphicspath{{./images/}}
\DeclareGraphicsExtensions{.pdf,.png,.jpeg,.jpg}

\usepackage{caption}
\usepackage[caption=false,font=small]{subfig}
\newcommand{\be}{\begin{equation}}
\newcommand{\ee}{\end{equation}}
\newcommand{\bea}{\begin{eqnarray}}
\newcommand{\eea}{\end{eqnarray}}
\def \Mpl {M_P}

\usepackage{float}
\usepackage{fixltx2e}
\usepackage{stfloats}
\fnbelowfloat

\usepackage{booktabs}
\usepackage{multirow}
\usepackage{lscape}
\usepackage{longtable}

\usepackage{fancyhdr}
\usepackage{cancel}
\usepackage{cleveref}
\usepackage{enumitem}
\usepackage{slashed}
\usepackage{algorithmic, algorithm}
\usepackage{diagbox}
\usepackage{pdflscape}

\usepackage[toc,page]{appendix}

\usepackage{natbib}
\usepackage{bibunits}
\code{D}
\edegree{Doctor}
\kdegree{박사}
\edept{Physics}
\kdept{물리학과}
\emajor{Particle Physics}
\kmajor{입자물리학}
\etitle{Towards self-consistent models for Inflation and Early Universe Cosmology}
\ktitle{우주 급팽창과 초기 우주의 일관적인 모형에 대한 연구}

\advisor{Hyun Min Lee}
\ename{Adriana G. Menkara}
\kname{아드리아나 멘카라}
\knamenospace{아드리아나멘카라}
\gcount{115}
\coveryear{2024}
\covermon{August}
\committeeA{Siyeon Kim}
\committeeB{Gungwon Kang}
\committeeC{Chang Hyon Ha}
\committeeD{Ki-Young Choi }

\renewcommand{\contentsname}{Table of Contents}
\renewcommand{\listtablename}{List of Tables}
\renewcommand{\listfigurename}{List of Figures}
\renewcommand{\bibname}{References}

\kabstractkname{국 문 초 록}
\kabstractename{국 문 초 록}
\eabstractkname{Abstract}
\eabstractename{Abstract}
\kackname{감 사 의 글}
\eackname{Acknowledgements}
\kvitaname{이 력 서}
\evitaname{Curriculum Vitae in Korean} 
\hypersetup{%
pdftitle = {Title},
pdfsubject = {Dissertation to get the degree of a Doctor of science},
pdfkeywords = {Keywords},
pdfauthor = {\textcopyright\ My Name, Lee Hyun Min},
}

\begin{document}

\newgeometry{layout=b5j,nohead,nofoot,nomarginpar,top=30mm,bottom=27mm,left=30mm,right=30mm}
\pdfbookmark[0]{Titlepage}{title}

\maketitle
\makesubmission
\clearpage\newpage
\restoregeometry

\newpage
\lhead{} \chead{\leftmark} \rhead{}
\doublespacing

\setcounter{page}{1}\pagenumbering{roman}

\newpage

\begin{eabstract}
Cosmic inflation stands as a pillar of modern Cosmology since it does not only resolve the flatness and horizon puzzles but it also provides
the seeds required for structure formation.  
In order to discern between the myriad of inflationary models, we need to look beyond the inflationary predictions and into the 
subsequent era of (p)reheating. 
In this thesis we take smalls steps towards linking our knowledge of the early universe with our knowledge of particle physics. The key questions we try to answer are: 
i) what is the nature of the inflaton,  ii) how can we use inflation and (p)reheating to shed light on the particle physics puzzles and iii) how to construct a natural model for inflation.

We first consider Higgs inflation, which is known to suffer from a unitarity problem. In order to recover unitarity up to the Planck scale, we propose several UV completions arising from higher curvature terms, discuss the implications of the extra degrees of freedom for vacuum stability and consider the possibility of producing Dark Matter 
during and after reheating. Furthermore, since at the high energies of the early universe quantum gravity becomes important, we provide an embedding of these models into Supergravity. We discuss the need to include higher curvature terms in order to cure the tachyonic instability induced by the extra fields inherent to supersymmetry and the phenomenology of SUSY breaking and its mediation to the visible sector.

Then, we consider a model for natural inflation with twin waterfalls, in which a $Z_2$ symmetry is responsible for the radiative stability of the potential.  
Including a separate $Z_2^\prime$ symmetry allows us to identify one of the waterfall fields with Dark Matter, that can be sufficiently produced during preheating.

Finally, we propose a model for inflation of the $\alpha-$attractor type, where perturbativity is ensured by construction.
We take conformal couplings and expand  around the pole of the Einstein-frame kinetic term. Using this “inflation at the pole" structure we can recover successful Higgs inflation without the need of a large non-minimal coupling. 
We also apply the setup to the Peccei-Quinn field responsible for the axion solution the Strong CP problem, and reproduce the observed Dark Matter relic abundance through the axion kinetic misalignment mechanism.


\end{eabstract}
\newpage
\pdfbookmark[0]{\contentsname}{toc} 
\tableofcontents
 \clearpage\newpage

\newpage
\pdfbookmark[0]{\listfigurename}{lof}

\cleardoublepage
\phantomsection
\addcontentsline{toc}{chapter}{\listtablename} 
\listoftables

\cleardoublepage
\phantomsection
\addcontentsline{toc}{chapter}{\listfigurename}
\listoffigures

\clearpage\newpage
{\Huge \textbf{Nomenclature}}

\vspace{1cm}

BBN / Big Bang Nucleosynthesis

CMB / Cosmic Microwave Backgrounds

DM / Dark Matter

EFT / Effective Field Theory

EW / Electroweak

FIDM / Freeze In Dark Matter

FIMP / Feebly interacting Massive Particles

FODM/ Freeze Out Dark Matter

LHC / Large Hadron Collider

LSS/ Large Scale Structure

$\Lambda$CDM / Cold Dark Matter plus a Cosmological Constant

MSSM / Minimal Supersymmetric Standard Model

NMSSM / Next to Minimal Supersymmetric Standard Model

QCD / Quantum Chromo Dynamics

SIDM /Self-Interacting Dark Matter

SM / Standard Model

SUSY/ SuperSymmetry

\clearpage\newpage

\lhead{}
\begin{doublespace}

\setcounter{page}{1}\pagenumbering{arabic}
     
\chapter{Introduction} \label{introduction}

\begin{small}
    ``Aristotle said a bunch of stuff that was wrong.
     Galileo and Newton fixed things up. 
     Then Einstein broke everything again. 
     Now, we've basically got it all worked out, except for small stuff, big stuff, hot stuff, cold stuff, 
     fast stuff, heavy stuff, dark stuff, turbulence, and the concept of time.”\\
    Zach Weinersmith
    \end{small}

    \vspace{5mm}

The Standard Model of Particle Physics is incomplete. 
The Standard Model of Big Bang Cosmology is also incomplete.
In particular, we don't understand the reason why the Higgs mass has the value it does(\textbf{Hierarchy Problem}), or why the neutron electric dipole moment is so tiny(\textbf{Strong CP Problem}). 
Within the Standard Model framework, we cannot explain
the origin of neutrino masses, which have been now probed by neutrino oscillations; nor can we
provide a suitable candidate for Dark Matter, which seems to comprise most of the matter content of our universe, and, surely, the quantization of gravity remains an elusive topic.
At the same time, Big Bang Cosmology alone cannot explain why our universe is so flat and homogeneous(\textbf{flatness problem}), or why does it look correlated on large scales(\textbf{horizon problem}).\\
Surely not all of these issues are in equal footing. 
The presence of Dark Matter, the horizon problem and the observation of neutrino oscillations all enforce us to extend the Standard Model based on direct evidence. 
On the other hand, the Hierarchy and Strong CP problems are based on a discrepancy between our expectations of what the parameters of our theories should be and what they are actually measured to be.
Nevertheless, each one of these questions alone is enough to conclude that there must be some physics yet undiscovered, and that the Standard Model is only an Effective Field Theory.

This simple statement, however, has huge implications. Let's think first about the Higgs Hierarchy problem.  At its heart, the question of why is the higgs mass observed to be $125$ GeV is the question of why the SM  seems to be insensitive to the UV physics.
The SM fermions are protected from getting mass corrections by their chiral symmetry, and the gauge bosons are protected by the gauge symmetry. However, any new physics that we introduce in order to solve the puzzles of the Standard Model will correct the Higgs mass.
In order to avoid getting large corrections, Supersymmetry, extra global symmetries or compositeness are usually assumed.
Between them, Supersymmetry has been particularly compelling not only because it is a theory free of quadratic divergences, but  because it can also lead to gauge coupling unification, provides a natural candidate for Dark Matter and can stabilize the inflationary potential.

All in all, towards the end of the 20th century, following the remarkable success of the recently consolidated Standard Model, there was a huge expectation that new particles, and in particular low-scale SUSY, would be discovered at the LHC. 
However, new particles haven't showed up after the finding of the Higgs boson, and particle physicists have since started looking for cues of new physics also in cosmology.
There's a good reason for this: the incredibly high energies of the early universe provide a unique opportunity to probe energy scales that we could never access through colliders.

In this thesis we take a phenomenological approach to inflation and early cosmology, from the particle physics perspective.
However, in contrast with particle physics, cosmology is an observational science, and so, it makes sense to start this thesis by trying to answer the question: \textit{what is it that we really know about our universe?}

Firstly, Planck data shows that our universe is remarkably consistent with a flat FLRW universe, with a curvature parameter

\begin{equation}
    \Omega_K = 0.007 \pm 0.0019.
\end{equation}

The fact that our universe is flat implies that it must also have been very homogeneous in its very early times. Otherwise, density fluctuations would have grown and the universe would have recollapsed. 
This intuition is confirmed by the observations of the \textbf{Cosmic Microwave Background}(CMB). 
The CMB  is an incredibly powerful tool that allows us to constrain our speculative models for the early universe. 
Thanks to the CMB we know that our universe was, at its early stages, exceptionally homogeneous, and that the primordial temperature fluctuations were small, scale invariant, gaussian, and adiabatic.
These observations are in perfect agreement with the most simple models of inflation, and they show no evidence of physics beyond the \textbf{$\Lambda$CDM} model of Standard Cosmology, and neither of isocurvature perturbations or cosmic defects.
The CMB also tells us how long inflation could have lasted. However, there is one thing the CMB can't really tell us anything about: the energy scale at which inflation took place.
Therefore, inflation model building remains a creative and active field of cosmology.
One of the big challenges of inflationary model building lies in the fact that we don't know what happened between the end of inflation and Big Bang Nucleosynthesis.
Current experiments simply cannot probe this part of the history of the universe, which typically falls into subhorizon scales and hence gets eventually washed out by inflation.

In this thesis, we delve into several aspects of inflation that are still not completely understood. The first question we try to address is 
whether it is really necessary to include new physics in order to accommodate an inflationary regime. 
In principle, the SM Higgs boson could drive the accelerated expansion. This would be an incredibly elegant solution, since it would provide a clear connection between two very different energy scales, as well as a link between the puzzles of particle physics and the structure of the SM, and the puzzles of cosmology, such as the nature of Dark Matter.
Unfortunately, the SM Higgs potential fails to reproduce the observable universe, leading to an overproduction of primordial energy fluctuations. 

It was first realized that a non-minimal coupling with gravity was necessary in order to renormalize the theory, and that adding this extra term could heal Higgs inflation given that the non-minimal coupling, $\xi$, were sufficiently large.
However, such a large coupling leads to a loss of unitarity, i.e, perturbativity is indeed broken at the energies required for inflation, making any prediction totally unreliable.
In \textbf{chapter \ref{chap_UV}}, we  address this issue by proposing several possible UV completions. These UV completions rely on extra scalar fields that change the dynamics of inflation, and arise from  adding general curvature terms that go beyond the $R^2$ term from Starobinsky inflation.
We call these UV complete models \textit{Higgs-sigma models}, in analogy with the sigma models of QCD.

The story, however, ought not to end here. Even if we manage to recover the predictive power of Higgs inflation, once inflation ends, the universe is cold and empty, containing only condensates of the inflaton field. 
It becomes then necessary to introduce a mechanism in order to transfer the energy stored in the inflaton condensates to the thermal plasma, realizing the necessary conditions for the hot Big Bang.
\textbf{Reheating} is a perturbative process in which, as the field oscillates around the minima of the potential, it decays into radiation.
However, during the first stages of reheating, collective effects of the inflaton condensates, such as Bose-enhancement, can lead to explosive particle production. This non-perturbative phenomenon is dubbed \textbf{preheating}.
We therefore continue \textbf{chapter \ref{chap_UV}} by performing an analysis of the reheating dynamics of these newly proposed UV completions.
We also know that Dark Matter needs to be produced at some point in the early universe in order to explain structure formation. Hence, we consider the possibility of producing Dark Matter during and after reheating, through a Freeze-In mechanism.
Although the Dark Matter we consider here is \textit{feebly interacting}, our approach could be extended to other Dark Matter candidates, such as WIMPs.

We know that our universe is not supersymmetric since we have seen no signals of superpartners, 
but nevertheless, SUSY could still provide a relaxation of the  Hierarchy problem if the symmetry breaking scale is below the Planck mass.
Supersymmetry must be broken, at its best, at the energies of our colliders, but it could also be broken at much higher energy scales that we could hope to probe through inflation.
Furthermore, when we gauge Supersymmetry, the SUSY algebra forces us to gauge the Poincare group. Thus, we obtain a theory that contains General Relativity as well: supergravity(SUGRA). We are trying to link the early universe to the Higgs boson, and, since at the high energies needed for inflation quantum gravity becomes important, it makes sense to continue our line of work by providing a supergravity embedding of the \textit{sigma models} for Higgs inflation.
However, the supersymmetric embedding includes new particles that could spoil the flatness of the inflationary potential, changing the inflationary predictions. 
In \textbf{chapter \ref{chap_sugra}} we study the stabilization of these new directions during inflation. In particular, in the Next-to-Mininimal SuperSymmetric Standard Model($NMSSM$) embedding of Higgs-$R^2$ inflation, the singlet scalar spectator fields develop a tachyonic instability that can be cured by adding higher curvature terms.
Moreover, we know that Supersymmetry must be broken by the vacuum energy that is driving inflation. However, after inflation the inflationary fields have a vanishing VEV, meaning that SUSY would be unbroken.
In \textbf{section \ref{sec:susybreak}} we consider possible extensions of the NMSSM in order to realize SUSY breaking and study its mediation to the visible sector.

Similarly to the Higgs boson, the inflaton field also suffers from its own naturalness issues. 
In particular, many models of inflation rely on extremely flat potentials that may be spoiled by radiative corrections, coming from the couplings of the inflaton to matter particles, unavoidably necessary for reheating. Moreover, in models where there are multiple fields present during inflation, the coupling of these fields to the inflaton may lead to corrections to the inflaton mass, raising the question of whether having a light inflaton is \textit{technically natural}.
In \textbf{chapter \ref{chapter_hybrid}} we discuss the case in which the inflaton is identified with a pseudo-Nambu Goldstone boson coming from a spontaneously broken global symmetry.
In principle, the inflaton field in a \textit{natural inflation} model could be identified with the Higgs boson, as well as with the \textit{QCD axion}.
In our case, we just consider a general axion and study the parameter space for successful inflation and subsequent reheating.

As it happens with the solutions to the Hierarchy Problem, in order to solve the Strong CP problem it is possible to introduce a new global symmetry. In the Peccei-Quinn solution, we introduce a new global $U(1)$ symmetry, that, when spontaneously broken, leads to the axion. The axion then rolls towards the minimum of the potential and dynamically relaxes the neutron EDM to zero.
However, in the same way that UV physics corrects the mass of the higgs boson, the effects of quantum gravity can spoil the axion solution. 
Gravity is believed to break any global symmetry, and so, even if we tried to impose the anomalous symmetry, instanton effects will add a mass term that is
not centered around a zero neutron EDM. In \textbf{section \ref{PQPole}}, we try to relate the Strong CP problem to inflation.
We extend the Standard model with the PQ field, a complex scalar field whose radial component drives inflation, while the angular part, identified with the QCD axion, can realize the observed amount of Dark Matter.
Gravitational effects lead to a non-zero velocity for the axion by the end of inflation, and Dark Matter is reproduced through kinetic misalignment, achieving a relatively low axion decay constant.
Axions with a non-zero initial velocity are rich in phenomenology and can lead to
axiogenesis \cite{Co:2019wyp,Co:2020jtv, Kawamura:2021xpu}, or provide signatures such as gravitational waves \cite{Gouttenoire:2021jhk} or axion fragmentation \cite{Eroncel:2022vjg}. 
Furthermore, in our model, the produced axions could become Dark Radiation at a level that may be detectable in the next generation of experiments.

Finally, in \textbf{chapter \ref{HiggsPole}} we explore the possibility of curing Higgs inflation not by adding a large non-minimal coupling to gravity, but instead by assuming conformal couplings. 
In this set up, that we dub \textit{Higgs-Pole inflation}, we take an expansion of the Higgs field around the pole for which the effective Planck mass in the Jordan frame vanishes, effectively decoupling gravity effects from the inflationary dynamics.
However, in order to realize successful inflation we find it necessary to impose a small running quartic coupling for the Higgs field during inflation, which calls for a nontrivial extension of the SM. 
Hence, we propose a supersymmetric embedding, similarly as we did in \textbf{chapter \ref{chap_sugra}}.  In the supergravity set-up, we can relax the Higgs quartic coupling to a small value along the D-flat direction during inflation.
The potential in the Higgs pole inflation set-up belongs to a class of inflation models containing a pole of the kinetic term, known as $\alpha$-attractors in supergravity.
We can apply the same \textit{inflation at the pole} set-up to the PQ field. In contrast with previous studies, which relied on a small non-minimal coupling \cite{Fairbairn:2014zta,Nakayama:2015pba,Ballesteros:2016euj,Ballesteros:2016xej}, applying a \textit{Pole} structure provides a possibility of realizing inflation without the need of relying on trans-Planckian values of the PQ field for inflation.

\subsection*{Outline}
Before jumping into the full discussion, we present here a succinct summary of every chapter.
\begin{itemize}

\item We start in \textbf{chapter \ref{inflation}} with a review of the basic concepts of inflation. We start by recalling the Big Bang puzzles, i.e, the flatness and horizon problem that lay behind the motivation for inflation, and compute how much should inflation last in order to solve them.
Then, we move to the simplest of models: slow roll single field inflation. We make a brief review of the CMB and how we can use it to rule out our inflationary models.
We then move to the study of particular models. We first review Higgs inflation and why it fails. We discuss the necessity of including a large non-minimal coupling with gravity and the issues of unitarity linked to this non-minimal coupling.
Finally, we review the original models of hybrid and natural inflation, that will serve as the basis for our work in \textbf{chapter \ref{chapter_hybrid}}.

\item In \textbf{chapter \ref{chap_partphys}} we make a review of the particle physics puzzles that are of concern for this thesis: Dark Matter and Supersymmetry.
We go through the Boltzmann equations for Dark Matter production and review three different mechanisms for dark matter production: Freeze-out, Freeze-in and the axion kinetic misalignment.
Because we will consider supergravity in several chapters of this thesis, we introduce a brief review of Supersymmetry.

\item Then, in \textbf{chapter \ref{chap_UV}} we delve into Higgs inflation. We introduced already the unitarity issues of this model in \textbf{chapter \ref{inflation}}, so we now introduce several UV completions.
 We start by the most simple case, Starobinsky inflation as a UV completion, and then, in the subsequent sections, we add general higher curvature terms.
 After we checked the perturbativity conditions, we move to the inflationary dynamics of these Higgs-sigma models, as well as the perturbative reheating dynamics.
We further add a singlet scalar, that serves as a dark matter candidate and study Dark Matter production from the reheating and post-reheating era, in both the cases where conformal couplings are assumed and when we allowed a small deviation and include a Higgs-portal coupling.

\item  In \textbf{chapter \ref{chap_sugra}} we provide a supergravity embedding of Higgs inflation. We study the effect of adding new spectator fields and the decoupling of heavy scalars. We propose adding higher curvature terms in order to cure the tachyonic instability from the singlet scalar directions.
The vacuum energy driving inflation breaks supersymmetry, so we study the phenomenology of SUSY breaking and its mediation to the visible sector. We can do this in two different ways: SUSY breaking from higher curvature terms favors high energy susy breaking, while
considering a renormalizable superpotential of the O’Raifeartaigh type it is possible to realize low scale SUSY breaking.

\item \textbf{Chapters \ref{chap_UV} and \ref{chap_sugra}} provide a complete picture of Higgs inflation. Now, in \textbf{chapter \ref{chapter_hybrid}} we will worry about the naturalness issues of inflationary model building.
We consider a model for hybrid inflation with twin waterfall fields. Adding a $Z_2$ symmetry in the waterfall sector protects the potential from radiative corrections. We first discuss the possible origin of this discrete symmetry and then move to the inflationary and (p)reheating dynamics.
If we assume an extra $Z_2'$ symmetry, we can further identify one of the waterfall fields with Dark Matter and hence study the amount of Dark Matter produced during (p)reheating.
Finally, we provided a microscopical dark QCD model from which our model can naturally arise.

\item In \textbf{chapter \ref{HiggsPole}} we construct a model for inflation at the pole, which relies on conformality.
These conformal couplings lead to the proximity of the effective Planck mass to zero in the Jordan frame during inflation, which corresponds to a pole in the kinetic term in the Einstein frame.
When we apply this set-up to the Higgs, we require a tiny quartic coupling. Hence, we need to include extra scalar or gauge fields in order to keep the running Higgs quartic coupling small during inflation. 
We provide first an extension with a singlet scalar field, and then a supergravity embedding of the Higgs pole inflation.
In \textbf{section \ref{PQPole}} we use the Pole structure to achieve axion Dark Matter by using the PQ field as the inflaton.

\item Finally, in \textbf{chapter \ref{chap:summary}} we summarize and include our final remarks.

\item This thesis contains two appendices. In \textbf{appendix I} we show how to derive the equations of motion when we have kinetic mixing in a FRW background. 
\textbf{Appendix II}, is the appendix I wish I had when I just started my PhD and jumped onto computing how the Ricci scalar changes under a conformal transformation.
There, we go step by step through the conformal transformation and provide as an explicit example the Lagrangian for PQ inflation in \textbf{Chapter \ref{PQPole}}.
\end{itemize}

The work completed during this Ph.D. is collected in the List of Publications at the end of this thesis, however, only material in [1–3, 7-10] is presented in this thesis.

\chapter{Inflation} \label{inflation} 

\begin{small}
    ``In the beginning, the Universe was created. This has made a lot of people very angry and been widely regarded as a bad move.”\\
    Douglas Adams
    \end{small} 
    \vspace{5mm}
\section{Why Inflation?}

Physics of the Early Universe is intriguing and fascinating,  but also full of challenges and speculation. We find ourselves in a  precision era, in which thanks to the observations made by BICEP Array\cite{Hui:2018cvg}, Simons Observatory\cite{Hensley:2021ydb}, CMB-S4 \cite{Abazajian:2019eic}, LiteBIRD\cite{Hazumi:2019lys} or PICO \cite{NASAPICO:2019thw}, we will be able to narrow down the possibilities for Cosmic Inflation.
Despite this unprecedented observational precision, we cannot know with any certainty what happened before reionization. However, we know that the Standard Big Bang Cosmology by itself cannot explain the universe as we observe it.
Why is the universe so flat, or so homogeneous? Why does the sky look correlated on large scales? 
The former question is known as the \textbf{flatness problem}. It encodes the fact that our universe seems to have started from very fine-tuned initial conditions.
The second question receives the name of the \textbf{horizon problem}. This is the puzzle that lies at the very core of any inflationary theory.

Initially, inflation was proposed by Alan Guth \cite{Guth:1997wk}. In Guth's theory of inflation, the scalar field stayed trapped in false minima of the potential. 
As long as the field remains trapped in the false minimum, its vacuum energy remains constant, $ H=\frac{\dot{a}}{a}$, implying that the scale factor evolves as 
\begin{equation*}
    a \propto e^{Ht},
\end{equation*}
 and thereby leading to a period of accelerated expansion.
However, Guth's old theory of inflation suffers from a catastrophic problem. Long story short, old inflation was explained by a delayed first order phase transition.
The tunneling from the false to the true vacuum, which is crucial for inflation to end, is unlikely to have happened simultaneously everywhere. 
Instead, different regions of spacetime would have undergone the phase transition at different times. Once in a region of spacetime the field decays, it forms bubbles that expand, as it happens with water droplets in a supercooled state.
The energy of those bubbles is stored within the bubble walls, while the interior of the bubbles itself remains empty.
Outside of the bubbles, the universe continues to inflate, separating the bubbles further and further away, making their collision unlikely. The result would be a very inhomogeneous universe that wouldn't match the one we live in.

Old inflation was then forgotten, and new inflation, proposed by Linde\cite{Linde:1982zj}, Albrecth and Steinhard\cite{Albrecht:1982wi}, took its place. 
In new inflation, it is common to have a scalar field rolling down a very flat potential.
Chaotic \cite{Linde:1983gd} and eternal inflation \cite{Guth:2000hz} have their roots here.
Now we do not need to worry about bubble formation and later collision, because our entire universe fits in just a portion of one of the bubbles.

In this chapter, we first explore the puzzles of Standard Big Bang Cosmology and show how inflation provides a common solution to all of them. 
Then, we jump into the simplest possible model for inflation: single field slow roll inflation.
This will set up the basis to understand the content in following chapters.  Then, we introduce some of the most relevant models for inflation, i.e, Higgs inflation, natural inflation and hybrid inflation. The last sections of this chapter cover a very brief review of Dark Matter and Supersymmetry.

\subsection{The Flatness Problem of the Universe}

The cosmological principle states that our universe is homogeneous.
Indeed, our universe is homogeneous on large scales today, but we have reasons to believe that it was already homogeneous at very early times.
The equilibrium between kinetic energy and gravity is a rather delicate matter.
Inhomogeneities are gravitationally unstable, and they eventually grow with time. If our universe were to contain any inhomogeneities at its early times, they should have grown and we should be able to observe them\footnote{Quantum fluctuations of the metric field will be responsible for the seeds that will lead to galaxy formation after inflation, but our analysis here is semiclassical.}.\\
Let's think of a slice $\Sigma$ of constant time of our universe. In this slice, we can define the values for all positions and velocities of particles.
If these velocities were slightly slow, the universe would collapse due to gravity. If instead they were slightly big, the universe will expand too fast, making structure formation impossible, leaving us in an empty universe.
So, within the framework of Big Bang Cosmology, we cannot do anything but accept that the initial conditions must be very fine tunned.
This situation is referred to as the flatness problem, since the difference between potential and kinetic energy defines the local curvature of the universe.
It is worth mentioning that there's nothing intrinsically wrong with the universe being flat, but as physicists, we would like to provide an explanation as to why it is so.\\

\subsection{The Horizon Problem}

In order to compute how much of the universe is in causal contact, we define the comoving particle horizon
\begin{equation}\label{parthor}
    \tau \equiv \int^t_0 \frac{dt'}{a(t')} = \int^a_0 \frac{da}{aH^{2}}= \int^a_0 d\ln\left(\frac{1}{aH}\right).
\end{equation}
$\tau$ is the maximum possible distance that light could have travel between an initial time that we set to $0$, and a later time $t$.
 If the distance that separates two regions is larger than $\tau$, these two regions could never have communicated.
Furthermore, we define the comoving Hubble radius as $(aH)^{-1}$. 
The comoving Hubble horizon is a measure of how far away particles can communicate once we take into account the expansion of the universe. 
It is important to notice that, if the distance separating this two particles is larger than $(aH)^{-1}$, it just means that particles are not in causal contact today, but they could have been in the past, and in particular, during the time of reionization.
If we now parametrize the evolution of the universe by an equation of state,

\begin{equation}\label{hubhor}
    \left(aH\right)^{-1} = H_{0}^{-1} a^{\frac{1}{2}\left(1 + 3\omega\right)},
\end{equation}

and substituting (\ref{hubhor}) in (\ref{parthor}), we find

\begin{equation}
    \tau \propto a^{\frac{1}{2}\left(1 + 3\omega\right)}.\label{tauu}
\end{equation}

Equation (\ref{tauu}) implies that the comoving horizon grows monotonically with time, and so, regions that are entering causal contact today were not in contact in the past \cite{Baumann:2009ds}. 

From the moment they last scattered, photons have been free-streaming without interacting any further.
We can then compare our present Hubble radius with the Hubble radius at the time of last scattering and obtain that, at the time of recombination, the universe was composed of $10^6$ disconnected patches \cite{Riotto:2018pcx}.
This is in clear contradiction with CMB observations \cite{Planck:2019evm}.

\subsection{Relic problem}

GUT theories with symmetry breaking predict heavy relics, such as magnetic monopoles, that could lead to overclosure of the universe.
Inflation solves this problem because this very heavy relics would have been washed away, together with any initial inhomogeneity.

\section{How much must inflation last?}

The curvature parameter $\Omega_K$ is defined as

\begin{equation}
    \Omega_K = - \frac{K}{a^2 H}\label{curvat}.
\end{equation}
During inflation, the Hubble rate $H$  is constant. Then, from equation (\ref{curvat})  we infer that the curvature must decrease as $a^{-2}$. 
This provides an explanation for the initial flatness, since any initial curvature would have been washed away.
But for this to work, inflation must last a sufficient amount of time. The only condition we need to require is that all the fluctuations that we observe today were inside the horizon at some point in the past,

\begin{equation}
    \left(a_0 H_0\right)^{-1} < (a_i H_i)^{-1}. \label{scalefac}
\end{equation}

In eq.~(\ref{scalefac}) we denoted by $a_0, a_i$ the scale factor today and at some initial time $t_i$, respectively.
We define an \textbf{e-folding} as the time that it takes for the universe to expand by a factor of $e$. The duration of inflation is then related to the total number of e-foldings
\begin{equation}
    N_{\mathrm{tot}}= \ln\left( \frac{a_e}{a_i}\right).
\end{equation}

The estimation of the number of e-foldings has some theoretical uncertainty associated to it, since it depends on the details of reheating, a period whose equation of state is unknown. We will go briefly through the reheating dynamics in \textbf{section \ref{sec:reheatingintro}}, but for the moment being, let's assume that inflation is followed by a period of radiation domination, denoted by a subscript R. Then, we can write

\begin{equation}
    \frac{a_0 H_0}{a_R H_R} = \frac{a_0}{a_R} \left(\frac{a_R}{a_0} \right)^2 = \frac{a_R}{a_0} \sim \frac{T_0}{T_R},
\end{equation}

where in the second step we have used that $H \sim a^{-2}$ during the radiation era.

Now let's make an approximate estimation of $N_{\mathrm{tot}}$. Then,

\begin{equation}
    \left(a_i H_i\right)^{-1 }  >\left(a_e H_e\right)^{-1} \sim 10^{-28} \left( \frac{T_R}{10^{15} \mathrm{GeV}}\right)\left(a_e H_e\right)^{-1},
\end{equation}

where we have introduced a reference temperature of $10^{15}$ GeV.

During inflation, the Hubble rate stays roughly constant, so $H_i = H_e$ and

\begin{equation}
    a_i^{-1} > 10^{28} \left(\frac{T_R}{10^{15} \mathrm{GeV}}\right) a_e^{-1},
\end{equation}
 so that
\begin{equation}
    \frac{a_e}{a_i} > 10^{28} \left(\frac{T_R}{10^{15} \mathrm{GeV}}\right),
\end{equation}

and the number of e-foldings is finally given by
\begin{equation}
    \ln \left( \frac{a_e}{a_i} \right) = \ln \left(10^{28} \frac{T_R}{10^{15} \mathrm{GeV}}\right) \simeq 64 + \ln\left(\frac{T_R}{10^{15} \mathrm{GeV}}\right). \label{efoldnum}
\end{equation}

We see from eq.~ (\ref{efoldnum}) that the minimum required number of e-folds can be lowered depending on the details of reheating.

\section{Slow roll inflation}\label{slowrollsec}

The simplest possible inflationary model is that of a single scalar field, whose action is given by
\begin{equation}\label{slowroll}
    S = \int \mathrm{d}^4 \sqrt{-g}\left[ \frac{1}{2}R + \frac{1}{2} g^{\mu \nu}\partial_{\mu}\phi \partial_{\nu}\phi - V(\phi)\right].
\end{equation}

Varying (\ref{slowroll}) with respect to the metric we find the stress energy tensor,
\begin{equation}
    T_{\mu \nu} \equiv - \frac{2}{\sqrt{-g}}\frac{\delta S_{\phi}}{\delta g^{\mu \nu}}.
\end{equation}

In a Friedmann-Robertson-Walker metric, the energy density and pressure are given by

\begin{align}
    \rho &= \frac{1}{2} \dot{\phi}^2 + V(\phi), \\
    p  &= \frac{1}{2} \dot{\phi}^2 + V(\phi).
\end{align}

Then, from energy conservation we get

\bea
    \dot{\rho} &=& -3H(\rho + p),\\
    \dot{\phi} \ddot{\phi} + V'(\phi)\dot{\phi} &=& - 3H \dot{\phi}^2,\\
    \ddot{\phi} + V'(\phi) + 3H \dot{\phi} &=& 0.\label{eq:inflatoneom}
\eea

During inflation the Hubble constant reads

\begin{equation} \label{eq:friedmann}
    H= \sqrt{\frac{8 \pi G \rho}{3}} = \sqrt{\frac{8 \pi G}{3}\left(\frac{1}{2} \dot{\phi}^2  + V(\phi)\right)}.
    \end{equation}

    Hence, the time derivative of $H^2$ leads to 
    \begin{equation}
        2 H \dot{H} = \frac{8\pi G}{3} \left( \dot{\phi}\ddot{\phi} + V'(\phi)\dot{\phi}\right) = -8 \pi G H \dot{\phi}^2.
    \end{equation}

The condition for exponential expansion
is $\left\vert  \dot{H} \right\vert \ll H^2$, so

\begin{equation}
    \vert \dot{H} \vert \ll H^2 = \frac{2}{3}\left(\frac{1}{2}\dot\phi + V(\phi)\right),
\end{equation}

and thus $\dot{\phi} \ll V(\phi)$.
Along this discussion, we have treated inflation classically. The reader may ask if it's justified to neglect quantum corrections coming from gravity.
In general, the energy densities that we are working with are much lower than the energy associated to the Planck scale. Even if the field $\phi$ were comparable to $M_P$ at the beginning of inflation (a common assumption in many inflation models), the potential is accompanied by 
small coupling constants that suppress gravitational effects. The necessity for very small couplings is known as the \textbf{naturalness problem of inflation}, and we will tackle this in \textbf{chapter \ref{chapter_hybrid}}. These coupling constants will also not appear in the flatness conditions for slow roll inflation, so we don't need to worry about spoiling the flatness of the potential and thus affecting the amount of time that inflation lasts.
On the other hand, small couplings in the potential are necessary in order to obtain a sufficient amount of inflation. However, in models with multi-field inflation, this small coupling constant can be avoided\cite{Linde:1993cn}.

For now, let's focus on the conditions for slow-roll inflation to happen and persist a sufficient amount of time.
From eqs.~(\ref{eq:inflatoneom}) and (\ref{eq:friedmann}), we can find the evolution of the Hubble parameter,

\bea
\dot{H} = -\frac{1}{2}\frac{\dot{\phi}^2}{M_p^2}. \label{eq:Hevolution}
\eea

Then, from eqs.(\ref{eq:friedmann}) and (\ref{eq:Hevolution}), we get the slow roll parameter

\bea 
\epsilon \equiv - \frac{\dot{H}}{H^2} = \frac{\frac{1}{2}\dot{\phi}^2}{M_P^2 H^2} = \frac{\frac{3}{2} \dot\phi^2}{\frac{1}{2}\dot\phi^2 + V}. \label{eq:epsilon}
\eea

Inflation occurs as long as the potential energy dominates over the kinetic energy, so that $\epsilon \ll 1$.
Inflation will continue as long as the acceleration of the field is small. This is parametrized through a second slow-roll parameter

\bea
\eta \equiv - \frac{\ddot{\phi}}{H \dot{\phi}}.\label{eq:eta}
\eea

If $\epsilon, |\eta| \ll 1$, we can take the so-called \textbf{slow-roll approximations}. 
$\epsilon \ll 1$ implies that the kinetic energy is very small and so we can approximate eq.~(\ref{eq:friedmann}) to

\bea
H^2 \approx \frac{V}{3 M_P^2}.
\eea

So, as long as slow-roll lasts, the Hubble rate can be approximated by the potential energy, which is almost constant. The second parameter simplifies the Klein-Gordon equation in eq.~(\ref{eq:inflatoneom}) to

\bea
3 H \dot\phi \approx - V',
\eea

so we have a direct relation between the slope of the potential and the speed of the inflaton field, with $H$ roughly constant. All in all, we can take an approximated form of the slow-roll parameters that depends exclusively on the form of the potential,

\bea
\epsilon_V &\equiv& \frac{M_P^2}{2}\left( \frac{V'}{V}\right)^2, \label{eq:slowrolleps}\\
\eta_V &\equiv& M_P^2 \frac{V''}{V}.\label{eq:slowrolleta}
\eea

Inflation will end when $\epsilon_V=1$ or $\eta_V=1$. In the rest of this thesis I will omit the subscript $V$ for cleaner notation, but whenever slow-roll parameters are mentioned, we will always be referring to the approximated form, and not to the ones in eqs.(\ref{eq:epsilon}) and (\ref{eq:eta}). Situations in which the approximations (\ref{eq:slowrolleps}) and (\ref{eq:slowrolleta}) are of order one but the non-approximated slow roll parameters are still small can happen, but they are rare and we don't encounter them in our scenarios.

\section{Reheating}\label{sec:reheatingintro}
After inflation the universe is cold and empty, and all of its energy is stored in the inflaton condensates. The energy stored in the inflaton field needs to be transferred somehow to the rest of particles of the SM and Dark Matter. The process through which this happens is dubbed \textbf{reheating}. Very briefly, once the inflaton reaches the minimum of the potential, it begins to oscillate around it.
Near the minimum, the potential can be approximated to a quadratic potential $V(\phi) \approx \frac{1}{2} m^2 \phi^2$, such that the equation of motion is that of a damped oscillator,

\bea
\ddot{\phi} + 3H\dot{\phi} = -m^2 \phi.
\eea

The energy density evolves accordingly to

\bea
\dot{\rho}_\phi + 3H\rho_\phi = - \Gamma_\phi \rho_\phi,
\eea

where $\Gamma_\phi$ is the inflaton decay rate. If the inflaton is only coupled to fermions, this reheating process is usually very slow due to the small couplings. However, if we include couplings to bosons, a much more efficient decay can take place. This rapid decay is called \textbf{preheating}. Preheating happens out of thermal equilibrium.

These particles arising from the decays of the inflaton collide with each other and eventually reach thermal equilibrium, leading to the initial conditions for the hot Big Bang.  The temperature at reheating completion is given by
\bea
\rho_{RH} = \frac{\pi^2}{30}g_*(T_RH)T_{RH}^4.\label{eq:reheatingtemp}
\eea

The longer it takes for reheating to complete, the less the final temperature $T_{RH}$ will be.
The minimum possible temperature has to be larger that 1 MeV, in order not to spoil BBN.
Reheating is a rich epoch that can lead to all kind of signatures to look for: isocurvature perturbations, that could arise in multifield models, such as the ones studied in \textbf{chapter \ref{chapter_hybrid}} and \textbf{chapter \ref{PQPole}}, gravitational waves, from phase transitions such as the waterfall phase transition introduced in \textbf{chapter \ref{chapter_hybrid}},
DM production included in \textbf{chapters \ref{chap_UV} ,\ref{chapter_hybrid}, \ref{HiggsPole}} or dark radiation in the case of axions in \textbf{chapter \ref{PQPole}}.

\section{Testing inflation}
In this section, we will briefly review what type of observables we can actually allow us to constrain and probe our models of inflation.
In practice, what observations measure are correlations in the Large Scale Structure (LSS) of the universe.
In this section, we study the correlations in the CMB. For that, we decompose the perturbations in their Fourier modes. That is because in Fourier space, every component will evolve independently and it's easy to predict. We will remain at linear order. 
\subsection{CMB}\label{sec:CMB}
\begin{figure}[t]
    \centering
    \includegraphics[width=0.85\textwidth,clip]{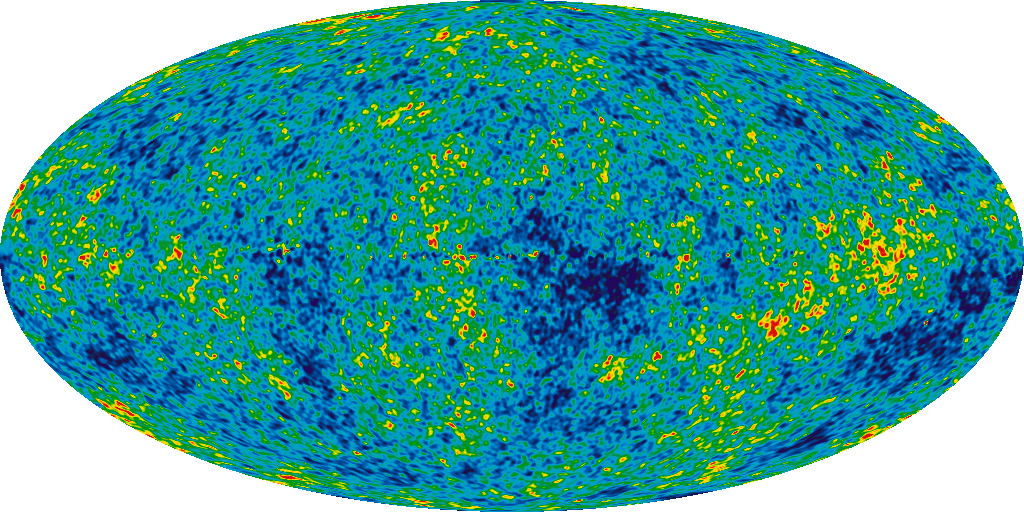}
    \caption{The CMB contains information about the temperature fluctuations, measured by the Planck satellite \cite{WMAP:2012fli}.} 
    \label{fig:cmb}
    \end{figure}

\begin{figure}[t]
        \centering
        \includegraphics[width=0.85\textwidth,clip]{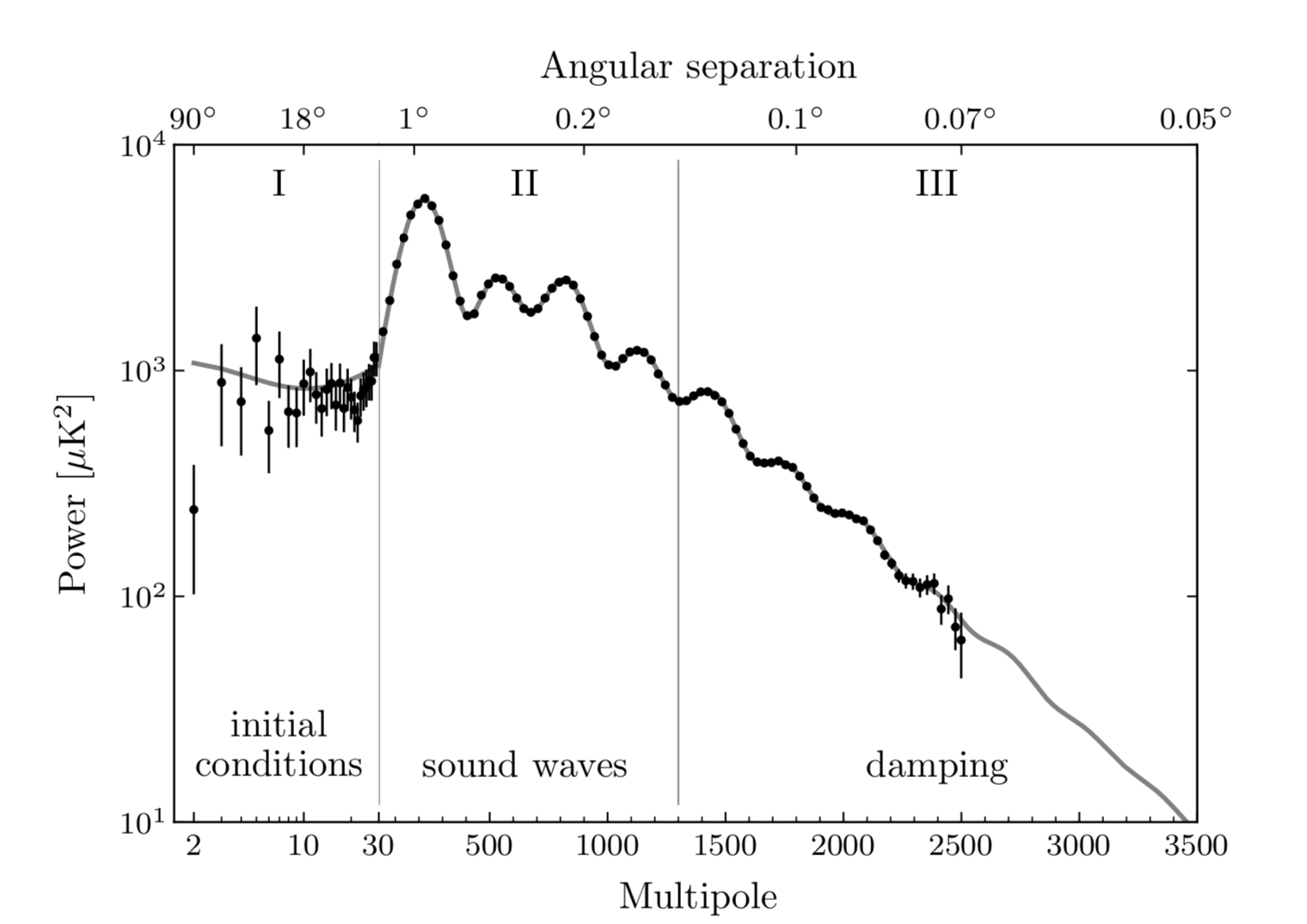}
        \caption{ Power spectrum of the CMB anisotropies as measured by the Planck satellite \cite{WMAP:2012fli}.} 
        \label{fig:cmbpower}
        \end{figure}

The CMB, showed in Fig. (\ref{fig:cmb}), is the first picture that we can obtain of our universe \cite{WMAP:2012fli}. It dates to the time when the universe had cooled down enough to allow hydrogen atoms to form. It is also the most powerful tool that we have in order to probe the early universe, not only because it is a window into the universe at it was into its very early times, but also because the perturbations imprinted on it are very small, allowing us to use perturbation theory accurately. 
The CMB contains the imprints left by the primordial sound waves at the time of photon-decoupling, or recombination. 
Different Fourier modes would have different amplitudes, and this fact reflects in the different peaks and the oscillatory behavior of the power spectrum, as shown in Fig. (\ref{fig:cmbpower}). The peaks that we observe from the CMB also provides us with information about the density of baryons and Dark Matter at the time of recombination, making it a powerful probe of structure formation.
Finally, we can also learn about the amount of neutrinos present in the early universe through the damping of the CMB peaks.\\

One of the most important features of the CMB is a temperature dipole of $3.36$ mK.
This is the largest anisotropy that we can observe in the CMB. However, it has a simple interpretation: it is caused by the relative movement of the solar system with respect to the rest frame of the CMB.
Being caused by the Doppler effect, it is simple to remove the dipole moment. What we are left with is a map of the primordial temperature fluctuations.
We are interested in studying the statistical correlations in these primordial temperature fluctuations.\\

\subsection*{Power spectrum and Harrison Zel'dovich }

It turns out that the evolution of primordial fluctuations is heavily dependent on the equation of state introduced in eq.~(\ref{tauu}).
 Thus, it becomes essential to define these perturbations so that do not depend on the equation of state, if possible. 
 This would make our predictions much more reliable, since we don't really know what is the equation of state between the end of inflation and BBN.
Taking this into account, we define the density contrast as 

\begin{equation}
    \delta \equiv \frac{\delta \rho}{\bar{\rho}},
\end{equation}

where  $\delta \rho$ is a small perturbation of the energy density and $\bar{\rho}$ its background density.
By definition, the mean value of the density perturbations is zero (they are random),

\begin{equation}
    \left \langle \rho \right \rangle = 0. 
\end{equation}

Thus, the first non trivial statistical quantity of interest is the two-point correlation function, $ \xi \left( \left \vert  \textbf{x} - \textbf{x}'\right \vert \right) $,

\begin{equation}
    \xi \left( \left \vert  \textbf{x} - \textbf{x}'\right \vert \right) \equiv \left \langle \delta(\textbf{x}, t) \delta(\textbf{x}', t) \right \rangle= \int \mathcal{D} \delta \mathbb{P}[\delta] \delta(\textbf{x},t) \delta(\textbf{x}', t).
\end{equation}

Here, $\mathbb{P}$ is the probability distribution for the given field configuration $\delta(\textbf{x})$. Because of isotropy and homogeneity, we can rewrite $ \xi \left( \left \vert  \textbf{x} - \textbf{x}'\right \vert \right)$ as simply $\chi(\textbf{r})$.

 In Fourier space, the two-point correlation function reads
\begin{align}
    \left \langle \delta(\textbf{k}) \delta^*(\textbf{k}) \right \rangle &= \int d^3 x d^3 x' e^{-i \textbf{k}\cdot \textbf{x}} e^{i \textbf{k}'\cdot \textbf{x}'} \left \langle \delta(\textbf{x}) \delta(\textbf{x}') \right \rangle \nonumber \\
    & = (2 \pi)^3 \delta_D\left(\textbf{k} - \textbf{k}'\right)\mathcal{P}(\textbf{k}).
\end{align}

The function $\mathcal{P}(\textbf{k})$ is called the \textbf{power spectrum}. It is nothing but the 3-dimensional Fourier transformation of the correlation function $\xi(\textbf{r})$.
 We notice that, because of isotropy (rotational invariance), the power spectrum only depends on the magnitude of the wavevector $k$.

We can show that the correlation function and the power spectrum form a Fourier pair such that
\begin{align*}
    \mathcal{P}(k) &= \int d^3 r e^{-ik\dot r} \xi(r) = \frac{4 \pi}{ k} \int^\infty_0 dr r \sin(kr)\xi(r),\\
    \xi(r) &= \int \frac{dk}{k} \frac{k^3}{2 \pi^2} \mathcal{P}(k)j_0(kr),
\end{align*}

with $j_0(x) = \sin(x)/x $.

Inflation predicts that the probability distribution of these primordial fluctuations follow a Gaussian distribution, as proved by observations of the CMB.

The primordial power spectrum can be written as a power law,

\begin{equation}
    \mathcal{P}\left( k, t_i\right) = A k^n,
\end{equation}

where n is called the \textbf{spectral index}. Harrison and Zel'dovich proposed that the universe is likely to have emerge from a primordial universe with $n=1$ \cite{Zeldovich:1972zz, Harrison:1969fb}.
The variance of the gravitational potential is given by 
\begin{equation} \label{eq:variancephi}
    \Delta _\Phi ^2 (k) \equiv \frac{k^3}{2\pi ^2} \mathcal{P}(k,t_i) \propto k^{n-1}
\end{equation}

From eq.~(\ref{eq:variancephi}) we clearly see that, for the harrison Zel'dovich spectrum, the power spectrum of the gravitational potential becomes independent of $k$. In other words, the field is scale invariant.
Indeed, observations of the CMB show 
\begin{equation} \label{eq:planckn} 
    n = 0.9667 \pm 0.0040,
\end{equation}

which is well in accordance with a scale invariant universe, and in agreement with the predictions of inflation.
The small deviation from the value $n=1$ can be understood from the fact that inflation is not exactly happening at a de-Sitter era. In particular, inflation must end, breaking the translation symmetry. This time dependent dynamics leads then to a small scale symmetry breaking, reflected on the value of n. 

The CMB normalization sets constrain the vacuum energy of inflation, setting a Hubble rate for inflation,

\bea
A_s=\frac{1}{24\pi^2} \frac{V_I}{\epsilon_* M^4_P}=2.1\times 10^{-9},\label{eq:CMBnormalization}
\eea

taken at the pivot scale $k_0 = 0.05 \mathrm{Mpc}^{-1}$.
The tensor-to-scalar ratio is constrained by the CMB to be

\begin{equation} \label{eq:planckr} 
    r <0.036
\end{equation}

Finally, we can infer the matter content of the universe by the scale dependence of the curvature perturbations.
Observations show that most of the matter in the universe is in the form of Cold Dark Matter (CDM),
\begin{equation}\label{eq:DMabundance}
    \Omega_c h^2 = 0.120 \pm 0.001.
\end{equation}
\section{Higgs inflation}\label{sec:higgsinflation}

So far, there has been no evidence of new physics in the LHC. Neither we have found proof for any significant amount of non-gaussianities or isocurvature perturbations from the inflationary period. 
This lack of evidence for new physics suggest that nature may be rather simple and it becomes natural to ask whether the Standard Model Higgs could actually be the inflaton. The possibility is also attractive because it would link the high energies of the primordial universe to our accessible electroweak physics data.
Unfortunately, it is not that simple to realize Higgs inflation. The first issue we encounter is that the quartic self coupling of the Higgs leads to an overproduction of primordial density perturbations, $\lambda \sim 10^{-3}$ \cite{Linde:1983gd}.
A possibility to save Higgs inflation is then to include a non-minimal coupling to gravity. In this section, we will briefly review the original model as proposed by \cite{Bezrukov:2007ep}.
\begin{equation}
    \delta S = \int \mathrm{d}^4 x \sqrt{- g} \xi H^{\dagger}H \mathrm{R} \label{eq:nonmin}
\end{equation}
In fact, adding the non-minimal coupling in eq.~(\ref{eq:nonmin}) has been shown to be necessary for the quantization of the SM in gravitational backgrounds \cite{Callan:1970ze, Birrell:1982ix}.

In the unitarity gauge, $H = \frac{1}{\sqrt{2}}\left(0,h\right)^{T}$, the action of Higgs inflation reads

\begin{align}
    S_J= \int \mathrm{d}^4 x \sqrt{-g}\left[\frac{M_P^2 + \xi h^2}{2}R - \frac{1}{2}\left(\partial h \right)^2 - U(h)\right],\label{higgsori}
\end{align}

with 

\begin{equation}
    U(h) = \frac{\lambda}{4} \left(h^2 - v_H^2\right)^2.
\end{equation}

We see that there are two dimensionful parameters in the action (\ref{higgsori}), corresponding to two very different scales. The first is the reduced Planck mass, $M_P\equiv \frac{1}{\sqrt{8 \pi G}} = 2.435 \times 10^{18}$ GeV. The second is the vacuum expectation value of the Higgs field, $v_H \simeq 250$ GeV. The relevant scale for inflation is $M_P$.

We can remove the non-minimal coupling by performing a conformal transformation of the metric, which has been introduced in \textbf{{appendix II}}.
In this case, the conformal function is defined to be
\begin{align}
    \Omega^2 = 1 + \frac{\xi h^2}{M_P^2}.
\end{align}

This will induce a non canonical kinetic term for the Higgs field. We can rewrite the action in the canonical form by making the field redefinition
\begin{equation}
     \odv[]{\chi}{h} = \sqrt{\frac{\Omega^2 + 6\xi^2 h^2/M_P^2}{\Omega^4}}.
\end{equation}

The resulting action is the Einstein frame action,

\begin{equation}
    S_E = \int \mathrm{d}4x \sqrt{- \hat{g}}\left[ - \frac{M_P^2 }{2}\hat{R} + \frac{\partial_\mu \chi \partial^\mu \chi}{2} - U\left(\chi\right)\right],
\end{equation}

with the Einstein potential
\begin{equation}\label{eins}
    U\left(\chi\right) = \frac{1}{\Omega\left(\chi\right)^4} \frac{\lambda}{4}\left( h\left(\chi\right)^2 - v_H^2\right)^2.
\end{equation}

For small field values, $h \simeq \chi$ and $\Omega \simeq 1$, recovering the Standard Model Higgs potential. However, in the large field limit, $h \gg M_P/\sqrt{\xi}$, 

\begin{equation}
    h \simeq \frac{M_P}{\sqrt{\xi}}\mathrm{exp}\left({\frac{\chi}{\sqrt{6}M_P}}\right). \label{largefield}
\end{equation}

Substituting (\ref{largefield}) in (\ref{eins}) we obtain the Einstein potential in terms of the canonical field $\chi$,

\begin{equation}\label{einstein_higgs}
    U\left(\chi\right) = \frac{\lambda M_P^4}{4 \chi^2}\left(1 + \mathrm{exp}\left(
        - \frac{2 \chi}{\sqrt{6}M_P}
    \right)\right)^{-2}.
\end{equation}

It is then for large values of the field, $\chi \gg \frac{M_P}{\sqrt{\xi}}$, where the potential is flat and chaotic inflation becomes possible.

\begin{figure}[h]
    \centering
    \includegraphics[width=0.5\textwidth]{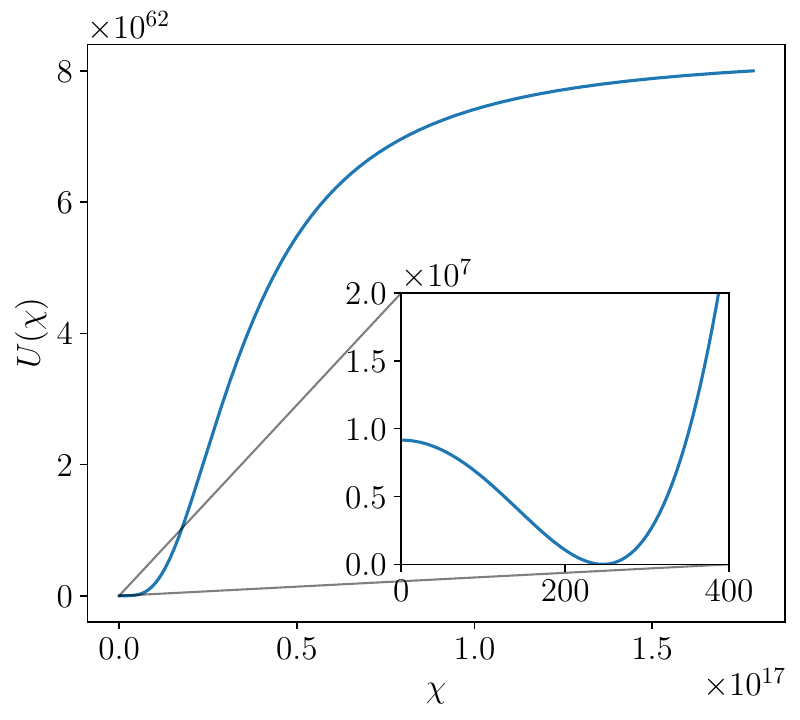}
    \caption{Potential for Higgs inflation. At large values, the potential is almost, but not exactly flat, allowing for slow-roll inflation to take place.}
    \label{fig:higgs_large}
\end{figure}
The slow-roll parameters for Higgs inflation are given by

\begin{align}
    \epsilon &\simeq \frac{4 M_P^4}{3 \chi^2 h^4},\\
    \eta &\simeq - \frac{4 M_P^2}{3 \chi h^2} .
\end{align}

The value of the Higgs field at the end of inflation is $h_{\mathrm{end}} \simeq M_P/\sqrt{\epsilon}$.
Then, we can compute the number of e-foldings, as given in eq.~(\ref{efoldnum}),

\begin{equation}
    N \simeq \frac{6}{8} \frac{h_0^2 - h_{\mathrm{end}}^2}{M_p^2/\xi}.
\end{equation}

This is an important result, since it shows that the non-minimal coupling, $\xi$, is correlated to the number of e-foldings.
Then, we can use the power-spectrum of the CMB

\begin{equation} \label{normalpower}
        \Delta^2_{\mathit{R}\big\vert_{k^*}} = 2.445 \pm 0.096 \times 10^{-9},
\end{equation}
to constrain the possible values of the non-minimal coupling.
Eq. (\ref{normalpower}) fixes the quotient $\lambda/ \xi^2$ to be $\lambda/ \xi^2 \simeq 4 \times 10^{-11}$. Then, in order to be consistent with observations, for $\lambda\simeq \mathcal{O}\left(1\right)$, the non minimal coupling must be large, $\xi \sim \mathcal{O}\left(10^4\right)$.

This large non-minimal coupling with gravity is unusual and in fact leads to unitarity violation in the context of Effective Field Theory, as has been shown in \cite{Burgess:2009ea, Burgess:2010zq, Barbon:2009ya, Hertzberg:2010dc}. 
Before discussing in detail the unitarity problem, two issues are worth mentioning.
 The first, is that the EW vacuum may not be stable at high energies. 
 The second is what are the possible sources of deviation from the flatness.
\subsection{Stability of the electroweak vacuum}\label{sec:ewstability}
When considering the SM Higgs to be the inflaton, we must worry about the stability of the electroweak vacuum at the high energies at which inflation takes place.
The Higgs self-coupling depends on the Yukawa coupling $y_t$. The value of $y_t$ that separates the (meta)stable from the unstable region is given by \cite{Bezrukov:2014ina}

\begin{equation}
    y_t = 0.9244 \pm 0.0012 \frac{m_H/ \mathrm{GeV}- 125.7}{0.4} + 0.0012 \frac{\alpha_s(m_Z)- 0.01184}{0.0007}.
\end{equation}

With today's experimental precision \cite{Butenschoen:2017ays, Espinosa:2016nld}, the Standard Model is compatible with both a (meta)stable and an unstable scenario.
The issue of the stability of the Higgs vacuum has been deeply discussed in \cite{Rubio:2018ogq}.

\subsection{Quantum corrections}\label{sec:quantumcorrections}
Slow-roll inflation, studied in \textbf{section \ref{slowrollsec}}, 
fits well within the classical regime.
However, quantum fluctuations are not completely negligible. In fact, it is precisely these quantum perturbations of the inflation field that leave imprints on the CMB, and that serve as seeds for structure formation.
When we consider more realistic models for inflation, which usually include new physics, loop contributions can effectively change the shape of the potential, spoiling the flatness and leading to a violation of the slow roll conditions.
It is important then to crosscheck that the Higgs inflationary potential is stable against this type of quantum corrections.
There are different types of quantum corrections that could affect the flatness of the potential:

    \begin{enumerate}[label=(\roman*)]
        \item Firstly, we have corrections coming from gravity.
        The size of these corrections is $U\left(\chi\right)/ M_p^4 \sim \lambda / \chi^2$, which is suppressed because of the large non-minimal coupling.
        \item Higher derivative interactions can also affect the shape of the potential, although these are suppressed by powers of $E/M_p$.
        \item Furthermore, we have to take into account that SM particles can couple to the inflaton through the Higgs, and they can radiatively contribute to the potential.
        \item Finally, there could be corrections due to the fact that the couplings, and in particular the non minimal coupling, are not constants but running parameters.
    \end{enumerate}

\subsection{Unitarity problem}\label{sec:unitprob}

Since inflation must end, time invariance must be a broken symmetry. This translates into a time dependency of the Hubble rate during inflation, related to the potential energy through
\begin{equation}
    H \simeq \frac{V^2}{M_p},
\end{equation}
with the inflationary scale, $\mu_\phi$,
\begin{equation}
    \mu_\phi = \frac{\dot{\phi}}{\phi},
\end{equation}
and the equation of motion for $\phi$
\begin{equation}
    \dot{\phi} = \frac{V'}{H}= \sqrt{\varepsilon}\nu^2.
\end{equation}
The slow-roll conditions $\varepsilon , \left| \eta \right|\ll  1$, are related to the inflationary parameters, 

\begin{equation}
    \dot{\phi}=\frac{V'}{H}\simeq\frac{M_p V'}{\sqrt{V}}\simeq \sqrt{\varepsilon V} \simeq \sqrt{\varepsilon}\nu^2.
\end{equation}

In the slow-roll type of models, $\phi$ typically varies between $M_p$ and $\nu$. For $\phi \simeq M_p$,
\begin{equation}
    \mu_\phi = \sqrt{\varepsilon}H,
\end{equation}

and for $\phi \simeq \nu$,

\begin{equation}
    \mu_\phi = \sqrt{\varepsilon} \nu.
\end{equation}

We can understand Higgs inflation as the EFT of a scalar field coupled to gravity. Thus, extra constraints will be required in order to ensure the validity of the EFT.
In particular, inflation must be an adiabatic process, which translates into an upper bound on the inflationary scale, $\mu_\phi, H \ll M$, where $M$ the mass of the lightest particle that has been integrated out.
If the process were not adiabatic, we wouldn't be able to separate the dynamics of the heavier and lighter fields. 

Furthermore, $H$ and $\mu_\phi$ are constrained by the temperature fluctuations of the CMB, which we studied in \textbf{section \ref{sec:CMB}}. These are given by
\begin{equation}
    \Delta^2_{\mathit{R}\big\vert k^*} = 2.445 \pm 0.096 \times 10^{-9}, \label{powersp}
\end{equation}

where $k^*$ is the pivot scale at horizon exit. 

The perturbations generated by quantum fluctuations in $\phi$ are measured by the parameter $\delta = \frac{H^2}{\dot{\phi}} $\cite{Giudice:2010ka}, and related to the temperature fluctuations through

\begin{equation}
    \delta = \left(24 \pi^2 \Delta^2_{\mathit{R}\big\vert k^*}\right)^{1/2}.
\end{equation}

We can then estimate $\delta$ to be
\begin{equation}\label{delt}
    \delta \simeq \frac{1}{\sqrt{\epsilon}}\left(\frac{\nu}{M_p}\right)^2 \simeq 7 \times 10^{-4}.
\end{equation}

This is a crucial result. Equation (\ref{delt}) provides a key relationship between the two characteristic scales of Higgs inflation,
\begin{equation}
    \frac{\nu}{M_p} \simeq 0.03 \varepsilon^{1/4}.
\end{equation}

In order for the EFT to remain valid, quantum corrections must be smaller (or of similar size) than the tree level contributions.
The quantum corrections affecting the shape of the potential will depend on the mass of the particles that have been integrated out.
In particular, for heavy fields, the scalar potential corrections can be dangerous \cite{Burgess:2003zw}.
On the other hand, the slow roll regime can be easily spoiled if the corrections are not suppressed by the Planck scale $M_p$, but instead by a lighter scale $\nu$ corresponding to the vacuum energy of the Higgs.

\begin{figure}[t]
    \centering
    \includegraphics[width=0.45\textwidth,clip]{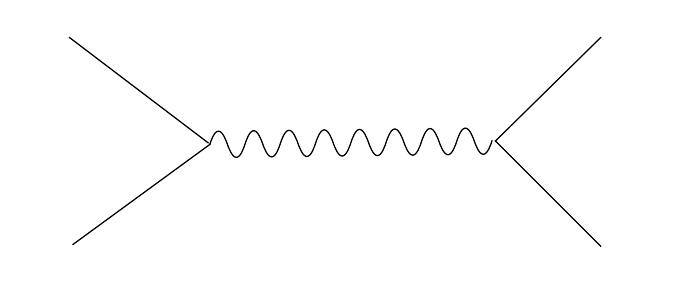}
    \caption{Scattering of Higgs through gravitons leading to the unitarity breakdown.} 
    \label{fig:scatteringhiggs}
    \end{figure}

In order to find the cut-off scale of Higgs inflation, we rely on the cross-sections of the processes $h g \rightarrow h g $ and $h h \rightarrow h h$, as shown in Fig. (\ref{fig:scatteringhiggs}).

Asking that the theory satisfies unitarity bounds, i.e., $\sigma \propto 1/E^2$, leads to

\begin{equation}
    E < E_{\mathrm{max}} \simeq \frac{M_p}{\xi}.
\end{equation}

For Higgs inflation, $H \simeq \sqrt{\lambda_H} M_p/\xi$. For $M \lesssim M_p /\xi$, we find $H/M \gtrsim \sqrt{\lambda_H}$.
This leaves a narrow window in which the EFT is valid.
\begin{equation}
    1 \gg H/M \gg \sqrt{\lambda_H}.
\end{equation}

\subsection{UV completions}\label{sec:uvcompletionshiggs}

We have seen that Higgs inflation with a non-minimal coupling, which naively could realize inflation,  suffers from a unitarity problem, and its EFT seems to be out of control for the values of the field that are relevant during inflation. 
Including new degrees of freedom can restore unitarity. It was recently noticed that the $R^2$ term from Starobinsky inflation \cite{Ema:2017rqn, Gorbunov:2018llf,He:2018mgb,Cheong:2020rao} can provide a UV completion of Higgs inflation, due to the presence of the scalaron field $\sigma$ associated with $R^2$.
In this section, we review how to realize Higgs inflation as a non-linear sigma model, taking the basis in which conformal invariance is manifest \cite{Ema:2020zvg,Ema:2020evi}.

\subsection{Realization as non-linear sigma models}\label{sec:nonlinearsigma}
In this section, we show how we can recast Higgs inflation into a non-linear sigma model.
We start from the now familiar Higgs inflation Lagrangian

\begin{equation}
    {\cal L}=\sqrt{-{\hat g}} \bigg[-\frac{1}{2}(1+\xi {\hat\phi}^{2}_i)  {\hat R}+\frac{1}{2} g^{\mu\nu} \partial_\mu {\hat\phi}_i \partial_\nu {\hat\phi}_i -\frac{\lambda}{4} ({\hat\phi}^2_i)^2 \bigg]. \label{Higgsinf}
\end{equation}

By performing a conformal transformation as in \textbf{appendix II} , we are introducing a new unphysical scalar degree of freedom, $\phi$, in the Lagrangian.

\begin{align}
{\cal L}&=\sqrt{-{g}}\, e^{4\varphi} \bigg[ -\frac{1}{2} e^{-2\varphi}(1+\xi {\hat\phi}^2_i)  {R}+ 3 (1+\xi{\hat\phi}^2_i)  e^{-3\varphi} \Box e^\varphi   +\frac{1}{2} e^{-2\varphi} (\partial_\mu{\hat\phi}_i)^2 -\frac{\lambda}{4} ({\hat\phi}^2_i)^2 \bigg] \nonumber\\
&=\sqrt{-{ g}}\, \bigg[ -\frac{1}{2} e^{2\varphi}(1+\xi {\hat\phi}^2_i)  {R}+ 3 (1+\xi{\hat\phi}^2_i)  e^{\varphi} \Box e^\varphi   +\frac{1}{2} e^{2\varphi} (\partial_\mu{\hat\phi}_i)^2 -\frac{\lambda}{4} \,e^{4\varphi}({\hat\phi}^2_i)^2 \bigg].
\end{align}

We now redefine the fields ${\phi}_i=e^{\varphi} {\hat\phi}_i$ and ${\Phi}=\sqrt{6}\,e^\varphi$.
\begin{align}
{\cal L}=&\sqrt{-{\hat g}} \bigg[  -\frac{1}{2} \Big(\frac{1}{6} {\Phi}^2+\xi {\phi}^2_i \Big) {R} -\frac{1}{2} (\partial_\mu{\Phi})^2 +3\xi\phi^2_i\, \frac{\Box {\Phi}}{{\Phi}} +\frac{1}{2}{\Phi}^2 \Big[\partial_\mu ({\Phi}^{-1}{\phi}_i)\Big]^2-\frac{\lambda}{4}({\phi}^2_i)^2  \bigg] \nonumber \\
=&\sqrt{-{\hat g}} \bigg[  -\frac{1}{2} \Big(\frac{1}{6} {\Phi}^2+\xi {\phi}^2_i \Big) {R} -\frac{1}{2} (\partial_\mu{\Phi})^2+\frac{1}{2} (\partial_\mu{\phi}_i)^2 +3\Big(\xi+\frac{1}{6} \Big)\phi^2_i\,\frac{\Box {\Phi}}{{\Phi}}-\frac{\lambda}{4}({\phi}^2_i)^2 \bigg]\label{Higgsinf2}
\end{align}

Identifying ${\Phi}=\phi+\sigma$,
\bea
\frac{\Box {\Phi}}{{\Phi}}=(\phi+\sigma)^{-1} (\Box \phi + \Box \sigma),
\eea

and taking the following combination of the fields,

\bea 
\frac{1}{6} (\phi+\sigma)^2 +\xi {\phi}^2_i =\frac{1}{6} \phi^2-\frac{1}{6} {\phi}^2_i -\frac{1}{6} \sigma^2,
\eea

we get a constrain equation for $\sigma$.
\bea\label{sigcon}
\sigma=\frac{1}{2} \bigg( \sqrt{\phi^2-12\Big(\xi+\frac{1}{6}\Big) {\phi}^2_i}-\phi\bigg). \label{constraint0}
\eea

Eq.~(\ref{sigcon}) links together the values of the sigma and the higgs field, which, in terms of the conformal field $\phi$, reads

\bea
{\Phi} =\frac{1}{2} \bigg( \sqrt{\phi^2-12\Big(\xi+\frac{1}{6}\big) {\phi}^2_i}+\phi\bigg).
\eea

Finally, using

\bea
\bigg[\frac{1}{2}{\Phi}^2+ 3\Big(\xi+\frac{1}{6} \Big){\phi}^2_i\bigg]\,\frac{\Box {\Phi}}{{\Phi}}&=&\frac{1}{2} (\phi^2-\sigma^2) \,\frac{\Box {\Phi}}{{\Phi}} \nonumber \\
&=& \frac{1}{2} (\phi-\sigma) (\Box \phi + \Box \sigma) \nonumber \\
&=&-\frac{1}{2} (\partial_\mu\phi)^2 + \frac{1}{2} (\partial_\mu  \sigma)^2,
\eea

we get the final from of the Lagrangian as

\bea
{\cal L}= \sqrt{-{g}} \bigg[  -\frac{1}{2} \Big(\frac{1}{6} {\phi}^2-\frac{1}{6} {\phi}^2_i-\frac{1}{6}\sigma^2 \Big) { R} -\frac{1}{2} (\partial_\mu{\phi})^2+\frac{1}{2} (\partial_\mu{\phi}_i)^2+ \frac{1}{2} (\partial_\mu \sigma)^2  -\frac{\lambda}{4}({\phi}^2_i)^2 \bigg]. \label{conformalL}
\eea.

Let's stop here and think about this result. We have performed a conformal transformation, removing the non-minimal coupling for the Higgs field, and found out that, when we do this, the kinetic term for the Higgs field becomes non-canonical, and we get a constrain equation for $\sigma$. This is just reflecting the fact that we have coupled the Higgs to the Ricci scalar, which contains derivatives of the metric. 
So far, the sigma field is only an auxiliary field.
The next step will be to study what happens once we promote $\sigma$ to be a dynamical variable.

\subsection*{Gauge fixing}\label{sec:gaugefixing}

Choosing the value for the conformal mode, $\phi=\sqrt{6}$, fixes the Planck scale.
\bea\label{fixed}
{\cal L}= \sqrt{-{g}} \bigg[  -\frac{1}{2} \Big(1-\frac{1}{6} {\phi}^2_i-\frac{1}{6}\sigma^2 \Big) { R} +\frac{1}{2} (\partial_\mu{\phi}_i)^2+ \frac{1}{2} (\partial_\mu \sigma)^2  -\frac{\lambda}{4}({\phi}^2_i)^2 \bigg]
\eea

The Lagrangian in eq.~(\ref{fixed}) shows conformal invariance, only broken by Einstein gravity. Thus, we call this the conformal frame.

The constraint equation 

\bea
f(\sigma,\phi_i)\equiv \bigg(\sigma+\frac{\sqrt{6}}{2}\bigg)^2 +3\Big(\xi+\frac{1}{6} \Big){\phi}^2_i-\frac{3}{2}=0 \label{constraint},
\eea

defines the vacuum structure. With this, we have proven that Higgs inflation can be regarded as a non-linear sigma model, subject to the constraint equation (\ref{constraint})

Now, we include the constraint equation (\ref{constraint}) as a Lagrange multiplier into our Lagrangian. This gives rise to our UV complete Lagrangian,

\bea\label{finallagrangiangague}
{\cal L}&=& \sqrt{-{g}} \bigg\{  -\frac{1}{2} \Big(1-\frac{1}{6} {\phi}^2_i-\frac{1}{6}\sigma^2 \Big) {R} +\frac{1}{2} (\partial_\mu{\phi}_i)^2+ \frac{1}{2} (\partial_\mu \sigma)^2  \nonumber \\
&&-\frac{\lambda}{4}({\phi}^2_i)^2-\frac{\kappa}{4}\left[\bigg(\sigma+\frac{\sqrt{6}}{2}\bigg)^2 +3\Big(\xi+\frac{1}{6} \Big){\phi}^2_i- \frac{3}{2}\right]^2 \bigg\}. \label{sigmamodels}
\eea

In equation (\ref{sigmamodels}), we can interpret $\kappa$ as a coupling. Again, we see that the kinetic terms, the non-minimal coupling and the Higgs self-coupling terms respect the conformal symmetry.
The conformal symmetry breaking comes from the Planck mass and the $\sigma$ potential only.
To ensure unitarity, we ask our theory to be perturbative. That implies that the couplings on eq.~(\ref{finallagrangiangague}) must be of order $\mathcal{O}\left(1\right)$,
\bea
\kappa \lesssim 1,\qquad \lambda+9\kappa \Big(\xi+\frac{1}{6} \Big)^2\lesssim 1, \qquad 6\kappa \Big(\xi+\frac{1}{6} \Big)\lesssim 1.
\eea

\section{Natural inflation}\label{sec:naturalinflation}

In order to realize sufficient inflation while still reproducing the CMB anisotropy measurements of density fluctuations, the potential
for the inflaton field must be very flat. In general, the potentials of models of slow-roll inflation, including new inflation \cite{Albrecht:1982wi, Linde:1981mu} and chaotic inflation \cite{Linde:1983gd}, must satisfy \cite{Adams:1990pn}

\bea
\frac{\Delta V}{(\Delta \phi)^4} \lesssim 10^{-6}-10^{-8},
\eea

so that the change in the potential, $\Delta V$ must be much smaller than the change in the field, $\Delta \phi$. The difference between this energy scales is a fine tuning problem of inflation.
This implies that the inflaton must be weakly self-coupled, leading to quartic couplings of the size of  $\lambda_\phi < 10^{-12}$.
There have been several approaches to the inflation fine tuning problem. For example, one may consider a tiny self quartic coupling ad-hoc, but then we need to make sure that radiative corrections and the effects of RG-running do not spoil inflation. But this approach doesn't really solve the naturalness problem of Cosmology, it just switches it with a particle physics naturalness problem instead.
Another solution is to try to protect the coupling by including new symmetries, as it can be \hyperref[subsec:intro_SUSY]{Supersymmetry}, studied in a later section. However, even in this case we rely on a tiny ratio of scales. The mass hierarchy in this case is stable but remains unexplained in the absence of a mechanism that can dynamically relax this ratio.
Then, natural inflation was proposed \cite{Freese:1990rb}. In this case the inflaton can be identified with a pNGb associated to a spontaneously broken global symmetry, and the flatness of the potential is guaranteed by a shift symmetry.
However, if this shift symmetry is exact, the potential is exactly flat and slow roll cannot happen. Hence, an extra symmetry breaking source is required.

The pNGb potential is of the form 

\bea
V(\phi)= \Lambda^4 \left(1 \pm \cos(N \phi/f )\right),
\eea

where f is the scale associated with the spontaneous symmetry breaking. For $T<f$, the global symmetry has been broken, and we can identify $\phi$ with the field oscillating in the bottom of a mexican hat potential.
In this case, Planck data constrains the axion decay constant to be $f> 0.6M_P$. Furthermore, the hight of the potential is around the GUT scale. Both scales are sufficiently low, so we don't have to worry about possible quantum gravity effects. 

\section{Hybrid inflation}\label{sec:hybrid}

In all of the models of inflation we have considered so far, inflation ends due to the violation of the slow-roll conditions.
In 1993, Linde proposed a model of inflation in which the inflaton fields gradually starts rolling faster and faster, and eventually ends due to a first order phase transition.
In this case, we need at least two fields: one driving inflation, and a second one that becomes tachyonic, dubbed the waterfall field, responsible for the end of inflation.

The potential for hybrid inflation is given by

\bea
\label{hybridpotential}
V(\sigma, \phi) = \frac{1}{4\lambda} \left(M^2 - \lambda \sigma^2\right)^2 + \frac{m^2}{2}\phi^2 + \frac{g^2}{2}\phi^2 \sigma^2
\eea
The effective mass of the water field is 

\bea
m_{\sigma,\mathrm{eff}} = - M^2 - g^2 \phi^2.
\eea

The dependence of this mass on the $\phi$ field is what eventually ends inflation. For $\phi > \phi_c = M/g$, the minimum of the potential is a $\sigma=0$ and is unique. However, for $\phi=\phi_c$, the phase transition happens and symmetry breaking occurs. Now, the motion in the $\phi$ direction is very fast and the field quickly rolls to the new minima in a time much shorter than the Hubble scale $H^{-1}$.

Under the condition $\phi< \phi_c$, the universe undergoes accelerated expansion. The first stages of inflation, according to \ref{hybridpotential}, are of the chaotic type. However, the last stages of inflation are  driving instead by the vacuum energy density, 

\bea
V(0,0) = \frac{M^4}{4\lambda},
\eea

as in the new inflation type of model. Hence, this model is dubbed \textbf{hybrid inflation}.
In hybrid inflation however, the coupling in the last term in eq.~(\ref{hybridpotential}), can receive dangerous radiative corrections, so it is important to ensure radiative stability in order to render inflation natural. 
 \chapter{Inflation and Particle Physics puzzles}\label{chap_partphys}
 \begin{small}
    ``The Universe is a poor's man accelerator.”\\
    Jakob Zeldovich
    \end{small} 
    \vspace{5mm}

    In the Early Universe, Dark Matter, very heavy or very weakly interacting particles could have been created.
    Many UV completions of the standard model expect incredibly heavy new particles, for example at the energies of GUTs, $10^{15}$ GeV. 
    Equally important, at very high temperatures phase transitions can happen, leading to gravitational waves  or cosmic defects that we can look for.
    In this thesis we will link inflation to Supersymmetry and Dark Matter production, so we provide here a review of Dark Matter production for the case of WIMPs, FIMPs and axions.
    We also provide a summary on the basic concepts of Supersymmetry.

 \section{Dark Matter}\label{sec:darkmatter}

The ordinary matter that we are used to (baryonic matter) composes only a tiny fraction of all the matter in our universe.
The vast majority of it is called Dark Matter, and we know about it because of its gravitational interaction. But even though we have plenty of evidence in favor of its existence, such as the rotation curves of galaxies or gravitational lensing, we know only one thing about it: its current energy density. 
So, when we write down our particular model for Dark Matter, we first write down our Lagrangian and then ask if our theory is able to reproduce the correct abundance of Dark Matter. 
For that, we rely uniquely on the cosmological evolution.

\subsubsection*{Equilibrium thermodynamics}

From the Friedmann-Robertson-Walker metric

\bea
ds^2 = dt^2 - a^2(t) \left(\frac{dr^2}{1-kr^2} + r^2 d\theta^2 + r^2 \sin^2 \theta d\phi^2\right),
\eea

we can derive the continuity equations

\bea
\left(\frac{\dot a}{a}\right)^2 + \frac{k}{a^2} &=& \frac{8 \pi G_N}{3}\rho = \frac{\rho}{3 M_P^2},\\
\frac{\ddot{a}}{a} &=& -\frac{4 \pi G_N}{3}\left(\rho +3p\right) = -\frac{1}{6M_P^2}\left(1 + 3\omega\right)p,\\
\dot \rho + 3H \left(\rho + p\right) &=& 0,
\eea
where $\omega$ is the thermodynamic equation of state, $\omega \equiv \rho/p$

We see then the relation between the energy density of the universe and the scale factor, $\rho \propto a^{-3(1+\omega)}$
The evolution of the universe at the rate of expansion will evolve depending on the dominant component of the universe, as summarized in Table \ref{domination}.

\begin{table}[h]
    \begin{center} 
    \begin{tabular}{ c c c c}
     & $\omega$ & $\rho$ & $a(t)$\\ 
     RD & 1/3 & $a^{-4}$ & $t^{1/2}$ \\  
     MD & 0 & $a^{-3}$ & $t^{2/3}$ \\
    $\Lambda$D & -1 & const & $e^{Ht}$ 
    \end{tabular} 
    \caption{Cosmological evolution according to the matter content of the universe.}
    \label{domination} 
    \end{center}
\end{table}


    In the early universe, particles were in a thermal bath \footnote{Dark Matter might have been at equilibrium with the rest of the plasma, as in the case of WIMPs, or it might have never been, as in the case of FIMPs.}.
    Due to the high temperatures, the rapid scatterings lead to equilibrium dynamics. We can describe the thermodynamical properties of the bath through the phase space distribution 

    \bea
    f\left(\vec{p}\right) = \frac{1}{e^{\left(E -\mu\right)/T} \mp 1}\label{bose},
    \eea

    with $E^2 = |\vec{p}|^2 +m^2$ and $\mu$ being the chemical potential, which is much smaller than the temperature in the early universe. In (\ref{bose}) we take the + sign for fermionic particles, leading to the \textbf{Fermi-Dirac distribution}, and the - sign for bosonic ones, leading to the \textbf{Bose-Einstein distribution}.
The number density of any given spices will be given by

    \bea
    n = \frac{g}{\left(2 \pi\right)^2}\int{d^3 p f(\vec{p})},
    \eea

    where $g$ describes the number of internal degrees of freedom. We have $g=1$ for a real scalar, $g=2$ for a complex scalar or a Majorana fermion and $g=4$ for a Dirac fermion. 

    Ignoring the chemical potential, the number density in equilibrium is given by

    \bea 
n_{eq} \approx \begin{dcases}
    \frac{g_{eff} \zeta(3)}{\pi^2} T^3 & \mathrm{for}\quad T \gg m,\label{eq:neq}\\ 
   g \left( \frac{mT}{2\pi}\right)^{3/2} e^{-m/T}& \mathrm{for}\quad T \ll m.
    \end{dcases}
    \eea

    Here, $g$ is the number of degrees of freedom, where for bosons, $g_{\mathrm{eff}}=g$ and for fermions $g_{\mathrm{eff}}= \frac{3}{4}g$.
    The energy density will be given by the integration over energies as

\bea
\rho = \frac{g}{(2 \pi)^3} \int d^3 p E(\vec{p}) f({\vec{p}}),
\eea

so, for relativistic species

\bea 
\rho_{eq} \approx \begin{dcases}
    \frac{g_* \pi^2}{30} T^4 & \mathrm{for}\quad T \gtrsim m,\\ 
   m n_{eq}& \mathrm{for}\quad T \lesssim m,
    \end{dcases}
    \eea

    where $g_*$ denotes the degrees of freedom of relativistic(massless) species only. 
Finally, the pressure is given by the integral 

\bea
P = \frac{g}{(2\pi)^3} \int d^3p f(\vec{p})\frac{|\vec{p}|}{3E}.
\eea

and the entropy density for relativistic species is given by

\bea
s =\frac{p+ \rho}{T} = \frac{2 \pi}{45}g_{*s}T^3,
\eea

The total entropy, $S=a^3s$ is conserved, so that $g_{*s}T^3$ is a constant.
If we assume radiation domination(Table \ref{domination}), the Hubble parameter becomes

\bea
\frac{1}{2t} = H = \sqrt{\frac{\rho_R}{3 M_P^2}}= 0.33 g_*^{1/2} \frac{T^2}{M_P}. \label{eq:ttoTconversion}
\eea

Eq. (\ref{eq:ttoTconversion}) is the key equation that allows us to translate between time and temperature.

\subsubsection*{Boltzmann equation}
As the temperature of the universe dropped, some of the species in the thermal bath departed from equilibrium. It is then crucial to understand the dynamics beyond equilibrium, described through the \textbf{Boltzmann equation}, which determines the amount of Dark Matter than can be produced,

    \bea
    E \frac{\partial f}{\partial t} - \frac{\dot a}{a} |\vec{p}|^2 \frac{\partial f}{\partial E} = C[f], \label{boltzmanneq}
    \eea
   
    where $C[f]$ is the collision operator, defined by the particular interactions of Dark Matter.
Taking the integral over momenta, we get

    \bea
    \dot n_X + 3 H n_X = \frac{g}{(2\pi)^3}\int \frac{d^3p}{E} C[f],
    \eea

    where the subscript X denotes the Dark Matter species. Then, for a general collision operator, we get

    \bea
    \dot{n}_X + 3 H n_X &=& - \int d\Pi_X d \Pi_a d \Pi_b \hdots d\Pi_i d \Pi_j \hdots \left(f_X f_a f_b \hdots - f_if_j \hdots \right) \nonumber\\
    &\times& \left(2 \pi\right)^4 \delta^4 \left(p_X + p_a + p_b + \hdots -p_i -p_j - \hdots\right)|\mathcal{M}|^2,
    \eea

where  $\Pi_i$ is the Lorentz Invariant Phase Space(LIPS) of the $i$-th species, 

\bea
d\Pi_i \equiv \frac{g_i}{(2 \pi)^3} \frac{d^3 p_i}{2E_i},
\eea

and the delta function enforces the conservation of energy and momenta.
Let's as an example consider the case in which the Dark Matter number density only changes due to direct and inverse annihilations. Then, the Boltzmann equation in (\ref{boltzmanneq}) becomes

\bea
    \dot n_X + 3H n_X = - n_X^2 \langle \sigma v \rangle_{XX} + n_\phi^2 \langle \sigma v\rangle_{\phi\phi}. \label{2to2}
\eea

The first term in the right-hand side of eq.~(\ref{2to2}) corresponds to DM annihilation into SM particles and the second term corresponds to SM particles annihilating into Dark Matter.
Using that a process and its backwards counterpart are related in equilibrium  (where the collision operator vanishes),

\bea
    (n^{eq}_X)^2 \langle  \sigma v\rangle_{XX} = (n^{eq}_\phi)^2 \langle \sigma v \rangle_{\phi\phi},
\eea

eq.~(\ref{2to2}) becomes
\bea
    \dot{n}_X + 3 H n_X = - \langle \sigma v \rangle_{XX} \left(n_X^2 - (n_{X}^{eq})^2\right) \label{wimpbol}.
\eea

Here, the angular brackets $\langle \hdots \rangle$ denote thermal average and $v$ is the velocity of the Dark Matter particles. The thermally averaged cross-section is defined as

    \begin{align}
    \langle \sigma v\rangle_{XY \rightarrow AB} \equiv \frac{1}{n^{eq}_X n^{eq}_Y}\int d\Pi_X d\Pi_X f_X^{eq} f_Y^{eq} \int d\Pi_A d\Pi_B (2 \pi)^4  \delta(p_X + p_Y - P_A - P_B)|\mathcal{M}|^2.
    \end{align}

    It is useful now to consider the limit of rapid annihilations (as compared to the Hubble rate). Then, the right-hand side of eq.~ (\ref{wimpbol}) is just zero and $n_X \approx n_X^{eq}$.
    In the opposite limit, annihilations are too slow and we get an inert species, $n_{X} = \mathrm{const}\times s$.

\subsection{Dark Matter Freeze-Out and WIMPS}\label{sec:wimps}
WIMP Freeze-Out relies on the idea that DM undergoes annihilations into SM particles through $X\bar{X}\leftrightarrow \phi\phi$.
 At high temperature, WIMPs are in thermal equilibrium and annihilations back and forward happen rapidly. 
 At low temperature, however, the number density dilutes as in eq. (\ref{eq:neq}),
so that annihilations are no longer efficient. We first define the yield as $Y = \frac{n}{s}$, allowing us to rewrite eq.(\ref{wimpbol}) as

\bea
\dot{n}_X + 3H n_X = s \dot Y_X = -\langle \sigma |v| \rangle s^2 \left(Y_X^2 - Y^2_{X,eq}\right).
\eea

Using the variable $x=\frac{m_X}{T}$, we can write 

\bea
\frac{dY}{dx}=-\frac{x \langle \sigma |v|\rangle s}{H(m)}\left(Y_X^2 - Y^2_{X,eq}\right) = -\frac{n_{X,eq}\langle \sigma |v|\rangle}{x H Y_{X,eq}}\left(Y_X^2 - Y^2_{X,eq}\right).
\eea

Freeze-out happens when $\Gamma_{ann}=n_X^{FO}\langle  \sigma v\rangle \approx H$. In order to be consistent with Large Scale Structure, Dark Matter needs to be cold, or non-relativistic. Then, taking $\langle  \sigma |v|\rangle = \sigma_0 x^{-n}$,

\bea
\frac{dY}{dx} = -1.329 \left(\frac{g_{*s}}{g_*^{1/2}}\right)M_P m_X \sigma_0 x^{-n-2} \left(Y^2_X - Y^2_{X,eq}\right)
\eea

Hence, we can estimate the total DM relic abundance as

\bea
\Omega_X h^2 = \frac{\rho}{\rho_c/h^2} \approx 0.2745  \left(\frac{m_X}{100 \mathrm{GeV}}\right)\left(\frac{Y_\infty}{10^{-11}}\right),
\eea

where we have used $\rho = m Y_\infty$, where $Y_\infty$ is defined at the yield at $T=0.$

\begin{figure}[t]
    \centering
    \includegraphics[width=0.85\textwidth,clip]{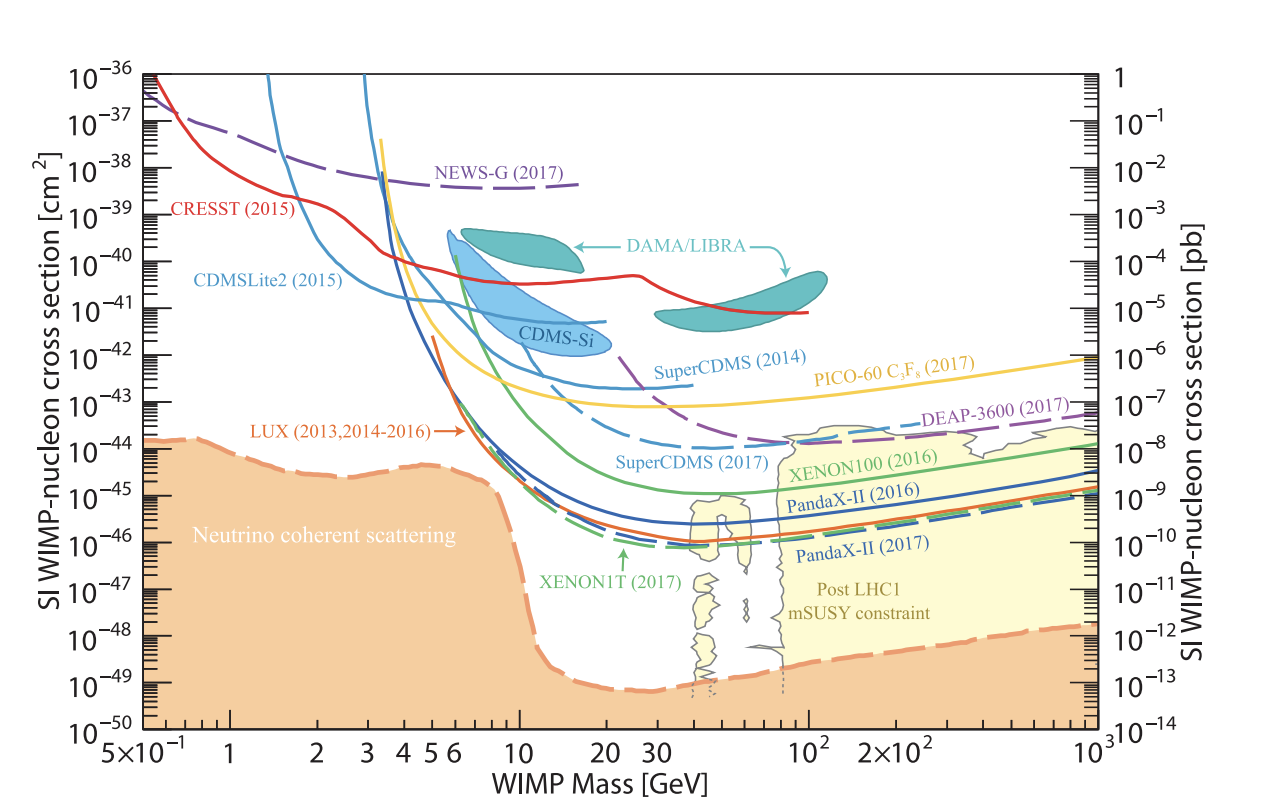}
    \caption{WIMP mass constraints based on Direct Detection \cite{ParticleDataGroup:2018ovx}.} 
    \label{directdet}
 \end{figure}

We notice that, in general, the larger the interactions are between the WIMPs, the lower the final relic abundance will be, due to a larger annihilation rate. The annihilation cross-section can be parametrized as 

\bea
\langle \sigma |v|\rangle \approx \frac{\alpha_{2\rightarrow 2}}{m_X^2}= \frac{g^2}{4 \pi} \frac{1}{m_X^2},
\eea

meaning that for couplings of order 1, in order to satisfy the observed energy density of DM, the masses of WIMPS must be around $100$ GeV. The appearance of the EW scale in this calculation has been referred as \textit{WIMP Miracle} in the literature.
However, the lack of evidence for WIMP Dark Matter and the current bounds by direct detection, as shown in Fig. \ref{directdet}, suggest that nature may have chosen another alternative for the nature of Dark Matter. 

\subsection{Freeze-In and Feebly Interacting Dark Matter}\label{sec:fimps}
The thermal relic that we studied in \textbf{section \ref{sec:wimps}} assumes that DM was once in thermal equilibrium with the SM bath. Hence, interactions are sufficient to deplete the number density until it reaches de observed energy density. However, if DM particles are very feebly interacting, it is very unlikely that they ever reached thermal equilibrium. 
We will assume a zero initial DM abundance.  FIMP DM can be produce by 2-to-2 scatterings from other particles in the thermal bath, or by oscillations of particles within the thermal bath. An example is given in the reheating dynamics of Higgs-sigma models in \textbf{chapter \ref{chap_UV}}.

Let's assume that FIMPs freeze out through the decay process $\psi \rightarrow X \phi$, where $\psi$ is a particle from the thermal bath, X the DM particle and $\phi$ a different particle.

\bea
\dot n_\chi + 3 H n_X \approx \langle \Gamma_\psi \rangle \left(n_\psi^{eq} - \frac{n_X}{n_X^{eq}}n_\psi^{eq}\right)\approx n_\psi^{eq}\langle \Gamma_\psi\rangle,
\eea
where we have assumed that $n_X \ll n_X^{eq}$, so that inverse processes can be neglected.
The thermally averaged decay rate is given by \cite{Giudice:2003jh},

\bea
\langle \Gamma_\psi \rangle = \frac{K_1(m_\psi/T)}{K_2(m_\psi/T)}\Gamma_\psi,
\eea

where $K_{1,2}$ are the modified Bessel functions of the second kind. It is interesting to see the behavior in the limits in which the temperature is very large or very low,

\bea 
\langle \Gamma_\psi \rangle \approx \begin{dcases}
    \frac{m}{2T}\Gamma_{\psi} & \mathrm{for}\quad T \gg m, \\ 
   \Gamma_\psi & \mathrm{for}\quad T \ll m.
    \end{dcases}
    \eea
    Then, for very low temperatures the averaged decay rate is just the usual decay rate, while for high temperatures we have a suppression factor of $m_\psi/T$.

    Rewriting the Boltzmann equation in terms of the yield,

    \bea
    \frac{d Y_X}{dx} = \frac{1 - \frac{1}{3} \frac{d \log{g_{*,s}}}{d\log{x}}}{s H x},
    \eea

    and defining the effective Hubble rate,

    \bea
    H_{\mathrm{eff}} = \frac{H}{1 - \frac{1}{3} \frac{d \log{g_{*,s}}}{d\log{x}}},
    \eea

    the solution to the Boltzmann equation, assuming $ H_{\mathrm{eff}} = H$, is

    \bea
    Y_X^{FI} = \int d \log{x} Y_\psi^{eq} \frac{\langle \Gamma_\psi\rangle}{H}.
    \eea

    The asymptotic behavior of the yield is given by

    \bea 
    \frac{Y_\psi^{eq}}{H} \langle \Gamma_\psi \rangle \propto \begin{dcases}
    \Gamma_{\psi} T^{-3}& \mathrm{for}\quad T \gg m \\ 
   e^{-m/T} & \mathrm{for}\quad T \ll m
    \end{dcases}
    \eea

    At high temperatures, the yield increases at $T$ drops, so this process is dominated by the IR contributions. After the temperature drops beyond the mass, the yield becomes exponentially suppressed, and Freeze-In is complete.
    We conclude that Freeze-in is dominated by the contribution $T \sim m_\psi$,

    \bea
    Y_X^{FI} = \int_0^\infty \frac{dx}{x} Y^{eq}_\psi \frac{\langle \Gamma_\psi\rangle}{H} \approx 0.33 \frac{g_\phi}{g_{*,s}\sqrt{g_*}}\frac{\Gamma_\psi M_P}{m_\psi^2}.
    \eea

\subsection{Strong CP problem, axions and the misalignment mechanism}

The axion is the pseudo-scalar particle that arises in the Peccei-Quinn solution \cite{Peccei:1977hh} to the Strong CP problem \cite{Wilczek:1977pj,Weinberg:1977ma}, which is, in short, the question of why the combination of CP violating phases of the standard model is so close to zero,

\bea
\theta_{SM}=\theta_{QCD} + \arg\left(\det Y_u Y_d\right) \leq 10^{-10}.
\eea

The axion couples to the $G^{\mu\nu} \tilde{G}_{\mu\nu}$ of the Standard Model, 

\bea
\Delta \mathcal{L}_{QCD} = \frac{a}{f_a} \frac{g_s^2}{32 \pi^2} G_{\mu\nu}\tilde{G}^{\mu\nu},\label{axiongluon}
\eea

so that when QCD becomes strongly coupled, i.e, at low energies, the QCD instantons generate the potential for the axion field,

\bea
V(a) = \mu^4\left(1 -\cos(a/f_a)\right).
\eea

If the axion falls into the minima of the potential, we have driven the theory to one with a CP conserving vacuum.
Hence, the neutron electromagnetic dipole moment is dynamically relaxed to zero.

In general, the axion field has a shift symmetry $a \rightarrow a+ 2\pi f_a$, which constrains the possible interactions terms,

\bea
\mathcal{L}_{\mathrm{int}} = c_1 \frac{\partial_\mu a}{f_a} \left(\bar{f}\gamma^\mu \gamma^5 f\right) + c_2 \frac{\alpha a}{8 \pi f_a} F_{\mu\nu}\tilde{F}^{\mu\nu}. \label{axionint}
\eea

In particular, we the axion will couple through derivative interactions to fermions, and to gauge bosons through the topological term as in eq.~(\ref{axionint}).

The QCD axion is also a good candidate for CDM, whose abundance is generated by the misalignment mechanism \cite{Preskill:1982cy, Abbott:1982af, Dine:1982ah}.
In the conventional misalignment mechanism, the axion velocity is initially set to zero. In this set up, the axion tends to be underproduced for masses heavier than $10\, \mu$eV. It was realized by \cite{Co:2019jts} that allowing for a non-zero initial velocity for the action lead to a broader window for the axion masses.
This non-zero velocity can be explained since the PQ symmetry will also be broken explicitly by quantum gravity effects. 

We note that the axion mass changes with temperature as

\bea \label{eq:axionmassintro}
 m_a(T) \sim m_{a,0} \left(\frac{\Lambda^4}{T^4}\right).
\eea
\subsubsection*{Conventional misalignment mechanism}
The misalignment mechanism is different from the rest of DM models introduced so far since axion abundance is not computed from the Boltzmann equation. In previous cases, the interactions of Dark Matter with the rest of the plasma determined the final relic abundance. However, the production through misalignment leads to large occupation numbers, and the axions produced can be treated as classical fields.

The axion evolves in the early universe following the equation of motion,

\bea
\ddot{\phi} + 3H\dot{\phi} + V'(\phi) = 0.
\eea

The conventional misalignment mechanism assumes $t_i\ll m_\phi^{-1}$, and the initial conditions for inflation, $\theta(t_i)= \theta_i$, $\dot\theta_i =0$.
If the universe is at this time dominated by radiation, where H scales as $H =\frac{1}{2t}$, the axion evolves as \cite{Co:2019jts}

\bea
\theta(t) \simeq \times 
    \begin{dcases}
    1 & \text{for $m_\phi t \ll 1$}\\
   2^{1/4} \Gamma\left(\frac{5}{4}\right) \sqrt{\frac{2}{\pi}} \left(\frac{1}{m_\phi t}\right)^{3/4}\cos(m_\phi t - \frac{3}{8}\pi) & \text{for $m_\phi t \gg 1$}
    \end{dcases}
\eea

then, we see that the axion stays frozen during the early stages but oscillations begin when the temperature drops below its mass, $3 H(T_*) = m_\phi$.
We define the abundance,

\bea
\Omega h^2 = \left(\frac{s_0}{\rho_c h^{-2}}\right) \frac{m_a^0 n_a}{s}.
\eea

The comoving number density, $\frac{n_a}{s}$ is constant, so we can evaluate it at the onset of oscillations,

\bea
\Omega h^2 = \left(\frac{s_{0}}{\rho_c h^{-2}}\right) \frac{m_{a}^0 n_{a*}}{s_*} \approx 0.12,
\eea

where we have assumed that axions compose the total amount of Dark Matter in the universe.
At the onset of oscillations, the amplitude of the axion oscillations is given by $\theta_0 f_a$, so the number density of axions is given by

\bea \label{eq:n*}
n_{a,*}= \frac{\langle \rho_*\rangle}{m_{a,*}}= \frac{1}{2} m_{a,*}\theta_0 f_a,
\eea

and the entropy,

\bea\label{eq:s*}
s_* = \frac{2\pi}{45}g_{*,s}T_*^3.
\eea

Using eqs.~(\ref{eq:n*}) and (\ref{eq:s*}), we obtain the abundance of Dark Matter as

\bea
\Omega h^2 \approx 3.9 \left(\frac{s_0}{\rho_c h^{-2}}\right)\frac{g_*^{3/4}}{g_*,s}\theta_0^2 \frac{m_{a,0}f_a^2}{(m_{a,*})^{1/2}M_P^{3/2}}
\eea

If we want to identify this axion with the QCD axion, we take the mass dependence in eq.~(\ref{eq:axionmassintro}), and the temperature $T_* \sim \sqrt{m_{a,*} M_P}$,

\bea
m_{a,*} \sim \frac{\Lambda^2}{f_a^{1/3}M_P^{2/3}},
\eea

leading to the final abundance form,

\bea\label{eq:axionabunintro}
\Omega h^2 \sim \theta_0^2 \frac{\Lambda}{T_{eq}}\left(\frac{f_a}{M_P}\right)^{7/6}.
\eea

From the abundance on eq.~(\ref{eq:axionabunintro}) we obtain the bound $f_a \lesssim 10^{12}$ GeV, in order not to overproduce Dark Matter.
Larger decay constants can only be achieved through very small misalignment angles, which conflicts with naturalness expectations. We would have just traded the strong CP problem for another naturalness problem in the axion decay constant window.


\subsection{Bounds on Dark Matter} \label{sec:darkmatterbounds}
It is also possible that  Dark Matter is depleted by 3-to-2 annihilations, instead of the 2-to-2 interactions we just studied for the WIMP paradigm. In this case, we dub the scenario Strongly Interacting Dark Matter(SIDM).
For both WIMPS and SIMPS, unitarity bounds set

\bea
(\sigma v)\simeq 3 \times 10^{-26} cm^3/s < \frac{16 \pi}{m^2_{DM}v},
\eea

which sets the Dark Matter masses to $m_{DM} \lesssim 100 \, \mathrm{TeV}$.
For FIMPs (Feebly Interacting Dark Matter), the temperature bounds 

\bea
\langle \sigma v\rangle \propto (\sigma v)\Big|_{T=m_{DM}}\times e^{-m_{DM}/T},
\eea
which sets the Dark Matter masses to be $m_{DM} \lesssim T_{RH}$.
Producing heavier Dark Matter calls for new mechanisms.

\section{Inflation and supersymmetry}
\label{subsec:intro_SUSY}

Supersymmetry plays an important role in many extensions of the Standard Model, and in particular, it can be responsible for the radiative stability of many inflationary models.
Furthermore, spontaneous breaking of SUSY can lead to unique phenomenological effects. In this section we review the basic concepts behind supersymmetry.

In particle physics, we associate symmetries to groups. The Coleman-Mandula's theorem states that it is impossible to combine Poincar\' e and internal symmetries in a way that is not trivial,\\ 
\begin{center}
    {Poincar\'e $\times$ Internal}.\\
\end{center}
There is, however, a caveat to the Coleman-Mandula's theorem: it only works for bosonic internal symmetries.
Supersymmetry relates bosons and fermions. It expands the Poincar\' e and internal symmetries with new spinor supercharges $Q_\alpha^i$. 
In this section we will consider the simplest case, supersymmetry with a single spinor charge $Q_\alpha$, or $\mathcal{N}=1$ supersymmetry.

The algebra of supersymmetry is a $Z_2$-graded algebra, that is, it has two classes of elements: even and odd. In physical terms, that would be bosons and fermions.
Besides the Poincar\' e algebra, the new (anti)commutation relations involving the supercharge $Q_\alpha$ are given by

\bea
\left[P^\mu, Q_\alpha \right] &=& \left[ P^\mu, \bar{Q}_{\dot{\alpha}} \right] = 0 \label{susy1} \\
\left\{Q_\alpha, Q_\beta \right\} &=& \left\{ \bar{Q}_{\dot{\alpha}}, \bar{Q}_{\dot{\beta}}\right\} = 0 \\
\left\{Q_\alpha, \bar{Q}_{\dot{\beta}}\right\} &=& 2\sigma^\mu_{\alpha \dot{\beta}}P_\mu \label{susytrans}
\eea

Eq (\ref{susytrans}) is really at the heart of supersymmetry. It implies that the composition of two supercharges is equivalent to a spacetime translation.
Furthermore, from eq. (\ref{susy1}) we see how supersymmetry relates boson to fermions.
Particles in a supersymmetric theory come in supermultiplets containing particles of spin $S$ and $S+1/2$.
Two examples of such multiplet representations are:

\begin{itemize}
    \item The chiral multiplet, containing a Majorana spin 1/2 fermion $\chi\left(x\right)$, and its superpartner, a complex spin 0 boson, $Z(x)$.
    \item The gauge multiplet, containing a massless spin 1 particle $A_\mu(x)$, and its spin 1/2 superpartner $\lambda(x)$, the gaugino, described by a Majorana spinor.
\end{itemize}

From a geometric perspective, superspace is obtained by adding the extra fermionic coordinates to the usual bosonic spacetime coordinates. 
Then, we can characterize the superspace by the set of coordinates $\{ x^\mu, \theta_\alpha, \bar\theta_{\dot{\alpha}}\}$, where the Grassmannian coordinates have dimension $\left[\mathrm{mass}\right]^{-1/2}$ and $x^\mu$ has the usual dimensions $\left[\mathrm{mass}\right]^{-1}$.
For the Grassmann variables, $\theta \theta =0$, meaning that any series expansion will be truncated at first order,

\bea
f(\theta, \bar\theta) = f_0 + f_1 \theta + f_2 \bar\theta.
\eea

Here, $f_0$ is even under the supersymmetry algebra, while $f_1$ and $f_2$ are odd. 
Each $\theta_\alpha$ and $\theta^\dagger_{\dot\alpha}$ has two different components, $\alpha, \dot\alpha =1,2$, so the general expansion of a superfield will be truncated at second order,

\bea
\bm{F}\left(x, \theta, \bar\theta \right)&=& f_0(x)+ \theta\phi(x) + \bar\theta \bar\chi(x) + f_1(x)\theta\theta + f_2(x)\bar\theta \bar\theta + \theta\sigma^\mu \bar\theta v_\mu(x) \nonumber\\
 &+& \bar{\theta}\bar{\theta}\theta \eta(x) + \theta\theta \bar{\theta}\bar{\lambda}(x) + f_3(x)\theta\theta \bar{\theta}\bar{\theta}\label{superfield}
\eea

Where $\theta\phi(x)\equiv \theta^{\alpha}\phi_{\alpha}(x)$, $\bar\theta \bar\chi(x)\equiv \bar\theta_{\dot\alpha} \bar\chi^{\dot\alpha}(x)$,  $\bar{\theta}\bar\lambda(x) \equiv \bar\theta_{\dot\alpha}\bar\lambda^{\dot{\alpha}}(x)$, etc. 
The scalar superfield in eq.~(\ref{superfield}) contains 4 complex scalars ($f_0$, $f_1$, $f_2$, $f_3$), two left-handed spinors($\phi$, $\eta$), two right-handed spinors ($\bar\chi$, $\bar\lambda$) and a vector ($\sigma^\mu_{\alpha\dot\alpha}\nu_\mu$).

We can define any group element of the super-Poincar\'e group as \footnote{$M^{\mu \nu}$ is the generator of Lorentz transformations, satisfying 
$\left[ M^{\mu\nu}, Q_{\alpha} \right] = \left(\sigma^{\mu \nu}\right)_\alpha^{\quad\beta} Q_\beta$ and $\left[ M^{\mu\nu}, \bar{Q}^{\dot\alpha} \right] = \left(\bar{\sigma}^{\mu\nu}\right)^{\dot\alpha}_{\quad\dot\beta} \bar{Q}^{\dot\beta}$.}

\bea
e^{i\left(x^\mu P_\mu + \theta^\alpha Q_\alpha + \bar \theta_{\dot\alpha} \bar{Q}^{\dot\alpha}\right)} e^{-\frac{1}{2} \omega^{\mu\nu} M_{\mu\nu}},
\eea

and denote
\bea
G(x,\theta, \bar{\theta}) = e^{i\left(x^\mu P_\mu + \theta^\alpha Q_\alpha + \bar \theta_{\dot\alpha} \bar{Q}^{\dot\alpha}\right)}.
\eea

The product of two group elements can be computed by making use of the Baker-Campbell-Hausdorff formula

\bea
e^{A}e^{B} = e^{A + B + \frac{1}{2}\left[A,B\right]}.
\eea

Hence, in order to see how a SUSY transformation acts on the coordinates $\left\{x^\mu, \theta, \bar{\theta}\right\}$, we only need to compute 

\bea
\left[ \xi Q, \bar{\theta} \bar{Q}\right] &=& \xi^\alpha Q_\alpha \bar{\theta}_{\dot\alpha} \bar{Q}^{\dot \alpha} - \bar{\theta}_{\dot \alpha} \bar{Q}^{\dot\alpha} \xi^\alpha Q_\alpha \nonumber\\
&=& -\xi^{\alpha}\bar{\theta}_{\dot \alpha}\left(Q_\alpha \bar{Q}_{\dot \alpha} + \bar{Q}_{\dot\alpha}Q_\alpha\right) \nonumber \\
&=&- \xi^\alpha \bar{\theta}_{\dot\alpha} \left\{Q_a, \bar{Q}^{\dot \alpha}\right\}\nonumber \\
&=& \xi^\alpha \bar{\theta}^{\dot\alpha} \left\{Q_a, \bar{Q}_{\dot \alpha}\right\} \nonumber \\
&=& 2\xi^\alpha \sigma_{\alpha \dot\alpha}^{\mu}\theta^{\dot\alpha} P_\mu,
\eea

where in the third line we used  $\psi^\alpha = \epsilon^{\alpha \beta}\psi_{\beta}$
and $\epsilon_{\dot \alpha \dot \beta} \epsilon^{\dot\beta \dot\gamma } = \delta_{\dot\alpha}^{\dot\gamma}$. Here, $\xi^\alpha$ and $\xi_{\dot \alpha}$ are Grassmann spinors, so we pick up minus signs when commuting them with odd variables such as $\theta$.
All together,

\bea
\label{susycoordtrans}
G\left(a, \xi, \bar\xi\right)G\left(x,\theta, \bar\theta\right) = G\left(x^\mu + a^\mu + i\xi\sigma^\mu \bar\theta - i\theta \sigma^\mu \bar\xi, \theta + \xi, \bar\theta + \bar\xi\right)
\eea

From eq (\ref{susycoordtrans}) we read

\bea
x^\mu &\rightarrow&  x^\mu + a^\mu + i\xi\sigma^\mu \bar\theta - i\theta \sigma^\mu \bar\xi\label{deltax}\\
\theta &\rightarrow& \theta + \xi \label{eq:st2}\\
\bar\theta &\rightarrow& \bar\theta + \bar\xi \label{eq:st3}
\eea

Then, the superfield in eq. (\ref{superfield}) transforms as

\bea
\bm{F}' \left(x, \theta, \bar\theta\right) = F\left(x + \delta x, \theta + \xi, \bar{\theta}+\bar{\xi}\right)
\eea

with the variations as defined in eqs. (\ref{deltax}) -(\ref{eq:st3}).
Then, an infinitesimal variation of the superfield $\bm{F}$ is then given by 

\bea
\delta F &=& \bm{F}' \left(x, \theta, \bar\theta\right) - \bm{F} \left(x, \theta, \bar\theta\right)\nonumber \\
&=& \bm{F} \left(x + a^\mu + i\xi \sigma^\mu \bar\theta -i\theta \sigma^\mu\bar\xi, \theta + \xi, \bar\theta + \bar\xi \right) - \bm{F} \left(x, \theta, \bar\theta\right) \nonumber \\
&=&\left(a^\mu + i\xi \sigma^\mu \bar\theta -i\theta \sigma^\mu\bar\xi \right) \partial_\mu\bm{F} + \xi^\alpha\partial_\alpha\bm{F} + \bar{\xi}_{\dot\alpha}\bar{\partial}^{\dot{\alpha}}\bm{F}
\eea

We notice that,  $\phi(x)\rightarrow \phi(x -a)$ can be equivalently written as 

\bea
\label{momentumoperator}
\delta \phi(x) = -a^\mu \partial_\mu \phi(x),
\eea
so $\delta \phi(x)= i a^\mu P_\mu \phi(x)$, and the transformation of the superfield becomes, at first order in $\xi$ and $\bar\xi$,

\bea
\delta \bm{F} = \left(a^\mu \partial_\mu + \xi^\alpha Q_\alpha + \bar{\xi}_{\dot\alpha}\bar{Q}^{\dot\alpha}\right).
\eea

The same transformation in eq.(\ref{susycoordtrans}) can be obtained from the differential operators

\bea
Q_\alpha &=& i\left( \frac{\partial}{\partial \theta^{\alpha}} -i \sigma^\mu_{\alpha \dot\alpha} \bar\theta^{\dot\alpha} \partial_\mu \right),\\
\bar{Q}^{\dot\alpha} &=& i\left( \frac{\partial}{\partial \bar\theta_{\dot\alpha}} - i \theta^{\alpha}\bar\sigma^\mu_{\alpha \dot\beta} \epsilon^{\dot\beta \dot\alpha} \partial_\mu \right).
\eea


Let's now act with the super-translation upon the superfield

\bea
&&e^{i\left(a^\mu P_\mu + \xi^\alpha Q_\alpha + \bar\xi_{\dot\alpha}\bar{Q}^{\dot\alpha}\right)}\bm{F}(x, \theta, \bar\theta)e^{-i\left(a^\mu P_\mu + \xi^\alpha Q_\alpha + \bar\xi_{\dot\alpha}\bar{Q}^{\dot\alpha}\right)} \nonumber\\
&=& \bm{F} \left(x + a^\mu + i\xi \sigma^\mu \bar\theta -i\theta \sigma^\mu\bar\xi, \theta + \xi, \bar\theta + \bar\xi \right) \nonumber \\
&=&\bm{F} + \delta \bm{F}
\eea

Noting that 
\bea
&&e^{i\left(a^\mu P_\mu + \xi^\alpha Q_\alpha + \bar\xi_{\dot\alpha}\bar{Q}^{\dot\alpha}\right)}\bm{F}(x, \theta, \bar\theta)e^{-i\left(a^\mu P_\mu + \xi^\alpha Q_\alpha + \bar\xi_{\dot\alpha}\bar{Q}^{\dot\alpha}\right)}\nonumber\\
&=& \bm{F}\left(x, \theta, \bar{\theta}\right) + i\left[a^{\mu}P_{\mu} + \xi^\alpha Q_\alpha + \bar\xi_{\dot\alpha}\bar{Q}^{\dot\alpha},\bm{F} \right],
\eea

we find that the superfield transforms as 

\bea
\label{susytr}
\delta_\xi \bm{F}\left(x, \theta,\bar\theta\right) = i \left[ \xi Q + \bar{\xi}\bar{Q}, \bm F\right] = \left(\xi Q + \bar{\xi} \bar{Q}\right)\bm{F}.
\eea

So far, we have worked with an arbitrary superfield as described by eq. (\ref{superfield}).
However, this representation the SUSY algebra can be further reduced. For that, we will define the chiral and anti-chiral superfields.
The chiral superfield satisfies\footnote{An anti-chiral superfield satisfies $D_{\alpha} \bm{\Psi} = 0$. If $\bm\phi$ is a chiral superfield,
then its conjugate, $\bar{\bm\phi}$ must be anti-chiral.
To see this, we only need to realize that while the field is a function of $x^\mu$ and $\theta$, its conjugate is a function of $\bar{x}^{\mu}$ and $\bar\theta$, but $D_{\alpha} x^{\mu}=D_{\alpha}\theta^{\dot\alpha}=0.$}

\bea
\overline{D}_{\dot\alpha}\bm\Phi =0,
\eea

where the covariant derivative $\overline{D}_{\dot\alpha}$ is defined as

\bea \label{eq:susycovariant}
\overline{D}_{\dot\alpha} = - \frac{\partial}{\partial \bar\theta^{\dot\alpha}} - i \theta^\alpha \sigma^\mu_{\alpha \dot\alpha}\partial_\mu.
\eea

Applying eq.~(\ref{eq:susycovariant}) to a chiral superfield, we get

\bea \label{eq:fieldchiral}
\bm \Phi &=& \phi(x) + i \bar\theta \bar{\sigma}^\mu \theta \partial_\mu \phi(x) + \frac{1}{4} \theta \theta \bar\theta \bar\theta \partial_\mu \partial^\mu \phi(x) + \sqrt{2}\theta \psi(x) \nonumber\\
&-& \frac{i}{\sqrt{2}}\theta\theta\bar\theta \partial_\mu \psi(x) + \theta \theta F(x).
\eea

Comparing eq.~ (\ref{superfield}) with eq.(\ref{eq:fieldchiral}), we see that $\bar{\chi} = \eta = f_2 =0$,  $f_1 =F$ and $f_0= \phi$.
The susy transformations in eqs.~(\ref{deltax}), (\ref{eq:st2}) and (\ref{eq:st3}) become,

\bea
\delta_\epsilon \phi &=& \epsilon \psi,\\
\delta_\epsilon \psi_\alpha &=& -i \left(\sigma^\mu \epsilon^\dagger\right)_\alpha \partial_\mu \phi + \epsilon_\alpha F,\\
\delta_\epsilon F &=& -i \epsilon^\dagger \bar{\sigma}^{\mu}\partial_\mu \psi.
\eea

From this transformation rules we see that $F$ actually transforms as a total derivative. Then, if the fields go to zero at infinity, this terms vanishes, so that the F-terms are SUSY invariants.
Furthermore, since $F$ is an auxiliary field, we can write a SUSY invariant Lagrangian using the equation of motion for F, $\frac{\partial \mathcal{L}}{\partial F} =0.$
Since $\theta^2 \bar\theta^2 $ and $\bm{\Phi}$ also transforms as total derivatives, we can write down the SUSY invariant Lagrangian
\begin{align}
\mathcal{L} = \int d^2 \bar\theta d^2 \theta \overline{{\bm\Phi}}_i \bm\Phi_i + \int d^2\theta \left(f_i \bm \Phi_i + \frac{1}{2}m_{ij}\bm\Phi_i \bm\Phi_j + \frac{1}{3}\lambda_{ijk}\bm \Phi_i \bm\Phi_j \bm\Phi_k + \lambda_i \bm\Phi_i\right) + \mathrm{h.c.} \label{eq:susylag},
\end{align}

where
\bea
\overline{\bm\Phi} \bm\Phi &= F^* F + \frac{1}{4}A^* \partial^2 A + \frac{1}{4}\partial^2 A^* A - \frac{1}{2}\partial_\mu A^* \partial^\mu A \bar\sigma^\mu \psi \nonumber\\
& + \frac{1}{2} \partial_\mu \bar\psi - \frac{i}{2} \psi \bar\sigma^\mu\partial_\mu \psi.
 \eea

 The equation of motion for F leads to
 \bea
 \frac{\partial \mathcal{L}}{\partial F_k} = F_k^* +\lambda_k + m_{ik} A_i + \lambda_{ijk}A_i A_j = 0,
 \eea

so that the F-term part of the potential is given by

 \bea
 V_F = \vert  \lambda_k + m_{ij}A_i + \lambda_{ijk}A_i A_j\vert^2.
\eea

Hence we see that the F-terms are always positive(or zero).

The first term in in eq. (\ref{eq:susylag}) is known as the K\"ahler potential, while the second is the superpotential.
 The chiral lagrangian can be written as 

\bea
 \mathcal{L}= \mathcal{L}_0 + \mathcal{L}_\psi - V_F,
 \eea

 with

 \bea
 V_F = \sum_i \Big\vert \frac{\partial W}{\partial \bm\Phi_i} \Big\vert^2
 \eea

 and 

 \bea
 \mathcal{L}_\psi = -\frac{1}{2} \sum_{i,j} \frac{\partial W}{\partial \bm\Phi_i \bm\Phi_j}\psi_i \psi_j + \mathrm{h.c.},
 \eea
 where the derivatives of the superpotential need to be evaluated at $\bm\Phi_i = A_i$. Here, we note that the superpotential is a function of only the un-conjugated fields, i.e, it is a holomorphic function.
 
 So far, we have constructed a supersymmetric theory containing scalars and fermions. If we want to include vectors, then we need to take into account that they satisfy $V= V^\dagger$.
 In the Wess-Zumino gauge, vector fields can be parametrized as

 \bea
 V= -\theta \sigma^\mu \bar\theta v_\mu(x) + i\theta\theta\bar\theta\bar\lambda(x) + \frac{1}{2}\theta\theta\bar\theta\bar\theta D(x),
 \eea

 where D is an auxiliary field that, like F, transforms as a total derivative under a SUSY transformation,

 \bea
 \delta_\epsilon v_\mu &= i \epsilon \sigma^\mu \bar \lambda + i\bar\epsilon \bar \sigma^\mu \lambda,\\
 \delta_\epsilon \lambda &= \sigma^{\mu\nu} \epsilon v_{\mu\nu} + i \epsilon D, \\
 \partial_\epsilon D &= \bar\epsilon \bar\sigma^\mu \partial_\mu \lambda - \epsilon \sigma^\mu \partial_\mu \bar\lambda.
 \eea
The Lagrangian for the vector part is given by

 \bea
 \mathcal{L}_V= \frac{1}{4} \left(W^\alpha W_\alpha\vert_{\theta\theta}\right) + \mathrm{h.c.}
 \eea

 with the field strength

 \bea
 W_\alpha = -\frac{1}{4}\bar D \bar D D_\alpha V.
 \eea

\subsection*{R-symmetry}
 It is usual to include an extra symmetry, the \textit{R-symmetry}. This is a global $U(1)$ symmetry for the $\theta$ field, defined by the commutation relations

 \bea
 \left[R, Q_\alpha\right] &= - Q_\alpha \\
 \left[R, \overline{Q}_{\dot\alpha}\right] &= \bar{Q}_{\dot{\alpha}}
 \eea
 The R-symmetry is the only one that distinguishes between the different components inside a multiplet. The spontaneous symmetry breaking of the R-symmetry is a sufficient condition for symmetry breaking.

 \subsection{SUSY breaking}

 When we say that a theory is spontaneously broken, what we mean is that the Noether charge Q doesn't act trivially upon the ground state

 \bea
 Q|0\rangle \ne 0.
 \eea

 Through the relation

 \bea
 H = P_0 = \frac{1}{4}\left(\left\{Q_1,\overline Q_{\dot{1}}\right\} + \left\{Q_2,\overline Q_{\dot{2}}\right\}\right),
 \eea

 we see that if $\langle 0 | H | 0 \rangle \ne 0$, 
 then necessarily  $Q_\alpha|0\rangle \ne 0$ 
 or $ \overline Q_{\dot{\alpha}}|0\rangle \ne 0$, meaning that SUSY is spontaneously broken. However, in order to break supersymmetry, the theory needs to overcome a topological obstruction: The Witten index.
 In supersymmetry we can have different grounds states, such as $(-1)^F |0\rangle = - |0\rangle$,
  corresponding to a fermionic state, and its bosonic counterpart $(-1)^F |0\rangle = + |0\rangle$. 
  The \textit{Witten index} takes account of the difference between the two,

  \bea
  \tr(-1)^F = \sum_E (n_B(E) - n_F(E)) = n_B(0) - n_F(0).
  \eea

Only a theory that has vanishing Witten index can undergo SSB.

Furthermore, we want to break SUSY while preserving Lorentz invariance, which means that the only fields that can get VEVs are Lorentz scalars.
 Hence, SUSY breaking can be achieved through the D of F terms. 
 In \textbf{chapter \ref{chap_sugra}} we will consider the case in which the F terms break SUSY, so we will make a brief review of its dynamics here.

 \subsection*{F-term breaking or O'Raifeataigh model}

 In this case, the superpotential is defined by
 \bea
 W = \mu^2 \Phi_0 + m \Phi_1 \Phi_2 + g \Phi_0 \Phi_1 \Phi_1.
 \eea

 The R-charges are +2, 0, and 2 for $\Phi_0, \Phi_1$ and $\Phi_2$ respectively.
The F term potential is given by 

\bea
V_F = |F_0|^2 + |F_1|^2 + |F_2|^2,
\eea

where the $F_i$'s are the derivatives of the super potential with respect to the $i$-th field.
Then

\bea
F_0 &=& \mu^2 + g\Phi_1\Phi_1,\\
F_1 &=& m\Phi_2 + 2g \Phi_0 \Phi_1,\\
F_2 &=& m\Phi_1.
\eea

For $\Phi_1=0$, $F_0 \ne 0$ and  $F_2 =0$. If we take  $\Phi_2 =0$, $F_1 =0$, and SUSY is broken spontaneously.
The non-zero F term shifts the masses of scalar fields by $m^2 + g \mu^2$, separating it from the fermion mass. 

For a non-zero F term, we get the SUSY transformation 

\bea
\delta_\epsilon \psi_0 = \sqrt{2} \epsilon F_0 = \sqrt{2}\epsilon \mu^2.
\eea

Then, we identify $\psi_0$ with the goldstino (the Goldstone boson associated to global SUSY breaking). In general, Goldstinos are problematic, but they get eaten by the gravitino when we gauge the theory (supergravity).


\chapter{UV completions of Higgs Inflation}\label{chap_UV}

\begin{small}
    ``The trick is to never stop looking. There's always another secret.”\\
    Brandon Sanderson
    \end{small} 

    \vspace{5mm}
In \textbf{section \ref{sec:higgsinflation}}, we went through the unitarity problem of Higgs inflation and showed that the Higgs kinetic terms can be identified with a non-linear sigma model in the Einstein frame.
In this chapter, we show how adding extra scalar degrees of freedom does indeed linearize Higgs inflation. We start by reviewing the simplest possible expansion, adding the $R^2$ term from Starobinsky inflation, and then propose a general class of UV completions by introducing general higher curvature terms $R^n$.

\section{Starobinsky as a UV completion}\label{uvstarobinsky}

We add the Starobinsky term $R^2$ to the Higgs inflation Lagrangian in (\ref{higgsori})

\bea
{\cal L}_{R2}=\sqrt{-{\hat g}} \bigg[-\frac{1}{2}(1+\xi {\hat\phi}^2_i)  {\hat R}+\frac{1}{2} g^{\mu\nu} \partial_\mu {\hat\phi}_i \partial_\nu {\hat\phi}_i -\frac{\lambda}{4} ({\hat\phi}^2_i)^2 +\alpha {\hat R}^2\bigg], \label{r2}
\eea
with 

\bea
\kappa = \frac{1}{36\alpha}. \label{r2-quartic}
\eea

We notice that $\kappa$ needs to be a tiny parameter in order to match the CMB observations, or equivalently, $\alpha$ is found to be very large. This is a known fine-tuning issue in Starobinsky inflation.

The dual theory for the Lagrangian (\ref{r2}) reads,
\bea
\frac{{\cal L}_{R2}}{\sqrt{-{\hat g}}}= -\frac{1}{2} {\hat R} (1+\xi {\hat\phi}^2_i + 4\alpha {\hat\chi}) -\alpha{\hat\chi}^2  + \frac{1}{2} (\partial_\mu{\hat\phi}_i)^2 -\frac{\lambda}{4} ({\hat\phi}^2_i)^2  \label{r2-dual},
\eea

where $\hat{\chi}$ is the dual scalar field.
After a conformal transformation, as given in eq. (\ref{eq:conformaltransformetric}), and taking the field redefinitions ${\hat\phi}_i=\Omega \phi_i$ and ${\hat\chi}=\Omega^2\chi$, the Lagrangian in equation (\ref{r2-dual}) becomes
\bea \label{midlad}
\frac{{\cal L}_{R2}}{\sqrt{-g}}&=&\Omega^{-4}\bigg[ \frac{1}{2}\Omega^2 \Big(-R+6(\partial_\mu\ln \Omega)^2-6\partial^2\ln\Omega\Big) (1+\xi\Omega^2\phi^2_i+4\alpha\Omega^2\chi ) \nonumber \\
&&-\alpha\Omega^4 \chi^2 +\frac{1}{2} \Omega^2 \Big( \partial_\mu(\Omega\phi_i)\Big)^2 -\frac{\lambda}{4} \Omega^4\phi^4_i \bigg] \nonumber \\
&=&-\frac{1}{2} (\Omega^{-2} +\xi\phi^2_i +4\alpha\chi) R +3\Omega^{-2} \Big((\partial_\mu\ln\Omega)^2-\partial^2\ln\Omega \Big) (1+\xi \Omega^2\phi^2_i+4\alpha\Omega^2\chi) \nonumber \\
&&-\alpha\chi^2 + \frac{1}{2} \Omega^{-2} (\Omega\partial_\mu\phi_i+\partial_\mu\Omega\,\phi_i)^2 - \frac{\lambda}{4} \phi^4_i . \label{r2a}
\eea

We can rewrite the derivative terms as
\bea
\Omega^{-2} (\Omega\partial_\mu\phi_i+\partial_\mu\Omega\,\phi_i)^2
&=&(\partial_\mu\phi_i)^2 +(\partial_\mu\ln\Omega)\, \phi_i\partial^\mu\phi_i +  (\partial_\mu\ln\Omega)^2 \phi^2_i \nonumber \\
&=& (\partial_\mu\phi_i)^2 +\phi^2_i \Big((\partial_\mu\ln\Omega)^2-\partial^2\ln\Omega \Big), 
\eea

where the total derivative term vanishes under integration. Then, eq.~(\ref{midlad}) becomes
\bea
\frac{{\cal L}_{R2}}{\sqrt{-g}}&=&-\frac{1}{2} (\Omega^{-2} +\xi\phi^2_i +4\alpha\chi) R +3\Big((\partial_\mu\ln\Omega)^2-\partial^2\ln\Omega \Big) \bigg(\Omega^{-2} +\Big(\xi+\frac{1}{6}\Big)\phi^2_i+4\alpha\chi\bigg) \nonumber \\
&&-\alpha\chi^2 + \frac{1}{2}  (\partial_\mu\phi_i)^2 - \frac{\lambda}{4} \phi^4_i. \label{r2b}
\eea

In this case, the conformal factor is given by 
\bea
\Omega^{-2}= \Big(1+\frac{\sigma}{\sqrt{6}} \Big)^2 \label{r2-conf},
\eea

and the constraint equation by

\bea
\Omega^{-2} +\xi\phi^2_i +4\alpha\chi= 1-\frac{1}{6} \phi^2_i -\frac{1}{6}\sigma^2. \label{r2-conf2} 
\eea

Rearranging (\ref{r2-conf2})
\bea
\Omega^{-2} +\Big(\xi+\frac{1}{6}\Big)\phi^2_i+4\alpha\chi= 1-\frac{1}{6}\sigma^2,
\eea

allows to rewrite the kinetic terms as

\bea
3\Big((\partial_\mu\ln\Omega)^2-\partial^2\ln\Omega \Big) \bigg(\Omega^{-2} +\Big(\xi+\frac{1}{6}\Big)\phi^2_i+4\alpha\chi\bigg)
&=&\frac{3}{\sqrt{6}}\Big(1-\frac{1}{6}\sigma^2\Big) \Omega\partial^2\sigma \label{kinterm} \nonumber \\
&=&-\frac{1}{2} \sigma \partial^2\sigma,
\eea

up to a total derivative. Putting together (\ref{kinterm}) and (\ref{r2b}) we obtain the lagrangian in the conformal frame,

\bea
\frac{{\cal L}_{R2}}{\sqrt{-g}}=-\frac{1}{2} R\Big(1-\frac{1}{6}\phi^2_i -\frac{1}{6} \sigma^2 \Big)+\frac{1}{2} (\partial_\mu\sigma)^2 +\frac{1}{2}(\partial_\mu\phi_i)^2 -\alpha \chi^2 -   \frac{\lambda}{4} \phi^4_i, \label{confinvlag}
\eea
with the constraint equation in terms of the new scalar field,
\bea
\chi=\frac{1}{4\alpha} \bigg[\frac{1}{2}-\frac{1}{3} \Big(\sigma+\frac{\sqrt{6}}{2}  \Big)^2-\Big(\xi+\frac{1}{6}\Big)\phi^2_i \bigg]. 
\eea

The Lagrangian in (\ref{confinvlag}) coincides with the one in (\ref{sigmamodels}), in which the term $R^2$ was not considered.
Thus, we have shown the equivalence between Starobinsky inflation and the linear-sigma models of Higgs inflation, making Starobinsky inflation a suitable UV completion.
The Einstein frame potential for the \textit{Higgs + Starobinsky inflation} is given by
\bea
U(\sigma,\phi_i) = \alpha \chi^2 = \frac{1}{16\alpha}\, \bigg[\frac{1}{2}-\frac{1}{3} \Big(\sigma+\frac{\sqrt{6}}{2}  \Big)^2-\Big(\xi+\frac{1}{6}\Big)\phi^2_i \bigg]^2. \label{r2-pot}
\eea

Similarly to our discussion in \textbf{section \ref{sec:gaugefixing}}, in order to ensure perturbativity, we ask that the coupling are weak,

\bea
\frac{1}{36\alpha} \lesssim 1, \qquad \lambda+ \frac{1}{4\alpha} \Big(\xi+\frac{1}{6} \Big)^2\lesssim 1,  \qquad \frac{1}{4\alpha}\,  \Big(\xi+\frac{1}{6} \Big)\lesssim 1.
\eea

\section{UV completions from higher curvature terms}\label{highercurvatureuv}
Now that we know that the $R^2$ term can provide a successful UV completion of Higgs inflation,
we move to the most general case, containing $R^{k+1}$ terms with $k>0$,

\begin{align}
{\cal L}_{\rm gen}=\sqrt{-{\hat g}} \bigg[-\frac{1}{2}(1+\xi {\hat\phi}^2_i)  {\hat R}+\frac{1}{2} g^{\mu\nu} \partial_\mu {\hat\phi}_i \partial_\nu {\hat\phi}_i -\frac{\lambda}{4} ({\hat\phi}^2_i)^2 +\sum_k\frac{2(-1)^{k+1}\alpha_k}{k+1}\, {\hat R}^{k+1}\bigg]\label{rn}
\end{align}

Each term $ R^{k+1}$ in eq.~(\ref{rn}) has an associated dual scalar field $\hat\chi_k$, such that

\bea
\frac{{\cal L}_{\rm gen}}{\sqrt{-{\hat g}}}= -\frac{1}{2} {\hat R} \Big(1+\xi {\hat\phi}^2_i + \sum_k 4\alpha_k {\hat\chi}_k\Big) -\sum_k 2\Big(\frac{k}{k+1}\Big)\,\alpha_k\,{\hat\chi}^{\frac{k+1}{k}}_k  + \frac{1}{2} (\partial_\mu{\hat\phi}_i)^2 -\frac{\lambda}{4} ({\hat\phi}^2_i)^2. \nonumber 
\eea

Similarly to what we did for the $R^2$ case, now we perform a conformal transformation, ${\hat g}_{\mu\nu}=\Omega^{-2} g_{\mu\nu}$, and redefine the fields as ${\hat\phi}_i=\Omega \phi_i$ and ${\hat\chi}_k=\Omega^{2}\chi_k$,
with $\Omega^{-2}=\big(1+\frac{\sigma}{\sqrt{6}}\big)^2$. The conformal Lagrangian is given by

\bea
\frac{{\cal L}_{\rm gen}}{\sqrt{-g}}&=&-\frac{1}{2} R\Big(1-\frac{1}{6}\phi^2_i -\frac{1}{6} \sigma^2 \Big)+\frac{1}{2} (\partial_\mu\sigma)^2 +\frac{1}{2}(\partial_\mu\phi_i)^2 \nonumber \\
&& -\sum_k \Omega^{-2+\frac{2}{k}}\Big(\frac{2k}{k+1}\Big)\,\alpha_k\,{\chi}^{1+\frac{1}{k}}_k   -   \frac{\lambda}{4} \phi^4_i \label{genL},
\eea

with the constraints
\bea
\sum_k 4\alpha_k\chi_k=\frac{1}{2}-\frac{1}{3} \Big(\sigma+\frac{\sqrt{6}}{2} \Big)^2-\Big(\xi+\frac{1}{6}\Big)\phi^2_i. \label{constraint-gen}
\eea

We can include the constrain equations (\ref{constraint-gen}) in the Lagrangian through
\bea
\frac{\Delta{\cal L}_{\rm gen}}{\sqrt{-g}}= y(x)\cdot \left[\sum_k 4\alpha_k\chi_k- \frac{1}{2}+\frac{1}{3} \Big(\sigma+\frac{\sqrt{6}}{2} \Big)^2+\Big(\xi+\frac{1}{6}\Big)\phi^2_i\right].
\eea

In this case, the dual scalar fields can be written in terms of the $y(x)$ couplings, as
\bea
\chi_k = 2^k\Omega^{2k-2} \,y^k \label{duals},
\eea

where $y(x)$ satisfies the algebraic equations

\bea
\sum_k 4\alpha_k \,2^k \Omega^{2k-2} \,y^k=\frac{1}{2}-\frac{1}{3} \Big(\sigma+\frac{\sqrt{6}}{2} \Big)^2-\Big(\xi+\frac{1}{6}\Big) \phi^2_i. \label{lagmul}
\eea
The scalar potential is given by 

\bea
U(\sigma,\phi_i)&=& \sum_k \Omega^{-2+\frac{2}{k}}\Big(\frac{2k}{k+1}\Big)\,\alpha_k\,{\chi}^{1+\frac{1}{k}}_k \nonumber \\
 &=&\sum_k \Big(\frac{2^{k+2} k}{k+1} \Big) \alpha_k (\Omega(\sigma))^{2k-2} (y(\sigma,\phi_i))^{k+1}.  \label{genpotg}
\eea

In order to decouple the sigma field from the rest of the scalars, we impose $\frac{\partial U}{\partial\sigma}=0$.

\bea
0= \frac{\partial U}{\partial\sigma}&=& \sum_k  2^{k+2} k  \alpha_k (\Omega(\sigma))^{2k-2} (y(\sigma,\phi_i))^k  \nonumber \\
&&+ \sum_k \Big(\frac{2^{k+2} k}{k+1} \Big) \alpha_k(2k-2) (\Omega(\sigma))^{2k-3} (y(\sigma,\phi_i))^{k+1}.
\eea

For $k \ge 1 $, $y=0$ will always present an extremum. 
At energies lower $m_\sigma$, we can integrate out the $\sigma$ field and recover the non-linear sigma model for Higgs inflation in eq.~(\ref{finallagrangiangague}).
It is a remarkable fact that this is the same non-linear sigma model that we obtained for the Starobinsky case in \textbf{section \ref{uvstarobinsky}}.

\subsection{Example A: A single curvature term}\label{sec:exampleA}

Let's consider a UV completion that contains a single term $R^{p+1}$.
The constraint equation in eq. (\ref{constraint-gen}) for this case simplifies to
\bea
\chi_p=\frac{1}{4\alpha_p} \bigg[\frac{1}{2}-\frac{1}{3} \Big(\sigma+\frac{\sqrt{6}}{2} \Big)^2-\Big(\xi+\frac{1}{6}\Big)\phi^2_i \bigg],
\eea

and the scalar potential is given by
\bea
U(\sigma,\phi_i)&=&\Omega^{-2+\frac{2}{p}}\Big(\frac{2p}{p+1}\Big)\, \alpha_p \chi^{1+\frac{1}{p}}_p \nonumber \\
&=& \frac{1}{3}\cdot 2^{-1-\frac{2}{p}}\Big(\frac{p}{p+1}\Big)\,\Big(\frac{1}{3\alpha_p}\Big)^{\frac{1}{p}}\Big(1+\frac{\sigma}{\sqrt{6}}\Big)^{2\big(1-\frac{1}{p}\big)} \nonumber \\
&&\times \left[ \frac{3}{2}-\bigg(\sigma+\frac{\sqrt{6}}{2}\bigg)^2 -3\Big(\xi+\frac{1}{6} \Big){\phi}^2_i\right]^{1+\frac{1}{p}} \nonumber \\
&\equiv& \frac{1}{4} \kappa_n (\sigma+\sqrt{6})^{4(1-n)}\left[-\sigma(\sigma+\sqrt{6})-3\Big(\xi+\frac{1}{6} \Big){\phi}^2_i\right]^{2n}  \label{genpot}
\eea

with $n=\frac{1}{2}\big(1+\frac{1}{p}\big)$. 
The coupling $\kappa_n$ in the scalar potential is proportional to $(\alpha_p)^{-1/|p|}$ for $p>0$, which corresponds to $n>\frac{1}{2}$.
For the case with $p<0$, $\kappa \propto (\alpha_p)^{1/|p|}$, corresponding to  $n<\frac{1}{2}$. 
Large value of $\alpha_p$ are favored for inflation with $p>0$.
The scalar potential in eq. (\ref{genpot}) has singular point at which $U=0$.
Given a non-integer $\frac{1}{p}$ (non-integer $2n$) and $ \alpha_p>0$, the possible values of the field are subject to the constraint
\bea
\sigma(\sigma+\sqrt{6}) +3\Big(\xi+\frac{1}{6} \Big){\phi}^2_i<0.
\eea

We also need to consider $\phi^2_i + \sigma^2<6$, so that the effective Planck mass from eq.~(\ref{genL}) remains positive.
In the case of a non-integer $\frac{1}{p}$, i.e., non-integer $2n$  
the potential becomes negative and cannot realize inflation.

\subsection{Example B: \texorpdfstring{$R^2 + R^3$}{R2 + R3}}
\label{sec:exampleB}

Now we consider the case in which only the $R^2$ and $R^3$ terms are non-zero, 
\bea
8(\alpha_1 y + 2\alpha_2 \Omega^2 y^2)=\frac{1}{2}-\frac{1}{3} \Big(\sigma+\frac{\sqrt{6}}{2} \Big)^2-\Big(\xi+\frac{1}{6}\Big) \phi^2_i\equiv -\frac{1}{3} f(\sigma,\phi_i). \label{r2r3}
\eea

Solving (\ref{r2r3}) for $y$, 
\bea
y=\frac{1}{4\alpha_2} \,\Omega^{-2} \Big(-\alpha_1+\sqrt{\alpha^2_1-\frac{1}{3} \alpha_2\Omega^2 f } \Big). \label{lagmul-sol}
\eea

Thus the dual scalar potential in eq.~(\ref{genpotg}) is given by
\bea 
U(\sigma,\phi_i) 
&=& 4\alpha_1 y^2 + \frac{32}{3} \alpha_2 \Omega^2 y^3 \nonumber \\
&=&\frac{1}{12\alpha^2_2}\,\Omega^{-4} \bigg[ 2\Big(\alpha^2_1-\frac{1}{3}\alpha_2\Omega^2 f\Big) \sqrt{\alpha^2_1-\frac{1}{3} \alpha_2\Omega^2 f} -\alpha_1 (2\alpha^2_1-\alpha_2\Omega^2 f) \bigg]\nonumber. 
\eea

To recover Starobinsky inflation, we set $\alpha_2 = 0$, so that
\begin{equation}
    U(\sigma,\phi_i)\approx \frac{1}{144\alpha_1} \, f^2
\end{equation}

which is the result we found in eq.~(\ref{r2-pot}).

\section{Inflation in Higgs-linear sigma models}\label{sec:inflationinhiggslinearsigmamodels}

For Higgs-$R^2$ inflation, our starting point is the Lagrangian in the conformal frame
\begin{align}
{\cal L}= \sqrt{-{g}} \bigg\{  -\frac{1}{2} \Big(1-\frac{1}{6} h^2-\frac{1}{6}\sigma^2 \Big) {R} +\frac{1}{2} (\partial_\mu h)^2+ \frac{1}{2} (\partial_\mu \sigma)^2 -\frac{\lambda}{4} h^4-U(\sigma, h) \bigg\} \label{sigmamodels2}
\end{align}
with the scalar potential

\bea
U(\sigma,h)=  \frac{\kappa_1}{4}\,\left[\sigma(\sigma+\sqrt{6})+3\Big(\xi+\frac{1}{6} \Big)h^2\right]^{2}. \label{sigmapot}
\eea

In the Einstein frame, the Lagrangian reads,

\bea
{\cal L}_E&=&  \sqrt{-{g_E}} \bigg\{  -\frac{1}{2}\,R(g_E)+\frac{3}{4\Omega^{\prime 4}} (\partial_\mu\Omega^{\prime 2})^2+\frac{1}{2\Omega^{\prime 2}} (\partial_\mu h)^2+ \frac{1}{2\Omega^{\prime 2}} (\partial_\mu \sigma)^2 -V(\sigma, h)\bigg\}  \nonumber \\
&=& \sqrt{-{g_E}} \bigg\{  -\frac{1}{2}\,R(g_E)+\frac{1}{2\Omega^{\prime 4}} \bigg[\Big( 1-\frac{1}{6}\sigma^2\Big)(\partial_\mu h)^2 +\Big( 1-\frac{1}{6} h^2\Big)(\partial_\mu \sigma)^2+\frac{1}{3} h\,\sigma\, \partial_\mu h \partial^\mu\sigma   \bigg]  \nonumber \\
&&\quad-V(\sigma,h) \bigg\} 
\label{general-einstein}
\eea
with
\begin{align}
&V(\sigma,h)&=\frac{1}{\Omega^{\prime 4}}\,\bigg( \frac{\lambda}{4} h^4+U(\sigma, h) \bigg) \nonumber \\
&=&\frac{1}{\big(1-\frac{1}{6}h^2 -\frac{1}{6}\sigma^2\big)^2} \bigg[\frac{1}{4} \kappa_1 \bigg(\sigma(\sigma+\sqrt{6})+3\Big(\xi+\frac{1}{6} \Big)h^2\bigg)^{2}+\frac{1}{4}\lambda h^4 \bigg]. \label{totalpotsigma}
\end{align}

From eq.~(\ref{totalpotsigma}) we identify the effective quartic coupling 
\bea
\lambda_{\rm eff}=\lambda+ 9 \kappa_1 \Big(\xi+\frac{1}{6} \Big)^2\lesssim {\cal O}(1). \label{lameff}
\eea

Once the Higgs field is integrated out, the effective quartic coupling is given by $\lambda$, which is smaller than $\lambda_{\rm eff}$ due to the threshold effects at tree-level \cite{Elias-Miro:2012eoi}. 
We have some freedom in the parameter space spanned by $\kappa_1$ and $\xi$, allowing us to tune the parameters to avoid vacuum instability.

During inflation, the Higgs boson is decoupled. We can minimize the potential to find the equation of motion for the Higgs,

\bea
h^2=\frac{\kappa_1\sigma (\sigma+\sqrt{6}) \big(\sigma-3\big(\xi+\frac{1}{6}\big)(\sigma-\sqrt{6}) \big)}{\lambda (\sigma-\sqrt{6})-3\kappa_1\big(\xi+\frac{1}{6}\big)\big(\sigma-3\big(\xi+\frac{1}{6}\big)(\sigma-\sqrt{6}) \big) }. \label{hmin}
\eea 

In the limit in which $\sigma \rightarrow - \sqrt{6}$, $h^2$ vanishes and the kinetic terms, including the kinetic mixing, disappear.
Substituting the equation of motion into the Lagrangian (\ref{general-einstein}) we find the effective Lagrangian valid during inflation

\bea \label{eq:einsframelag}
\frac{{\cal L}_{\rm eff}}{\sqrt{-{g_E}} } =  -\frac{1}{2}\,R(g_E)+ \frac{ (\partial_\mu \sigma)^2 }{2(1-\sigma^2/6)^2}- V_{\rm eff}(\sigma), 
\eea

with the inflationary potential 
\bea
V_{\rm eff}(\sigma)= 9\lambda\,\kappa_1\sigma^2 \bigg[\lambda (\sigma-\sqrt{6})^2+\kappa_1 \bigg(\sigma-3\Big(\xi+\frac{1}{6}\Big)(\sigma-\sqrt{6}) \bigg)^2  \bigg]^{-1}.
\eea

We can canonicalize the lagrangian by the following field redefinition

\bea \label{cansigmafield}
\sigma=-\sqrt{6} \tanh \Big(\frac{\chi}{\sqrt{6}}\Big),
\eea

leading to an exponentially flat potential 

\bea
V_{\rm eff}(\chi) = \frac{9\kappa_1}{4} \Big(1-e^{-2\chi/\sqrt{6}} \Big)^2 \left[1+\frac{\kappa_1}{4\lambda} \Big(6\xi+e^{-2\chi/\sqrt{6}} \Big)^2 \right]^{-1}. \label{effpot}
\eea

There are two interesting limits for the potential (\ref{effpot}). In the case that $9\kappa_1 \xi^2\ll \lambda$, we recover the pure sigma field inflation. On the opposite side, for  $9\kappa_1 \xi^2\gg \lambda$, we recover the original model of Higgs inflation.
Therefore, our parameter space allows us to interpolate between the two models.

\bea
V_{\rm eff}(\chi) \approx \left\{ \begin{array}{c} \frac{9\kappa_1}{4} \Big(1-e^{-2\chi/\sqrt{6}} \Big)^2 \quad \mathrm{for} \quad 9\kappa_1 \xi^2\ll \lambda,  \\  \frac{\lambda}{4\xi^2} \Big(1-e^{-2\chi/\sqrt{6}} \Big)^2 \quad \mathrm{for} \quad 9\kappa_1 \xi^2\gg \lambda. \end{array}\right. 
\eea

Inflation is driven by the vacuum energy density, which for $\chi\gg 1$ reads
\bea \label{Inf_scale}
V_0=  \frac{9\kappa_1\lambda}{4 (\lambda+9\kappa_1 \xi^2)}.
\eea

\subsection{Inflationary predictions} \label{sec:inflationarypredsigma}
We compute the slow-roll parameters, from eqs. (\ref{eq:slowrolleps}) and (\ref{eq:slowrolleta}) as

\bea
\epsilon &=&  \frac{1}{3} \frac{(2\lambda+3\kappa_1\xi(1+6\xi))^2}{(\lambda+9\kappa_1\xi^2)^2}\,e^{-4\chi/\sqrt{6}} , \\
\eta &=& -\frac{2}{3} \, \cdot \frac{2\lambda+3\kappa_1\xi(1+6\xi)}{\lambda+9\kappa_1\xi^2}\, e^{-2\chi/\sqrt{6}}  \nonumber \\
&&\qquad\qquad\quad\,\,+ \frac{2\kappa_1}{3}\, \cdot\frac{(-\lambda+12\lambda\xi+18\kappa_1\xi^2(1+6\xi))}{(\lambda+9\kappa_1\xi^2)^2}\, e^{-4\chi/\sqrt{6}}.
\eea 

The number of e-foldings, given in eq.~(\ref{efoldnum}), can be computed in terms of the canonical field $\chi$ as
\bea
N&=& \frac{3}{2}\,\cdot\frac{\lambda+9\kappa_1\xi^2}{2\lambda+3\kappa_1\xi(1+6\xi)}\, \Big(e^{2\chi_*/\sqrt{6}} -e^{2\chi_e/\sqrt{6}}\Big),
\eea

where $\chi_*, \chi_e$ are the inflaton field values at the horizon exit and  at the end of inflation, respectively.
Taking the approximation $e^{2\chi_*/\sqrt{6}} \gg e^{2\chi_e/\sqrt{6}}$ 
the spectral tilt in eq. (\ref{eq:planckn}),  and the tensor to scalar ratio in eq.({\ref{eq:planckr}}) become
\bea
n_s &=&  1-\frac{2}{N} -\frac{9}{2N^2} + \frac{3\kappa_1}{N^2}\, \frac{(-\lambda+12\lambda\xi+18\kappa_1\xi^2(1+6\xi))}{(2\lambda+3\kappa_1\xi(1+6\xi))^2},\\
r &=& \frac{12}{N^2}. 
\eea

The difference with pure Starobinsky or pure Higgs inflation is small, and contained in the terms $1/N^2$. These contributions were expected to be small since we remain in the perturbativity regime, where $\lambda\sim \kappa_1\xi^2\lesssim 1$.
Therefore, the predictions for linear higgs sigma models are almost degenerate with those of Higgs\cite{Bezrukov:2007ep} and Starobinsky \cite{Starobinsky:1980te} inflation.
From the CMB normalization in eq.~(\ref{eq:CMBnormalization}), our parameters are related through

\bea \label{eq:cmbnormhiggssigma}
\frac{\sqrt{\lambda+9\kappa_1\xi^2}}{\sqrt{\kappa_1\lambda}} = 1.5\times 10^5.
\eea

\begin{figure}[t]

    \begin{center}
     \includegraphics[width=90mm]{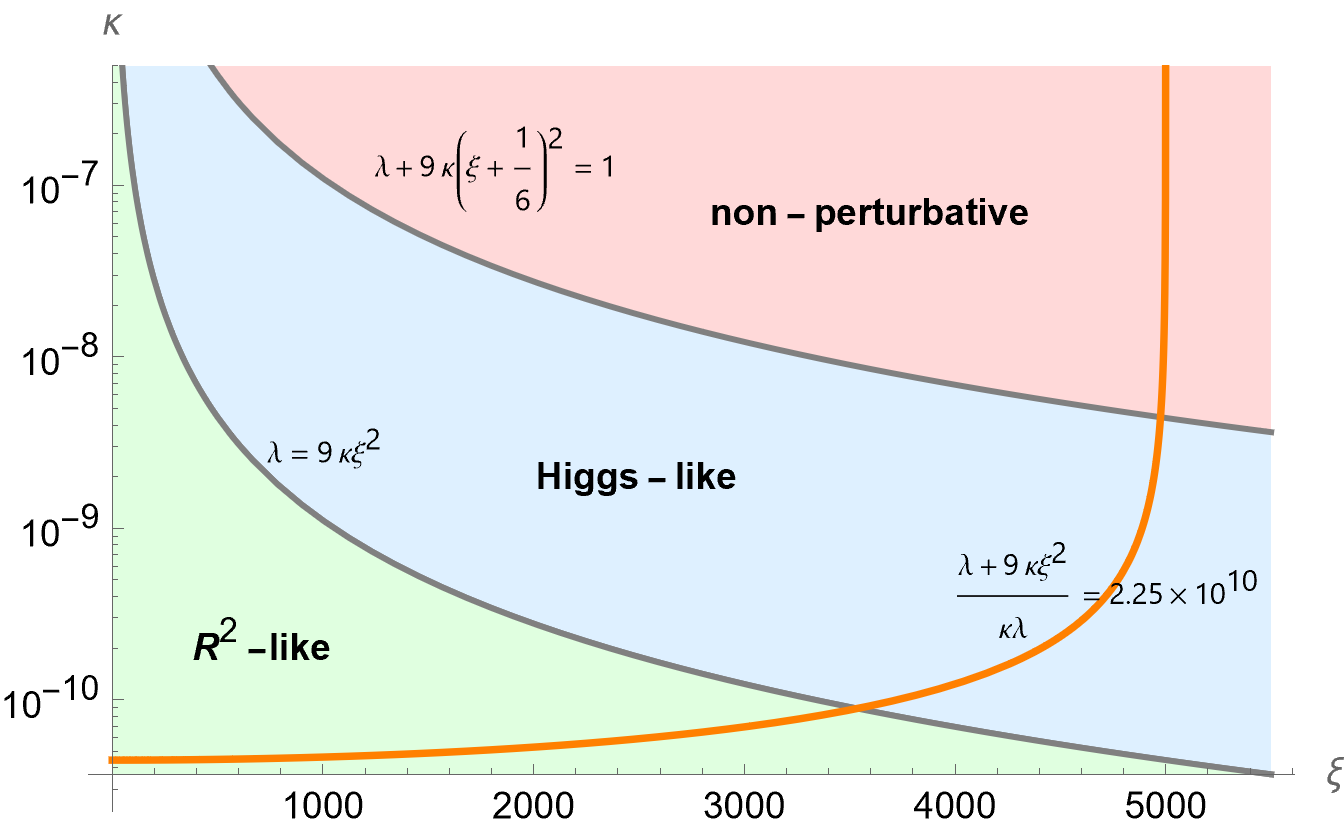}
    \end{center}
  \caption{Consistent inflation for the parameter space in $(\xi,\kappa)$.}
    \label{fig:kappaxi}
   
  \end{figure}

The parameter space available for Higgs-$R^2$ reheating is shown in Fig.(\ref{fig:kappaxi}). The blue and green regions correspond to the Higgs-like inflation  and the $R^2$-like inflation, respectively. The orange line is given by the CMB normalization and we took the conventional choice of $\lambda=0.01$.

\section{Perturbative reheating in the Higgs-\texorpdfstring{$R^2$}{R2} model}

In order to study the reheating dynamics, we need to solve the Boltzmann and Friedmann equations, as introduced in \textbf{section \ref{sec:darkmatter}}.
In this particular case, the system reads
\bea
&&\ddot{\sigma}+\frac{\sigma}{3\Omega^{2}\Mpl^2}\dot{\sigma}^{2}+\frac{h}{3\Omega^{2}\Mpl^2} \dot{\sigma} \dot{h}+(3H+\Gamma_{\sigma_0})\dot{\sigma} \nonumber \\
&&+\frac{2 \sigma}{3\Omega^{2}\Mpl^2} U+\frac{1}{\Omega^{2}}\left(1-\frac{\sigma^{2}}{6\Mpl^2}\right) U_{\sigma}-\frac{h \sigma}{6\Omega^{2}\Mpl^2} U_{h}=0,\label{EOM_sigma}  \\
&&\ddot{h}+\frac{h}{3\Omega^{2}\Mpl^2}\dot{h}^{2}+\frac{\sigma}{3\Omega^{2}\Mpl^2} \dot{\sigma} \dot{h}+(3H+\Gamma_{h_{\rm{osc}}})\dot{h}
\nonumber \\
&&+\frac{2 h}{3\Omega^{2}\Mpl^2} U+\frac{1}{\Omega^{2}}\left(1-\frac{h^{2}}{6\Mpl^2}\right) U_{h}-\frac{h \sigma}{6\Omega^{2}\Mpl^2} U_{\sigma}=0, \\
&&\dot{\rho}_r+4H\rho_r-\frac{\Gamma_{\sigma_0}}{\Omega^4}\left[\left(1-\frac{h^{2}}{6\Mpl^2}\right)\dot{\sigma}^2+\frac{h\sigma}{6\Mpl^2}\dot{\sigma}\dot{h}\right] \nonumber \\
&&-\frac{\Gamma_{h_{\rm{osc}}}}{\Omega^4}\left[\left(1-\frac{\sigma^{2}}{6\Mpl^2}\right)\dot{h}^2+\frac{h\sigma}{6\Mpl^2}\dot{\sigma}\dot{h}\right]=0,\label{rho_r} \\
&&3 H^{2}\Mpl^2 =\rho_{\sigma+h}+\rho_r \label{Friedmann}
\eea

where $U\equiv V\Omega^4$ and the subscripts denote the derivative with respect to that field.
The total energy density of the system is given by,
\begin{align}
\rho_{\sigma+h}\equiv \frac{1}{2\Omega^4}\left[\left(1-\frac{h^{2}}{6\Mpl^2}\right) \dot{\sigma}^{2}+\left(1-\frac{\sigma^{2}}{6\Mpl^2}\right) \dot{h}^{2}+\frac{h \sigma}{3\Mpl^2} \dot{\sigma} \dot{h}+2U\right], 
\end{align}
with $\rho_r$ being the radiation energy density, and $\Gamma_{\sigma_0}$ and $\Gamma_{h_{\rm{osc}}}$ the decay rates of the sigma and Higgs condensates, respectively.
The equation of state parameter $w$, as given in eq.~(\ref{tauu})
\begin{align}
w=\frac{p_{\sigma+h}+p_{r}}{\rho_{\sigma+h}+\rho_{r}}=\frac{p_{\sigma+h}+\rho_{r}/3}{\rho_{\sigma+h}+\rho_{r}},    
\end{align}

where we have assumed that the universe is radiation dominated in the last equality. The total pressure is given by 

\begin{align}
p_{\sigma+h}\equiv \frac{1}{2\Omega^4}\left[\left(1-\frac{h^{2}}{6\Mpl^2}\right) \dot{\sigma}^{2}+\left(1-\frac{\sigma^{2}}{6\Mpl^2}\right) \dot{h}^{2}+\frac{h \sigma}{3\Mpl^2} \dot{\sigma} \dot{h}-2U\right].
\end{align}

\subsection{Background field evolution after inflation} \label{BG_after_inflation}

At the end of inflation, $\sigma$ settles at the minimum and starts oscillating around it.
The Higgs squared mass depends on the value of the $\sigma$ field, so, once $\sigma$ start oscillating, it triggers a tachyonic instability in the Higgs direction, releasing it from its initial background value in eq.~(\ref{hmin}).
We now divide the fields into a background part, denoted by the subscript ``0", and a time dependent perturbation.
Furthermore, we divide the background evolution of $h$ into a slowly oscillating part $h_0\left(\sigma_0\right)$, which depends on the sign of $\sigma_0$, and a rapidly oscillating part $h_{\rm{osc}}$ \cite{Fan:2019udt,He:2020qcb}.

\bea
\sigma(t)&=&\sigma_0(t), \\
h(t)&=&  h_0(\sigma_0)+h_{\rm{osc}}(t),  \label{h_0} 
\eea

The Higgs is related to the sigma condensate through the inflationary condition in eq.~(\ref{hmin}), which for for $\sigma_0<0$ and  $|\sigma_0|/\Mpl\ll 1$ becomes

\bea
(h_0(\sigma_0))^2
\simeq -\frac{3\sqrt{6}\kappa\tilde{\xi}}{\lambda+9\kappa\tilde{\xi}^2}\Mpl\sigma_0. \label{hv}   
\eea

where we defined the effective non-minimal coupling $\tilde{\xi}\equiv \xi+1/6$.

However, when $\sigma_0>0$, we find $h_0=0$. This sudden change in behavior, in which the Higgs background changes accordingly on the sign of $\sigma$, 
can be understood from the cubic coupling  $\kappa \tilde{\xi}\sigma h^2$, present in the scalar potential $\eqref{totalpotsigma}$.
When $\sigma<0$, the Higgs field gets a tachyonic mass and develops a non-zero VEV. 
However, when $\sigma>0$, the $h_0$ part of the Higgs condensate vanishes and stabilizes at the minimum.
Therefore, during reheating, we are alternating between a phase in which the EW symmetry is broken and a phase in which it is restored.

Expanding the Lagrangian (\ref{general-einstein}) around the background field values, $|\sigma|= |\sigma_0|\ll M_{\rm Pl}$ and $h=h_0$, we get the masses for $\sigma_0$ and $h_{\rm{osc}}$ as
\begin{align}
m_{\sigma }^{2}= \begin{cases}3\kappa\Mpl^2 \equiv m^2_{\sigma,+} & \quad \mathrm{for} \quad  \sigma_0>0, \\ \frac{3\kappa\lambda\Mpl^2}{\lambda+9\kappa\tilde{\xi}^2}\equiv m^2_{\sigma,-}& \quad \mathrm{for} \quad  \sigma_0<0 \end{cases}, \label{m_s}   
\end{align}
and 

\begin{align}
m_{h }^{2}= \begin{cases}3\sqrt{6}\kappa \tilde{\xi} \Mpl\sigma_0\equiv m^2_{h,+} & \quad \mathrm{for} \quad \sigma_0>0, \\ 6\sqrt{6}\kappa \tilde{\xi} (-\Mpl\sigma_0)\equiv m^2_{h,-}& \quad \mathrm{for} \quad \sigma_0<0.
\end{cases}  \label{m_h}  
\end{align}

The interaction between the inflaton fields makes their masses time dependent.
The masses for the sigma condensate for $\sigma_0<0$ in eq.~(\ref{m_s}) and the masses for the Higgs condensate in eq.~(\ref{m_h}) are only valid in the regime ${\tilde\xi}|\sigma_0|/M_{\rm Pl}\gg 1$. 
In order to ensure this condition, the inflaton field must not go far away from its value at the end of inflation. 
Since ${\tilde\xi}|\sigma_e|/M_{\rm Pl}\simeq 20-700\gg 1$ for ${\tilde\xi}=100-4000$, we don't need to worry about being out of the valid regime, and we can use the perturbative results  in eqs.~(\ref{m_s}) and (\ref{m_h}).

\begin{figure}[t]
    \begin{minipage}{0.5\hsize}
     \begin{center}
      \includegraphics[width=70mm]{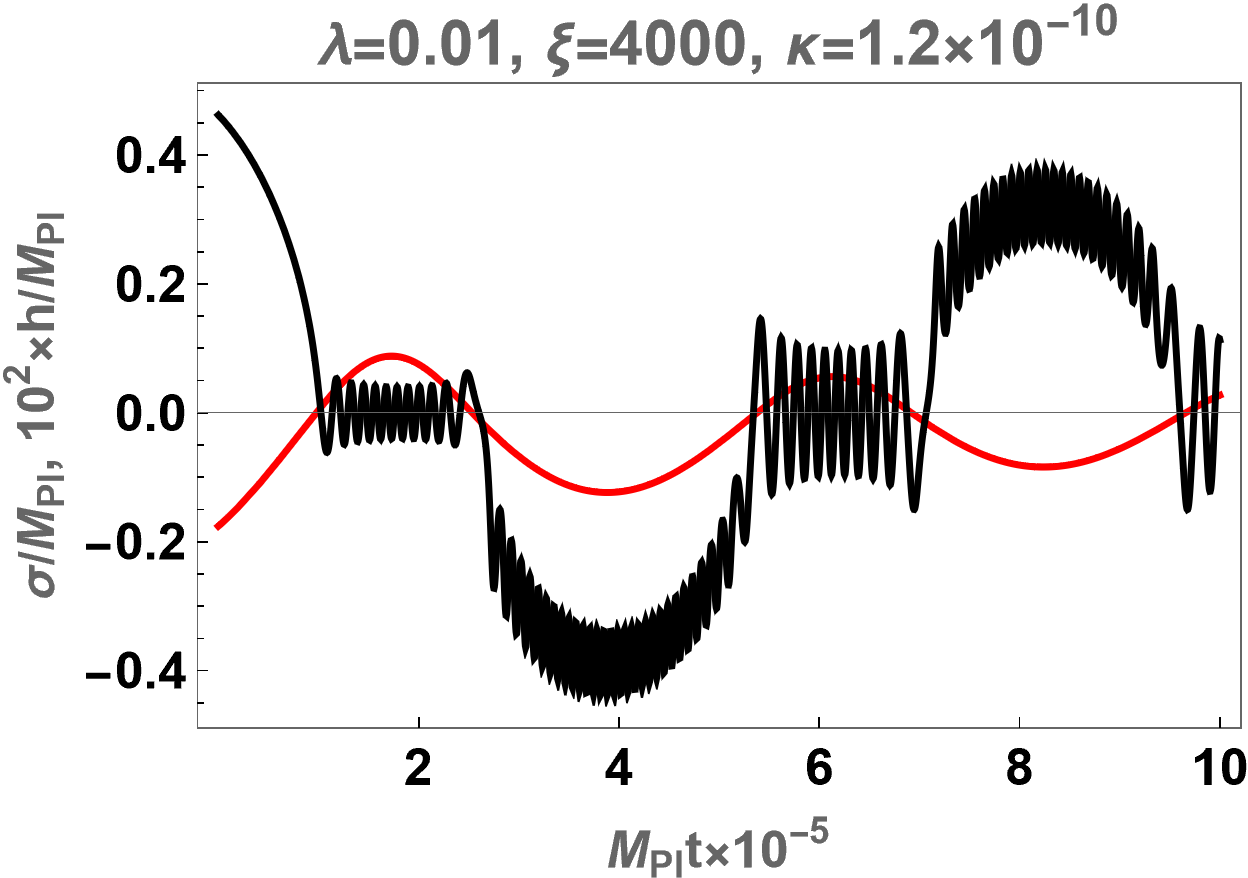}
     \end{center}
    \end{minipage}
    \begin{minipage}{0.5\hsize}
     \begin{center}
      \includegraphics[width=70mm]{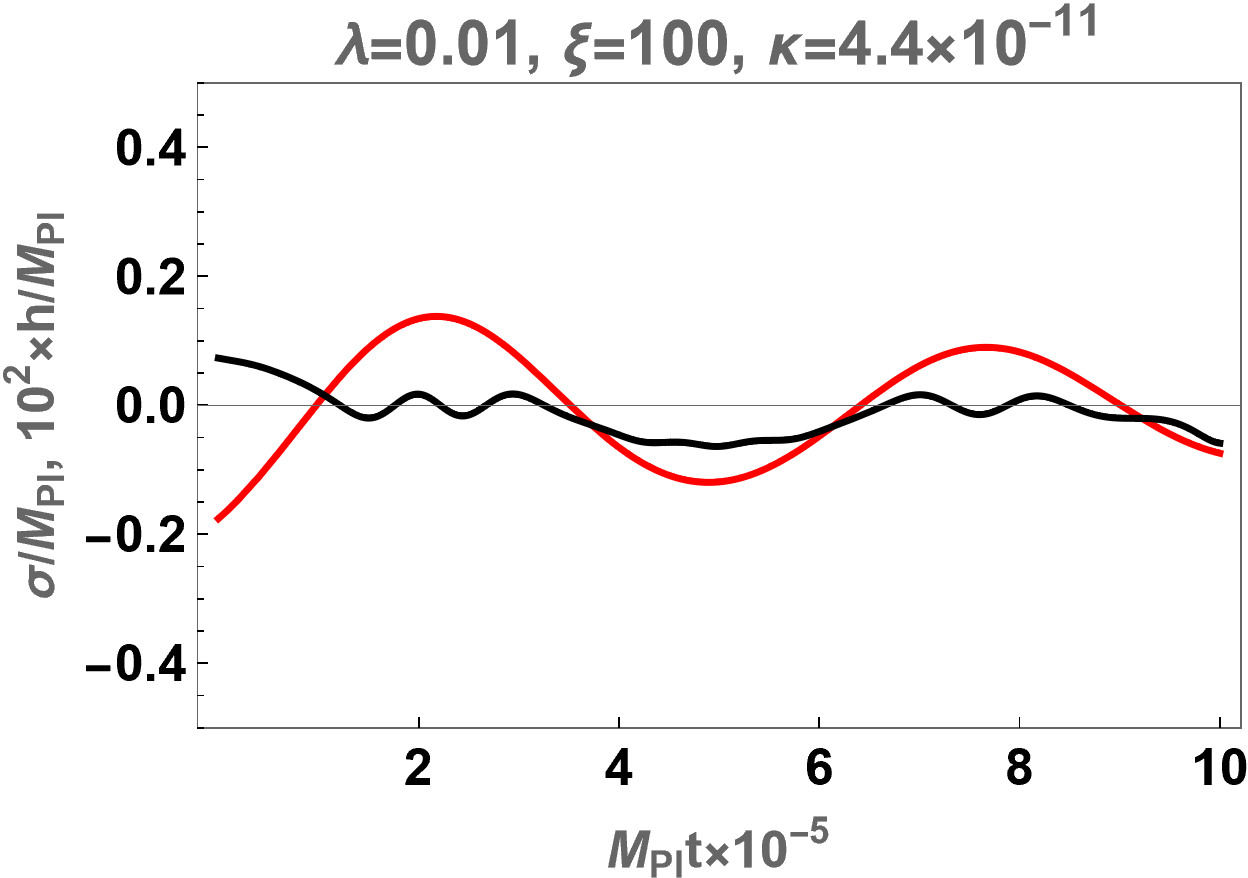}
     \end{center}
    \end{minipage}
   \caption{Time evolution of inflaton condensates, $\sigma$ and $h$, during reheating, in red and black lines, respectively. We took $\xi=4000$ ($100$) on left (right) plots.}
     \label{fig:osc}
   \end{figure}

   In Fig.~(\ref{fig:osc}), we show the numerical solution to eqs.~$\eqref{EOM_sigma}$-$\eqref{Friedmann}$ for the background evolution of $\sigma$ and $h$.
   We take $\Gamma_{\sigma_0}=\Gamma_{h_0}=\Gamma_{h_{\rm{osc}}}= 0$ at the onset of oscillations, since at early times the inflaton fields haven't still decayed.
We can compare now our approximate solutions to the numerical one. We find that the time evolution of the Higgs condensate is well approximated by eq.~(\ref{hv}) for a large $\tilde{\xi}$, as seen in Fig.~\ref{fig:osc}.
However, the approximation breaks for $\tilde{\xi}\gtrsim 100$, and thus, eq. (\ref{hv}) is no longer valid. 
We also see from the numerical result that the rapidly oscillating part $h_{\rm osc}$ of the Higgs condensate is prominent for $\sigma_0>0$, when the Higgs background $h_0$ becomes zero.

\subsection{Decay rates of inflaton condensates}\label{sec:decayratesofinflatoncondensates}

In order to obtain an approximate analytical result, we define $\sigma_0(t)  \sim \sin (m_{\sigma} t)$ and $ h_{\rm{osc}}(t)\sim \sin(m_h t)$, with the constant masses given in eqs.~$\eqref{m_s}$ and $\eqref{m_h}$.
We also neglect the expansion of the universe. 
It's important to notice the limits of this approach: this won't work when the dynamics of the fields becomes non-linear ~\cite{He:2020qcb}.
We divide now $\sigma$ and $h$ into their backgrounds, $\sigma_0$ and $h_0 + h_{\mathrm{osc}}$, and their quantum fluctuations, $\delta \sigma$ and $\delta h$.
\begin{align}
    &\sigma=\sigma_0(t)+\delta \sigma,\label{def_fl_s} \\ &h=h_0(\sigma_0)+h_{\rm{osc}}(t)+\delta h. \label{def_fl_h}    
\end{align}

\subsubsection*{Decay rates of the sigma condensate}

In order to find the dominant contributions to the Lagrangian~$\eqref{eq:einsframelag}$, we expand around the minima, using eqs.~$\eqref{def_fl_s}$ and $\eqref{def_fl_h}$.
Then, we find that $\sigma_0$ will mainly decays through
\bea
\mathcal{L}\supset c\sigma_0(\delta h)^2,
\eea
where the constant $c$ is defined as
\bea
c= \begin{cases}-\frac{3}{2}\sqrt{6}\kappa\tilde{\xi}\Mpl  & \mathrm{for} \quad\sigma_0>0, \\ 3\sqrt{6}\kappa\tilde{\xi}\Mpl&  \mathrm{for} \quad\sigma_0<0.\end{cases}    
\eea

The background field $\sigma_0$ only couples to the SM particles through the conformal factors, $\Delta$ and $\Omega^2$, and its thus suppressed by the Planck scale.
Then, using the standard result for the two body decay ~\cite{Ichikawa:2008ne,Nurmi:2015ema,Kainulainen:2016vzv}, we obtain  

\bea
\Gamma_{\sigma_{0} \rightarrow \delta h\delta h} 
= \begin{cases}\frac{9\sqrt{3}}{16 \pi}\Mpl\kappa^{3/2}\tilde{\xi}^2\left(1-4\sqrt{6}\tilde{\xi}\frac{\sigma_0}{\Mpl}\right)^{1/2}  & \mathrm{for}\,\,\,\sigma_0>0, \\ \frac{9\sqrt{3}}{4 \pi}\Mpl\kappa^{3/2}\tilde{\xi}^2\sqrt{\frac{\lambda_{\rm eff}}{ \lambda}}\left(1+8 \sqrt{6}\tilde{\xi}\frac{\lambda_{\rm eff}}{\lambda} \frac{\sigma_{0}}{\Mpl}\right)^{1 / 2}& \mathrm{for}\,\,\,\sigma_0<0. \end{cases}  \label{Gamma_s}  
\eea

We realize then, that in both cases, $\sigma >0$ and $\sigma<0$, the decay mode $\sigma_0 \rightarrow \delta h \delta h$ is kinematically blocked due to the large non-minimal coupling.

\subsubsection*{Decay rates of the Higgs condensate}

We remember that the background part of the higgs field is divided as 

\begin{equation}
h(t) = h_0(\sigma_0)+h_{\rm{osc}}(t),
\end{equation}

where $h_{\rm{osc}}$ is only present for $\sigma_0>0$.
The dominant decay channel for the Higgs is into a top quark pair,

\begin{align}
    \mathcal{L}\supset -\frac{y_t}{\sqrt{2}}h\bar{t}t=-\frac{y_t}{\sqrt{2}}(h_0(\sigma_0)+h_{\rm{osc}})\bar{t}t, \label{int_h_t}
\end{align}
due to the rapid oscillations of the Higgs field and the large Yukawa coupling.

We can identify the time dependent mass of the top quark from the interaction Lagrangian
\bea
m_t
= \begin{cases}
    \sqrt{\frac{3\sqrt{6}\kappa}{2\lambda_{\rm eff}}\tilde{\xi} (-\Mpl\sigma_0)}  & \mathrm{for}\,\,\,\sigma_0<0, \\ 
    0& \mathrm{for}\,\,\,\sigma_0>0, \end{cases}  
\eea

leading to the decay rates

\begin{align}
    \Gamma_{h_{\rm{osc}} \rightarrow t \bar{t}}= \begin{cases} \frac{3 y_{t}^{2}}{16 \pi}\Mpl\left(3 \sqrt{6}\kappa \tilde{\xi}\frac{\sigma_0}{\Mpl}\right)^{1 / 2}  & \mathrm{for} \quad\sigma_0>0, \\ \frac{3 y_{t}^{2}}{16 \pi}\Mpl\left(-6 \sqrt{6}\kappa \tilde{\xi}\frac{\sigma_0}{\Mpl}\right)^{1 / 2} \left(1-\frac{y_t^2}{\lambda_{\rm eff}}\right)^{3 / 2}& \mathrm{for}\quad \sigma_0<0. \end{cases}  \label{Gamma_h}  
\end{align}

For $\sigma_0 <0$, the channel is kinematically allowed for  $\tilde{\xi} \gtrsim 5000$,
with  $y_t = 0.5$ at the inflationary scale.
In the case with $\sigma_0 >0$, the channel is always open. Thus, reheating happens mainly through decays of $h_{\mathrm{osc}}$ into a pair of $t \bar{t}$.
We notice that the Higgs can decay into other SM particles, such as $h_{\mathrm{osc}} \rightarrow WW, ZZ, b\bar{b}$. However, these channels are subdominant.

The slowly oscillating part of the Higgs condensate, $h_0$, only contributes when $\sigma_0>0$.
Thus, the mode $h_0\to t{\bar t}$ is open for $|\sigma_0|/M_P\lesssim 0.1( \lambda/y^2_t) \tilde{\xi}^{-1}$, 

\begin{equation}
    \Gamma_{h_0\to t{\bar t}}\sim y^2_t m_\sigma.
\end{equation}

However, as compared with the channel $\sigma_{0} \rightarrow \delta h\,\delta h$, the decay $h_0\to t{\bar t}$ is kinematically blocked for a wider field range of $\sigma_0$, given  $\lambda \ll y^2_t$.  
We then conclude that, for $\tilde{\xi} |\sigma_0|/M_{\rm Pl}\gtrsim 1$, the decay modes of the sigma and $h_0$ condensates are kinematically blocked, and the dominant decay channel for reheating is $h_{\rm osc}\to t{\bar t}$, with $\sigma_0>0$.

\subsection{Analytic and numerical solutions for reheating}\label{RT_MT}
We now use the decay rates from the previous section to solve the Boltzmann equations in eqs.~(\ref{EOM_sigma})-(\ref{Friedmann}). 
This allows us to study the evolution of the inflaton and the radiation energy densities, as well as the reheating and maximum temperatures and the equation of state.

\subsubsection*{Analytic solutions}

The dynamics of the $h$ and $\sigma$ fields is intertwined during reheating, which makes it difficult to solve the system analytically.
Nevertheless, we propose an approximate analytical solution, and compare the results with the numerical results. 
We hope the analytic expressions give us some intuition of the physics underlying reheating.
We recommend the references ~\cite{Battefeld:2008bu,Choi:2008et,Battefeld:2009xw,Braden:2010wd,Meyers:2013gua,Elliston:2014zea,Hotinli:2017vhx,Leung:2012ve,Huston:2013kgl,Leung:2013rza,Watanabe:2015eia,DeCross:2015uza,DeCross:2016cbs,DeCross:2016fdz,Schimmrigk:2017jwa,Gonzalez:2018jax,Martin:2021frd} for a detailed analysis in multifield reheating models.
Our first assumption is that we can separate the energy densities of the $\sigma$ and $h$ fields,

\begin{align}
    \rho_{\sigma+h}&\simeq \rho_{\sigma}+\rho_{h},\\
    p_{\sigma+h}&\simeq p_{\sigma}+p_{h}.
\end{align}

When $\sigma_0 >0$, we can write

\begin{align}
    &\rho_{\sigma}=\frac{1}{2}\dot{\sigma}^2+\frac{1}{2}m^2_{\sigma,+}\sigma^2, \ \  \rho_{h}= \frac{1}{2}\dot{h}^2+\frac{1}{2}m^2_{h,+} h^2+\frac{\lambda_{\rm{eff}}}{4}h^4,\\
    &p_{\sigma}=\frac{1}{2}\dot{\sigma}^2-\frac{1}{2}m^2_{\sigma,+}\sigma^2, \ \  p_{h}= \frac{1}{2}\dot{h}^2-\frac{1}{2}m^2_{h,+} h^2-\frac{\lambda_{\rm{eff}}}{4}h^4,
\end{align}

 where

\begin{align}
    &m^2_{\sigma,+}=3\kappa \Mpl^2, \label{ms1} \\
    &m^2_{h,+}=3\sqrt{6}\kappa \tilde{\xi} \Mpl\sigma_0.  \label{mh1}
\end{align}

However, when $\sigma<0$, the Higgs condensates become tachyonic, developing a non-zero vev and abruptly changing the masses to $m^2_{\sigma,-}$ and $m^2_{h,-}$,

\begin{align}
  m^2_{\sigma,-} &= \frac{3\kappa\lambda\Mpl^2}{\lambda+9\kappa\tilde{\xi}^2} , \\
  m^2_{h,-} &= 6\sqrt{6}\kappa \tilde{\xi} (-\Mpl\sigma_0).
\end{align}

With these approximations, we can approximate Boltzmann equations to
\begin{align}
    &\dot{\rho}_{\sigma}+3H(\rho_{\sigma}+p_{\sigma})+\Gamma_{\sigma}(\rho_{\sigma}+p_{\sigma})=0,\label{CE_sigma}\\
    &\dot{\rho}_{h}+3H(\rho_{h}+p_{h})+\Gamma_{h}(\rho_{h}+p_{h})=0,\label{CE_h}\\
    &\dot{\rho}_r+4H\rho_r-\Gamma_{\sigma}(\rho_{\sigma}+p_{\sigma})-\Gamma_{h}(\rho_{h}+p_{h})=0,\label{Rapp}\\
    &3\Mpl^2H^2=\rho_{\sigma}+\rho_{h}+\rho_r,\label{H2app}
\end{align}

where we have neglected the higher order terms suppressed by the Planck scale.

The initial conditions needed to solve the Boltzmann equations can be derived from the end of inflation.
Inflation ends when the slow-roll parameter $\varepsilon =1$. This implies $\sigma_e \simeq -0.18 M_p$, and $h_e$ given by eq.~(\ref{hv}) with $\sigma = \sigma_e.$  
We also have the kinetic conditions ${\dot\sigma}^2_e=\frac{1}{2}m^2_{\sigma,-} \sigma^2_e$  and ${\dot h}^2_e\simeq h^2_e\cdot\frac{{\dot \sigma}^2_e}{4\sigma^2_e}=\frac{1}{8} m^2_{\sigma,-} h^2_e$. 
Using these results, we can rewrite the total pressure and energy densities as

\bea
\rho_\sigma + p_\sigma &\simeq&  \frac{1}{2}m^2_{\sigma,-} \sigma^2_e, \\
\rho_h + p_h &\simeq& \frac{1}{8} m^2_{\sigma,-} h^2_e.
\eea

We define $\lambda_{\mathrm{eff}}=\lambda+9\kappa\tilde{\xi}^2$, so that the ratio between $\sigma_e$ and $h_e$,

\bea
\frac{h^2_e}{\sigma^2_e}\simeq \frac{3\sqrt{6} \kappa \tilde{\xi}}{0.18 \lambda_{\rm eff}}. 
\eea

The CMB normalization restricts our parameter space, $\kappa\tilde{\xi}\lesssim 0.02\lambda_{\rm eff}$, for which $\sigma_e\gtrsim h_e $.
This implies that $\rho_\sigma + p_\sigma \gtrsim \rho_h + p_h$ at the onset of oscillations. However, as we discussed in \textbf{section \ref{sec:decayratesofinflatoncondensates}},
the decay modes of $\sigma$ are not efficient.
Only when $\sigma>0$ the rapidly oscillating field $h_{\mathrm{osc}}$ becomes effective.
We take the ansatz $h_{\rm osc}(t)=A\, \cos(m_{h,-} t)$ and take the approximate energy conservation for the Higgs field as  $\rho_h\sim m^2_{\sigma,-} h^2_e\sim A^2 m^2_{h,-}$.
Then, the amplitude at the onset of the oscillations can be written as  $h_{\rm osc}$ as $A\sim (m_{\sigma,-}/m_{h,-})h_e$.

Our next assumption is that we can neglect the partial pressures during reheating, $p_{\sigma}=p_h=0$, and that at the very beginning of reheating there has been no particle production, i.e, $\Gamma_{\sigma} = \Gamma_{h} = 0$.
Then, we can follow the evolution of the energy densities from eqs.~(\ref{CE_sigma}) and (\ref{CE_h}), as  

\begin{align}
    \rho_{\sigma}=\rho_{\sigma,\rm{end}}\left(\frac{a}{a_{\rm{end}}}\right)^{-3} ,\ \ \rho_{h}=\rho_{h,\rm{end}}\left(\frac{a}{a_{\rm{end}}}\right)^{-3}.  \label{inflatonc}
\end{align}

We can substitute eq.~(\ref{inflatonc}) into eq.~$\eqref{Rapp}$, and including the decay rates for the inflaton condensates, we find $\rho_r$ as a function of $a$,
\begin{align}
\rho_{r}=\frac{2\sqrt{3}}{5}\Mpl \frac{\Gamma_{\sigma}\rho_{\sigma,\rm{end}}+\Gamma_{h}\rho_{h,\rm{end}}}{\sqrt{\rho_{\rm{end}}}}\left(\left(\frac{a}{a_{\text {end }}}\right)^{-\frac{3}{2}}-\left(\frac{a}{a_{\text {end }}}\right)^{-4}\right),   \label{sol_rhor} 
\end{align}

with $\rho_{\rm{end}}\equiv \rho_{\sigma,\rm{end}}+\rho_{h,\rm{end}}$. From eq.~(\ref{eq:reheatingtemp}), $\rho_r=\frac{\pi^2g_{\rm{reh}}}{30}T^4$. 
Reheating will be complete (at least in a good approximation) when $\rho_{\sigma}(a_{\rm{reh}})+\rho_{h}(a_{\rm{reh}})=\rho_{r}(a_{\rm{reh}})$.
Then, 

\begin{align}
    \left(\frac{a_{\rm{reh}}}{a_{\rm{end}}}\right)^3= \frac{25}{12}\frac{\rho_{\rm{end}}^3}{\Mpl^2\left(\Gamma_{\sigma}\rho_{\sigma,\rm{end}}+\Gamma_{h}\rho_{h,\rm{end}}\right)^{2}},   
\end{align}

where we took $a_{\rm{reh}}\gg a_{\rm{end}}$. 
Then, the reheating temperature
becomes
\begin{align}
T_{\rm{reh}}^4= \frac{72\Mpl^2}{5\pi^2g_{\rm{reh}}}\left(\frac{\Gamma_{\sigma}\rho_{\sigma,\rm{end}}+\Gamma_{h}\rho_{h,\rm{end}}}{\rho_{\sigma,\rm{end}}+\rho_{h,\rm{end}}}\right)^2.    
\end{align}

In order to find the maximum temperature reached during reheating, we find the maximum value $a_{\rm{max}}=(8/3)^{2/5}\,a_{\rm{end}}$, leading to

\begin{align}
    T_{\rm{max}}^4=   \frac{12\sqrt{3}}{\pi^2g_{\rm{reh}}}\left(\frac{3}{8}\right)^{\frac{3}{5}} \Mpl\frac{\Gamma_{\sigma}\rho_{\sigma,\rm{end}}+\Gamma_{h}\rho_{h,\rm{end}}}{\sqrt{\rho_{\rm{end}}}}.
 \end{align}

 These results are a generalization of the single-field cases in ~\cite{Chung:1998rq,Giudice:2000ex,Ellis:2015pla,Ellis:2015jpg,Garcia:2017tuj}.

 \subsubsection*{Numerical solutions}

In this section we present the numerical solutions to the Boltzmann equations.
Inflation ends when $\phi_e/M_p \simeq 0.18$.

\begin{figure}[t]
    \begin{minipage}{0.5\hsize}
     \begin{center}
      \includegraphics[width=70mm]{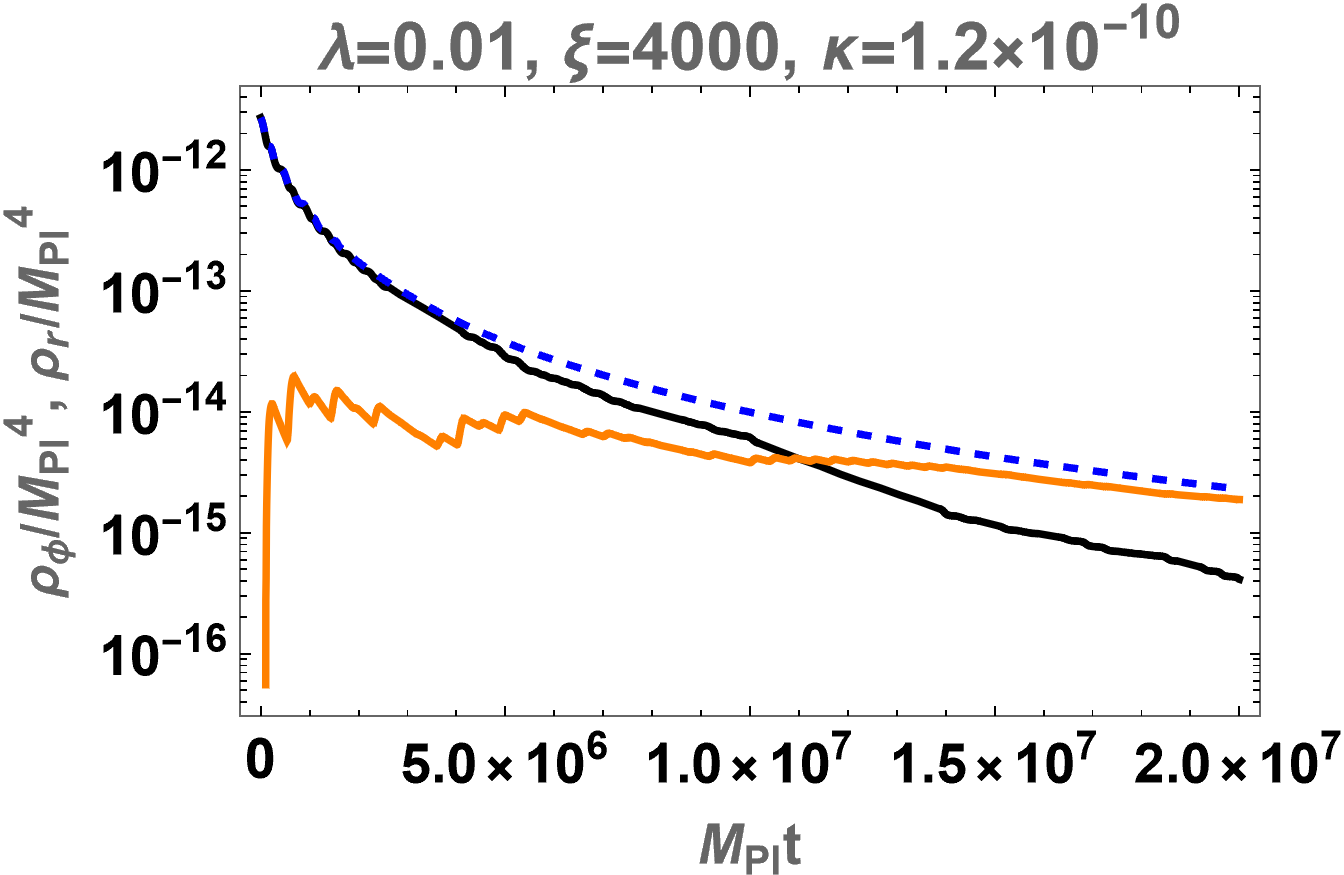}
     \end{center}
    \end{minipage}
    \begin{minipage}{0.5\hsize}
     \begin{center}
      \includegraphics[width=70mm]{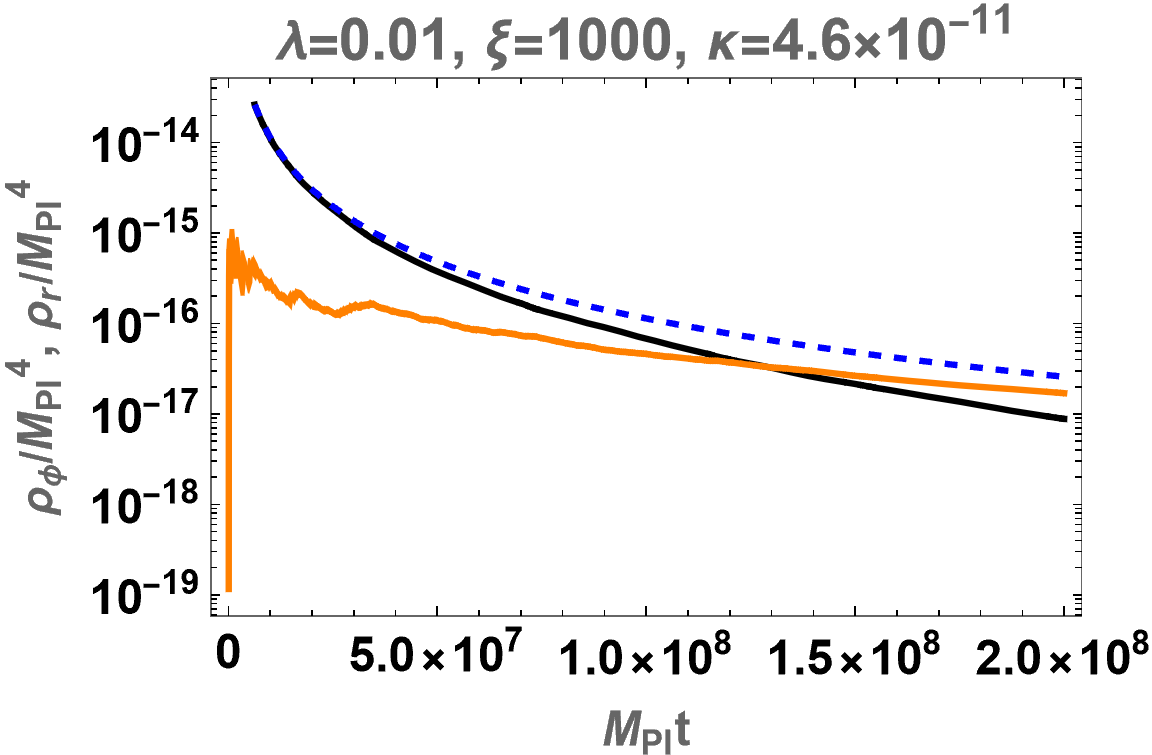}
     \end{center}
    \end{minipage}
    
    \begin{minipage}{0.5\hsize}
     \begin{center}
      \includegraphics[width=70mm]{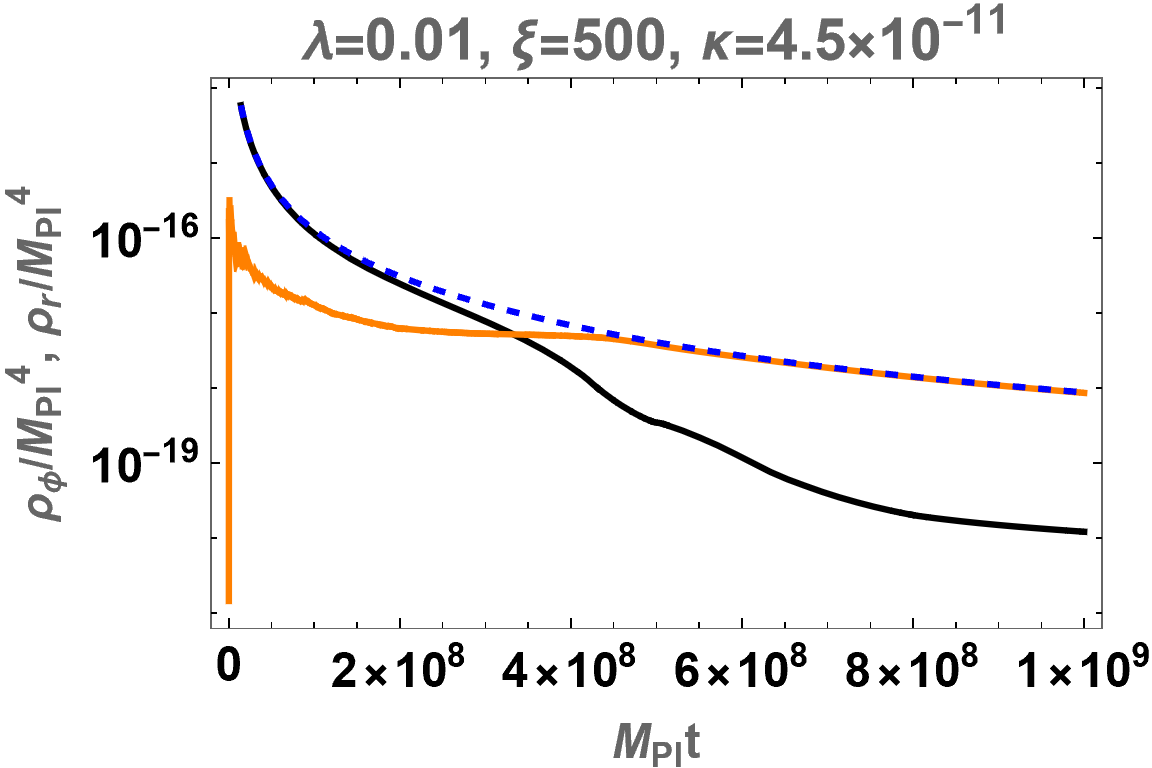}
     \end{center}
    \end{minipage}
    \begin{minipage}{0.5\hsize}
     \begin{center}
      \includegraphics[width=70mm]{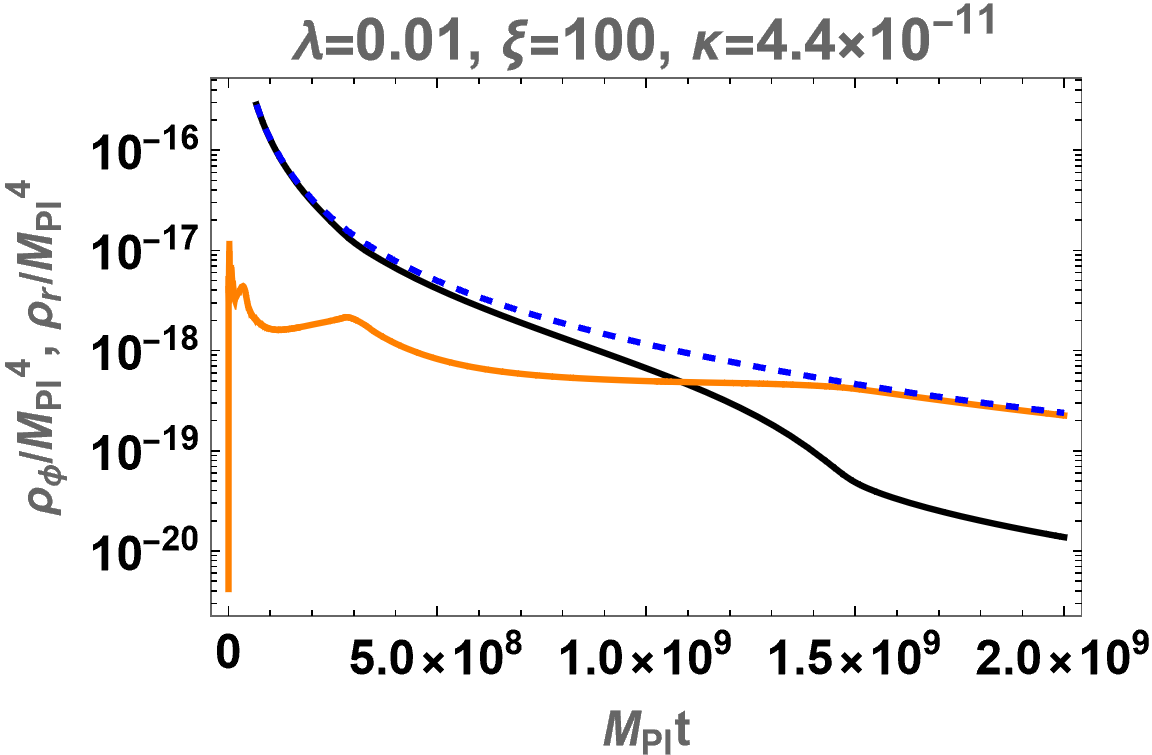}
     \end{center}
    \end{minipage}
    
    \caption{Time evolution of the energy densities during reheating. The inflaton energy density, $\rho_\phi\equiv \rho_{\sigma+h}$, the radiation energy density, $\rho_r$, and the total energy density, $\rho_{\sigma+h}+\rho_r$, are shown in black, orange and blue dashed lines, respectively. 
    }
     \label{fig:rhophi_vs_rhor}
   \end{figure}

In Fig.~\ref{fig:rhophi_vs_rhor}, we depict the numerical results for the time evolution of the inflaton energy density, $\rho_{\sigma+h}$ (black), and the radiation energy density, $\rho_r$ (orange). We take four different values of the non-minimal coupling, $\xi=4000, 1000, 500, 100$.
The parameter $\kappa$ in eq.~(\ref{r2-quartic}), must satisfy eq.~(\ref{eq:cmbnormhiggssigma}) with $\lambda =0.01$.
We denoted in the blue dashed line the total energy density $\rho_{\sigma+h}+\rho_r$, which is decaying due to the Hubble expansion. 
Reheating is complete at  $\rho_{\sigma+h}=\rho_r$.
From Fig.~\ref{fig:rhophi_vs_rhor}, we see a tendency to delay reheating completion when we lower the non-minimal coupling.

We assume that the radiation produced from the inflation decays thermalizes immediately. 
Temperature evolves as $T=\big(\frac{30}{\pi^2g_{\rm reh}}\rho_r\big)^{1/4}$, with $g_{\rm reh}=106.75$ during reheating.

 \begin{figure}[t]
    \begin{minipage}{0.5\hsize}
     \begin{center}
      \includegraphics[width=70mm]{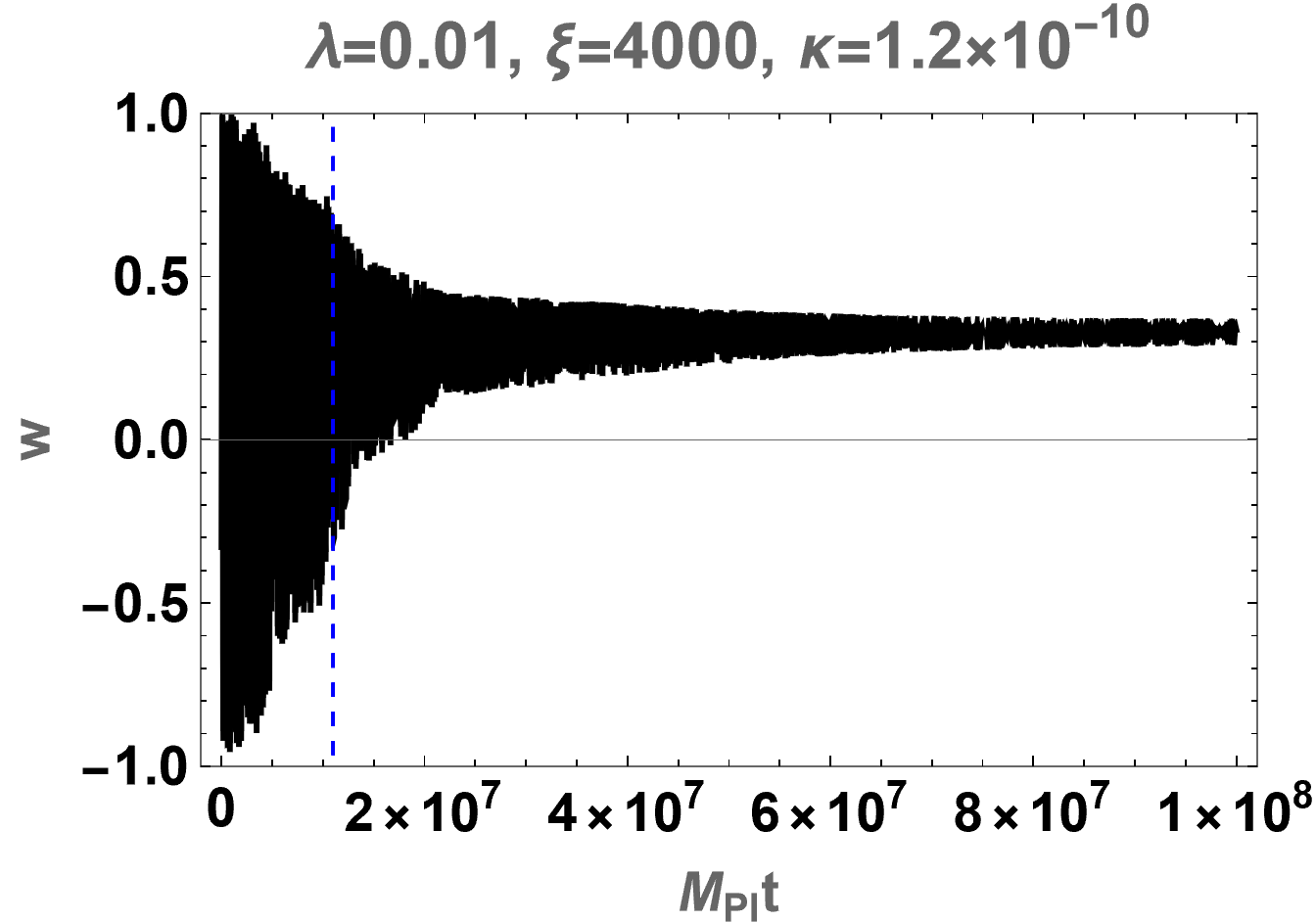}
     \end{center}
    \end{minipage}
    \begin{minipage}{0.5\hsize}
     \begin{center}
      \includegraphics[width=70mm]{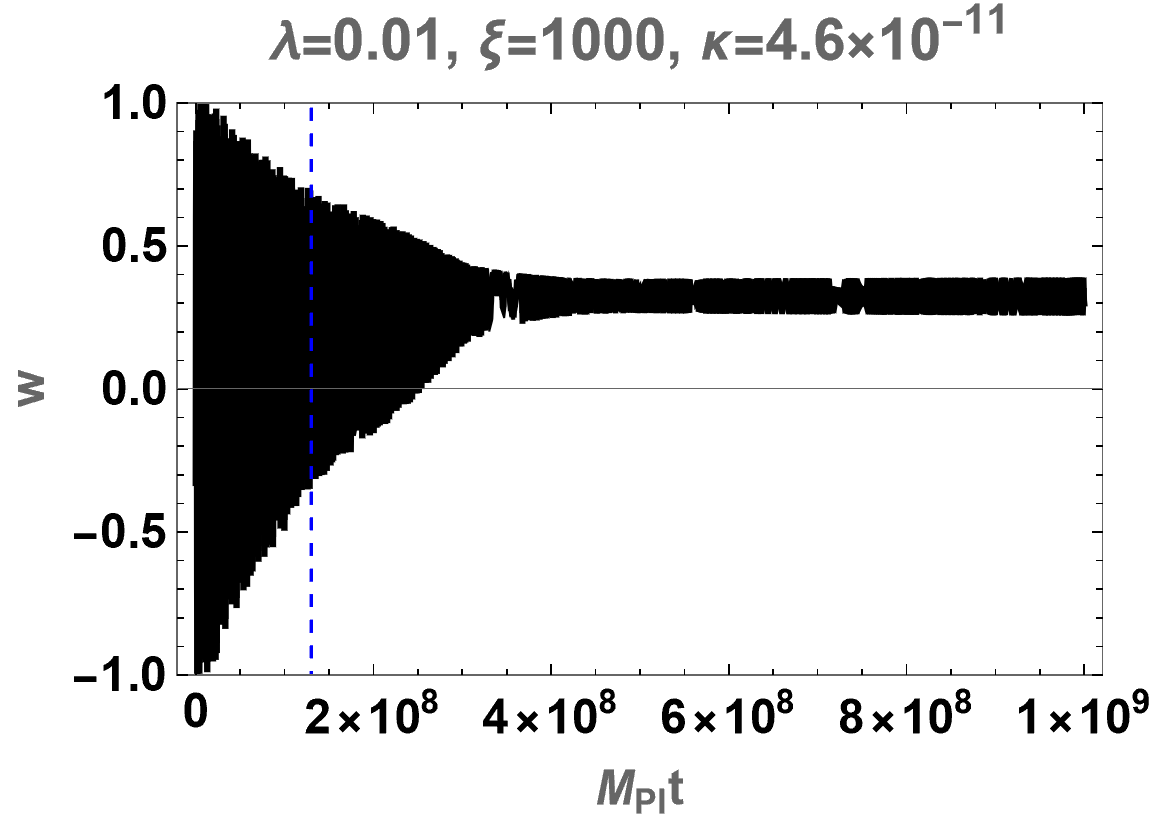}
     \end{center}
    \end{minipage}
    
    \begin{minipage}{0.5\hsize}
     \begin{center}
      \includegraphics[width=70mm]{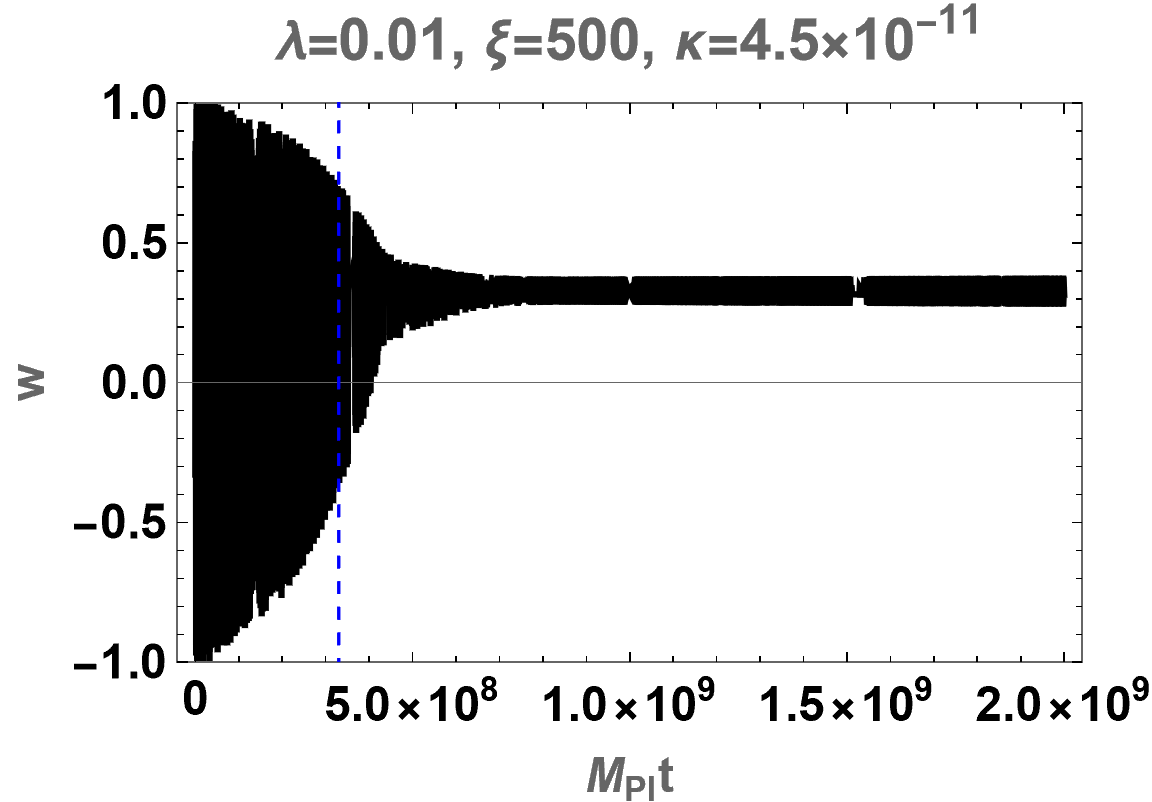}
     \end{center}
    \end{minipage}
    \begin{minipage}{0.5\hsize}
     \begin{center}
      \includegraphics[width=70mm]{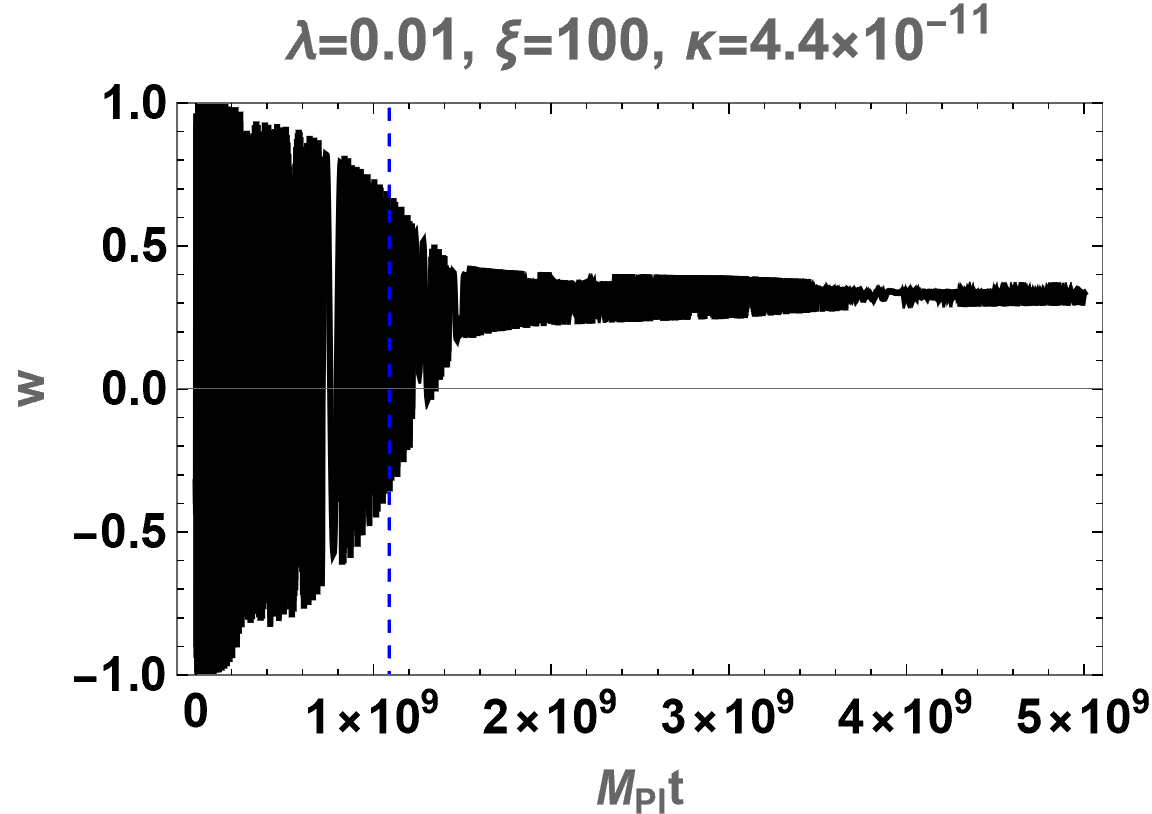}
     \end{center}
    \end{minipage}
    
    \caption{Time evolution of the equation of state $w$ during reheating. The blue dashed line denotes reheating completion.
    }
     \label{fig:w}
   \end{figure}

\begin{figure}[t]

    \begin{center}
     \includegraphics[width=90mm]{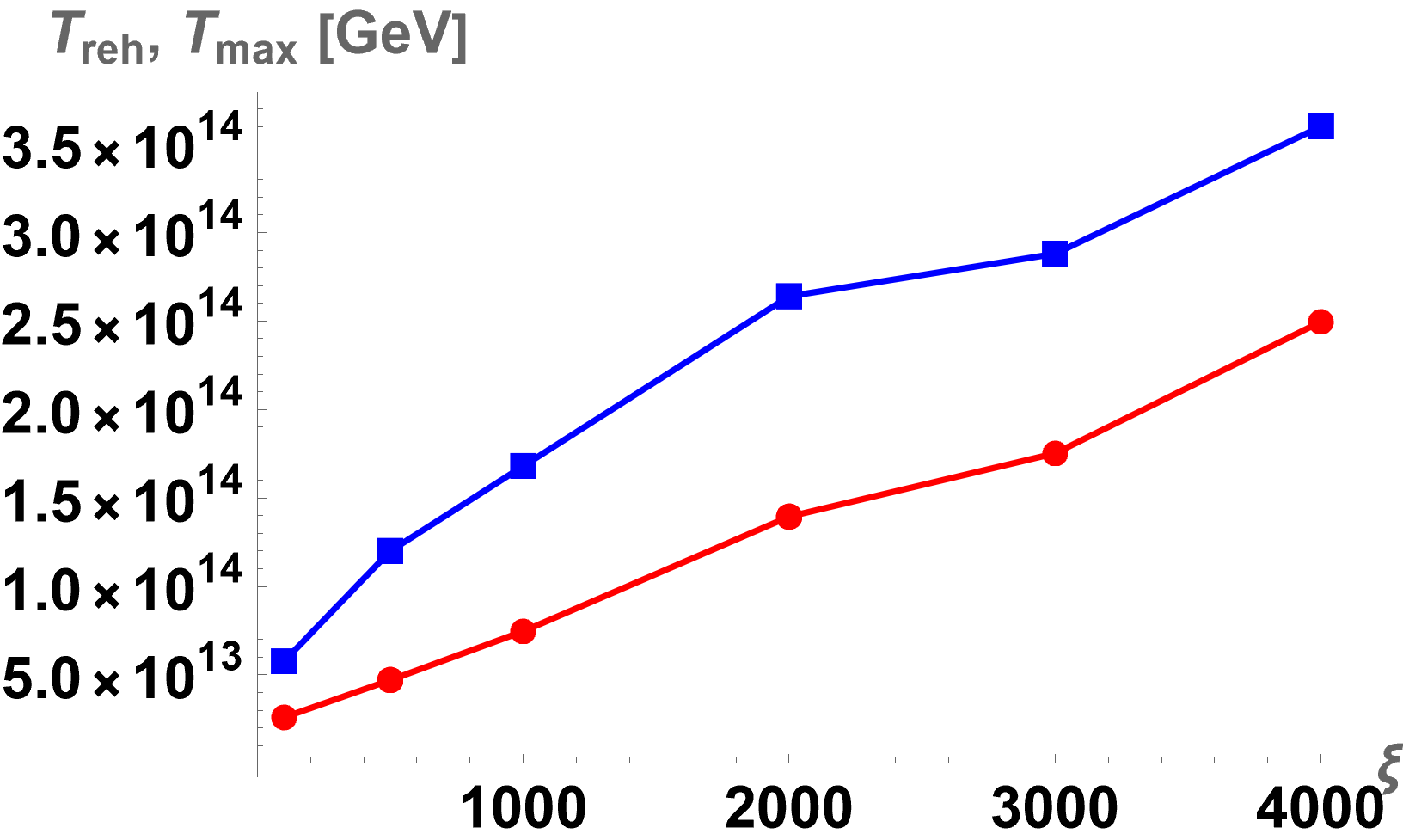}
    \end{center}
   \caption{Reheating temperature(red) $T_{\rm{reh}}$ and maximum temperature(blue) $T_{\rm{max}}$ as a function of the Higgs non-minimal coupling $\xi$.
   }
    \label{fig:sum_T}
  \end{figure}

In Fig.~\ref{fig:sum_T}, we show the reheating temperature $T_{\rm{reh}}$ and the maximum temperature $T_{\rm{max}}$ achieved for the different values of the  non-minimal coupling.
We see that the difference between $T_{\rm{reh}}$ and $T_{\rm{max}}$ is not significant, and that the results are not really sensitive to the change in $\xi$.
Then, in the range $100\leq \xi\leq 4000$, the reheating temperature is given by $2.6\times 10^{13}\,{\rm{GeV}}\leq T_{\rm{reh}}\leq 2.5\times 10^{14}\,{\rm{GeV}}$, while the maximum temperature reached during reheating varies between $5.8\times 10^{13}\,{\rm{GeV}}\leq T_{\rm{max}}\leq 3.6\times 10^{14}\,{\rm{GeV}}$.

Finally, Fig.~\ref{fig:w} follows the evolution of the equation of state, through the parameter $w$.
The results are consistent with our expectations:  the average value of $w$ can be well approximated as $\langle w\rangle=0$ , corresponding to a matter dominated universe during reheating. Then, when reheating completes, it averages to $w=1/3$, corresponding to a radiation dominated universe. 
The blue dashed line represents precisely the moment of reheating completion.

Using our numerical results for the reheating temperature $T_{\rm reh}$ and the averaged equation of state, $\langle w\rangle=0$,
the number of e-foldings gets modified by eq.~(\ref{efoldnum}). Given the pivot scale, $k=0.05\,{\rm Mpc}^{-1}$, the number of e-foldings is given by
\bea
N=53.2-54.0,
\eea

for $T_{\rm reh}=2.6\times 10^{13}-2.5\times 10^{14}\,{\rm GeV}$.

The spectral index~(\ref{eq:planckn}) and the tensor-to-scalar ratio~(\ref{eq:planckr}) become
\bea
n_s&=&0.9608-0.9614, \\
r&=&0.0041-0.0042,
\eea

where we took $V_{\rm end}\simeq V_I\simeq 3M^2_{\rm Pl} H_k^2$. 
We conclude that our results are consistent with Planck data {Planck:2018jri}, and with the Planck/BICEP/Keck limit on the tensor-to-scalar ratio at 95\% CL \cite{BICEP:2021xfz}.
The correction to the number of e-foldings due to the delayed reheating is given by $-\Delta N=0.88-1.6$, which amounts to  $-\Delta n_s=0.00064-0.0012$.

\section{Freeze-In Dark Matter}

In order to study Dark Matter production, we extend Higgs-$R^2$ inflation by including a singlet scalar Dark Matter field.
We consider the case of Freeze-In DM studied in \textbf{section \ref{sec:fimps}}.

\subsection{A model for scalar Dark Matter}
\label{DM_int}

We extend the Jordan frame Lagrangian by including a scalar DM field, $\hat{X}$, such that

\begin{align}
\nonumber \mathcal{L} / \sqrt{-g_J}=\ &\frac{1}{2}(\Mpl^2+\xi \hat{h}^2+\eta \hat{X}^2)R_J -\frac{1}{2}(\partial_{\mu} \hat{h})^{2}-\frac{1}{2}(\partial_{\mu} \hat{X})^{2}-\tilde{V}(\hat{h},\hat{X})+\alpha R_J^2+{\cal L}_{\rm{SM}}\\
=\ &\frac{1}{2}(\Mpl^2+\xi \hat{h}^2+\eta \hat{X}^2+4\alpha\hat{\chi})R_J -\frac{1}{2}(\partial_{\mu} \hat{h})^{2}-\frac{1}{2}(\partial_{\mu} \hat{X})^{2}-\tilde{V}(\hat{h},\hat{X})-\alpha\hat{\chi}^2+{\cal L}_{\rm{SM}},
\end{align}

where, in the second line, we introduced the auxiliary field $\hat{\chi}$. The parameter $\eta$ is the non-minimal coupling for $\hat{X}$ and the scalar potential is given by 

\begin{align}
    \tilde{V}(\hat{h},\hat{X})=\frac{\lambda}{4}\hat{h}^4+\frac{m_X^2}{2}\hat{X}^2+\frac{\lambda_X}{4}\hat{X}^4+\frac{\lambda_{hX}}{4}\hat{h}^2\hat{X}^2. \label{V_til}
\end{align} 

The scalar potential in eq. (\ref{V_til}) respects the $Z_2$ symmetry required for the stability of DM.
Now, we move the non minimal couplings to the scalar potential by performing a conformal transformation~(\ref{eq:conformaltransformetric}), with the field redefinitions
$\hat{h}= \Delta h$, $\hat{\chi}= \Delta^2 \chi$ and $\hat{X}= \Delta X$,  
with 

\begin{align}
&\Delta^{-2} \equiv  \left(1+\frac{\sigma}{\sqrt{6}\Mpl}\right)^{2},\\
&\left(1+\frac{\sigma}{\sqrt{6}\Mpl}\right)^{2}+\xi \frac{h^2}{\Mpl^2}+\eta \frac{X^2}{\Mpl^2}+4\alpha \frac{\chi}{\Mpl^2} =1-\frac{h^2}{6\Mpl^2}-\frac{X^2}{6\Mpl^2}-\frac{\sigma^2}{6\Mpl^2},\\
&\Omega^{ 2}=1-\frac{h^{2}}{6\Mpl^2}-\frac{X^{2}}{6\Mpl^2}-\frac{\sigma^{2}}{6\Mpl^2}.
\end{align}

Then, the Einstein frame Lagrangian is given by

\begin{align}
    \nonumber \mathcal{L} / \sqrt{-g_E}=&\ \frac{\Mpl^2}{2}R_E -\frac{1}{2\Omega^4}\left(1-\frac{h^{2}}{6\Mpl^2} -\frac{X^{2}}{6\Mpl^2} \right)\left(\partial_{\mu} \sigma\right)^{2}-\frac{1}{2\Omega^4}\left(1-\frac{\sigma^{2}}{6\Mpl^2} -\frac{X^{2}}{6\Mpl^2} \right)\left(\partial_{\mu} h\right)^{2}\\
    \nonumber &-\frac{1}{2\Omega^4}\left(1-\frac{\sigma^{2}}{6\Mpl^2} -\frac{h^{2}}{6\Mpl^2} \right)\left(\partial_{\mu} X\right)^{2}-\frac{h X}{6\Mpl^2\Omega^4}\partial_{\mu}h\partial^{\mu}X-\frac{h \sigma}{6\Mpl^2\Omega^4}  \partial_{\mu} h \partial^{\mu} \sigma\\
    &-\frac{ X \sigma}{6\Mpl^2\Omega^4} \partial_{\mu} X \partial^{\mu} \sigma-V+\frac{1}{(\Omega\Delta)^4}{\cal L}_{\rm{SM}},  \label{L_E_DM} 
\end{align}

where the Einstein frame potential reads

\begin{align}
    \nonumber V= &\ \frac{1}{\Omega^4}\Biggl[\frac{1}{4} \kappa\left(\sigma(\sigma+\sqrt{6}\Mpl)+3\tilde{\xi} h^{2}+3\tilde{\eta}X^2\right)^{2}\\
    &+\frac{\lambda}{4}h^4+\frac{m_X^2\Delta^{-2}}{2}X^2+\frac{\lambda_X}{4}X^4+\frac{\lambda_{hX}}{4}h^2X^2\Biggr].  \label{V_E_DM}
\end{align}

where we defined the effective non-minimal coupling,

\begin{align}
    \tilde{\eta}\equiv \eta+\frac{1}{6}.\label{def_delta}
\end{align}

Thus, $\tilde{\eta}=0$ corresponds to the conformal case.
We learn from the Lagrangian~$\eqref{L_E_DM}$ with eq.~$\eqref{V_E_DM}$, that in our model, Dark Matter $X$ couples feebly to $\sigma$ and $h$ with gravitational interactions, for conformality, $|\tilde{\eta}|\ll 1$, and a vanishing Higgs-portal coupling, $|\lambda_{hX}|\ll 1$.
We can produce Dark Matter during or after reheating. In each case the equation of state describing the universe is different, so we proceed to study them separately.
We also notice that DM could be produced thermally, that is, by scattering with the SM radiation (~\cite{Garcia:2017tuj,Chowdhury:2018tzw,Kaneta:2019zgw,Anastasopoulos:2020gbu,Brax:2020gqg,Kaneta:2021pyx}) 
or directly through decays of the inflaton condensates, which we dub non-thermal production (~\cite{Ellis:2015jpg,Garcia:2017tuj,Dudas:2017rpa,Kaneta:2019zgw,Garcia:2020eof,Garcia:2020wiy,Mambrini:2021zpp,Clery:2021bwz}). 

\subsection{Dark Matter Freeze-In after reheating}

In this case, reheating is complete and the universe is already within its radiation domination era. Then, the only contribution to the Dark Matter abundance is the thermal scattering with SM particles.
When DM is decoupled from the SM plasma, the DM number density~$n_X$ follows the Boltzmann equation, where the production reaction rate, $R(T)$, given as in reference ~\cite{Hall:2009bx},

\begin{align}
\dot{n}_X+3Hn_X=R(T).\label{DMeq}
\end{align}

For the thermal process $i_1(p_1)+i_2(p_2)\rightarrow X(p_3)+X(p_4)$, with the amplitude~$|\mathcal{M}|^2_{i_1+i_2\rightarrow X+X}$, the reaction rate reads ~\cite{Hall:2009bx,Edsjo:1997bg}
\begin{align}
R=\frac{T}{2^{11}\pi^6}\int_{4m_X^{2}}^{\infty} ds \,d\Omega\,  K_{1}\left(\frac{\sqrt{s}}{T}\right)\sqrt{s-4m_X^{2}}\,\overline{\left|\mathcal{M}_{i_1+i_2\rightarrow X+X}\right|^2},\label{R_T}
\end{align}

where we denoted the SM radiation by $i_{1,2}$,  $d\Omega\equiv 2\pi d\cos \theta_{13}$ is the solid angle spanned by ${\bf{p}}_1$ and ${\bf{p}}_3$, and $K_1(z)$ is the first modified Bessel function of the second kind.
The bar in the amplitude means that the symmetric factors coming from the initial and final states have already been included.

During radiation domination the temperature scales as $T\propto a^{-1}$, thus $\dot{T}=-HT$ and $H=\sqrt{\frac{g_{\rm{reh}}\pi^2}{90}}\frac{T^2}{\Mpl}$. 
Then, the Boltzmann equation~$\eqref{DMeq}$ becomes
\begin{align}
\frac{d Y}{d T}=-\frac{1}{HT^4}R(T)
=-\sqrt{\frac{90}{\pi^{2} g_{\rm{reh}}}}\frac{\Mpl}{T^{6}}R(T),\label{DMeq2}  
\end{align}
where $Y\equiv n_XT^{-3}$ is the DM relic abundance.

\subsubsection*{Thermal production from the contact terms}

In the post-reheating period, the inflaton fields have stopped oscillating and finally settled around the origin.
Then, we can neglect the VEVs for $\sigma_0$ and $h_0$ and follow the dynamics of the quantum fluctuations $\delta \sigma, \delta h,$ and $\delta X$.
For simplicity, we denote these quantum fluctuations  by  $\sigma, h,$ and $ X$, respectively.
The interaction Lagrangian is given by
\bea
\mathcal{L}_X/\sqrt{-g}&=&-\frac{X^2}{12\Mpl^2}(\partial_{\mu}\sigma)^2-\frac{X^2}{12\Mpl^2}(\partial_{\mu}h)^2-\frac{h^2}{12\Mpl^2}(\partial_{\mu}X)^2 \nonumber\\
&&-\frac{\sigma^2}{12\Mpl^2}(\partial_{\mu}X)^2  -\frac{X\sigma}{6\Mpl^2} \partial_{\mu}X\partial^{\mu}\sigma
-\frac{h X}{6\Mpl^2} \partial_{\mu}h\partial^{\mu}X \nonumber \\
&&+c_{\sigma XX}\sigma X^2 +c_{\sigma\sigma XX}\sigma^2 X^2+c_{h h XX} h^2 X^2  \nonumber\\
 &&+\frac{X^2}{12\Mpl^2}g^{\mu\nu}T^{\rm{SM}}_{\mu\nu},\label{DM_coupling}
\eea
where
\begin{align}
 c_{\sigma XX} &=-\frac{m^2_X}{\sqrt{6}\Mpl} -\frac{3}{2}\sqrt{6}\kappa \tilde{\eta}M_P,\\
 c_{\sigma\sigma XX} &=-\frac{m_X^2}{4\Mpl^2}-\frac{1}{2}\kappa( 3\tilde{\eta} +1),\\
c_{hh XX} &=-\frac{m_X^2}{6\Mpl^2}-\frac{9}{2}\kappa \tilde{\xi}\tilde{\eta}-\frac{\lambda_{hX}}{4}.\label{c_3}
\end{align}

The energy-momentum tensor of the SM particles is given by 

\begin{equation}
T^{\rm{SM}}_{\mu\nu}\equiv -\frac{2}{\sqrt{-g}}\frac{\delta\left(\sqrt{-g}{\cal L}_{SM}\right)}{\delta g^{\mu\nu}},
\end{equation}

where the Higgs contribution was extracted. There are also additional Dark Matter self-interactions, like $X^4$ or $X^2(\partial_{\mu}X)^2$, but these are irrelevant for our discussion.
We note that the DM couplings coming from $g^{\mu\nu}T^{\rm{SM}}_{\mu\nu}$  vanish exactly because all the SM fermions and gauge bosons are massless during reheating
Therefore, there is no direct coupling between DM and the SM particles at tree level, except for the coupling to the Higgs. 
There are extra non-zero effective couplings between DM and SM gauge bosons coming from the trace anomaly, but these terms are loop suppressed \cite{Watanabe:2010vy,Choi:2019osi}.
Since the reheating is complete at this stage, we only need to consider scattering with the SM plasma. During radiation domination we can set $m_h=0$,
so the Higgs field couples directly to DM via the derivative interactions and the mixing quartic coupling, as shown in eq.~$\eqref{DM_coupling}$, resulting in the scattering amplitude 
\begin{align}
\mathcal{M}_{h+h\rightarrow X+X}=-\frac{s+2 m_{X}^{2}}{6 M_{\mathrm{Pl}}^{2}}-18 \kappa \tilde{\eta}\tilde{\xi}-\lambda_{h X},   \label{M_h}
\end{align}
where $s$ is the Mandelstam parameter denoting the center of mass energy.

\subsubsection*{Thermal production from the graviton exchanges}

There will be an extra contribution to Dark Matter production coming from graviton exchanges~\cite{Garny:2015sjg,Garny:2017kha,Tang:2017hvq,Bernal:2018qlk,Barman:2021ugy,Clery:2021bwz,Mambrini:2021zpp,Haque:2022kez,Haque:2021mab}. 
We first expand the metric around flat space $g_{\mu\nu}\simeq \eta_{\mu\nu}+2h_{\mu\nu}/\Mpl$ and ignore the mixing quartic terms and other higher order terms,
\begin{align}
\mathcal{L}\supset \ &\frac{1}{\Mpl}h^{\mu\nu}\left(T^{\rm{SM}}_{\mu\nu}+T^{h}_{\mu\nu}+T^{\sigma}_{\mu\nu}+T^{X}_{\mu\nu}\right),\label{int_h}
\end{align}
 where
\begin{align}
T^{\phi}_{\mu\nu}=\partial_{\mu}\phi\partial_{\nu}\phi-\frac{1}{2}\eta_{\mu\nu}\eta^{\rho\sigma}\partial_{\rho}\phi\partial_{\sigma} \phi -\frac{1}{2}\eta_{\mu\nu}m_{\phi}^2\phi^2 \,\, \mathrm{for} \ \ \phi=h, \sigma, X. 
\end{align}

From eq.~$\eqref{int_h}$,  we get the scattering amplitude for the process $h+h\rightarrow X+X$ with graviton exchanges by
\begin{align}
\mathcal{M}^G_{h+h\rightarrow X+X}=-\frac{1}{\Mpl^2}\frac{(t-m_X^2)(s+t-m_X^2)}{s},    
\end{align}

where we used the Mandelstam variable  $t=\frac{s}{2}\left(\sqrt{1-\frac{4m_X^2}{s}}\cos \theta_{13}-1\right)+m_X^2$.
Adding this effect to eq.~$\eqref{M_h}$, we find the total squared amplitude,
\begin{align}
|\mathcal{M}^{\rm{total}}_{h+h\rightarrow X+X}|^2=\left(\frac{s+2 m_{X}^{2}}{6 M_{\mathrm{Pl}}^{2}}+18 \kappa \tilde{\eta}\tilde{\xi}+\lambda_{h X}+\frac{(t-m_X^2)(s+t-m_X^2)}{s\Mpl^2}\right)^2.\label{M_hGX}
\end{align}
The contributions comingn from the other SM particles through graviton exchange are given by
\begin{align}
&|\mathcal{M}^G_{f+f\rightarrow X+X}|^2=\frac{-1}{2 \Mpl^{4} s^{2}}\left(s+2 t-2 m^{2}_X\right)^{2}\left(\left(t-m^{2}_X\right)^{2}+s t\right),\label{M_fGX}\\
& |\mathcal{M}^G_{V+V\rightarrow X+X}|^2=\frac{2}{\Mpl^{4} s^{2}} \left(m_{X}^{4}-2 m^{2}_X t+t(s+t)\right)^{2}.\label{M_VGX}
\end{align}

Now, in order to integrate the Boltzmann equation, we have to calculate the reaction rate $R(T)$ from eq.~$\eqref{R_T}$ with eqs.~$\eqref{M_hGX}$, $\eqref{M_fGX}$ and $\eqref{M_VGX}$.
After integrating eq.~$\eqref{DMeq2}$ from $T_{\rm{reh}}$ to $T_*$ with $T_* \ll m_X\ll T_{\rm{reh}}$, we find that the asymptotic value 
of $Y(T_*)$ is independent of $T_*$,
\begin{align}
Y(T_*) \simeq Y(T_{\rm{reh}}) +\frac{\sqrt{10}}{20480\pi^4g_{\mathrm{reh}}^{1 / 2}}\frac{4m_X^4+45\Mpl^4\left(\lambda_{hX}+18\kappa \tilde{\eta} \tilde{\xi}\right)^2}{m_X\Mpl^3}+\frac{209 \sqrt{10}}{240 \pi^{6} g_{\mathrm{reh}}^{1 / 2}} \frac{T_{\mathrm{reh}}^{3}}{M_{\mathrm{Pl}}^{3}}\label{Y_T*}
\end{align}
where $g_{\rm{reh}}$ is treated as a constant. 
The second term on the right-hand side is IR dependent, while the third term encodes its UV counterpart.
The last term, which is proportional to the reheating temperature, $Y(T_{\rm{reh}})$, depends on the details of reheating. We will proceed to determine this value in the next section.

\subsection{Dark Matter Freeze-In during reheating}
 During reheating, the universe is matter-dominated, as seen in Fig.~\ref{fig:w}. 
The temperature $T$ and the scale factor $a$ evolve according to the non-trivial relation~$\eqref{sol_rhor}$.
In this case, we need to take into account both the production from the scattering with SM particles and the decays of inflaton condensates. 
We find out that, in our model, both the sigma  and the Higgs condensates are responsible for the non-thermal production, while
the SM radiation is responsible for the thermal production.
 Thus, we divide the total DM abundance into a thermal and a non-thermal contribution,
\begin{align}
Y(T_{\rm{reh}})=Y_{\rm{thermal}}(T_{\rm{reh}})+Y_{\rm{non-thermal}}(T_{\rm{reh}}).
\end{align}

\subsubsection*{Thermal production from the SM plasma}

In this case, all the SM particles except for the Higgs contribute through graviton exchanges from eqs.~$\eqref{M_fGX}$ and $\eqref{M_VGX}$.
The temperature scales as $T\propto a^{-3/8} $, so that $\dot{T}=-\frac{3}{8}HT$  and $H= \sqrt{\frac{\pi^{2} g_{*}}{90}}\frac{T^{4}}{T_{\text {reh }}^{2}}$.
Then, the Boltzmann equation~$\eqref{DMeq}$ reads

\begin{align}
\frac{d}{dT}(n_XT^{-8})=-\frac{8}{3HT^9}R(T)= -\frac{8}{3}\sqrt{\frac{90}{\pi^{2} g_{\rm{reh}}}}\frac{\Mpl T_{\rm{reh}}^2}{T^{13}}R(T).   
\end{align}

Integrating the above equation from $T_{\rm{reh}}$ to $T_{\rm{max}}$ and taking the limit $m_X\ll T_{\rm{reh}}\ll T_{\rm{max}}$ we obtain

\begin{align}
Y_{\rm{thermal}}(T_{\rm{reh}})\simeq \frac{69\sqrt{10}}{40\pi^6g_{\rm{reh}}^{1/2}} \frac{T_{\rm{reh}}^3}{\Mpl^3}.\label{Y_Treh_th}   
\end{align}

\subsubsection*{Non-thermal production from inflaton condensates}
 
For the non-thermal contribution, we use $a$ as the dynamical variable, instead of $T$~\cite{Clery:2021bwz}.

\begin{align}
\frac{d }{d a}(n_Xa^3)=\frac{a^2R(a)}{H}
\simeq\sqrt{\frac{3}{\rho_{ \rm{end}}}}\Mpl a^2\left(\frac{a}{a_{\rm{end}}}\right)^{3/2}R(a), \label{DMeq1}  
\end{align}

where we used $3\Mpl^2 H^2\simeq \rho_{\sigma}+\rho_{h}$ for the matter domination era.
The reaction rate can be written as ~\cite{Clery:2021bwz,Garcia:2020wiy} 
\begin{align}
R=\frac{1}{8\pi}\sum_{n=1}^{\infty}\left|\mathcal{M}_n\right|^2\sqrt{1-\frac{4m_{X,{\rm{eff}}}^2}{n^2\omega^2}},  \label{R_a} 
\end{align}
with $\mathcal{M}_n$ being the transition amplitude for the inflaton condensate with a Fourier mode $n$ and $\omega$ being the frequency of the two-particle final state.
The equation of state during reheating is general, and only in the case in which $n=1$ the inflationary potential is quadratic.
Then, the effective DM mass during reheating, $m_{X,{\rm{eff}}}^2$ is generally different from the bare mass $m^2_X$ from eq.~$\eqref{V_til}$. 

For the non-thermal scattering we need to take into account the VEVs of the inflaton fields during reheating.
From now on we take the case in which Dark Matter is conformally coupled to  gravity, i.e, $\tilde{\eta}=0$ and $\lambda_{hX}=0$.
 In this particular case, we find that the dominant interactions come from both the non-derivative couplings
\begin{align}
\mathcal{L}\supset \begin{cases}-\frac{\kappa}{2}\sigma_0^2X^2 & \mathrm{for} \quad\sigma_{0}>0, \\ -\frac{\kappa}{2}\frac{\lambda}{\lambda+9\kappa \tilde{\xi}^2}\sigma_0^2X^2 & \mathrm{for} \quad \sigma_{0}<0,\end{cases}    \label{s_0^2X^2}
\end{align}
and the derivative couplings
\begin{align}
\mathcal{L}\supset -\frac{1}{12\Mpl^2}X^2(\partial_{\mu}\sigma_0)- \frac{1}{6\Mpl^2}X\sigma \partial_{\mu}X\partial^{\mu}\sigma_0-\frac{1}{12\Mpl^2}\sigma_0^2(\partial_{\mu}X)\label{der_s_0^2X^2}.
\end{align}

Thus, the contact interaction terms contribute to the scattering process, $\sigma_0+\sigma_0\rightarrow X+X$, with an oscillating background~$\sigma_0$. 
In eq.~$\eqref{s_0^2X^2}$, the resulting DM interactions do not differ much for $\sigma_0>0$ and $\sigma_0<0$ if $\tilde{\xi}$ is relatively small. Here, the effective mass of DM can be taken to $m_{X,{\rm{eff}}}^2=\kappa \sigma_0^2$ from eq.~$\eqref{s_0^2X^2}$.
The contributions from eq.~(\ref{s_0^2X^2}) and (\ref{der_s_0^2X^2}) to the scattering amplitude are

\begin{align}
&\mathcal{M}_{1}^{\rm{non-der}}=-\frac{\kappa}{4}\sigma_e^2,\label{nt_nd}\\
&\mathcal{M}_{1}^{\rm{der}}=-\frac{\kappa}{8}\sigma_e^2\left(1-\frac{\sigma_0^2}{3\Mpl^2}\right),\label{nt_d}
\end{align}

where $\sigma_{e}$ is the oscillation amplitude of the sigma field at the end of inflation, 

\begin{equation}
    \rho_{\sigma}=\sigma_{e}^2m^2_{\sigma}/2=3\sigma_{e}^2\kappa\Mpl^2/2.
\end{equation}

We also have a contribution coming from the graviton exchanges with $\eqref{int_h}$ ~\cite{Mambrini:2021zpp,Clery:2021bwz}, as follows,
\begin{align}
\mathcal{M}_1^{G}=\frac{3}{8} \kappa \sigma_e^{2}\left(1+\frac{\sigma_{0}^{2}}{6 \Mpl^{2}}\right). \label{nt_G}    
\end{align}

Then, from ~$\eqref{nt_nd}$, $\eqref{nt_d}$, and $\eqref{nt_G}$, we obtain the total scattering amplitude as 
\begin{align}
\mathcal{M}_1^{\rm{total}}= \frac{5}{48}\kappa\sigma_e^2\frac{\sigma_{0}^{2}}{ \Mpl^{2}}.  \label{M_1_total} 
\end{align}

Remarkably, or maybe as a result of the conformal limit, the leading contributions proportional to $\kappa \sigma_e^2$ cancel out. 
Then, the final total amplitude is suppressed by a factor  $\sigma_0^2/\Mpl^2\sim 10^{-2}$ relative to the contact interactions. 

Finally, from eq.~$\eqref{R_a}$ with eq.~$\eqref{M_1_total}$, the reaction rate reads
\begin{align}
R_{\rm{scatter}}(a)\simeq \frac{25}{248832 \pi }\frac{\rho_{\sigma}^4}{\kappa^2\Mpl^{12}}=\frac{25}{248832 \pi }\left(\frac{a}{a_{\rm{end}}}\right)^{-12}\frac{\rho_{\sigma,{\rm{end}}}^4}{\kappa^2\Mpl^{12}},\label{R_condensate}
\end{align}

where we have extracted the leading term with respect to $\sigma_0^2/\Mpl^2$, and taken the average for the sigma condensate $\left\langle\sigma_{0}^{4}\right\rangle=3\sigma_e^4/8=\rho_{\sigma}^2/6\kappa^2\Mpl^4$.

Even for $\tilde{\eta}=0$, there exists a three-point coupling $\sigma_0 X^2$ in addition to eq.~$\eqref{s_0^2X^2}$, for $m_X\ll m_{\sigma}$.
\begin{align}
\mathcal{L}\supset \begin{cases}-\frac{m_X^2}{\sqrt{6}\Mpl}\sigma_0X^2 & \mathrm{for} \quad \sigma_{0}>0, \\ -\frac{m_X^2}{\sqrt{6}\Mpl}\frac{\lambda-3\kappa\tilde{\xi}+9\kappa\tilde{\xi}^2}{\lambda+9\kappa \tilde{\xi}^2}\sigma_0X^2 & \mathrm{for} \quad \sigma_{0}<0,\end{cases}    \label{s_0X^2}
\end{align}
which leads to the decay of the sigma condensate.
But this decay rate is given by 
\begin{align}
R_{\rm{decay}}(a)\simeq \frac{1}{72\pi}\frac{m_X^4\rho_{\sigma}}{\kappa\Mpl^4},   
\end{align}
which is subdominant since
\begin{align}
R_{\rm{decay}}/ R_{\rm{scatter}}\sim \left(\frac{\Mpl}{\sigma_e}\right)^6\left(\frac{m_X}{m_{\sigma}}\right)^4 \ll 1,
\end{align}

Using eq.~$\eqref{R_condensate}$ into the Boltzmann eq. ~$\eqref{DMeq1}$, and integrating between $a_{\rm{reh}}$ and $a_{\rm{end}}$, 
\begin{align}
Y_{\rm{non-thermal}}(T_{\rm{reh}})\simeq \frac{\sqrt{3}\pi g_{\rm{reh}}}{2239488}\frac{T_{\rm{reh}}}{\kappa^2\Mpl^{11}} \frac{\rho_{\sigma,{\rm{end}}}^4}{\rho_{\rm{end}}^{3/2}} \label{Y_Treh}
\end{align}
with $n_X(a_{\rm{end}})=0$.
Thus, this is the total amount of Dark Matter produced during reheating.

 \subsubsection*{Conformal couplings for Dark Matter}

We derive the total DM abundance in the case of conformal and non conformal coupling with gravity.
Using the yield found in ~$\eqref{Y_T*}$, together with eqs.~$\eqref{Y_Treh_th}$ and~$\eqref{Y_Treh}$, we obtain the DM relic abundance at present 
\begin{align}
\nonumber\Omega h^2  
=&\ 1.6\times 10^8 \left(\frac{m_X}{1\rm{GeV}}\right)\left(\frac{g_0}{g_{\rm{reh}}}\right)Y(T_*),\\
\nonumber \simeq&\ 1.6\times 10^8 \left(\frac{m_X}{1\rm{GeV}}\right)\left(\frac{g_0}{g_{\rm{reh}}}\right)\Biggl[\frac{\sqrt{3}\pi g_{\rm{reh}}}{2239488}\frac{T_{\rm{reh}}}{\kappa^2\Mpl^{11}} \frac{\rho_{\sigma,{\rm{end}}}^4}{\rho_{\rm{end}}^{3/2}}+ \frac{623\sqrt{10}}{240\pi^6g_{\rm{reh}}^{1/2}}  \frac{T_{\rm{reh}}^3}{\Mpl^3}\\
&+\frac{\sqrt{10}}{20480\pi^4g_{\mathrm{reh}}^{1 / 2}}\frac{4m_X^4+45\Mpl^4\left(\lambda_{hX}+18\kappa \tilde{\eta} \tilde{\xi}\right)^2}{m_X\Mpl^3}\Biggr],\label{Omegah^2}
\end{align}

where $g_0=3.91$ is the number of the effective relativistic degrees of freedom today. 

\begin{figure}[t]

  \begin{center}
   \includegraphics[width=90mm]{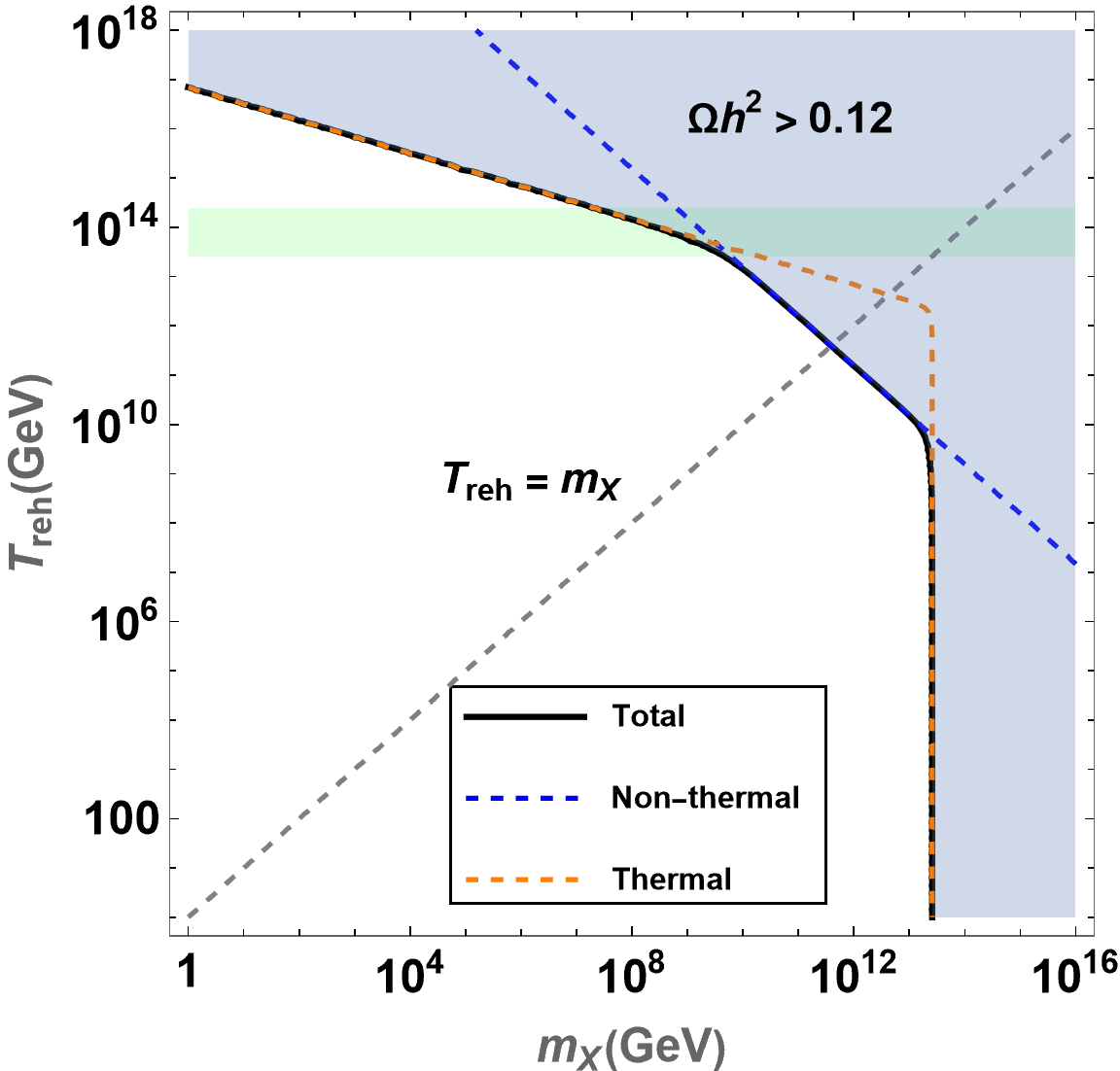}
  \end{center}
 \caption{Parameter space for the Dark Matter mass and the reheating temperature. The blue region corresponds to the case where we overproduce Dark Matter $\Omega h^2>0.12$. 
 The green band shows the predicted reheating temperature in the Higgs-$R^2$ model.}
  \label{fig:DM}
\end{figure}

In Fig.~\ref{fig:DM}, we show the available parameter space for $(m_X,T_{\rm{reh}})$, in the conformal case, i.e, with~$\tilde{\eta}=\lambda_{hX}=0$. We assume that our candidate $X$ composes the total amount of Dark Matter present in the universe, from~(\ref{eq:DMabundance}).
We also set $\rho_{{\rm{end}}}=9\times 10^{61}\,{\rm{GeV}}^4$, which holds almost the same for $100\leq \xi\leq 4000$, and $\rho_{\sigma,{\rm{end}}}=\rho_{{\rm{end}}}$. 
We show in the orange dashed line the result for the thermal production during and after reheating, i.e, the contribution from the second and third terms in eq.~$\eqref{Omegah^2}$; and in the blue dashed line, the one from the non-thermal production during reheating, i.e, from the first term in eq.~$\eqref{Omegah^2}$.
The total relic DM abundance is depicted in the black line. 
The blue shaded region shows the region where we are overproducing Dark Matter.
The green band corresponds to the reheating temperature obtained in the Higgs-$R^2$ model as shown in Fig.~\ref{fig:sum_T}, in our case, $2.6\times 10^{13}\,{\rm{GeV}}\leq T_{\rm{reh}}\leq 2.5\times 10^{14}\,{\rm{GeV}}$, for $100\leq \xi\leq 4000$. 
In order to be consistent with our integration assumptions, we need to stay above the gray dashed line, for which $T_{\rm{reh}}\gg m_X$.

In summary, we find that we can explain the observed amount of Dark Matter if the scalar field has a mass  of $2.1\times 10^{7}\,{\rm{GeV}}\leq m_X\leq 4.6\times 10^{9}\,{\rm{GeV}}$. 
Having a mass $m_X< 2.1\times 10^{7}\,{\rm GeV}$ is also possible, but then the scalar dark matter is less abundant and we would need an extra production mechanism.
Finally, we remark that for the range of DM masses that are consistent with the observed relic density, the velocity of the DM particles is sufficiently small at the time of recombination.
Quantitatively, if DM is produced dominantly from the inflaton scattering during reheating, its velocity would be $v_X=p_X/E_X$ with $p_X=m_\sigma (a_{\rm reh}/a_{\rm rec})$. 
Using $a_{\rm rec}/a_{\rm reh}=T_{\rm reh}/T_{\rm rec}$ and taking $m_\sigma\sim 10^{13}\,{\rm GeV}$ and $T_{\rm reh}\sim 10^{13}\,{\rm GeV}$ as in our model, the velocity has a value of $v_X\sim {\rm eV}/m_X\sim 10^{-18}$ for $m_X\sim 10^{9}\,{\rm GeV}$, which is small enough to be consistent with the Lyman-$\alpha$ constraint \cite{Bringmann:2016din}.

\subsubsection*{Non-conformal couplings for Dark Matter}

In this case we allow for deviations from conformality and a non-zero Higgs-Portal coupling $\lambda_{hX}$.
Including the Higgs portal translates into a new coupling $\kappa \tilde\eta \tilde\xi$.
However, in order to not over produce Dark Matter, these deviations from conformality must be small. 
The upper limits are roughly $|\lambda_{hX}|\lesssim 10^{-12}$ and $|\tilde{\eta}| \lesssim 10^{-6}$ for $\tilde{\xi} \kappa \sim 10^{-7}$. 
Around these values, the thermal production with $\lambda_{hX}\neq 0$ or $\tilde{\eta}\neq 0$ affects the total DM abundance.
Furthermore, new terms enter in the interaction Lagrangian, 

\begin{align}
    \mathcal{L}\supset \begin{cases}-3 \sqrt{\frac{3}{2}} \Mpl \tilde{\eta} \kappa \sigma_{0} X^{2}   & \mathrm{for} \quad\sigma_{0}>0, \\ -3 \sqrt{\frac{3}{2}}\Mpl \kappa \frac{\tilde{\eta} \lambda-\lambda_{hX}\tilde{\xi}/2}{\lambda+9\kappa \tilde{\xi}^2}\sigma_{0}X^2 & \mathrm{for} \quad \sigma_{0}<0.\end{cases}    
\end{align}
    
For $\sigma_0>0$, the reaction rate from the decay of the inflaton condensate becomes

    \begin{align}
    R_{{\rm{decay}}, \tilde{\eta}}(a)\simeq \frac{9}{8\pi}\tilde{\eta}^2 \kappa\rho_{\sigma},  
    \end{align}

    which is smaller than the one from the inflaton scattering in the conformal case in eq.~$\eqref{R_condensate}$. 
    To put some numbers, in the case of  $\tilde{\eta}\sim 10^{-6}$ $R_{{\rm{decay}}, \tilde{\eta}}/R_{\rm{scatter}}\sim 10^{-2} \ll 1$.
    \begin{figure}[t]

        \begin{center}
         \includegraphics[width=90mm]{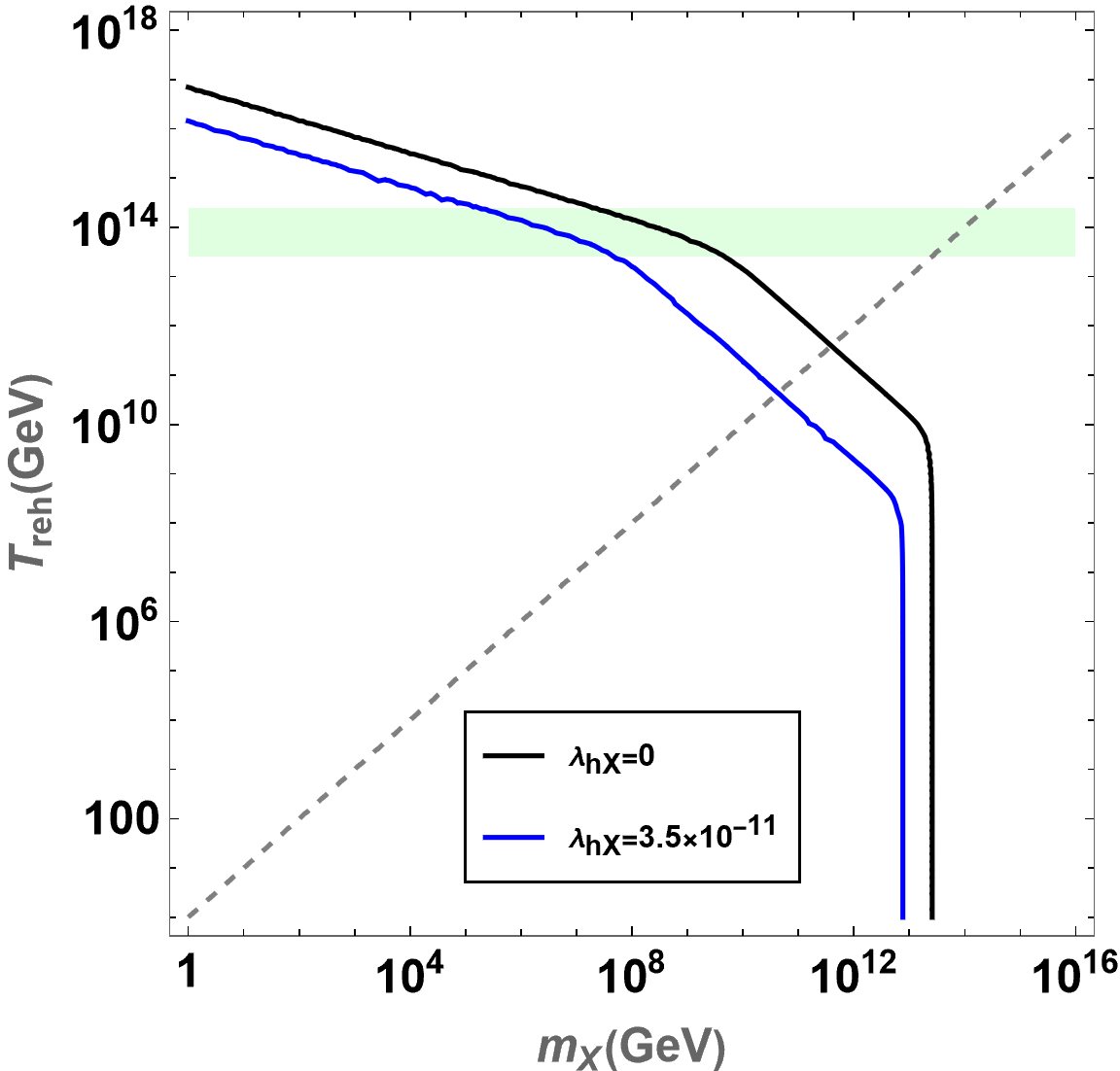}
        \end{center}
      
       \caption{Dark Matter relic energy density with a non-zero Higgs-portal coupling, $\lambda_{hX}=3.5\times 10^{-11}$, depicted in blue line. 
       For comparison, we show in the black line the case with conformal couplings. }
        \label{fig:DM_lambda}
      \end{figure}
    In Fig.~\ref{fig:DM_lambda}, we took  $\lambda_{hX}=3.5\times 10^{-11}$. 
    In this particular case, we recover the correct DM relic density, achieving a smaller DM mass than in the conformal limit.
  
\chapter{Embedding of Higgs Inflation into SUGRA}
\label{chap_sugra} 

\begin{small}
'If I could remember the names of these particles, I would have been a botanist.'\\
-Enrico Fermi
    \end{small}
    \vspace{5mm}

At the high energies of inflation, quantum gravity becomes important, so it is natural to extend any inflationary model into the local version of SUSY, supergravity.
However, constructing such an embedding is not a trivial task. The SUSY breaking potential term, 
which is precisely driving inflation, generally spoils the flatness of an inflaton potential. For a review on how to circumvent the problems of supergravity embeddings of inflation see \cite{Yamaguchi:2011kg}.

In this chapter, we provide an embedding of Higgs inflation into supersymmetry.
For that, we construct a Next-to-Minimal-Supersymmetric Standard Model (NMSSM) model, which extends Higgs inflation into the context of $R^2$-supergravity.

\section{\texorpdfstring{$R^2$}-supergravity and dual description}\label{S1}

\subsection{\texorpdfstring{$R^2$}-supergravity and the NMSSM}

We start by the superconformal action \cite{Kaku:1978nz,Kaku:1978ea,Townsend:1979ki,Kugo:1982cu,Kugo:1983mv}
\begin{align}
S=[|X^0|^2\tilde{\Omega}(z^{\alpha},\bar{z}^{\bar{\beta}})]_D+[(X^0)^3\tilde{W}(z^{\alpha})]_F+[f_{AB}(z^{\alpha})\bar{\mathcal{W}}^A\mathcal{W}^B]_F+[\alpha \bar{\mathcal{R}}\mathcal{R}]_D,   \label{S_HD} 
\end{align}
where $[...]_{D,F}$ denotes the superconformal $D$ and $F$ terms, corresponding to the real and chiral multiplets with (Weyl weight, chiral weight)$=(2,0)$ and $(3,3)$, respectively. 
We introduce a $X^0$ chiral compensator multiplet in order to obtain a theory that is manifestly conformal. This $X^0$ supermultiplet plays the same role as the conformal model $\varphi$ in the non-supersymmetric case.
The weights of $X^0$ are $(1,1)$, and the ones of its complex conjugate, $\bar{X}^{\bar{0}}$, are $(1,-1)$.
We denote the chiral and anti-chiral matter multiplets as $z^{\alpha}$ and $\bar{z}^{\bar{\beta}}$, with weights $(0,0)$, which correspond to the NMSSM chiral superfields. 
The conformal factor, $\tilde{\Omega} $, must be a real function of the matter multiplets $z^{\alpha}$ and $\bar{z}^{\bar{\beta}}$. We will refer to it as the \textbf{frame function}.
The K\"ahler potential then is related to the frame function through

\begin{equation}
\tilde{\mathcal{K}}=-3\log \left(-\frac{\tilde{\Omega}}{3} \right).
\end{equation}

The superpotential is defined by $tilde{W}(z^{\alpha})$, while $\mathcal{W}^A$ is the gauge field strength, and f is the gauge kinetic function.
We added a curvature multiplet, $\mathcal{R}$~\cite{Kugo:1983mv} with weights $(1,1)$,
\begin{align}
\mathcal{R}=(X^0)^{-1} \Sigma (\bar{X}^{\bar{0}}),
\end{align}
where $\Sigma $ is a chiral projection operator. 
The coefficient $\alpha$ accompanying the $\mathcal{R}$ multiplet, is a real number and must remain positive in order to keep the stability of the theory.
In the limiting case where $\alpha=0$, the action $\eqref{S_HD} $ reduces to the \textbf{standard superconformal action}, up to second derivatives. 
In our setup, we take a non-zero $\alpha$, thereby introducing new dynamical degrees of freedom ~\cite{Cecotti:1987sa}. 
These extra degrees of freedom are responsible for the UV completion of Higgs inflation.

The compensator multiplet $X^0$ is unphysical, and it can be eliminated by fixing a gauge.
We impose the \textbf{dilatation gauge condition}, $X^0=1$, on the lowest scalar component of $X^0$-multiplet.
The bosonic Lagrangian in the Jordan frame, after integrating out the auxiliary fields, reads
\begin{align}
\nonumber\mathcal{L}/\sqrt{-g}=&-\tilde{\Omega}_{\alpha\bar{\beta}}\partial_{\mu} z^{\alpha} \partial^{\mu} \bar{z}^{\bar{\beta}}+(-i\tilde{\Omega}_{\alpha}\partial_{\mu} z^{\alpha} \mathcal{A}^{\mu}+{\rm{c.c.}})+\tilde{\Omega}(-\mathcal{A}^{2}+\left|F^{0}\right|^{2})+(3F^0\tilde{W}+{\rm{c.c.}})\\
\nonumber&+\left(-\frac{\tilde{\Omega}}{6}+\frac{\alpha}{6}|F^0|^2+\frac{\alpha}{3}\mathcal{A}^2\right)R+\frac{\alpha}{36}R^2+\alpha\left(\mathcal{A}^{2}+\left|F^{0}\right|^{2}\right)^{2}+\alpha(\nabla_{\mu}  \mathcal{A}^{\mu})^{2}\\
&-\alpha\left|\partial_{\mu} F^{0}-3 i \mathcal{A}_{\mu} F^{0}\right|^{2}-\tilde{\Omega}^{\alpha\bar{\beta}}(\tilde{\Omega}_{\alpha}\bar{F}^{\bar{0}}+\tilde{W}_{\alpha})(\tilde{\Omega}_{\bar{\beta}}F^{0}+\bar{\tilde{W}}_{\bar{\beta}})\nonumber \\ 
&-\frac{1}{2}({\rm{Re}}f)^{-1 AB}\tilde{\Omega}_{\alpha}k_A^{\alpha}\tilde{\Omega}_{\bar{\beta}}k_B^{\bar{\beta}},\label{comp_HD}
\end{align}
where $\mathcal{A}^2=\mathcal{A}_{\mu}\mathcal{A}^{\mu}$ and $F^0$ is the $F$ term component of $X^0$. 
 Although $F^0$ is just an auxiliary field in standard supergravity, in our case it is a dynamical degree of freedom.
 The subindices, as in $\tilde{\Omega}_\alpha$, denote the derivatives with respect to $z^{\alpha}$ or $\bar{z}^{\bar{\alpha}}$, and $\tilde{\Omega}^{\alpha\bar{\beta}}\equiv (\tilde{\Omega}_{\alpha\bar{\beta}})^{-1}$. 
 The covariant derivatives on $z^{\alpha}$ generally include gauge connections, but we omit them for simplicity in our discussion.
Also, $k_A^{\alpha}$ are the Killing vectors defined by the gauge transformations of chiral superfields, $\delta z^{\alpha}=\theta^Ak_A^{\alpha}$.
In the second line of eq.~$\eqref{comp_HD}$, we include the non-minimal couplings between the matter chiral multiplets and the Ricci scalar, $\tilde{\Omega} (z^{\alpha},\bar{z}^{\bar{\alpha}})R$, as well as the $R^2$ term necessary for the UV completion of Higgs inflation. 

Let's now take a look at the Higgs sector of the NMSSM,
\begin{align}
 z^{\alpha}=\{S,H_{u},H_{d}\}.
\end{align}
Here, $S$ is a singlet chiral superfield, while $H_u$ and $H_d$ correspond to the $SU(2)_L$ Higgs doublets,
\begin{align}
H_{u}=\left(\begin{array}{c}
H_{u}^{+} \\
H_{u}^{0}
\end{array}\right), \quad H_{d}=\left(\begin{array}{c}
H_{d}^{0} \\
H_{d}^{-}
\end{array}\right).    
\end{align}
Following \cite{Einhorn:2009bh,Ferrara:2010yw,Ferrara:2010in},  we choose the frame function and the superpotential as
\begin{align}
&\tilde{\Omega}(z^{\alpha},\bar{z}^{\bar{\beta}})=-3+|S|^2+|H_{u}|^2+|H_{d}|^2+\left(\frac{3}{2}\chi H_u\cdot H_d +{\rm{h.c.}}\right),\label{O_NMSSM}\\
&\tilde{W}(z^{\alpha})=\lambda S H_{u} \cdot H_{d}+\frac{\rho}{3} S^{3},\label{W_NMSSM}
\end{align}
where $|H_{u}|^2=H_{u}^{\dagger} H_{u}$, $H_{u} \cdot H_{d} \equiv-H_{u}^{0} H_{d}^{0}+H_{u}^{+} H_{d}^{-}$, and 

\begin{equation}
\tilde{\Omega}(z^{\alpha},\bar{z}^{\bar{\beta}})=-3\,{\rm exp}\big(-\tilde{\mathcal{K}}(z^{\alpha},\bar{z}^{\bar{\beta}})/3\big)
\end{equation}
 We choose $\chi, \lambda$ and $\rho$ to be real parameters. 

Thus, eq.~$\eqref{comp_HD}$ describes a supergravity embedding for the mixed Higgs-$R^2$ model.
 Expanding the kinetic terms of the NMSSM Lagrangian we get
\begin{align}
\nonumber \mathcal{L}/\sqrt{-g}=&\left\{\frac{1}{2}-\frac{1}{6}|S|^2-\frac{1}{6}|H_u|^2-\frac{1}{6}|H_d|^2+\left(-\frac{1}{4}\chi H_u\cdot H_d +{\rm{h.c.}}\right)\right\}R\\
&-|\partial_{\mu}S|^2- |\partial_{\mu}H_u|^2- |\partial_{\mu}H_d|^2+\frac{\alpha}{36}R^2+\cdots ,
\end{align}
where the ellipsis includes the  $\mathcal{A}_{\mu}$ and $F^0$ terms, as well as the scalar potential.
As a result of the supergravity embedding, we find several new scalar fields, including $S$ and the multi-Higgs multiplets. 
We will study in later sections the stability of the inflationary potential in this new directions.
\subsection{Dual-scalar Lagrangian}\label{sec:duallagrangian}

In \textbf{chapter \ref{chap_UV}}, we used the scalar-dual description of Higgs inflation in order to obtain the extra degrees of freedom from the higher curvature terms.
In this chapter we will also derive the dual Lagrangian for $\eqref{S_HD}$, promoting the higher curvature terms, such as $R^2$, to dynamical scalar fields.
For the time being we won't fix the gauge, but later on we will study the different frames corresponding to different gauge fixings. 
In order to derive the \textbf{master action}, we follow ~\cite{Cecotti:1987sa}. 
We rewrite the last term in eq.~$\eqref{S_HD}$ as 

\begin{align}
[\alpha \bar{\mathcal{R}}\mathcal{R}]_D=[\alpha\bar{C}C]_D+[T(C-\mathcal{R})]_F,  \label{S_HD2}
\end{align}
where $T$ and $C$ are the chiral multiplets, with weights $(2,2)$ and $(1,1)$, respectively. 
In order to see the correspondence with eq.~$\eqref{S_HD2}$, we use the equation of motion for $T$,

\begin{equation}
    C=\mathcal{R}.
\end{equation} 

From the second term in the right-hand side of eq.~$\eqref{S_HD2}$, we get
\begin{align}
\nonumber [T(C-\mathcal{R})]_F&=[TC-\Sigma(T(X^0)^{-1}\bar{X}^{\bar{0}})]_F\\
&=[TC]_F-[T(X^0)^{-1}\bar{X}^{\bar{0}}+{\rm{c.c.}}]_D,
\end{align}
up to total derivative. Then, the dual action becomes
\begin{align}
S=[|X^0|^2(\tilde{\Omega}+\alpha\bar{C}C-(T+\bar{T}))]_D+[(X^0)^3(\tilde{W}+TC)]_F+[f_{AB}(z^{\alpha})\bar{\mathcal{W}}^A\mathcal{W}^B]_F,  \label{S_HD3}
\end{align}
where we have performed the redefinitions
\begin{align}
    T &\rightarrow T(X^0)^2,\\
    C &\rightarrow CX^0,
\end{align}
so that now our fields $T$ and $C$ are weightless.
In order to define the superpotential, we compare our result in eq.~$\eqref{S_HD3}$ with standard supergravity, 

\begin{align}
S=[|X^0|^2\Omega (z^I,\bar{z}^{\bar{J}})]_D+[(X^0)^3W(z^{I})]_F+[f_{AB}(z^{I})\bar{\mathcal{W}}^A\mathcal{W}^B]_F.  \label{S_sugra}
\end{align}

We see that we can define the frame function~$\Omega$ and the superpotential~$W$ as
\begin{align}
\nonumber \Omega(z^{I},\bar{z}^{\bar{J}})&\equiv\tilde{\Omega}(z^{\alpha},\bar{z}^{\bar{\beta}})+|C|^2-(T+\bar{T})\\
&=-3+|S|^2+|H_{u}|^2+|H_{d}|^2+\left(\frac{3}{2}\chi H_u\cdot H_d +{\rm{h.c.}}\right)+|C|^2-(T+\bar{T}),\label{tilde_O}\\
W(z^{I})& \equiv  \tilde{W}(z^{\alpha})+\frac{1}{\sqrt{\alpha}}TC=\lambda S H_{u} \cdot H_{d}+\frac{\rho}{3} S^{3}+\frac{1}{\sqrt{\alpha}}TC,\label{tilde_W}\\
f_{AB}(z^{I})&=f_{AB}(z^{\alpha}),\label{f}
\end{align}

where we took $C\rightarrow C/\sqrt{\alpha}$, and took the K\"ahler potential as  
\begin{equation*}
    \Omega(z^{I},\bar{z}^{\bar{J}})=-3\,{\rm exp}\big(-\mathcal{K}(z^{I},\bar{z}^{\bar{J}})/3\big).
\end{equation*} 

We have removed the higher derivative term $\alpha \bar{\mathcal{R}}\mathcal{R}$, and as it happen for the non-supersymmetric case, we have instead two additional chiral superfields, $T$ and $C$. 
For  $\alpha=0$, we recover the original NMSSM inflationary model in~\cite{Einhorn:2009bh,Ferrara:2010yw,Ferrara:2010in}.
In that case, $C$ appears only in the superpotential and $T=0$.

After fixing the gauge,  
and integrating out the auxiliary fields, the bosonic lagrangian becomes

\begin{align}
\nonumber \mathcal{L}/\sqrt{-g}=& -\frac{1}{6}(X^0)^2\Omega R-\Omega (\partial_{\mu}X^0)^2 -X^{0} \partial^{\mu} X^{0}\left(\Omega_{I} \partial_{\mu} z^{I}+\Omega_{\bar{I}} \partial_{\mu} \bar{z}^{\bar{I}}\right)+\left(X^{0}\right)^{2}\Omega\mathcal{A}_{\mu}^2\\
&-(X^0)^2 \Omega_{I\bar{J}} \partial_{\mu} z^{I} \partial^{\mu} \bar{z}^{\bar{J}}-V ,\label{master}
\end{align}
where the gauge field is defined as,
\begin{align}
\mathcal{A}_{\mu}=  -\frac{\mathrm{i}}{2 \Omega}\left(\partial_{\mu} z^{I} \Omega_{I} -\partial_{\mu} \bar{z}^{\bar{I}} \Omega_{\bar{I}} \right). \label{sol_A}
\end{align}

The scalar potential, composed by the $F$ and $D$ terms, $V=V^F+V^D$ is given by
\begin{align}
\nonumber V^F=&\left(X^{0}\right)^{4} \left(\Omega_{I \bar{J}}-\frac{\Omega_{I} \Omega_{\bar{J}}}{\Omega}\right)^{-1}\left(W_{I}-\frac{3 \Omega_{I}}{\Omega} W\right)\left(\bar{W}_{\bar{J}}-\frac{3\Omega_{\bar{J}}}{\Omega} \bar{W}\right) +\frac{9}{\Omega}\left(X^{0}\right)^{4}|W|^{2} \\
=& \left(X^{0}\right)^{4}e^{\mathcal{K}/3}\left[ \mathcal{K}^{I\bar{J}} \left(W_{I}+\mathcal{K}_IW\right)\left(\bar{W}_{\bar{J}}+\mathcal{K}_{\bar{J}}\bar{W}\right)-3 |W |^2\right],\\
V^D=&\frac{\left(X^{0}\right)^{4}}{2}({\rm{Re}}f)^{-1 AB}\Omega_{\alpha}k_A^{\alpha}\Omega_{\bar{\beta}}k_B^{\bar{\beta}},
\end{align}

where $\mathcal{K}^{I\bar{J}}$ is the inverse of K\"ahler metric.
The scalar potential and the bosonic Lagrangian are generic and do not depend on the choice in Eqs.~$\eqref{tilde_O}$-$\eqref{f}$.

For the particular choice of the NMSSM in the dual-scalar description of $R^2$-supergravity, we use the following notations,
\begin{align}
&z^I=\left\{S,H_{u},H_{d},C,T\right\},\\
&z^{i}=\{S,H_{u},H_{d},C\},\label{z^i}\\
&z^{\alpha}=\{S,H_{u},H_{d}\}.
\end{align}

 Eqs.~(\ref{tilde_O}) and (\ref{tilde_W}), allows us to write the K\"ahler metric  as

\begin{align}
\mathcal{K}_{I\bar{J}}= -\frac{3}{\Omega}\left(\begin{array}{cc}
\delta_{i \bar{j}}-\frac{\Omega_i\Omega_{\bar{j}}}{\Omega} & \frac{\Omega_i}{\Omega} \\
\frac{\Omega_{\bar{j}}}{\Omega}  & - \frac{1}{\Omega}
\end{array}\right), \ \   \mathcal{K}^{I\bar{J}}=-\frac{\Omega}{3} \left(\begin{array}{cc}
 \delta^{i\bar{j} }& \delta^{i\bar{k}}\Omega_{\bar{k}}\\
\delta^{\bar{j}\ell}\Omega_{\ell}& -\Omega+\delta^{k\bar{\ell}} \Omega_{k} \Omega_{\bar{\ell}}
\end{array}\right).\label{metric}
\end{align}

Where we have made use of  $\Omega_{i \bar{j}}=\delta_{i \bar{j}}$ and $\Omega_{T \bar{T}}=0$.
Then, the $F$-term  of the scalar potential can be written as
\begin{align}
V^F=\left(X^{0}\right)^{4}\left[\delta^{i \bar{j}} W_{i} \bar{W}_{\bar{j}}+\left(W_i\delta^{i \bar{j}} \Omega_{\bar{j}}  \bar{W}_{\bar{T}}-3W_{T} \bar{W}+{\rm{c.c}}\right)-\left(\Omega-\delta^{i \bar{j}}{\Omega}_{i} \Omega_{\bar{j}}\right)|W_{T}|^{2}\right].\label{SP_1}
\end{align}

Now we choose the Jordan frame i.e, $X^0=1$. We remark that, in the absence of the chiral superfields,  $T$ and $C$, we recover the F-term for NMSSM with global SUSY. 

\section{Dual-scalar supergravity for NMSSM} \label{sec:dualscalarsugra}

Different choices of the gauge will lead to different frames. Some aspects of the underlying physics are clearer in a specific frame.
In this section, we will introduce three different frame, i.e,  the Jordan frames,  the Einstein frame and linear sigma frame. 

\subsection{Jordan frames}\label{CJF}

 The Jordan frame corresponds to setting $X^0=1$ in Eq.~$\eqref{master}$.
 The bosonic Lagrangian in this case reads 
\begin{align}\label{jor}
\mathcal{L}_J/\sqrt{-g}=& -\frac{1}{6}\Omega R- \Omega_{I\bar{J}} \partial_{\mu} z^{I} \partial^{\mu} \bar{z}^{\bar{J}}-V_J,
\end{align}

We have omitted the gauge part of the Lagrangian, because it vanishes during inflation.
Eq. (\ref{jor}) is completely general. Then, we take the frame function to be
\begin{align}
\Omega=-3e^{-\frac{\mathcal{K}}{3}}= -3 +\delta_{I\bar{J}}z^I\bar{z}^{\bar{J}}+J(z)+\bar{J}(\bar{z}), \label{CJ}   
\end{align}

where $J$ is an arbitrary holomorphic function and
the scalar fields have canonical kinetic terms~\cite{Ferrara:2010yw,Ferrara:2010in}. 
Global supersymmetry corresponds to $J=0$ and a superpotential that contains only cubic terms. 
This is usually called as ``\textbf{canonical superconformal supergravity}"~\cite{Ferrara:2010in}.
There is a subtle difference between Eq.~$\eqref{CJ}$ and Eq.~$\eqref{tilde_O}$.  In $\Omega$, the dual scalar field $T$ appears as the combination $T+\bar{T}$ , so that T is not dynamical in the Jordan frame. 
The $T$-dependence in the frame function in Eq.~$\eqref{CJ}$ is a characteristic feature of  $R^2$ supergravity.

In the Jordan frame, from Eqs.~$\eqref{tilde_O}$ and $\eqref{tilde_W}$,  we obtain the bosonic Lagrangian as
\begin{align}
\nonumber \mathcal{L}_J/\sqrt{-g}=&\left\{\frac{1}{2}-\frac{1}{6}|S|^2-\frac{1}{6}|H_u|^2-\frac{1}{6}|H_d|^2-\frac{1}{6}|C|^2+\left(-\frac{1}{4}\chi H_u\cdot H_d +{\rm{h.c.}}\right)+\frac{1}{3}{\rm{Re}}T\right\}R\\
&- |\partial_{\mu}S|^2- |\partial_{\mu}H_u|^2- |\partial_{\mu}H_d|^2- |\partial_{\mu}C|^2+\Omega\mathcal{A}_{\mu}^2-V_J.\label{dual2}
\end{align}
Where the scalar potential $V_J=V_J^F+V_J^D$ reads
\begin{align}
\nonumber V_{J}^F=&\left|\lambda H_u\cdot H_d+\rho S^{2}\right|^{2}+\lambda^2\left| S \right|^{2}(|H_u|^2+|H_d|^2)+\frac{1}{\alpha}|T|^2 \nonumber \\
&+\frac{3}{2}\frac{\chi\lambda}{\sqrt{\alpha}} (S\bar{C}+\bar{S}C)(|H_u|^2+|H_d|^2) \nonumber \\
&+\frac{1}{\alpha}|C|^2\left\{3+\frac{3}{2}\chi( H_u\cdot H_d+{\rm{c.c.}})+\frac{9}{4}\chi^2(|H_u|^2+|H_d|^2)-2{\rm{Re}}T\right\},\\\label{SP_1explicit}
V_{J}^{D}=&\frac{g^{\prime 2}}{8}\left(\left|H_{u}\right|^{2}-\left|H_{d}\right|^{2}\right)^{2}+\frac{g^{2}}{8}\left(\left(H_{u}\right)^{\dagger} \vec{\tau} H_{u}+\left(H_{d}\right)^{\dagger} \vec{\tau} H_{d}\right)^{2}.  
\end{align}

We have introduced the Pauli matrices, ${\tau_i} (i=1,2,3)$, and defined $f_{AB}=\delta_{AB}$.

Then, eq. (\ref{SP_1explicit}) represents a  generalized NMSSM inflation model in the Jordan frame, where two additional complex fields, $C$ and $T$, appear due to the $R^2$ term. 
We also realize that there is a non-minimal coupling only for the real part of $T$, but not for ${\rm{Im}}T$.

We redefine $C\rightarrow\sqrt{\alpha}C$ and $T\rightarrow\sqrt{\alpha}T$. Then, taking the limit  $\alpha\rightarrow 0$, the frame function becomes independent of $C$ and $T$.
Using the equations of motion for the auxiliary fields,
\begin{align}
T=0, \ \ C=\frac{\frac{3}{2} \chi \lambda  S\left(|H_u|^{2}+|H_d|^{2}\right)}{-3+\left(-\frac{3}{2}\chi H_{u}\cdot H_{d}+{\rm{c.c.}}\right)-\frac{9}{4} \chi^{2}\left(|H_u|^{2}+|H_d|^{2}\right)},  
\end{align}

we can integrate them out.

Using this result in eq.~$\eqref{SP_1explicit}$, we get the scalar potential.
\begin{align}
\nonumber V_{J}^F|_{\alpha\rightarrow 0}=&\left|\lambda H_u\cdot H_d+\rho S^{2}\right|^{2}\\
&+\lambda^{2}|S|^{2}\left(|H_u|^{2}+|H_d|^{2}\right) \frac{-3+\left(-\frac{3}{2} \chi H_{u}\cdot H_{d}+{\rm{c.c.}}\right)}{-3+\left(-\frac{3}{2}\chi H_{u}\cdot H_{d}+{\rm{c.c.}}\right)-\frac{9}{4} \chi^{2}\left(|H_u|^{2}+|H_d|^{2}\right)},
\end{align}
which reproduces the result of refs.~\cite{Ferrara:2010yw,Ferrara:2010in}.
In the case in which $\alpha$ is sizable, $C, T$ become dynamical and they affect the inflationary dynamics. 
Considering a conformal coupling for the Higgs fields, i.e. $\chi=0$, the NMSSM sector decouples from the scalaron, recovering pure Starobinsky inflation in supergravity~\cite{Cecotti:1987sa}. 
Varying the size of the non-minimal coupling, we can interpolate between Higgs and Starobinsky inflation models in supergravity.  

In the Einstein frame with $X^0=\sqrt{-3/\Omega}=e^{\mathcal{K}/6}$, eq.~$\eqref{master}$ leads to the dual-scalar Lagrangian
\begin{align}
\mathcal{L}_E/\sqrt{-g}=\frac{1}{2}R-\mathcal{K}_{I \bar{J}} \partial_{\mu} z^{I} \partial^{\mu} \bar{z}^{\bar{J}}-V_E, \label{E-frame}  
\end{align}
where the scalar potential~$V_E$ is related to $V_{J}$ through a conformal transformation,

\begin{align}
\nonumber V_E=&\frac{9}{\Omega^2}V_{J}.\label{V_E}
\end{align}

In this case, the kinetic term for $T$ is present in eq.~$\eqref{E-frame}$.

\subsection{Sigma-model frames}\label{LSF}

The ``linear sigma frame" is interesting because it makes obvious the unitarity restoration of Higgs inflation after including the extra degrees of freedom. 
From the scalar potential it will be clear to read the perturbativity conditions.

In this case, we define the matter multiplets as
\begin{align}
\hat{z}^i\equiv X^0 z^i, \ \ \hat{T}\equiv  (X^0)^2 T, 
\end{align}

with $z^{i}=\{S,H_{u},H_{d},C\}$. The frame function and the superpotential in the Einstein frame become
\begin{align}
|X^0|^2\Omega (z^I,\bar{z}^{\bar{J}})
&= -3 |X^0|^2+|\hat{S}|^2+|\hat{H}_{u}|^2+|\hat{H}_{d}|^2+|\hat{C}|^2 \nonumber \\
&\quad+\frac{3\chi}{2}\left( \frac{\hat{H}_u\cdot \hat{H}_d\bar{X}^{\bar{0}}}{X^0} +{\rm{h.c.}}\right)-\left( \frac{\hat{T}\bar{X}^{\bar{0}}}{X^0} +{\rm{h.c.}}\right),\\
(X^0)^3W(z^I)&=\lambda \hat{S} \hat{H}_{u} \cdot \hat{H}_{d}+\frac{\rho}{3} \hat{S}^{3}+\frac{1}{\sqrt{\alpha}}\hat{T}\hat{C}.
\end{align}

As previously, we impose the gauge conditions, except for the dilatation, and integrate out the auxiliary fields.
Then, from eq.~$\eqref{S_HD3}$ we get the bosonic Lagrangian
\begin{align}
 \mathcal{L}_{LS}/\sqrt{-g}&=\bigg\{\frac{(X^0)^2}{2}-\frac{1}{6}|\hat{S}|^2-\frac{1}{6}|\hat{H}_u|^2-\frac{1}{6}|\hat{H}_d|^2-\frac{1}{6}|\hat{C}|^2 \nonumber \\
&\qquad+\left(-\frac{1}{4}\chi \hat{H}_u\cdot \hat{H}_d +{\rm{h.c.}}\right)+\frac{1}{3}{\rm{Re}}\hat{T}\bigg\}R \nonumber \\
&\quad+\left(\left(\partial \log X^0\right)^{2}+\Box \log X^0\right)\left(-3(X^0)^2+\left(\frac{3}{2}\chi \hat{H}_u\cdot \hat{H}_d +{\rm{h.c.}}\right)-2{\rm{Re}}\hat{T}\right) \nonumber  \\
&\quad- |\partial_{\mu}\hat{S}|^2- |\partial_{\mu}\hat{H}_u|^2- |\partial_{\mu}\hat{H}_d|^2- |\partial_{\mu}\hat{C}|^2+\Omega\mathcal{A}_{\mu}^2-V_{LS},
\end{align}

where
\begin{align}
\nonumber V_{LS}^F=&|\lambda \hat{H}_u\cdot \hat{H}_d+\rho \hat{S}^{2}|^{2}+\lambda^2| \hat{S} |^{2}(|\hat{H}_u|^2+|\hat{H}_d|^2)+\frac{1}{\alpha}|\hat{T}|^2 \nonumber \\
&+\frac{3}{2}\frac{\chi\lambda}{\sqrt{\alpha}} (\hat{S}\bar{\hat{C}}+\bar{\hat{S}}\hat{C})(|\hat{H}_u|^2+|\hat{H}_d|^2) \nonumber \\
&-\frac{1}{\alpha}|\hat{C}|^2\left\{-3(X^0)^2-\frac{3}{2}\chi( \hat{H}_u\cdot \hat{H}_d+{\rm{c.c.}})-\frac{9}{4}\chi^2(|\hat{H}_u|^2+|\hat{H}_d|^2)+2{\rm{Re}}\hat{T}\right\},\label{V_LS} \\
V_{LS}^{D}=&\frac{g^{\prime 2}}{8}\left(|\hat{H}_{u}|^{2}-|\hat{H}_{d}|^{2}\right)^{2}+\frac{g^{2}}{8}\left((\hat{H}_{u})^{\dagger} \vec{\tau} \hat{H}_{u}+(\hat{H}_{d})^{\dagger} \vec{\tau} \hat{H}_{d}\right)^{2}.
\end{align}

The gauge field in the sigma-model frame is given by
\begin{align}
\mathcal{A}_{\mu}= -\frac{i}{2\Omega}(X^{0})^2\left(\Omega_{\hat{i}}\partial_{\mu}((X^{0})^{-1}\hat{z}^{\hat{i}})+X^0\Omega_{\hat{T}}\partial_{\mu}((X^{0})^{-2}\hat{T})-{\rm{c.c.}}\right). \label{Amu}
\end{align}

Now we can fix the dilation gauge, $X^0=1+\frac{1}{\sqrt{6}} \sigma$, where $\sigma$ is a function of $\hat{z}^I$ and $\bar{\hat{z}}^{\bar{J}}$.
 The $\sigma$ field satisfies a similar constraint equation as in the non supersymmetric case ~\cite{Lee:2021dgi},

\begin{align}
\nonumber &\frac{(X^0)^2}{2}-\frac{1}{6}|\hat{S}|^2-\frac{1}{6}|\hat{H}_u|^2-\frac{1}{6}|\hat{H}_d|^2-\frac{1}{6}|\hat{C}|^2+\left(-\frac{1}{4}\chi \hat{H}_u\cdot \hat{H}_d +{\rm{h.c.}}\right)+\frac{1}{3}{\rm{Re}}\hat{T}\\ &=\frac{1}{2}-\frac{1}{6}|\hat{S}|^2-\frac{1}{6}|\hat{H}_u|^2-\frac{1}{6}|\hat{H}_d|^2-\frac{1}{6}|\hat{C}|^2-\frac{1}{12}\sigma^2, \label{frameredef}
\end{align}

which can be rewritten as 
\begin{align}
\left(1+\frac{1}{\sqrt{6}} \sigma\right)^2 +\left(-\frac{1}{2}\chi \hat{H}_u\cdot \hat{H}_d +{\rm{h.c.}}\right) +\frac{2}{3}{\rm{Re}}\hat{T} =1-\frac{1}{6}\sigma^2.\label{def_sigma}
\end{align}

Now we use the constraint equation  (\ref{def_sigma}) to remove ${\rm{Re}}\hat{T}$, promoting $\sigma$ to be dynamical. 

\begin{align}
\nonumber \mathcal{L}_{LS}/\sqrt{-g}=&\frac{1}{2}\left(1-\frac{1}{3}|\hat{S}|^2-\frac{1}{3}|\hat{H}_u|^2-\frac{1}{3}|\hat{H}_d|^2-\frac{1}{3}|\hat{C}|^2-\frac{1}{6}\sigma ^2\right)R\\
&- |\partial_{\mu}\hat{S}|^2- |\partial_{\mu}\hat{H}_u|^2- |\partial_{\mu}\hat{H}_d|^2- |\partial_{\mu}\hat{C}|^2-\frac{1}{2}(\partial_{\mu}\sigma)^2 +\Omega\mathcal{A}_{\mu}^2-V_{LS},\label{LS_frame}
\end{align}

where the scalar potential~$\eqref{V_LS}$ reads
\begin{align}
\nonumber V_{LS}^F=&|\lambda \hat{H}_u\cdot \hat{H}_d+\rho \hat{S}^{2}|^{2}+\lambda^2| \hat{S} |^{2}(|\hat{H}_u|^2+|\hat{H}_d|^2) \nonumber \\
&+ \frac{1}{4\alpha}\left(\sigma^2+\sqrt{6}\sigma-\left(\frac{3}{2}\chi \hat{H}_u\cdot \hat{H}_d +{\rm{h.c.}}\right)\right)^2 \nonumber \\
\nonumber &+ \frac{1}{\alpha}({\rm{Im}}\hat{T})^2+\frac{3}{2}\frac{\chi\lambda}{\sqrt{\alpha}} (\hat{S}\bar{\hat{C}}+\bar{\hat{S}}\hat{C})(|\hat{H}_u|^2+|\hat{H}_d|^2) \nonumber \\
&+\frac{1}{\alpha}|\hat{C}|^2\left\{3+2\sqrt{6}\sigma+\frac{3}{2}\sigma^2+\frac{9}{4}\chi^2(|\hat{H}_u|^2+|\hat{H}_d|^2)\right\}.\label{SP_2explicit}
\end{align}

Once we consider the extra kinetic terms, coming from $\mathcal{A}^2_{\mu}$, with eq.~(\ref{Amu}) in eq.~(\ref{LS_frame}), we find that all the scalar fields including ${\rm Im}\,{\hat T}$ are dynamical.
Thus, there is no unitarity violation up to the Planck scale. 

We recover similar results as in \textbf{chapter \ref{chap_UV}}. Firstly, the scalaron $\sigma$ plays the role of the sigma field, which unitarizes Higgs inflation.
Also, as it was the case in the non-supersymmetric case,  the lagrangian shows local conformal invariance, except for the Planck mass and the scalar potential. 
The kinetic terms for the angular component of the complex scalar fields arise from the term $\Omega A^2_\mu$ in the Lagrangian (\ref{LS_frame}),

\bea
\Omega {\cal A}^2_\mu &=&-\frac{1}{4\Omega}\, (X^0)^4 \bigg[(X^0)^{-1} (\Omega_{\hat i} \partial_\mu {\hat z}^{\hat i}+\Omega_{\hat T}\partial_\mu {\hat T})+(\partial_\mu (X^0)^{-1}) (\Omega_{\hat i} {\hat z}^{\hat i} +2 \Omega_{\hat T}{\hat T})-{\rm c.c.}\bigg]^2 \nonumber \\
&=&-\frac{1}{4\Omega}\, \bigg[ (X^0)^{-1}\Big(({\bar {\hat C}}\partial_\mu {\hat C} +{\bar {\hat S}}\partial_\mu {\hat S}+{\bar {\hat H}}_u \partial_\mu {\hat H}_u+{\bar {\hat H}}_d \partial_\mu {\hat H}_d-{\rm c.c}) -2i \partial_\mu b \Big) \nonumber \\
&&\quad-4ib\,\partial_\mu (X^0)^{-1} \bigg]^2.
\eea

We can define a new real scalar field, $b$,
\bea
{\hat T}-{\bar {\hat T}}-\frac{3}{2}\chi ({\hat H}_u \cdot {\hat H}_d -{\bar {\hat H}}_u\cdot {\bar {\hat H}_d}) =2ib,
\eea 

where we still make use of the constraint equation (\ref{frameredef}),
\bea
\Omega=-6(X^0)^{-2} \bigg(\frac{1}{2}-\frac{1}{6}|\hat{S}|^2-\frac{1}{6}|\hat{H}_u|^2-\frac{1}{6}|\hat{H}_d|^2-\frac{1}{6}|\hat{C}|^2-\frac{1}{12}\sigma^2\bigg).
\eea

We can read the mass of the ${\rm Im {\hat T}}$ field from the scalar potential (\ref{SP_2explicit}),
\bea
V_{LS}^F\supset \frac{1}{\alpha}({\rm{Im}}\hat{T})^2=  \frac{1}{\alpha}\bigg(b-\frac{3}{4}i \chi  ({\hat H}_u\cdot {\hat H}_d -{\bar {\hat H}}_u \cdot {\bar {\hat H}_d}) \bigg)^2.
\eea

In the linear-sigma model frame we have removed the non minimal coupling from the kinetic part of the Lagrangian, and instead, we have introduced the new field $b$ in the scalar potential.
Thus, in order to ensure that the theory remains in the perturbative regime, we require

\bea
\frac{\chi}{\alpha}\lesssim 1, \qquad  \frac{\chi^2}{\alpha}\lesssim 1.
\eea
 
These angular components, that are the responsible for linearizing Higgs inflation, are decoupled during inflation and do not affect the dynamics.

\section{Higgs-Sigma inflation}\label{sec:susyhiggsinflation}
The supergravity set-up requires the inclusion of new scalar fields that could alter the inflationary dynamics.
In this section we derive the effective action for slow-roll inflation and study under which conditions the potential remains stable.

\subsection{Effective action for inflation}

We start from the effective supergravity lagrangian in the Jordan frame, as we described in \textbf{section~\ref{CJF}}.
The only degrees of freedom that are relevant for inflation are the scalaron ${\rm{Re}}T$ and the neutral Higgs component, $h$, from $H_u^{0}\rightarrow \frac{1}{2}h$ and $H_d^{0}\rightarrow \frac{1}{2}h$.
The Lagrangian in the Einstein frame is then
\begin{align}
\nonumber \mathcal{L}/\sqrt{-g}=&\frac{1}{2}R-\frac{1}{2}\frac{\big(1+\xi (1+6\xi)h^2+\frac{2}{3}{\rm{Re}}T\big)}{(1+\xi h^2+\frac{2}{3}{\rm{Re}}T)^2}(\partial_{\mu}h)^2 -\frac{1}{3}\frac{1}{(1+\xi h^2+\frac{2}{3}{\rm{Re}}T)^2}(\partial_{\mu}{\rm{Re}}T)^2\\
&-\frac{2\xi h}{(1+\xi h^2+\frac{2}{3}{\rm{Re}}T)^2}\partial_{\mu}h\partial^{\mu}{\rm{Re}}T-V(h,{\rm{Re}}T),
\end{align}

where we defined the effective non-minimal coupling, 
\begin{align}
\xi \equiv -\frac{1}{6}+\frac{\chi}{4},
\end{align}

 and Einstein frame potential is given by 
\begin{align}
V(h,{\rm{Re}}T)=\frac{1}{(1+\xi h^2+\frac{2}{3}{\rm{Re}}T)^2}\left(\frac{1}{16} \lambda^{2} h^{4}+\frac{1}{\alpha}({\rm{Re}}T)^{2}\right).    
\end{align}

The Einstein frame Lagrangian is equivalent to the one in the non-supersymmetric Higgs-$R^2$ inflation in refs.~\cite{Ema:2017rqn,Ema:2020zvg,Lee:2021dgi}. 
In order to compare the two models, we take $\alpha \rightarrow 1/\kappa_1$ and $\lambda \rightarrow 2 \sqrt{\lambda}$.
In the case of the non supersymmetric case, $g=g'=0$.
In the linear-sigma frame, with the redefinition of the fields as 

\begin{align*}
    {\hat z}^i&=X^0 z^i,\\
     {\hat T}&=(X^0)^2 T,\\
     \hat{h}&=X^0h=\left(1+\frac{1}{\sqrt{6}} \sigma\right)h,
\end{align*}

and using eq.~$\eqref{def_sigma}$, we find the Einstein frame Lagrangian as 

\begin{align}
    \nonumber \mathcal{L}/\sqrt{-g}=&\frac{1}{2}R-\frac{1}{2}\frac{1}{\left(1-\frac{1}{6}\hat{h}^2-\frac{1}{6}\sigma^2\right)^2}\biggl[\left(1-\frac{\sigma^2}{6}\right)(\partial_{\mu}\hat{h})^2+\left(1-\frac{\hat{h}^2}{6}\right)(\partial_{\mu}\sigma)^2\\
    &+\frac{1}{3}\hat{h}\sigma \partial_{\mu}\hat{h}\partial^{\mu}\sigma\biggr]-V(\hat{h},\sigma),\label{L_inf_hat}
\end{align}

with the scalar potential is the same as in eq.~(\ref{effpot}). Hence, even if we have extra scalar fields due to SUSY, the inflationary dynamics of this model is the same as we studied in \textbf{chapter \ref{chap_UV}}.

\subsection{Decoupling of heavy scalars}

So far we have assume that the only fields playing a role during inflation are  the radial component of the neutral Higgs fields and the scalaron. However, it is necessary to analyze the full Einstein potential, in order to check that the heavy fields are indeed decoupled from the inflation field and do not affect the inflationary trajectory.
For this purpose, we parametrize the MSSM Higgs fields as follows,
\begin{align}
H_u^0=\frac{1}{\sqrt{2}}h\,{\rm{cos}}\beta \, e^{i\delta_1},\ \ H_d^0=\frac{1}{\sqrt{2}}h\, {\rm{sin}}\beta\,  e^{i\delta_2} 
\end{align}
where we take $h,\beta,\delta_{1,2}$ to be real parameters. Then, the scalar potential in Einstein frame becomes 
\begin{align}
V_E=\frac{V_{LS}}{\left(1-\frac{1}{3}|S|^2-\frac{1}{6} h^{2}-\frac{1}{3}|H_u^+|^2-\frac{1}{3}|H_d^-|^2-\frac{1}{3}|C|^2-\frac{1}{6} {\sigma}^{2}\right)^2}. \label{full_V}   
\end{align}

Expanding the \textit{linear sigma frame} potential, we get
\begin{align}
\nonumber V_{LS}=&\left|-\frac{1}{4} \lambda h^{2} \sin 2 \beta e^{i \gamma}+\lambda H_{u}^{+} H_{d}^{-}+\rho S^{2}\right|^{2}+\lambda^{2}|S|^{2}\left(\frac{1}{2} h^{2}+\left|H_{u}^{+}\right|^{2}+\left|H_{d}^{-}\right|^{2}\right)\\
\nonumber&+\frac{1}{4\alpha}\left(\sigma^2+\sqrt{6}\sigma+\frac{3}{4}\chi h^2 \sin 2\beta \cos \gamma -\frac{3}{2}\chi (H_u^+H_d^-+{\rm{c.c.}})\right)^2+\frac{1}{\alpha}({\rm{Im}}T)^2\\
\nonumber&+\frac{3}{2} \frac{\chi \lambda}{\sqrt{\alpha}}(S\bar{C}+{\rm{c.c.}})\left(\frac{1}{2} h^{2}+\left|H_{u}^{+}\right|^{2}+|H_d^-|^{2}\right)\\
\nonumber&+\frac{1}{ \alpha} |C|^{2}\left\{3+2\sqrt{6}\sigma +\frac{3}{2}\sigma^2+\frac{9}{4} \chi^{2}\left(\frac{1}{2} h^{2}+\left|H_{u}^{+}\right|^{2}+|H_d^-|^{2}\right)\right\}\\
\nonumber &+\frac{g^{2}+g^{\prime 2}}{8}\left(\frac{1}{2}h^{2} \cos 2 \beta+\left|H_{u}^{+}\right|^{2}-|H_d^-|^{2}\right)^{2}\\
&+\frac{g^2}{4} h^{2}\left\{\left|H_{u}^{+}\right|^{2}\sin^2 \beta+\left|H_{d}^{-}\right|^{2}\cos^2 \beta+\left(\frac{1}{2}H_u^+H_d^-\sin 2\beta e^{-i\gamma}+{\rm{c.c.}}\right)\right\},\label{full_VJ}
\end{align}

where $\gamma \equiv \delta_1+\delta_2$.
The scalar potential depends on the phases only through the combination, $\gamma = \delta_1+\delta_2$, so it is independent of the would-be neutral Goldstone boson.

We consider $h$ and $\sigma$ to be slowly varying, and thus we take then to be background fields.
Then, the potential ~$\eqref{full_V}$, with eq.~$\eqref{full_VJ}$, is minimized at

\begin{align}
H_u^+=H_d^-=S=C={\rm{Im}}T=\gamma=0, \ \ {\rm{and}}\ \   \beta=\pi/4. \label{SC}
\end{align}

In order to check the stability of these extrema, we expand the fields up to quadratic order.
\begin{align}
\beta=\frac{\pi}{4}+\tilde{\beta},\ \   X=0+\tilde{X}, \ \ {\rm{with}}\ \ X=\{H_u^+,H_d^-,S,C,{\rm{Im}}T,\gamma \}   
\end{align}

We denote fluctuations of the field with a tilde. Now we proceed to discuss every field individually.

\subsubsection*{Stabilization of $\beta$}

The Lagrangian for $\beta$ is given by

\begin{align}
    - \frac{h^2}{2\Delta}(\partial_{\mu} \tilde{\beta})^2- \frac{1}{2}V_{\beta\beta}\tilde{\beta}^2,\label{beta}
\end{align}
    where, denoting second derivatives as $V_{\beta\beta}$, 
\begin{align}
    &V_{\beta\beta}=\left[-\frac{\lambda^2}{4}h^4-\frac{3\chi h^2}{4\alpha}\left(\sigma^2+\sqrt{6}\sigma+\frac{3\chi h^2}{4}\right)+\frac{g^{\prime 2}+g^2}{8}h^4\right]\frac{2}{\Delta^2},\\
    &\Delta\equiv 1-\frac{h^2}{6}-\frac{\sigma^2}{6}.
\end{align}

We now use the explicit form of the background fields (\ref{hmin}) and (\ref{cansigmafield}),

\bea
\sigma &\simeq -\sqrt{6}\left(1-2e^{-\frac{2}{\sqrt{6}}\phi}\right), \label{sigma_app}\\
h^2 &\simeq \frac{144\frac{\xi}{\alpha}}{\lambda^2+6\frac{\xi}{\alpha}(6\xi+1)}e^{-\frac{2}{\sqrt{6}}\phi}\label{h_app}.
\eea

and find the canonically normalized mass of $\beta$,
\begin{align}
m_{\beta}^2 = \frac{3\frac{\lambda^2}{\alpha}+9(g^{\prime 2}+g^2)\frac{\xi}{\alpha}}{\lambda^2+36\frac{\xi^2}{\alpha}} = 4H^2\left(1+\frac{3\xi}{\lambda^2}(g^{\prime 2}+g^2)\right),
\end{align}
where we used eq.~$\eqref{Inf_scale}$.
Then, we see that mass of $\beta$ is larger than the Hubble scale for $\xi (g^{\prime 2}+g^2)/\lambda^2\gg 1$, and thereby it is stabilized and decoupled during inflation.

\subsubsection*{Stabilization of the charged Higgs}

The quadratic Lagrangian for the charged Higgs sector is given by
\begin{align}
-\frac{1}{\Delta}|\partial_{\mu} \tilde{H}_{u}^{+}|^{2}-\frac{1}{\Delta}|\partial_{\mu} \tilde{H}_{d}^{-}|^{2}    -\left(\tilde{H}_{u}^{+*}, \tilde{H}_{d}^{-}\right)\left(\begin{array}{ll}
V_{++} & V_{+-} \\
V_{+-} & V_{--}
\end{array}\right)\left(\begin{array}{c}
\tilde{H}_{u}^{+} \\
\tilde{H}_{d}^{-*}
\end{array}\right),
\end{align}
where the matrix elements are defined as
\begin{align}
&V_{++} =V_{--}=\left[\frac{2V}{3}\Delta+\frac{g^2}{8}h^2\right]\frac{1}{\Delta^2}, \\
&V_{+-}=\left[-\frac{\lambda^2}{4}h^2-\frac{3\chi }{4\alpha}\left(\sigma^2+\sqrt{6}\sigma+\frac{3\chi h^2}{4}\right)+\frac{g^2}{8}h^2\right]\frac{1}{\Delta^2},
\end{align}
and $V$ is the effective scalar potential given in eq.~$\eqref{totalpotsigma}$,
\begin{align}
V=\left[\frac{\lambda^2}{16}h^4+\frac{1 }{4\alpha}\left(\sigma^2+\sqrt{6}\sigma+\frac{3\chi h^2}{4}\right)^2\right]\frac{1}{\Delta^2}.    
\end{align}

Diagonalizing the mass matrix we find the mass eigenvalues as
\begin{align}
&\left\{0,\ \  \frac{3\frac{\lambda^2}{\alpha}+9g^2\frac{\xi}{\alpha}}{\lambda^2+36\frac{\xi^2}{\alpha}}=4H^2\left(1+\frac{3g^2\xi}{\lambda^2}\right)\right\}.
\end{align}

where the massless field, i.e., the would-be Goldstone boson, gets eaten by the charged gauge boson.
Then, the massive charged Higgs gets a mass of order the Hubble scale or beyond for $g^2\xi\lambda^2\gg 1$, decoupling during inflation.

\subsubsection*{Stabilization of $\gamma$ and ${\rm{Im}}T$}

In the case of $\gamma$ and ${\rm{Im}}T (\equiv \tau)$ we have to take into account their kinetic mixing,
\begin{align}
-\frac{1}{2}\left(\partial_{\mu} \tilde{\gamma}, \partial_{\mu}\tilde{\tau}\right)\left(\begin{array}{ll}
a & c \\
c & b
\end{array}\right)\left(\begin{array}{l}
\partial^{\mu} \tilde{\gamma} \\
\partial^{\mu} \tilde{\tau}
\end{array}\right) -\frac{1}{2}V_{\gamma\gamma}\tilde{\gamma} ^2-\frac{1}{2}V_{\tau\tau}\tilde{\tau}^2,\label{gamma_tau}   
\end{align}
where 
\begin{align}
&a=\frac{h^2}{4\Delta}\left[1+\frac{h^2}{6\Delta}\left(1-\frac{3\chi}{2}\right)^2\right], \ \ b=\frac{2}{3\Delta^2},\ \ c=-\frac{h^2}{6\Delta^2}\left(1-\frac{3\chi}{2}\right),\\
&V_{\gamma\gamma}=-\frac{3\chi h^2}{8\alpha}\left(\sigma^2+\sqrt{6}\sigma+\frac{3\chi h^2}{4}\right)\frac{1}{\Delta^2}, \ \ V_{\tau\tau}=\frac{2}{\alpha \Delta^2}.
\end{align}
Since $ab-c^2=\frac{h^2}{6\Delta^3}>0$ and $a+b>0$, there is no ghost mode for the background fields.

Next, we canonicalize the kinetic matrix of eq.~$\eqref{gamma_tau}$, 
\begin{align}
\left(\begin{array}{c}
\tilde{\gamma} \\
\tilde{\tau}
\end{array}\right)=\frac{1}{\sqrt{2}}\left(\begin{array}{ll}
\left(\frac{\sqrt{\frac{b}{a}}}{\sqrt{ab}+c}\right)^{1/2} & -\left(\frac{\sqrt{\frac{b}{a}}}{\sqrt{ab}-c}\right)^{1/2} \\
\left(\frac{\sqrt{\frac{a}{b}}}{\sqrt{ab}+c}\right)^{1/2}  & \left(\frac{\sqrt{\frac{a}{b}}}{\sqrt{ab}-c}\right)^{1/2} 
\end{array}\right)\left(\begin{array}{c}
\tilde{\gamma}^{\prime} \\
\tilde{\tau}^{\prime}
\end{array}\right)\equiv \mathcal{M}\left(\begin{array}{c}
\tilde{\gamma}^{\prime} \\
\tilde{\tau}^{\prime}
\end{array}\right).
\end{align}

For $\sqrt{ab}-c>0$, we get
\begin{align}
-\frac{1}{2}\left(\partial_{\mu}\tilde{\gamma}^{\prime}\right)^{2}-\frac{1}{2}\left(\partial_{\mu}{\tilde{\tau}}^{\prime}\right)^{2} -\frac{1}{2}\left(\tilde{\gamma}^{\prime} , \tilde{\tau}^{\prime}\right)\mathcal{M}^{T}\left(\begin{array}{ll}
V_{\gamma\gamma} & 0 \\
0 & V_{\tau\tau}
\end{array}\right)\mathcal{M}  \left(\begin{array}{c}
\tilde{\gamma}^{\prime} \\
\tilde{\tau}^{\prime}
\end{array}\right) .
\end{align}

Diagonalizing the mass terms, we find the mass eigenvalues
\begin{align}
m^2_{\pm}&=  \frac{1}{2} \frac{V_{\gamma\gamma} b+V_{\tau\tau} a}{a b-c^{2}} \pm \frac{1}{2} \bigg[\bigg(\frac{c}{a b-c^{2}}\bigg)^{2}\bigg(V_{\gamma\gamma} \sqrt{\frac{b}{a}}+V_{\tau\tau} \sqrt{\frac{a}{b}}\bigg)^{2} \nonumber \\
&\qquad+\frac{1}{a b-c^{2}}\bigg(-V_{\gamma\gamma} \sqrt{\frac{b}{a}}+V_{\tau\tau} \sqrt{\frac{a}{b}}\bigg)^{2}\bigg]^{1/2}.  \label{mass_tau_gamma}
\end{align}

We now make use of ~$\eqref{sigma_app}$ and $\eqref{h_app}$,
\begin{align}
&m^2_+= \frac{3+18\xi}{\alpha}=4(6\xi+1)\left(1+\frac{36\xi^2}{\alpha\lambda^2}\right)H^2,\\
&m^2_-=\frac{3 \lambda ^2}{\alpha  \lambda ^2+36 \xi ^2}=4H^2.
\end{align}

Again, both of mass eigenvalues in the $\gamma$ and ${\rm{Im}}T$ sector are larger than the Hubble scale.
Thus, the directions from $\gamma$ and ${\rm{Im}}T$ are stable too.

\subsubsection*{Stabilization of $S$ and $C$}

Lastly, we analyze the $S$-$C$ direction. In this case, the Lagrangian contains mass-mixing for the $S$ and $C$ fields, 
\begin{align}
-\frac{1}{\Delta}|\partial_{\mu}\tilde{S}|^2-\frac{1}{\Delta}|\partial_{\mu}\tilde{C}|^2-V_{S\bar{S}}|\tilde{S}|^2-\frac{1}{2}V_{SS}(\tilde{S}^2+{\rm{c.c.}})-V_{S\bar{C}}(\tilde{S}\tilde{C}^*+{\rm{c.c.}}) -V_{C\bar{C}}|\tilde{C}|^2,   
\end{align}

with
\begin{align}
&V_{S\bar{S}}= \left[\frac{\lambda^2}{2}h^2+\frac{2V}{3}\Delta\right]\frac{1}{\Delta^2},\ \ V_{SS}=-\frac{\lambda \rho}{2\Delta^2}h^2,\ \ V_{S\bar{C}}=\frac{3\chi \lambda }{4\sqrt{\alpha}\Delta^2}h^2,\label{V_ss}\\
&V_{C\bar{C}}=\left[\frac{1}{\alpha}\left(3+2\sqrt{6}\sigma+\frac{3}{2}\sigma^2+\frac{9\chi^2 h^2}{8}\right)+\frac{2V}{3}\Delta\right]\frac{1}{\Delta^2}.\label{V_cc}
\end{align}

We follow the same steps as for the previous cases and introduce the canonically normalized fields, $\tilde{S}'$ and $\tilde{C}'$.
Then, we divide them into their real and imaginary components, 
\begin{align}
\frac{1}{\sqrt{\Delta}}\tilde{S}=\tilde{S}'=\frac{1}{\sqrt{2}}({\rm{Re}}\tilde{S}'+i{\rm{Im}}\tilde{S}'), \ \ \frac{1}{\sqrt{\Delta}}\tilde{C}=\tilde{C}'=\frac{1}{\sqrt{2}}({\rm{Re}}\tilde{C}'+i{\rm{Im}}\tilde{C}').      
\end{align}
In the above basis, the mass matrix is given by
\begin{align}
-\frac{1}{2}\left({\rm{Re}}\tilde{S}^{\prime}, {\rm{Re}}\tilde{C}^{\prime}\right)\left(\begin{array}{ll}
V_{S\bar{S}}+V_{SS} & V_{S\bar{C}} \\
V_{S\bar{C}} & V_{C\bar{C}}
\end{array}\right)\Delta\left(\begin{array}{c}
{\rm{Re}}\tilde{S}^{\prime} \\
{\rm{Re}}\tilde{C}^{\prime}
\end{array}\right),\label{Re}
\end{align}

and

\begin{align}
-\frac{1}{2}\left({\rm{Im}}\tilde{S}^{\prime}, {\rm{Im}}\tilde{C}^{\prime}\right)\left(\begin{array}{ll}
V_{S\bar{S}}-V_{SS} & V_{S\bar{C}} \\
V_{S\bar{C}} & V_{C\bar{C}}
\end{array}\right)\Delta\left(\begin{array}{c}
{\rm{Im}}\tilde{S}^{\prime} \\
{\rm{Im}}\tilde{C}^{\prime}
\end{array}\right).\label{Im}
\end{align}

We diagonalize the mass matrices in eqs.~$\eqref{Re}$ and $\eqref{Im}$, and obtain the following eigenvalues
\begin{align}
&m^2_{1,2}=\frac{\Delta}{2}\left[V_{S\bar{S}}+V_{SS}+V_{C\bar{C}} \pm \sqrt{\left(V_{S\bar{S}}+V_{SS}-V_{C\bar{C}}\right)^{2}+4 V_{S\bar{C}}^{2}}\right],\label{m12}
\end{align}
and
\begin{align}
m^2_{3,4}=\frac{\Delta}{2}\left[V_{S\bar{S}}-V_{SS}+V_{C\bar{C}} \pm \sqrt{\left(V_{S\bar{S}}-V_{SS}-V_{C\bar{C}}\right)^{2}+4 V_{S\bar{C}}^{2}}\right].\label{m34}
\end{align}

We can explicitly write these masses as 
\begin{align}
&m_{1,2}^2=\frac{18\xi\left( 6 \xi  (6 \xi +1)+\alpha  \lambda  (\lambda -\rho )\right)\pm 3f(\lambda,\alpha,\xi,\rho)}{2 \alpha  \left(\alpha  \lambda ^2+36 \xi ^2\right)},    \\
&m_{3,4}^2=m_{1,2}^2 \ \ {\rm{with}}\ \ \rho\rightarrow -\rho,
\end{align}
where
\begin{align}
f(\lambda,\alpha,\xi,\rho)  &=\Big[\alpha ^2 \lambda ^2 (6 \lambda  \xi +\lambda -6 \xi  \rho )^2+72 \alpha  \lambda  (6 \xi +1) \xi ^2 (6 \lambda  \xi +\lambda +6 \xi  \rho ) \nonumber \\
&\qquad+1296 (6 \xi +1)^2 \xi ^4\Big]^{1/2} . 
\end{align}

\begin{figure}[t]
    \begin{minipage}{0.5\hsize}
     \begin{center}
      \includegraphics[width=70mm]{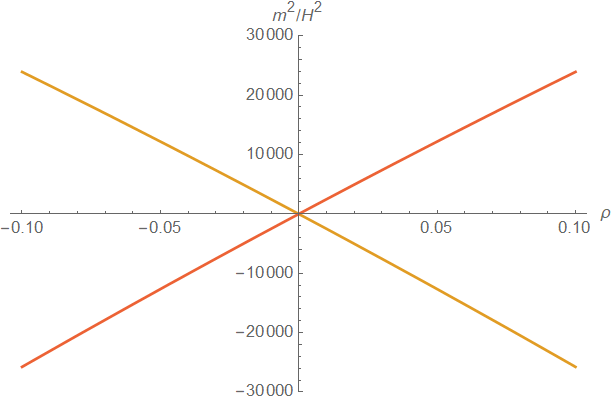}
     \end{center}
    \end{minipage}
    \begin{minipage}{0.5\hsize}
     \begin{center}
      \includegraphics[width=70mm]{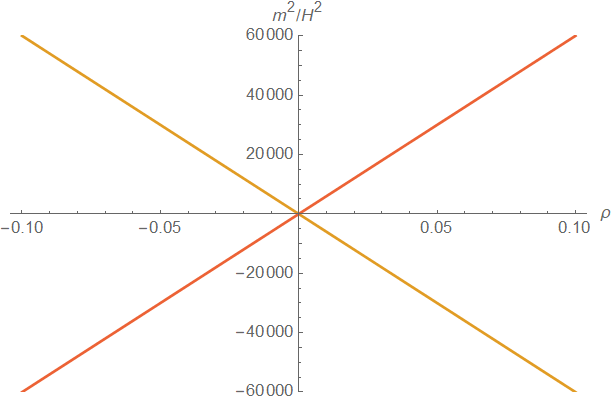}
     \end{center}
    \end{minipage}
     \caption{We plot the squared mass eigenvalues for the lighter states of singlet scalars, $S$ and $C$, as a function of the self-coupling $\rho$: $m^2_2$ (yellow) and $m^2_4$ (orange).
     We took $(\lambda,\xi,\alpha)=(0.5,10^4,10^{10})$ for $R^2$-like inflation on the left plot and  $(\lambda,\xi,\alpha)=(4\times 10^{-5},1,10^{2})$ for Higgs-like inflation on the right plot.
     }
     \label{fig:SC_mass}
   \end{figure}

In Fig.~\ref{fig:SC_mass}, we show the behavior of the mass eigenvalues in the $S$-$C$ sector. 
We take two benchmark examples, one representative of $R^2$-like inflation and one for Higgs-like inflation.
We observe that the heavier mass eigenvalues, in this case, $m^2_1$ and $m^2_3$, always remain positive. However, the lighter states,  $m^2_2$ in yellow and $m^2_4$ in orange, can take negative values.
This result is independent of the choice of the parameter $\rho$, confirming  a tachyonic instability that spoils the inflationary trajectory.

This tachyonic instability problem is a well-known problem arising from the supergravity embedding of both Higgs and Starobinsky inflation~\cite{Ferrara:2010yw,Ferrara:2010in,Kallosh:2013lkr}. 
A possible solution is to add quartic couplings for  $S$ and $C$ to the frame function, which pushes the tachyonic direction above the Hubble scale~\cite{Lee:2010hj,Ferrara:2010in,Kallosh:2013lkr}. 
Then,
\begin{align}
\Delta \Omega=-\zeta_s|S|^4-\zeta_c|C|^4\label{zeta_c}-\zeta_{sc}|S|^2|C|^2  
\end{align}
where $\zeta_s$, $\zeta_c$ and $\zeta_{sc}$ are real parameters.
In particular,the term $-\zeta_c|C|^4$ leads to the dual term
\begin{align}
-[\zeta_c\alpha^2 |X^0|^{-2}(\bar{\mathcal{R}}\mathcal{R})^2]_D.
\end{align}
These extra couplings can naturally arise from the renormalizable couplings of $S$ or $C$ to vector-like heavy multiplets \cite{Lee:2010hj}. 
Then, the scalar potential get corrections to the mass terms for $S$ and $C$ \cite{Kallosh:2010ug,Kallosh:2010xz,Kallosh:2011qk}.
\begin{align}
&\Delta V_{S\bar{S}}=\frac{\zeta_s\lambda^2}{4} \frac{h^4}{\Delta^2} \left(1+\frac{\sigma}{\sqrt{6}}\right)^{-2}, \\
&\Delta V_{C\bar{C}}=\frac{\zeta_c}{\alpha}\frac{1}{\Delta^2} \left(1+\frac{\sigma}{\sqrt{6}}\right)^{-2}\left(\sigma^2+\sqrt{6}\sigma+\frac{3\chi h^2}{4}\right)^2, \\
&\Delta V_{S\bar{C}}=\frac{\zeta_{sc}\lambda}{2\sqrt{\alpha}} \frac{1}{\Delta^2}\left(1+\frac{\sigma}{\sqrt{6}}\right)^{-2}h^2\left(\sigma^2+\sqrt{6}\sigma+\frac{3\chi h^2}{4}\right).
\end{align}

\begin{figure}[t]
    \begin{minipage}{0.5\hsize}
     \begin{center}
      \includegraphics[width=70mm]{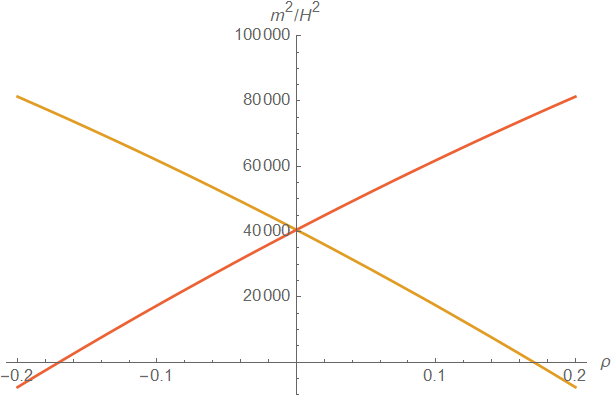}
     \end{center}
    \end{minipage}
    \begin{minipage}{0.5\hsize}
     \begin{center}
      \includegraphics[width=70mm]{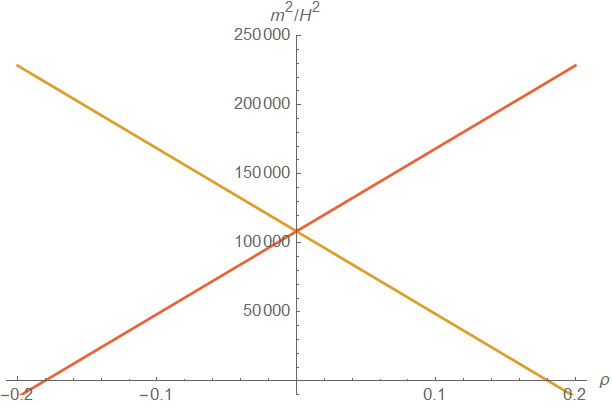}
     \end{center}
    \end{minipage}
       \caption{Squared mass eigenvalues for the lighter states of singlet scalars after including extra couplings in the frame function. }
     \label{fig:five}
   \end{figure}

In Fig.~\ref{fig:five}, we show how the situation changes once we introduced the extra couplings on the frame function.
The states that before were tachyonic now keep a positive mass squared during inflation.
We took the particular values  $(\zeta_s,\zeta_c)=(3,0.4)$ and $\zeta_{sc}=0$.

Now, we study the allow parameter space once the extra couplings have been added.
We see that a large value of $\zeta_{sc}$ is not favored, since it tends to lower the eigenvalues $m^2_{2}$ in eq.~$\eqref{m12}$ and $m^2_{4}$ in eq.~$\eqref{m34}$.
For this reason, we restrict to including only a nonzero $\xi_s$ in the frame function.
Still, a nonzero $\zeta_c$ should be included in order to achieve Supersymmetry breaking, realizing a sufficiently small cosmological constant. 

\begin{figure}[t]
 
    \begin{minipage}{0.5\hsize}
     \begin{center}
      \includegraphics[width=70mm]{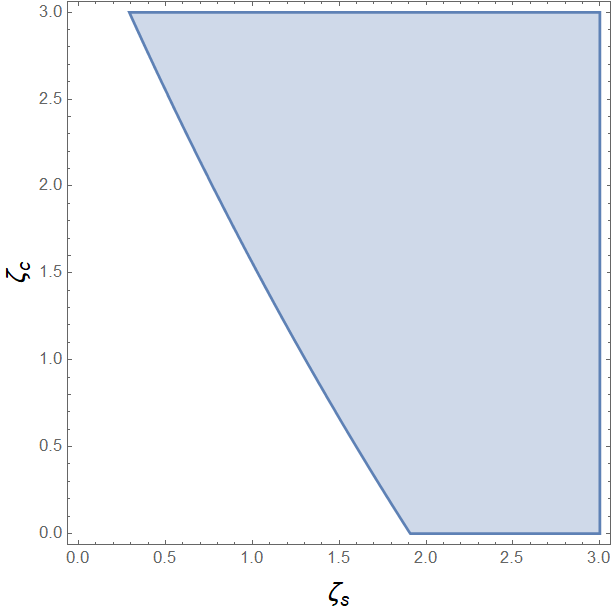}
     \end{center}
    \end{minipage}
    \begin{minipage}{0.5\hsize}
     \begin{center}
      \includegraphics[width=70mm]{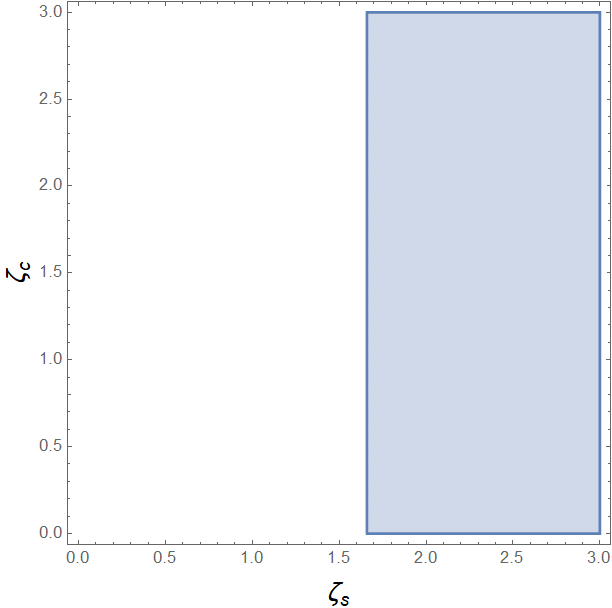}
     \end{center}
    \end{minipage}
       \caption{Parameter space for $\zeta_s$ and $\zeta_c$ satisfying $m_2>H$. We took  $(\lambda,\xi,\alpha)=(0.5,10^4,10^{10})$ for $R^2$-like inflation on left and  $(\lambda,\xi,\alpha)=(4\times 10^{-5},1,10^{2})$ on right for Higgs-like inflation. In both figures, we fixed $\rho=0.1$ and $\phi=10$.}
     \label{fig:six}
   \end{figure}

\section{Supersymmetry breaking}\label{sec:susybreak}
In this section, we study the phenomenology of SUSY breaking. In our model, once inflation ends, 
the VEVs of the sigma and Higgs fields vanish in the vacuum, and SUSY is unbroken. Thus, we need to extend our model
in order to accommodate Supersymmetry breaking. Firstly, we introduce extra curvature terms, which allows us to break SUSY without the introduction of a hidden sector.
Then, we propose a different alternative in which we introduce an extra chiral singlet.

\subsection{Higher curvature terms for SUSY breaking}

Following the idea of~\cite{Dalianis:2014aya}, we introduce extra curvature terms
\begin{align}
S=[|X^0|^2f(\mathcal{R}/X^0,\bar{\mathcal{R}}/\bar{X}^{\bar{0}})]_D ,  
\end{align}
where
\begin{align}
f=-3+  \alpha|\mathcal{R}/X^0|^2 -\gamma_c \alpha \left(\mathcal{R}/X^0+{\rm{c.c.}}\right)-\zeta_c\alpha^2|\mathcal{R}/X^0|^4. \label{def_f}
\end{align}
Including the term $\zeta_c$ was motivated in the previous section for the stability of the inflationary potential.
For SUSY breaking, we also need to include a linear term, proportional to $\gamma_c$, in $\mathcal{R}$. 
The frame function and the superpotential in the $C$ and $T$ sector are now given by
\bea
 \Omega &=&-3+(T+{\bar T})+|C|^2-\gamma_c (C+\bar{C})-\zeta_c |C|^4,\label{dual_2} \\
W&=&\frac{1}{\sqrt{\alpha}}\, TC.
\eea

The local Minkowski minimum with SUSY breaking \cite{Dalianis:2014aya} is defined by  
\bea
\langle C \rangle &=&\frac{1}{54\zeta_c} \,\Big(1+\sqrt{1+324 \zeta_c} \Big)\equiv c_0, \\ 
\langle T \rangle &=& \gamma_c c_0 + c^2_0 (1-6\zeta_c c^2_0)\equiv t_0,
\eea
together with the constraint
\bea
\gamma_c=-c_0 + \frac{2}{c_0} \bigg(1+\frac{1}{3} c^2_0\bigg)
\eea
and $9-36 \zeta_c c^2_0>0$.  The latter condition requires $\zeta_c<0.48$, while we need $\zeta_c>0.15$ for $c_0<1$. 
Then, the linear coupling  must satisfy $1.7<\gamma_c<2.5$.
We obtain the $F$-terms as
\bea
F^C &=& -e^{K/2} K^{C{\bar C}} (D_C W)^\dagger - e^{K/2} K^{C{\bar T}} (D_T W)^\dagger, \\
F^T &=&  -e^{K/2} K^{T{\bar T}} (D_T W)^\dagger - e^{K/2} K^{T{\bar C}} (D_C W)^\dagger,
\eea
with
\bea
D_C W &=& \frac{1}{9\sqrt{\alpha} }\, (3+c^2_0) \bigg(\frac{246-192c_0+55c^2_0+27 c^3_0}{66-5c^2_0}\bigg), \\
D_T W &=&  \frac{1}{9\sqrt{\alpha} }\, c_0\bigg(\frac{39-14c^2_0}{66-5c^2_0}\bigg),
\eea
and the gravitino mass is given by
\bea
m_{3/2}= \frac{243}{8}\sqrt{\frac{6}{\alpha}}\,\cdot \frac{c_0(3+c^2_0)^{1/2}}{(66-5c^2_0)^{3/2}}. 
\eea
Here, $c_0$ is subject to $c_0<\sqrt{66/5}$. 
Then, the F-terms are of order $F^C\sim F^T\sim M_P m_{3/2}$ and the gravitino mass satisfies $m_{3/2}\sim M_P/\sqrt{\alpha}$.
The perturbativity constraints from Higgs inflation lead to $\alpha\lesssim 10^{10}$, meaning that $m_{3/2}\gtrsim 10^{13}\,{\rm GeV}$.
Hence, for this scenario, SUSY is predicted to be broken at high energies.

\subsection{O'Raifeartaigh model for SUSY breaking}

In this subsection, we include an extra singlet chiral scalar $\Phi$.
The frame function and the superpotential are of the O'Raifeartaigh type introduced in \textbf{chapter \ref{chap_partphys}},
\bea
\Omega&=& -3-(T+{\bar T})+|C|^2+|\Phi|^2 -\gamma\, |\Phi|^4, \label{framephi} \\
W &=& \frac{1}{\sqrt{\alpha}}\, TC + \kappa\, \Phi + g\,  \Phi C^2 +\lambda \Phi^3 +\kappa' C + g' \Phi^2 C + \lambda' C^3 \label{Osuper}
\eea
We note that an $Z_{4R}$ R-symmetry can ensure the above form of the superpotential, given the charge assignments, $R[\Phi]=R[C]=+2$ and $R[T]=0$.
The dual scalar superfield $T$ is neutral under the $Z_{4R}$ R-symmetry.

Let's study the simpler case with $\lambda=\kappa'=\lambda'=g'=0$.
Then, the minima will be at $C=T=0$, with a non-zero F term $F_\Phi=\kappa$. 
The  $\Phi$  direction can be stabilized by either loop corrections \cite{Intriligator:2007py, Shih:2007av} or by including higher order terms \cite{Kitano:2006wz} in the K\"ahler potential. 
So, for $\gamma\neq 0$, the squared mass $m^2_\Phi=4\gamma\kappa^2/M^2_P$.
The coupling between $\Phi$ and $C$ gives rise to a mass splitting in the $T$ and $C$ sector, 
\bea
m^2_{s,\pm} &=& \frac{M^2_P}{\alpha} \pm 2g\kappa, \label{simplescalar} \\
m_f &=&  \frac{M_P}{\sqrt{\alpha}}. 
\eea
We see that the SUSY breaking effects are controlled by $\kappa$, so, in contrast to the previous section,
we can get now get a low-scale for SUSY breaking.
In order to be consistent with the smallness of the cosmological constant, the gravitino mass must be given by  $m_{3/2}=|F_\Phi|/(\sqrt{3}M_P)=\kappa/(\sqrt{3}M_P)$.

We can further add general couplings in the superpotential~(\ref{Osuper}).
This would shift the local minimum to $C=0$, $T=-\sqrt{\alpha}\,\kappa'$, due to a nonzero $\kappa'$.
However, the pseudo-flat direction can be still stabilized at $\Phi=0$ due to the extra quartic term for $\Phi$ in the frame function.
Then, the F-terms are given by $F_C=F_T=0$ and $F_\Phi=\kappa$, as it was for $\lambda=\kappa'=\lambda'=g'=0$.
Taking the couplings in the superpotential~(\ref{Osuper}) to be real and including the quartic correction for $\Phi$ in the frame function, we find 
\bea
M^2_R =\left(\begin{array}{cc} \frac{M^2_P}{\alpha}+2g \kappa  & 2g' \kappa \\ 2g' \kappa & 6\lambda \kappa+m^2_\Phi \end{array}\right),  \quad 
M^2_I =\left(\begin{array}{cc} \frac{M^2_P}{\alpha}-2g \kappa & -2g' \kappa  \\ -2g' \kappa & -6\lambda \kappa+m^2_\Phi \end{array}\right)
\eea 
with
\bea
m^2_\Phi\equiv \frac{4\gamma \kappa^2}{M^2_P}.
\eea
The result in eq.~(\ref{simplescalar}) can be recovered by taking $g'=\lambda=0$ and $\gamma=0$.
For general extra couplings, however, we obtain 
\begin{align}
m^2_{s1,s2} &=&\frac{1}{2} \bigg[\frac{M^2_P}{\alpha}+2(g+3\lambda)\kappa+m^2_\Phi \pm \sqrt{\Big(\frac{M^2_P}{\alpha}+2(g-3\lambda)\kappa-m^2_\Phi \Big)^2+16g^{\prime 2}\kappa^2} \bigg], \\
m^2_{s3,s4} &=&\frac{1}{2} \bigg[\frac{M^2_P}{\alpha}-2(g+3\lambda)\kappa+m^2_\Phi  \pm \sqrt{\Big(\frac{M^2_P}{\alpha}-2(g-3\lambda)\kappa-m^2_\Phi \Big)^2+16g^{\prime 2}\kappa^2} \bigg].
\end{align}
In order to keep $m^2_{s3,s4}$ positive, we must satisfy the conditions
\bea
 \frac{M^2_P}{\alpha}>2g \kappa, \quad m^2_\Phi > 6\lambda \kappa, \label{stability1}
\eea
and
\bea
\bigg(\frac{M^2_P}{\alpha}-2g \kappa\bigg)\Big(m^2_\Phi-6\lambda \kappa  \Big)> 4g^{\prime 2} \kappa^2.
\eea
This also leads to positive $m^2_{s1,s2}$. 
In particular, the second condition in  eq.~(\ref{stability1}) corresponds to
\bea
\frac{\gamma}{\lambda}> \frac{3M^2_P}{2\kappa}.
\eea
Therefore, under these conditions, the local minimum for SUSY breaking remains stable.

\subsection{Comments on soft masses in the visible sector}

The non-minimal coupling and the dual super fields from the $R^2$ terms give extra contributions to the $\mu$ term,
proportional to the gravitino mass \cite{Lee:2010hj},
\bea
\mu=\lambda\langle {\tilde S}\rangle + \frac{3}{2}\,\chi m_{3/2} -\frac{1}{2}\chi K_{\bar I} {\bar F}^{\bar I}. \label{muterm}
\eea
Where we have rescaled the superfields by ${\tilde H}_{u,d}=e^{{\cal K}/6} H_{u,d}$ and ${\tilde S}=e^{{\cal K}/6} S$, etc.
The second term in eq.~(\ref{muterm}) arises due to the presence of the non-minimal coupling $\chi$ ~\cite{Lee:2010hj},  while the third term is called the Giudice-Masiero contribution \cite{Giudice:1988yz}.
Therefore, the $\mu$ term has a naturally small value, due to not only the VEV of the NMSSM singlet $S$, but also for consistency with the gravitino mass and the SUSY breaking scale,
since, for large $\chi$, the $\mu$ term is typically much larger than the gravitino mass \cite{Lee:2010hj}.

In our  Jordan frame construction of supergravity, the visible sector and the hidden sector composed of $T, C$ and $\Phi$ are both sequestered in the frame function \cite{Randall:1998uk}.
Although the soft masses in NMSSM vanish at tree level, anomaly mediation can happen at loop level \cite{Randall:1998uk,Giudice:1998xp}.
Hence, we introduce gravity mediation by adding the contact terms between the visible and the hidden sector in the frame function,

\bea
\Omega_{\rm contact} = C_{{\bar\alpha}{\beta}} X^\dagger X z^\dagger_{\bar \alpha}  z_\beta +{\rm c.c}
\eea

where $X=C,\Phi$, and $z_\alpha$ are NMSSM superfields, and $ C_{{\bar\alpha}{\beta}} $ are the coupling parameters.
If $C_{{\bar\alpha}{\beta}}\neq \delta_{{\bar\alpha}{\beta}}$, gravity mediation would introduce dangerous flavor problems \cite{Falkowski:2005zv},
so in that case, SUSY breaking effects must translate to the visible sector by another mechanism, such as gauge or $U(1)'$ mediation.

\chapter{Inflation and Graceful Exit with twin waterfalls}
\label{chapter_hybrid} 
\begin{small}
  “Everything’s got to end sometime. Otherwise, nothing would ever get started.”\\
The Eleventh Doctor
  \end{small}

  \vspace{5mm}

\section{Pseudo-Nambu-Goldstone inflation with twin waterfalls}

We consider a pseudo-Nambu-Goldstone boson $\phi$, that acts as the inflaton, and two real scalar fields $\chi_1, \chi_2$, which are the waterfall fields responsible for the end of inflation.

The inflationary potential is divided into
\bea
V(\phi,\chi_1,\chi_2,H) =V_I(\phi)+V_W(\phi,\chi_1,\chi_2) +V_{\rm RH}(\chi_1,\chi_2,H).
\eea
Here,  $V_I(\phi)$ only depends on the inflaton field $\phi$ and $V_W(\phi,\chi_1,\chi_2)$ is the waterfall field part as in the hybrid inflation models introduced in section (\ref{sec:hybrid}). Finally,
$V_{\rm RH}(\chi_1,\chi_2,H)$ contains the couplings with the Higgs responsible for reheating.

The most general potential respecting the $Z_2$ discrete symmetry
\bea
Z_2:\quad \phi\rightarrow-\phi,  \qquad  \chi_1\leftrightarrow \chi_2,
\eea 
is given by

\bea
V_I(\phi) &=& V_0 +A(\phi), \label{inflatonpottwin} \\
V_W(\phi,\chi_1,\chi_2)&=&B(\phi) (\chi^2_1-\chi^2_2) +\frac{1}{2} m^2_\chi (\chi^2_1+\chi^2_2) -
\alpha^2 \chi_1\chi_2\nonumber \\
&&+\beta (\chi^3_1+\chi^3_2) + \gamma (\chi^2_1\chi_2+\chi_1\chi^2_2) \nonumber \\
&&+\frac{1}{4} \lambda_\chi (\chi^4_1+\chi^4_2) + \frac{1}{2} {\bar\lambda}_\chi \chi^2_1 \chi^2_2+ \frac{1}{3} \lambda'_\chi(\chi^3_1\chi_2+\chi_1\chi^3_2),  \label{fullpottwinpre} \\
V_{\rm RH}(\chi_1,\chi_2,H)&=& \kappa_1 (\chi^2_1+\chi^2_2)|H|^2 +\kappa_2 \chi_1\chi_2 |H|^2.
\eea

Here, $V_0$ is the constant vacuum energy during inflation, while $A(\phi)$ and $B(\phi)$ are arbitrary functions of $\phi$, satisfying $A(-\phi)=A(\phi)$ and $B(-\phi)=-B(\phi)$. 
We choose $A(\phi)=-\frac{1}{2}m^2_\phi \phi^2$ and $B(\phi)=-g\phi$.
We could also have introduced the renormalizable waterfall field couplings $\phi^2 (\chi^2_1+\chi^2_2)$, but the shift symmetry for $\phi$ naturally suppresses them.  

If we further assume a separate $Z'_2$ symmetry that acts only on $\chi_2$, 

\bea
Z'_2:\quad \phi\to\phi,  \qquad  \chi_1\to \chi_1, \qquad \chi_2\to -\chi_2, 
\eea

we can set $\alpha=\beta=\gamma=\lambda'_\chi=\kappa_2=0$, making $\chi_2$ a possible Dark Matter candidate. In this case, our free parameters are $\lambda_\chi, {\bar\lambda}_\chi $, and $\kappa_1$.

We now take $A(\phi)$ and $B(\phi)$ to be
\bea
A(\phi)&=&\Lambda^4 \cos\Big(\frac{\phi}{f}\Big), \label{fA} \\
B(\phi)&=&-\frac{1}{2} \mu^2 \sin\Big(\frac{\phi}{2f} \Big). \label{fB}
\eea

Then, the inflaton potential is given by
\bea
V_I(\phi) &=& V_0 + \Lambda^4 \cos\Big(\frac{\phi}{f} \Big), \label{inflaton} 
\eea

where the vacuum energy density during inflation is chosen to be $V_0\gtrsim \Lambda^4$. 

At the same time, the potential for the waterfall fields becomes
\bea
V_W(\phi,\chi_1,\chi_2)&=&-\frac{1}{2}\mu^2\sin\Big(\frac{\phi}{2f} \Big) (\chi^2_1-\chi^2_2) +\frac{1}{2} m^2_\chi (\chi^2_1+\chi^2_2) -
\nonumber \\
&&+\frac{1}{4} \lambda_\chi (\chi^4_1+\chi^4_2) + \frac{1}{2} {\bar\lambda}_\chi \chi^2_1 \chi^2_2.  \label{full}
\eea

The waterfall fields don't get a vev during inflation, but there would be mass mixing between the waterfall fields for $\alpha\neq 0$ in eq.~(\ref{full}).

\bea
m^2_1(\phi) &=& m^2_\chi - \sqrt{ \mu^4 \sin^2\Big(\frac{\phi}{2f} \Big)+\alpha^4},  \label{chi1mass}\\
m^2_2(\phi)&=& m^2_\chi +\sqrt{ \mu^2 \sin^2\Big(\frac{\phi}{2f} \Big)+\alpha^4}, \label{chi2mass}
\eea

and the mixing angle $\theta$ between the waterfall fields depends on the inflaton field by
\bea
\sin2\theta(\phi)=\frac{2\alpha^2}{m^2_2(\phi)-m^2_1(\phi)}.
\eea

These waterfall masses depend on the value of the inflaton field.
For $\alpha=0$, there is no mixing between the waterfall fields, so we can just keep track of the waterfall field $\chi_1$ to determine the end of inflation.

We define $\phi_c=2f\arcsin(\sqrt{m^4_\chi-\alpha^4}/\mu^2)$ with $\sqrt{m^4_\chi-\alpha^4}<\mu^2$ and $\alpha<m_\chi$. Then, as long as $\phi<\phi_c$ we are in a slow-roll inflation regime.
During inflation, the waterfall fields are heavy enough for $m_\chi>H_I$, with $H_I$ being the Hubble scale during inflation, so they can be decoupled.
At $\phi=\phi_c$ the waterfall field with mass $m_1$ starts becoming unstable, ending the inflation even if the slow-roll condition for the inflaton direction is not violated.  

\subsection{Origin of discrete symmetries}
We comment upon the origin of the discrete symmetries for the inflaton and the waterfall fields. Suppose that a $U(1)$ global symmetry is broken to a $Z_4$ symmetry, under which the inflaton $\phi$ and a complex scalar field $\Phi$, transform by
\bea
Z_4:\quad \phi\to \phi, \qquad \Phi\to -i \Phi.
\eea
Moreover, we take the CP symmetry in the dark sector as
\bea
{\rm CP}:\quad \phi\to -\phi, \qquad \Phi\to \Phi^*.
\eea
As a result of combining $Z_4\times {\rm CP}$, we get
\bea
Z_4\times {\rm CP}:\quad \phi\to -\phi, \qquad \Phi\to i\Phi^*.
\eea
In this case, writing $\Phi=\frac{1}{\sqrt{2}}(\chi_1+i\chi_2)$, we can realize $\phi\to-\phi$ and $\chi_1\leftrightarrow \chi_2$ under $Z_4\times {\rm CP}$, as required for the $Z_2$ symmetry, thus providing the potential for hybrid inflation in our model.
On the other hand, the separate $Z'_2$ symmetry for $\chi_2$ corresponds to
\bea
Z'_2:\quad \phi\to \phi,\qquad \Phi\to \Phi^*.
\eea

\subsection{UV corrections}
Due to the couplings of the waterfall fields to the inflaton, the inflaton potential receives loop corrections. The one-loop Coleman-Weinberg potential for the inflaton is given in cutoff regularization with cutoff scale $M_*$, as follows,
\bea
V_{\rm CW} &=&\frac{1}{64\pi^2}\sum_{i=1,2} \left[2m^2_{\chi_i}M^2_* -m^4_{\chi_i} \ln\bigg( \frac{e^{\frac{1}{2}}M^2_* }{m^2_{\chi_i}}\bigg)  \right] \nonumber \\
&\simeq&\frac{1}{16\pi^2}\,m^2_\chi M^2_* -\frac{1}{64\pi^2}\bigg[m^4_\chi +\mu^4\sin^2\Big(\frac{\phi}{2f}\Big)+\alpha^4\bigg] \ln  \frac{M^2_*}{m^2_\chi}.  \label{CW}
\eea 
Then, the constant vacuum energy proportional to $M^2_*$ must be renormalized in order to get the correct vacuum energy density.
The quadratic divergences cancel due to the $Z_2$ discrete symmetry, and the logarithmically divergent terms of the inflaton potential can be ignored during inflation as far as $\mu^2\lesssim 8\pi \Lambda^2$ is satisfied.  

Consequently,  as far as the mass parameters in the waterfall field sector satisfies $H^2_I\lesssim \mu^2\sim m^2_\chi\lesssim 8\pi \Lambda^2$, the waterfall fields remain decoupled and the inflation mass doesn't get corrections from these couplings.
Moreover, for a QCD-like phase transition with  $\Lambda\ll f$ and the inflaton mass,  $\sqrt{|m^2_\phi|}= \frac{\Lambda^2}{f} $,  we maintain a natural hierarchy of scales in our model,
\bea
\sqrt{|m^2_\phi|} \ll H_I\ll \mu\sim m_\chi\ll \sqrt{8\pi} \Lambda \ll f.  
\eea 

As we vary the inflationary scale $H_I$, we can also scale the rest of the mass parameters subject to the above hierarchy. 
We remark that the bound on the waterfall fields couplings is consistent with the condition that a QCD-like phase transition occurs during inflation, $\Lambda>H_I$.

\section{Inflationary predictions}
From the inflaton potential in eq.~(\ref{inflatonpottwin}), the slow-roll parameters are given by
\bea
\epsilon &=& \frac{M^2_P \Lambda^8 \sin^2(\phi/f)}{2f^2 (V_0+\Lambda^4\cos(\phi/f))^2}, \\
\eta &=& -\frac{M^2_P \Lambda^4 \cos(\phi/f)}{f^2 (V_0+\Lambda^4\cos(\phi/f))},
\eea
and the number of e-foldings is also obtained as
\bea
N = \frac{f^2}{2M^2_P \Lambda^4}\, \bigg[V_0 \ln \Big(\tan^2\Big(\frac{\phi_c}{2f}\Big)\Big)+\Lambda^4\ln \Big(\sin^2\Big(\frac{\phi_c}{f}\Big)\Big) \bigg] -(\phi_c\to \phi_*) 
\eea
where $\phi_*,\phi_c$ are the inflaton field values at the horizon exit and at the end of inflation, respectively.

Considering $V_0\gg \Lambda^4$, we can approximate the slow-roll parameters to
\bea
\eta_* &\simeq& -\frac{M^2_P \Lambda^4}{f^2V_0 }\, \cos(\phi_*/f), \\
\epsilon_* &\simeq & \frac{M^2_P \Lambda^8}{2f^2V^2_0}\,  \sin^2(\phi_*/f), \\
N &\simeq&  \frac{f^2 V_0}{M^2_P \Lambda^4}\,\ln \Big(\frac{\tan(\phi_c/(2f))}{\tan(\phi_*/(2f))} \Big). \label{Nefold}
\eea

As a result, the spectral index and the tensor-to-scalar ratio can be determined by
\bea
n_s &=&1+2\eta_*-6\epsilon_*,  \label{ns}\\
r&=&16\epsilon_*. \label{r}
\eea

The CMB normalization, $A_s=\frac{1}{24\pi^2} \frac{V_0+\Lambda^4}{\epsilon_* M^4_P}\simeq 2.1\times 10^{-9}$, leads to
\bea
r=3.2\times 10^7\,\cdot\frac{V_0}{M^4_P}.  \label{cmb}
\eea

In order to get the spectral index consistent with Planck data, we need $2\eta_*\simeq -0.0033$, because  $ \epsilon_*\ll |\eta_*|$ in this case. The critical value $\phi_c$ of the inflaton should satisfy $ \phi_*\lesssim \phi_c\lesssim f$ for the number of e-foldings $N=50-60$ to solve the horizon problem. 
Moreover, the Planck bound on the tensor-to-scalar ratio, $r<0.036$, gives rise to the upper bound on $H_I<4.6 \times 10^{13}\,{\rm GeV}$. 

From the Hubble scale during inflation, $H^2_I\simeq V_0/(3M^2_P)$ and eqs.~(\ref{ns}), (\ref{Nefold}) and (\ref{cmb}),
\bea
n_s&\simeq& 1+2\eta_*\simeq 1- \frac{|m^2_\phi|}{3H^2}\, \cos(\phi_*/f), \quad |m^2_\phi|= \frac{\Lambda^4}{f^2}, \\
N&=& \frac{\cos(\phi_*/f)}{|\eta_*|}\,  \ln \Big(\frac{\tan(\phi_c/(2f))}{\tan(\phi_*/(2f))} \Big), \\
\frac{H_I}{f}&=&2.9\times 10^{-4}\,\Big|\eta_*  \tan(\phi_*/f)\Big|, \label{cmb2}
\eea

together with the condition determining the end of inflation,
\bea
\sin\Big(\frac{\phi_c}{2f}\Big) = \frac{\sqrt{m^4_\chi-\alpha^4}}{\mu^2}. 
\eea
Eq.~(\ref{cmb2}) shows a correlation between the axion decay constant and the Hubble scalar during inflation.
Taking $|\eta_*|=0.033/2$ and $\cos(\phi_*/f)=0.95$, we get $f\simeq 6.4\times 10^5 H_I$, which is much larger than the Hubble scale. Therefore, the global symmetry responsible for the PNG inflaton is already broken during inflation.  

\begin{figure}[t]
    \centering
    \includegraphics[width=0.40\textwidth,clip]{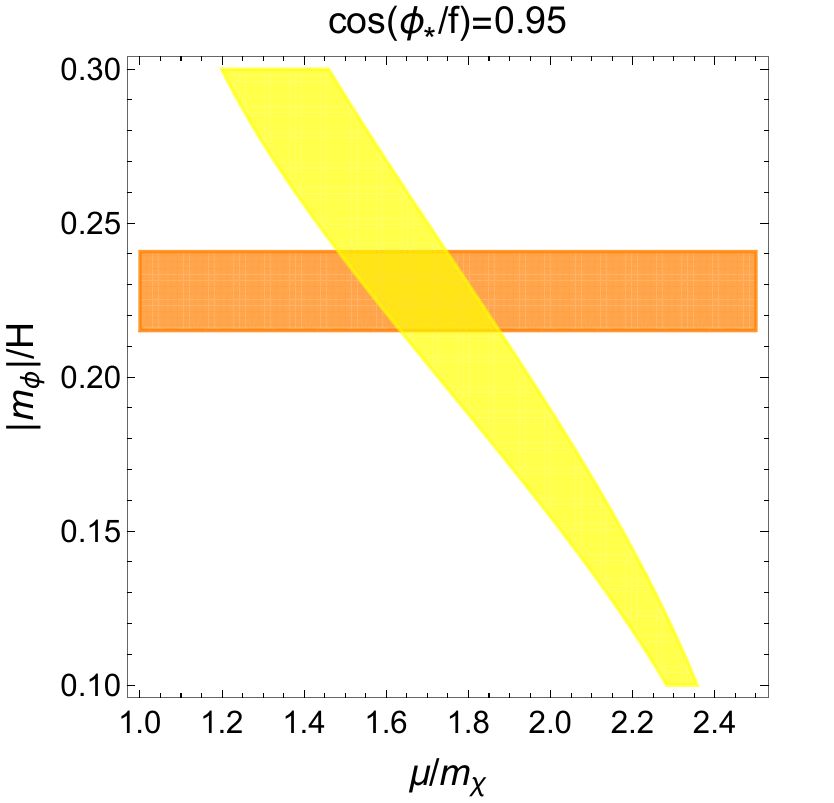}\,\,\,\,\,\,
    \includegraphics[width=0.40\textwidth,clip]{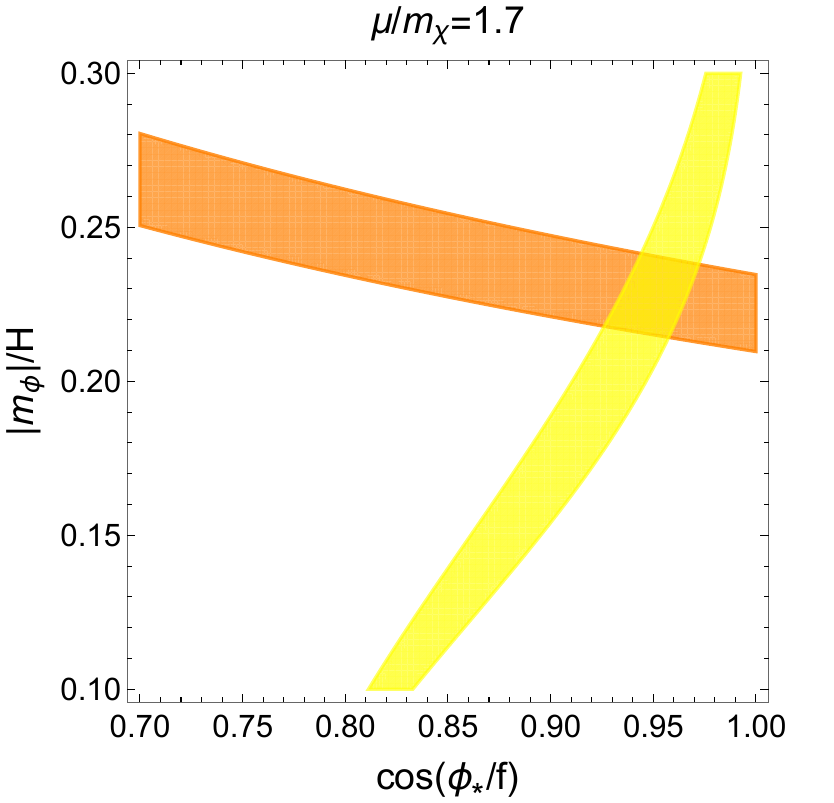}
    \caption{Parameter space for successful inflation. We plot $|m_\phi|/H$ vs $\mu/m_\chi$ on the left and $|m_\phi|/H$ vs $\cos(\theta_*/f)$ on the right. 
    The orange region is consistent with Planck data within $1\sigma$ and  the number of e-foldings between $N=40-60$ is shown in the yellow region.} 
    
    \label{fig:inf1}
    \end{figure}

In Fig.~\ref{fig:inf1}, we show the available parameter space for the successful inflation.
The orange region is consistent with Planck data within $1\sigma$ and the yellow region is the one for $40$ to $60$ e-foldings.
Thus, successful inflation will correspond to the overlap between the two regions. 
We took $\cos(\theta_*/f)=0.95$ on left and $\mu/m_\chi=1.7$ on the right. 
Although we chose $\alpha=0$ in Fig.~\ref{fig:inf1}, we only have to replace $m^2_\chi$  by $\sqrt{m^4_\chi-\alpha^4}$ for a nonzero $\alpha$. 
As we increase $\mu/m_\chi$, the waterfall transition would take place at a lower value of the inflaton field, $\phi_c$, so inflation needs to start closer to the origin in order to satisfy the correct number of e-foldings in eq.~(\ref{Nefold}).

\begin{figure}[t]
\centering
\includegraphics[width=0.45\textwidth,clip]{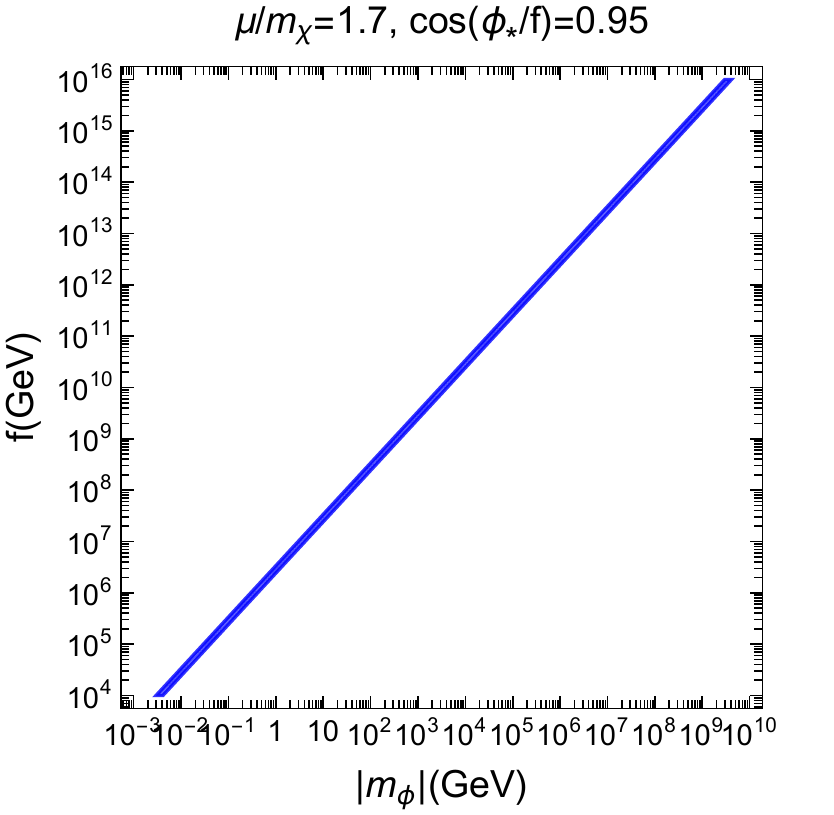}
\caption{Parameter space for inflation in $f$ vs $|m_\phi|$. 
}
\label{fig:inf2}
\end{figure}

In Fig.~\ref{fig:inf2}, we show the parameter space for the axion decay constant $f$ vs $|m_\phi|$,  consistent with Planck data within $1\sigma$. 
We took  $\mu/m_\chi=1.7$ and $\cos(\theta_*/f)=0.95$. Thus, we find that for $f=10^4\,{\rm GeV}-10^{16}\,{\rm GeV}$, the successful inflation is achieved for the values of $|m_\phi|=\Lambda^2/f$ between $3.5\times 10^{-3}\,{\rm GeV}$ and $3.5\times 10^9\,{\rm GeV}$, which correspond to the range for the QCD-like condensation scale, $5.9\,{\rm GeV}\lesssim \Lambda\lesssim 5.9\times 10^{12}\,{\rm GeV}$.

For the benchmark point in  Fig.~\ref{fig:inf2}, the Hubble scale $H_I$ varies in the range between $0.016\,{\rm GeV}$ and $1.6\times 10^{10}\,{\rm GeV}$. From eq.~(\ref{cmb}) with $V_0\simeq 3M^2_P H^2_I$, this means we get a tiny tensor to scalar ratio,

\bea
4.1\times 10^{-33}\lesssim r\lesssim 4.1 \times 10^{-9}.
\eea

The scale of inflation is given by $2.6\times 10^8\,{\rm GeV}\lesssim V^{1/4}_0\lesssim 2.6 \times 10^{14}\,{\rm GeV}$, which is much larger than the QCD-like condensation scale $\Lambda$. 
Quantum corrections coming from the waterfall field couplings won't be spoiled provided that $\mu\ll \sqrt{8\pi} \Lambda=30\,{\rm GeV}-3.0\times 10^{13}\,{\rm GeV}$.
Hence, we find a wide parameter space realizing successful for inflation, satisfying $H_I\ll \mu\ll \sqrt{8\pi} \Lambda$, and the waterfall fields can be light.

\section{Waterfall transition}

After inflation, reheating takes place through the decay or scattering of the waterfall fields.
The effective inflaton mass is given by

\bea
m^2_{\phi, {\rm eff}} =  -\frac{\Lambda^4}{f^2} \cos\Big(\frac{\langle\phi\rangle}{f}\Big) +\frac{\mu^2}{8f^2}( \langle\chi^2_1\rangle - \langle\chi^2_2\rangle)\sin\Big(\frac{\langle\phi\rangle}{2f}\Big) \label{effinfmass}
\eea

The first term is a tachyonic mass, while the second is the contribution from the waterfall fields.
When $\phi=\phi_c$, the first waterfall field starts rolling fast and develops a non zero vev that modifies the inflation mass in eq~.(\ref{effinfmass}).
The inflaton settles at the minimum of the potential, at $\phi/f= \pi$.
The waterfall masses don't change after inflation, and they are still given by  eqs.~(\ref{chi1mass}) and (\ref{chi2mass}).

\subsection{The vacuum structure}

Taking $\alpha=0$, the minimum of the potential  

\bea
v_\phi &=& \pi f, \\
 v_\chi&=& \sqrt{\frac{\mu^2-m^2_\chi}{\lambda_\chi}}. \label{chivev}
\eea

where $\langle\phi\rangle=v_\phi$,  $\langle\chi_1\rangle=v_1\equiv v_\chi $ and $\langle\chi_2\rangle=0$.
Since one of the waterfall fields gets a vev, the $Z_2$ symmetry is broken by the vacuum.

The cosmological constant in the true vacuum is given by

\bea
V_{\rm eff}(\chi_1=v_\chi,\chi_2=0) = V_0-\Lambda^4- \frac{1}{4}\lambda_\chi v^4_\chi \simeq 0, \label{zerocc}
\eea

Expanding around the inflaton vev, $\phi=v_\phi+a$,  we get
\bea
m^2_{a} = \frac{1}{f^2} \Big(\Lambda^4+\frac{1}{8} \mu^2 v^2_\chi\Big), \label{inflatonmass} 
\eea

and similarly, for the waterfall fields, $\chi_1=v_\chi+{\tilde\chi}_1$, 
\bea
m^2_1&=&  2\lambda_\chi v^2_\chi \nonumber \\
&=&2(\mu^2-m^2_\chi),  \\
m^2_2 &=& \mu^2 + m^2_\chi +{\bar\lambda}_\chi v^2_\chi \nonumber  \\
&=& \mu^2+m^2_\chi +\frac{{\bar\lambda}_\chi }{\lambda_\chi}(\mu^2- m^2_\chi). 
\eea

Vacuum stability requires $\lambda_\chi>0$ and $\lambda_\chi+{\bar\lambda}_\chi>0$ for ${\bar\lambda}_\chi<0$. 

For $\mu^2/m^2_\chi<3+2({\bar\lambda}_\chi/\lambda_\chi)/(1-{\bar\lambda}_\chi/\lambda_\chi)$, the waterfall field $\chi_2$ is heavier than ${\tilde\chi}_1$. Otherwise, the waterfall field $\chi_2$ is lighter than ${\tilde\chi}_1$. 
$m_2>m_1$ is favored by the number of e-foldings, so that the decay ${\tilde\chi}_1\to \chi_2\chi_2$ is not open for reheating.

The interaction terms are given by

\bea
{\cal L}_{\rm int} = \frac{\mu^2}{8f^2}\,v_\chi a^2 {\tilde\chi}_1+ \frac{\mu^2}{16f^2} a^2 ({\tilde\chi}^2_1-{\chi}^2_2)+\cdots. \label{inflaton-int}
\eea

So, even if there are no quadratic divergences due to the presence of the $Z_2$ symmetry, we still have logarithmic divergences due to the coupling of the inflaton to the waterfall, i.e., the first term in eq. (\ref{inflaton-int})

For $V_0\gg \Lambda^4$, eqs.~(\ref{chivev}) and (\ref{zerocc}) give rise to $4V_0/\lambda_\chi\sim v^4_\chi\sim \mu^4/\lambda_\chi^2$. 
Then, the quartic coupling and the VEV for the waterfall fields are related to the dimensionful parameters of the inflation, as follows,
\bea
\lambda_\chi\sim \frac{\mu^4}{4V_0}, \qquad v^2_\chi\sim \frac{4V_0}{\mu^2}. \label{relation}
\eea

Then, from the decoupling condition during inflation, $\mu\gtrsim H_I$, we can constrain the parameter space for the waterfall fields.
For $f\simeq 6.4\times 10^{5}  H_I$ and $\cos(\phi_*/f)=0.95$,

\bea
\lambda_\chi\sim \mu^4/(4V_0)\simeq 1.4\times 10^{-20}(\mu/(100H_I))^4 (H_I/10^{5}\,{\rm GeV})^2 
\eea
 and 
 \bea
 v_\chi\sim \sqrt{4V_0/\mu^2}\simeq 0.035 M_P(100H_I/\mu)
 \eea 

From eqs.~(\ref{inflatonmass}) and eq.~(\ref{relation}), we find that $\mu^2 v^2_\chi\sim 4V_0\gg \Lambda^4$ and
 $m^2_a\sim \mu^2 v^2_\chi/(8 f^2)$. 

\section{Preheating}

After the end of inflation, the equations of motion for the scalar fields $\psi=H$ and $\chi_2$, can be rewritten in terms of the rescaled field, $\varphi=a\,\psi$, 
\bea
\varphi^{\prime\prime} -\partial^2_i\varphi+\Big(m^2_{\psi,{\rm eff}} a^2+(6\zeta_\psi-1)\,\frac{a^{\prime\prime}}{a}  \Big)\varphi=0 \label{spert}
\eea
where $\zeta_\psi$ is the non-minimal coupling of the scalar field  $\psi$ and we defined the conformal time, $\eta=\int dt/a(t)$.

For a conformal non-minimal coupling, $\zeta_\psi=\frac{1}{6}$, we get the total number density of produced particles at the end of the waterfall transition by
\bea
n_\psi= \frac{1}{2\pi^2}\, \int^\infty_0 dk\, k^2 n^\psi_k, \label{numberd}
\eea
where
\bea
n^\psi_k =- \frac{g_\psi}{2} +\frac{1}{2\omega_k} \Big(|v'_k|^2 +\omega^2_k |v_k|^2 \Big), \quad \omega^2_k=k^2+m^2_{\psi,{\rm eff}}.
\label{modenumber}
\eea
Here, $v_k$ is the mode function for the scalar field, 
and $g_{\psi}$ is the number of degrees of freedom for $\psi=H,\chi_2$.

During the waterfall transition, we take the effective scalar masses \cite{Garcia-Bellido:2001dqy} as
\bea
m^2_{\psi,{\rm eff}}(t) = m^2_{\psi,0}+ \frac{1}{4} c_\psi v^2_\chi (1+\tanh \lambda(t-t_*))^2 \label{effmass}
\eea
where $\psi \equiv H, \chi_2$. Here, $m^2_{\psi,0}$ are the bare masses and $c_\psi$ are the couplings with $\chi_1$, given by $c_\psi=\kappa_1, {\bar\lambda}_\chi$ for $H$ and $\chi_2$, respectively.
Here, $\lambda$ is the strength of the waterfall transition, given by $\lambda=m_1/2$, with $m_1=\sqrt{\lambda_\chi} v_\chi$ being the tachyonic mass of the waterfall field $\chi_1$.  The duration of the waterfall transition is given by $t_*\simeq \ln(32\pi^2/\lambda_\chi)/(2m_1)$ \cite{Garcia-Bellido:2001dqy}.
for $\lambda_\chi=10^{-20}-10^{-10}$, the duration is given by $t_*\simeq 14/m_1-26/m_1$ .
 As $m_1\gg H_{\rm end}$, the waterfall transition takes place less than one Hubble time and finishes rapidly. Thus, the Hubble expansion can be ignored during the waterfall transition, allowing use to set the scale factor to $a=1$.

The number density of produced particles with momentum $k$ is given by 
\bea
 n^\psi_k &=&\frac{ \cosh\Big(\pi (w_{\psi,2}-w_{\psi,1})/\lambda\Big)+\cosh(2\pi\delta) }{ 2\sinh(\pi \omega_{\psi,1}/\lambda)\sinh(\pi \omega_{\psi,2}/\lambda)} \label{npsi}
 \eea
 where $w^2_{\psi,1}=k^2+m^2_{\psi,0}$,  $w^2_{\psi,2}=k^2+m^2_{\psi,0}+c_\psi v^2_\chi$ and $\delta=\frac{1}{2}\sqrt{\frac{c_\psi v^2_\chi}{\lambda^2}-1}$.
 Therefore, we obtain the energy density of the produced particles at the end of the waterfall transition by
 \bea
 \rho_\psi(t_*)=\frac{g_\psi}{2\pi^2}\, \int^\infty_0 dk\, k^2 n^\psi_k w_{\psi,2}.
 \eea

 As a consequence, the ratio of the produced energy density to the initial vacuum energy density, $V_0\simeq \frac{1}{4}\lambda_\chi v^4_\chi=\frac{m^4_1}{4\lambda_\chi}$, is given by
\bea
\frac{\rho_\psi(t_*)}{V_0} =\frac{2g_\psi\lambda_\chi}{\pi^2} \int^\infty_0 d\kappa \,\kappa^2 n^\psi_k  \sqrt{\kappa^2+\frac{m^2_{\psi,0}}{m^2_1}+\alpha_\psi} \label{analytic}
\eea
with $\kappa=k/m_1$ and $\alpha_\psi=c_\psi/\lambda_\chi$. 
In order to compare our results with the lattice calculations \cite{Garcia-Bellido:2001dqy}, for $m_{\psi,0}=0$, we quote the numerical solution for $\frac{\rho_\psi(t_*)}{V_0}$ in terms of a fitting function, $f(\alpha,\gamma)=\sqrt{\alpha+\gamma^2}-\gamma$ \cite{Garcia-Bellido:2001dqy}, as follows,
\bea
\frac{\rho_\psi(t_*)}{V_0} =2\times 10^{-3} g_\psi\lambda_\chi f(\alpha_\psi,1.3). \label{preheating}
\eea

The numerical result for the number density at the waterfall transition is given in \cite{Garcia-Bellido:2001dqy} by 
\bea
n_\psi(t_*)=  10^{-3} g_\psi m^3_1f(\alpha_\psi,1.3)/\alpha_\psi.
\eea

\begin{figure}[!t]
\begin{center}
\includegraphics[width=0.45\textwidth,clip]{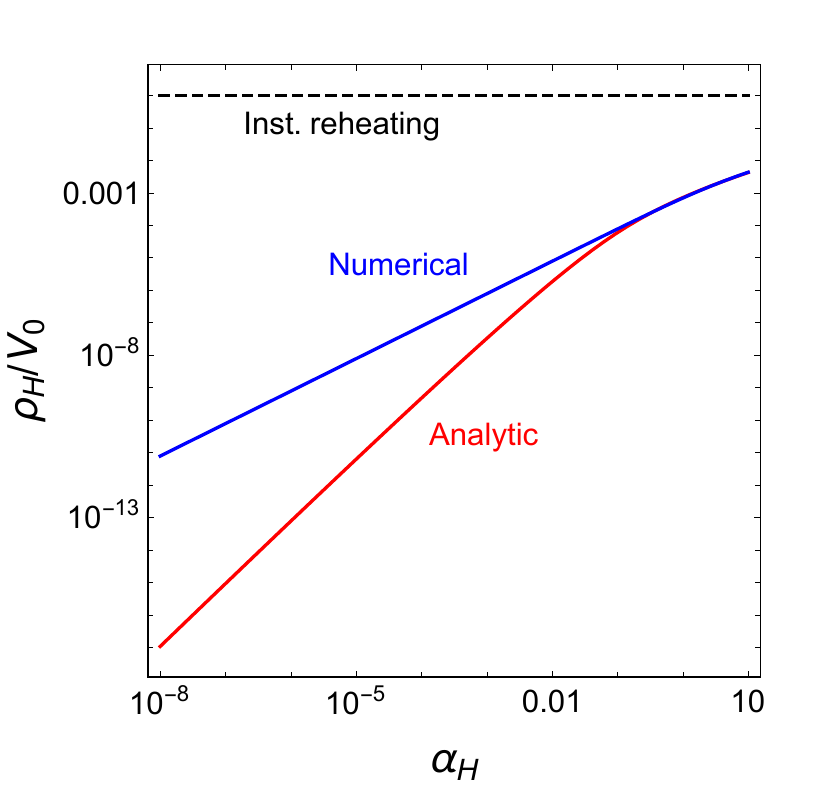}
\end{center}
\caption{Analytic and numerical solutions for $\rho_H/V_0$ from preheating as a function of $\alpha_H=\kappa_1/\lambda_\chi$ in red and blue lines, respectively. }
\label{fig:preheating}
\end{figure}

In Fig.~\ref{fig:preheating}, we depict the analytic and numerical results for $\rho_H/V_0 $ for the Higgs as a function of $\alpha_H=\kappa_1/\lambda_\chi$ in red and blue lines, respectively.
 Here, we took $m_{H,0}=0$. 
The analytic result is based on the formula in eq.~(\ref{analytic}), while the numerical result is taken from eq.~(\ref{preheating}). 
For $\alpha_H\gtrsim 0.1$, both the analytic and the numerical results agree. For comparison, we included the case in which reheating is instantaneous in the black line.

\section{Reheating}

After preheating, the fields settle at the bottom of the inflationary potential and start oscillating. If reheating hasn't completed by preheating, we can determine the reheating temperature by the perturbative decays of the fields.

The dynamics of the fields during reheating follow from the Boltzmann equations studied in section \ref{sec:darkmatter},
\bea
{\ddot \phi}+3H {\dot \phi}  &=&-\Gamma_\phi {\dot\phi} +\frac{\Lambda^4}{f} \sin\Big(\frac{\phi}{f}\Big)+ \frac{\mu^2}{4f} \cos\Big(\frac{\phi}{2f} \Big) (\chi^2_1-\chi^2_2),  \\
{\ddot\chi}_1 + 3H {\dot\chi}_1  &=&-\Gamma_{\chi_1} {\dot\chi}_1+\mu^2\sin\Big(\frac{\phi}{2f} \Big) \chi_1-m^2_\chi \chi_1-\alpha^2 \chi_2-\lambda_\chi \chi^3_1 -{\bar\lambda}_\chi \chi_1 \chi^2_2, \\
{\ddot\chi}_2 + 3H {\dot\chi}_2 &=&-\Gamma_{\chi_2} {\dot\chi}_2 -\mu^2\sin\Big(\frac{\phi}{2f} \Big) \chi_2-m^2_\chi \chi_2-\alpha^2\chi_1-\lambda_\chi \chi^3_2 -{\bar\lambda}_\chi \chi^2_1 \chi_2, \\
{\dot\rho}_R + 4H \rho_R &=&\Gamma_{\phi} {\dot\phi}^2 +\Gamma_{\chi_1}{\dot\chi}^2_1  +\Gamma_{\chi_2}{\dot\chi}^2_2 ,
\eea

together with the Friedmann equation
\bea
H^2= \frac{\rho_I+\rho_R}{3M^2_P},
\eea

where $\rho_I$ is the sum of energy densities for the inflaton and the waterfall fields,
\bea
\rho_I = \frac{1}{2} {\dot\phi}^2 + \frac{1}{2} {\dot\chi}^2_1 + \frac{1}{2} {\dot\chi}^2_2 +V(\phi,\chi_1,\chi_2).
\eea

Taking $\alpha=0$ and $\chi_2=0$, we can just focus on the dynamics of $\chi_1$ during reheating.
Since $\Lambda^4\ll V_0$ during inflation, we can ignore the contributions from the inflaton field.

Then, the reheating temperature will be determined from the decay of $\chi_1$,
\bea
T_{\rm RH}= \bigg(\frac{90}{\pi^2 g_{\rm RH}} \bigg)^{1/4} \sqrt{M_P \Gamma_{\chi_1}}  \label{RH}
\eea
where $g_{\rm RH}$ is the number of relativistic degrees of freedom at the time of reheating completion and $ \Gamma_{\chi_1}$ is the decay rate for $\chi_1$.

If reheating is delayed, the number of e-foldings required to solve the horizon problem \cite{Choi:2016eif, Aoki:2022dzd} is modified to
\bea
N=61.1 +\Delta N -\ln \bigg(\frac{V^{1/4}_0}{H_k} \bigg) -\frac{1}{12} \ln \bigg(\frac{g_{\rm RH}}{106.75} \bigg)
\eea

where the correction to the number of e-foldings due to the non-instantaneous reheating is given by
\bea
\Delta N= \frac{1}{12} \bigg(\frac{3w-1}{w+1} \bigg) \ln\bigg(\frac{45 V_0}{\pi^2 g_{\rm RH} T^4_{\rm RH}} \bigg). \label{DN2}
\eea

Here, $H_k$ is the Hubble parameter evaluated at the horizon exit for the Planck pivot scale, $k=0.05\,{\rm Mpc}^{-1}$, and $w$ is the averaged equation of state during reheating. 

We can also introduce the Higgs portal couplings,

\bea
{\cal L}_{H} = - \kappa_1 (\chi^2_1+\chi^2_2)|H|^2 -\kappa_2 \chi_1\chi_2 |H|^2.
\eea

Then, taking the waterfall field to $\chi_1= v_\chi+\chi_c(t)$, where  $\chi_c(t)$ denotes the waterfall condensate, we get the decay rate of $\chi_1$ as





\bea
\Gamma_{\chi_1}= \Gamma_{\chi_1\to hh}+ \Gamma_{\chi_1\to\chi_2\chi_2}, \label{decay}
\eea
with
 
\bea
 \Gamma_{\chi_1\to hh}&=&\frac{\kappa_1^2 v^2_\chi }{2\pi m_{\chi_1}} \sqrt{1-\frac{4m^2_H}{m^2_{\chi_1}}}, \\
  \Gamma_{\chi_1\to\chi_2\chi_2} &=&\frac{{\bar\lambda}^2_\chi v^2_\chi }{2\pi m_{\chi_1}} \sqrt{1-\frac{4m^2_{\chi_2}}{m^2_{\chi_1}}}.
\eea
Then,  ignoring the Higgs and $\chi_2$ masses, we obtain the reheating temperature approximately by 
\bea
T_{\rm RH}\simeq \bigg(\frac{90}{\pi^2 g_{\rm RH}} \bigg)^{1/4} \bigg( \frac{\kappa^2_1+{\bar\lambda}^2_\chi}{4\pi \lambda_\chi}\bigg)^{1/2} \sqrt{M_P m_{\chi_1}}. 
\eea
Therefore, for $\Gamma_{\chi_1}\ll  H_I\sim m_{\chi_1}$, namely,  $\kappa^2_1+{\bar\lambda}^2_\chi\ll 4\pi \lambda_\chi$, and taking $H_I\lesssim 1.6\times 10^{10}\,{\rm GeV}$ for $f\lesssim 10^{16}\,{\rm GeV}$, we get the reheating temperature as $T_{\rm RH}\ll  10^{14}\, {\rm GeV}$.

The waterfall transition takes place faster than the Hubble expansion, so preheating does not affect the number of e-foldings.
However,reheating can be delayed, so that eq.~(\ref{DN2}) becomes
\bea
\Delta N= \frac{1}{12} \bigg(\frac{3w-1}{w+1} \bigg) \ln\bigg(\frac{45 \rho_{\chi_1}(t_*)}{\pi^2 g_{\rm RH} T^4_{\rm RH}} \bigg).\label{DN}
\eea
Here, $\rho_{\chi_1}(t_*)=V_0-\rho_R(t_*)$ is the energy density of the waterfall field at the end of the waterfall transition and $w$ is the equation of state for the waterfall condensate.
Here, $H_k$ is the Hubble parameter evaluated at the horizon exit for the Planck pivot scale, $k=0.05\,{\rm Mpc}^{-1}$, and $w$ is the averaged equation of state during reheating. 
Then, taking $w=0$ for $V_0\gtrsim \rho_R(t_*)$ and $g_{\rm RH}=106.75$ in eq.~(\ref{DN2}), we obtain the number of efoldings  as
\bea
N=51.3-\frac{1}{3}\ln\bigg(\frac{H_I}{1.6\times 10^{10}\,{\rm GeV}}\bigg)-\frac{1}{12}\ln \bigg(1-\frac{\rho_R(t_*)}{V_0}\bigg)+\frac{1}{3}\ln\bigg(\frac{T_{\rm RH}}{10^{14}\,{\rm GeV}}\bigg).
\eea
As a result, we conclude that there is a wide range of the parameter space for a successful inflation, and a sufficiently large reheating temperature is achieved due to the decay of the waterfall field.

\subsection{Time evolution of radiation energy density}

We make a concrete discussion of the time evolution of the waterfall and radiation energy densities after preheating and determine the reheating temperature concretely. 

After preheating, we consider the Boltzmann equations in the presence of the coherent oscillation of the waterfall field, as follows,
\bea
{\dot\rho}_R + 4H \rho_R&=&\Gamma_{\chi_1\to hh}\rho_{\chi_1}, \\
{\dot\rho}_{\chi_1} + 3H \rho_{\chi_1} &=&-\Gamma_{\chi_1}\rho_{\chi_1},
\eea
with
\bea
H^2= \frac{\rho_R+\rho_{\chi_1}}{3M^2_P}. \label{Hubbleeq}
\eea
We take the initial condition for the waterfall energy density at preheating as $\rho_{\chi_1}(t_*)=V_0-\rho_R(t_*)$, where $\rho_R(t_*)=\rho_H$ is the radiation energy density from preheating. For a constant decay rate, $\Gamma_{\chi_1}\simeq \Gamma_{\chi_1\to hh}$, we obtain the analytic solutions to the above Boltzmann equations by
\bea
\rho_{\chi_1}(t) &=& \rho_{\chi_1}(t_*) \bigg(\frac{a(t)}{a_*}\bigg)^{-3}\, e^{-\Gamma_{\chi_1} (t-t_*)}, \\
\rho_R(t)&=& \rho_{\chi_1}(t_*) \bigg(\frac{a(t)}{a_*}\bigg)^{-4}\,\bigg[\frac{\rho_R(t_*)}{ \rho_{\chi_1}(t_*)}+\int^{u}_{u_*} \bigg(\frac{a(u)}{a_*}\bigg) e^{u_*-u} \,du \bigg]. \label{totalrad}
\eea 
Here, $u=\Gamma_{\chi_1}t$ and $u_*=\Gamma_{\chi_1}t_*$.

Combining eq.~(\ref{Hubbleeq}) and the effective continuity equation for the total energy density, $\rho=\rho_{\chi_1}+\rho_R$, given by
\bea
{\dot\rho} + 3H \rho (1+w(t))=0,
\eea
with $w(t)$ being the effective equation of state,  we find the total energy density as
\bea
\rho(t)=V_0 \bigg(1+\sqrt{\frac{3}{4} V_0}\, (1+{\overline w}) \,\frac{t-t_*}{M_P}\bigg)^{-2}.
\eea
where
\bea
{\overline w}(t) = \frac{1}{t-t_*} \int^t_{t_*} dt'\, w(t').
\eea
Here, we used $\rho_{\chi_1}+\rho_R\simeq V_0$ at the waterfall transition.  Then, for a slowly-varying ${\overline w}$, we also obtain the scale factor as a function of time,
\bea
\frac{a(t)}{a_*} \simeq \bigg( 1+\sqrt{\frac{3}{4} V_0} \, (1+{\overline w}) \,\frac{t-t_*}{M_P}\bigg)^{\frac{2}{3} \frac{1}{1+{\overline w}}}.
\label{scalefactor}
\eea

Setting ${\overline w}=0$ for the waterfall field oscillation near the quadratic potential and using the results in eqs.~(\ref{totalrad}) and (\ref{scalefactor}), we get the total radiation energy density after preheating, as follows,
\bea
\rho_R(t)= \rho_{\chi_1}(t_*) \bigg(1+\frac{v}{A}\bigg)^{-8/3}\,\bigg[ \frac{\rho_R(t_*)}{\rho_{\chi_1}(t_*)}+\int^v_0 \bigg(1+\frac{v'}{A}\bigg)^{2/3} e^{-v'} \,dv' \bigg] \label{rad}
\eea
where $v\equiv \Gamma_{\chi_1}(t-t_*)$ and
\bea
A&=&\frac{\Gamma_{\chi_1}}{m_{\chi_1}}\,\bigg(\frac{3}{4} \frac{V_0}{m^2_{\chi_1}M^2_P}\bigg)^{-1/2} \nonumber \\
&=& 4\sqrt{\frac{2}{3}} \frac{\Gamma_{\chi_1}}{ m_{\chi_1}} \bigg(\frac{v_\chi}{M_P} \bigg)^{-1}. \label{Apara}
\eea
Here, we used $V_0=\frac{m^4_1}{4\lambda_\chi}$ and $m^2_1=\lambda_\chi v^2_\chi=m^2_{\chi_1}/2$ in the second line of eq.~(\ref{Apara}).

\begin{figure}[!t]
\begin{center}
 \includegraphics[width=0.45\textwidth,clip]{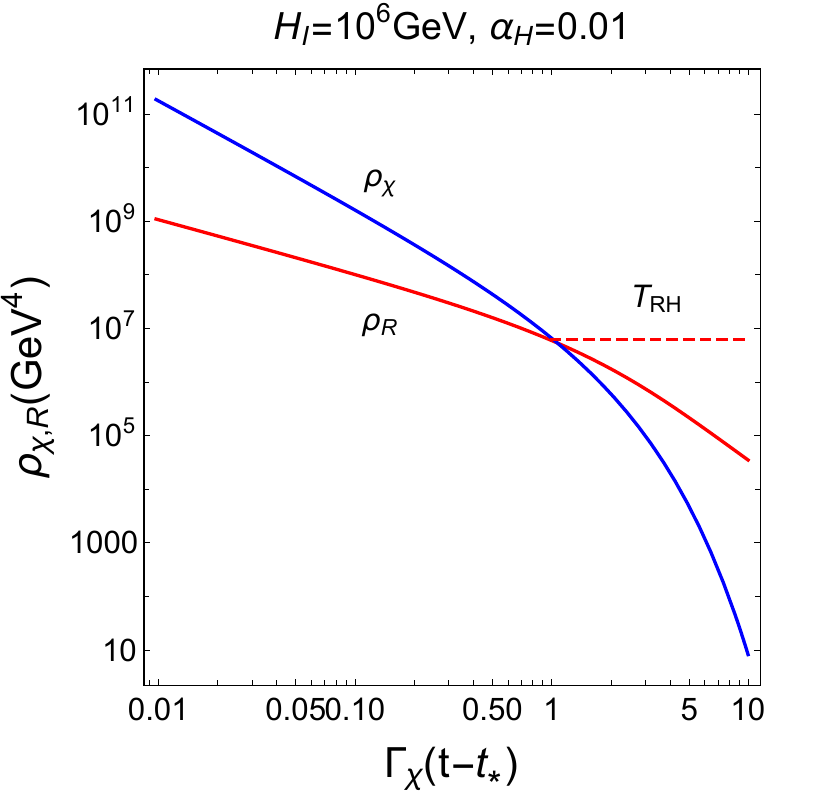}\,\,  \includegraphics[width=0.45\textwidth,clip]{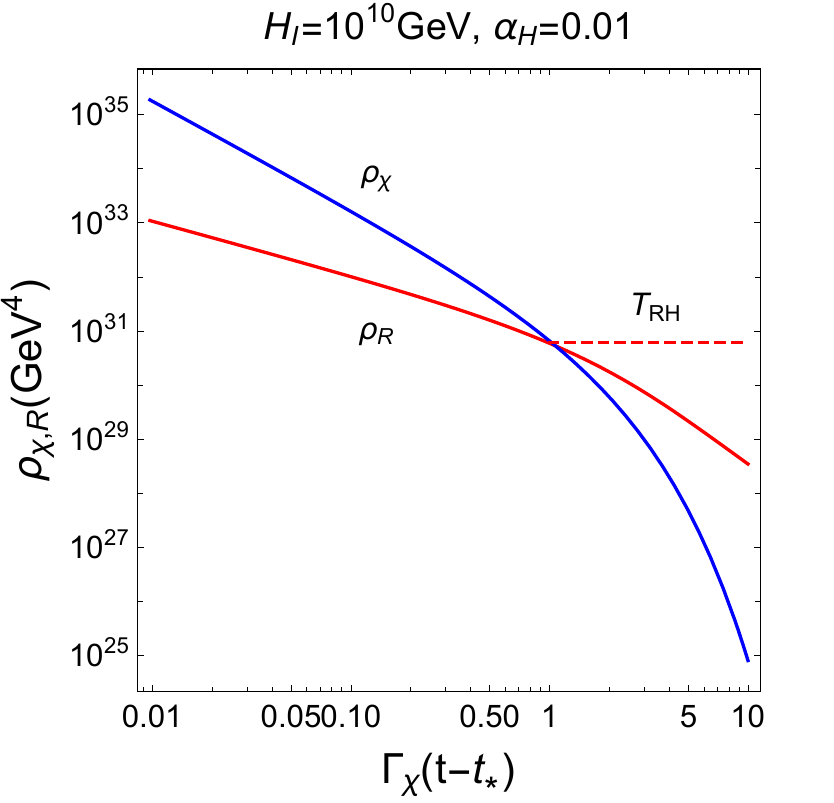}  
 \end{center}
\caption{$\rho_\chi, \rho_R$ as a function of $\Gamma_{\chi_1}(t-t_*)$ in blue and red solid lines, respectively. Red dashed lines correspond to the reheating completion. We took  $\alpha_H=0.01$ and $m_1=100\,{\rm H}_I$ for both plots, and $H_I=10^6,\, 10^{10}\,{\rm GeV}$ in left and right plots, respectively. }
\label{fig:evolution}
\end{figure}

In Fig.~\ref{fig:evolution}, we depict the time evolution of $\rho_{\chi_1}, \rho_R$ as a function of  $\Gamma_{\chi_1}(t-t_*)$ in blue and red solid lines, respectively, for $H_I=10^6\, 10^{10}\,{\rm GeV}$ in left and right plots. We find that the reheating temperature $T_{\rm RH}$ is determined from $\rho_R=\rho_{\chi_1}$, for which $\Gamma_\chi(t-t_*)\sim 1$. Thus, the results are consistent with the identification of the reheating temperature by eq.~(\ref{RH}).

The second term from reheating in eq.~(\ref{rad}) is maximized at $v_{\rm max}=\Gamma_{\chi_1}(t_{\rm max}-t_*)=0.80 A$, resulting in the maximum radiation energy,
\bea
\rho_R(t_{\rm max})= 1.8^{-8/3} \bigg(\rho_R(t_*)+1.0 A \rho_{\chi_1}(t_*) \bigg).
\eea  
Therefore, for $\rho_R(t_*)\lesssim V_0$ and $\rho_{\chi_1}(t_*)\simeq V_0$, we get the maximum ratio of reheating to  preheating contributions to the radiation energy density by
\bea
R_{\rm rh}\equiv \frac{1.0A\rho_{\chi_1}(t_*)}{\rho_R(t_*)}&\simeq& 1600\, \frac{\Gamma_{\chi_1}/m_{\chi_1}}{\lambda_\chi f(\alpha_H,1.3)}\,\bigg(\frac{v_\chi}{M_P}\bigg)^{-1} \nonumber \\
&\simeq& 130\,\frac{\alpha^2_H}{f(\alpha_H,1.3)}\,\bigg(\frac{v_\chi}{M_P}\bigg)^{-1}. \label{ratio}
\eea
Here, we used eq.~(\ref{Apara}) and the numerical result for preheating with the Higgs coupling, $\alpha_H=\kappa_1/\lambda_\chi$, in eq.~(\ref{preheating}), and took eq.~(\ref{decay}) in the second equality.
Thus, from eq.~(\ref{ratio}) with eq.~(\ref{chivev}), we find that 
\bea
R_{\rm rh}\simeq 3700\, \frac{\alpha^2_H}{f(\alpha_H,1.3)}\,\bigg(\frac{m_1}{100H_I}\bigg).
\eea

\begin{figure}[!t]
\begin{center}
 \includegraphics[width=0.45\textwidth,clip]{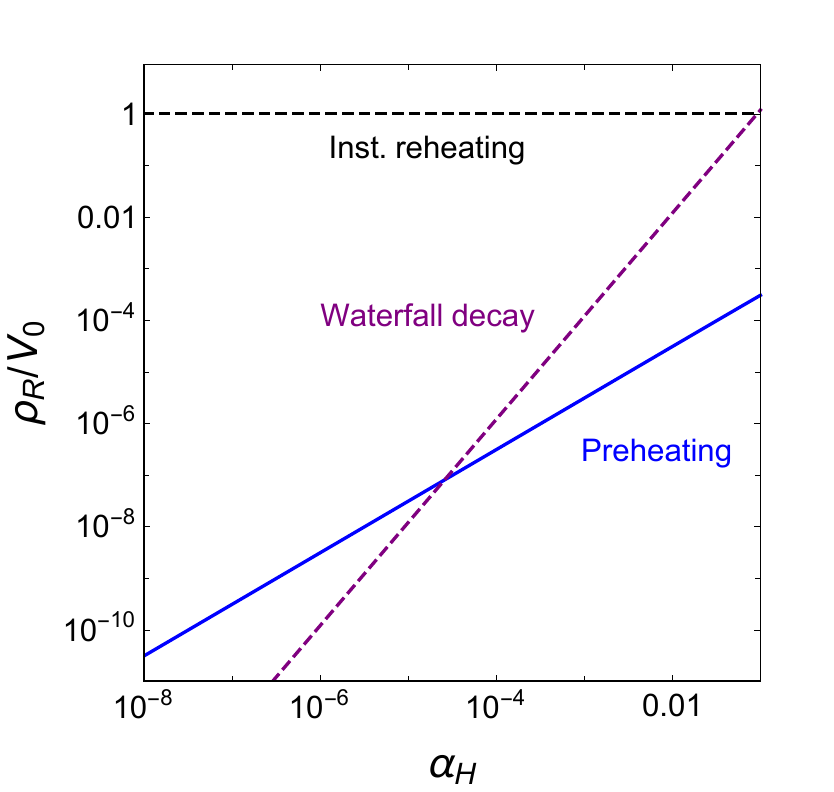}
\end{center}
\caption{$\rho_R/V_0$ as a function of $\alpha_H=\kappa_1/\lambda_\chi$, from preheating in blue solid line, and its maximum value from decay of waterfall field  in purple dashed line. We took $m_1=100\,{\rm H}_I$.  Black dashed lines correspond to the instantaneous reheating. }
\label{fig:comp}
\end{figure}

In Fig.~\ref{fig:comp}, we draw  the fraction of the radiation energy in the waterfall energy density at the end of inflation, $\rho_R/V_0$, as a function of $\alpha_H=\kappa_1/\lambda_\chi$, for preheating in the blue solid line and for the decay of the waterfall field in the purple dashed line. For comparison, we also show the case for instantaneous reheating, $\rho_R/V_0=1$, in the black dashed line.
Then, for $m_1=100 H_I$, we get $R_{\rm rh}\gtrsim 1$ for $\alpha_H\gtrsim 2.6\times 10^{-5}$, so the decay of the waterfall field is dominant for reheating.

For $R_{\rm rh}\gtrsim 1$ and $A\ll v\ll 1$, we can approximate eq.~(\ref{rad}) to
\bea
\rho_R\simeq \frac{4}{5} (\Gamma_{\chi_1}M_P)^2 v^{-1}=\frac{\pi^2}{30} g_* T^4,
\eea
and rewrite the scale factor in eq.~(\ref{scalefactor}) as a function of the radiation temperature $T$ as
\bea
\bigg(\frac{a}{a_*}\bigg)^3\simeq \bigg(\frac{v}{A}\bigg)^2=\bigg(\frac{24\Gamma^2_{\chi_1}M^2_P}{g_* \pi^2 A T^4}\bigg)^2. \label{scaleRH}
\eea
So, choosing $T=T_{\rm RH}$ in the above result, we can take into account the red-shift factor from the waterfall transition to the reheating completion.

\begin{figure}[!t]
\begin{center}
\includegraphics[width=0.325\textwidth,clip]{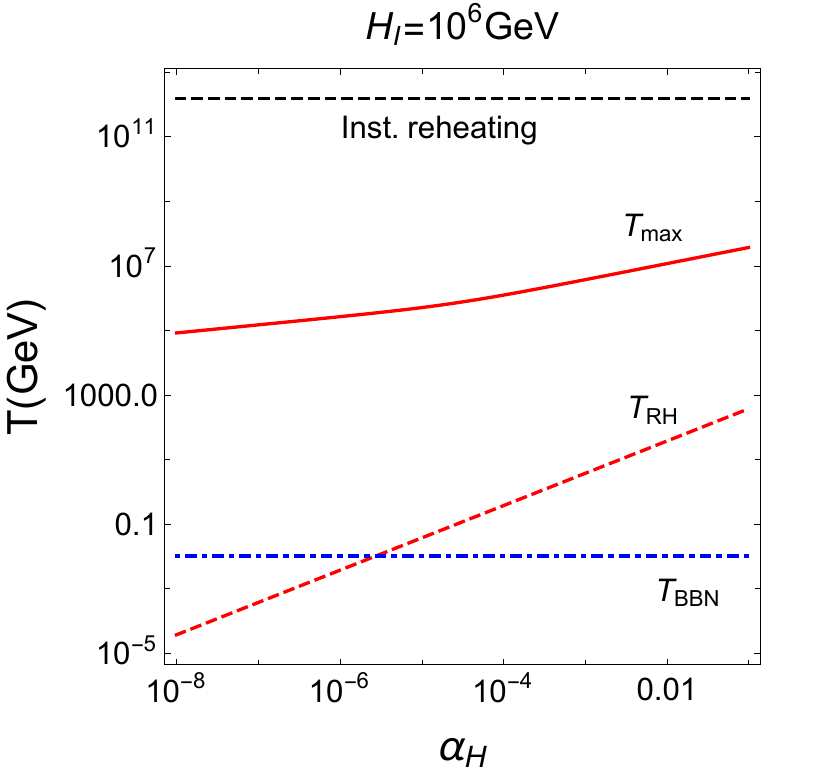}  \includegraphics[width=0.325\textwidth,clip]{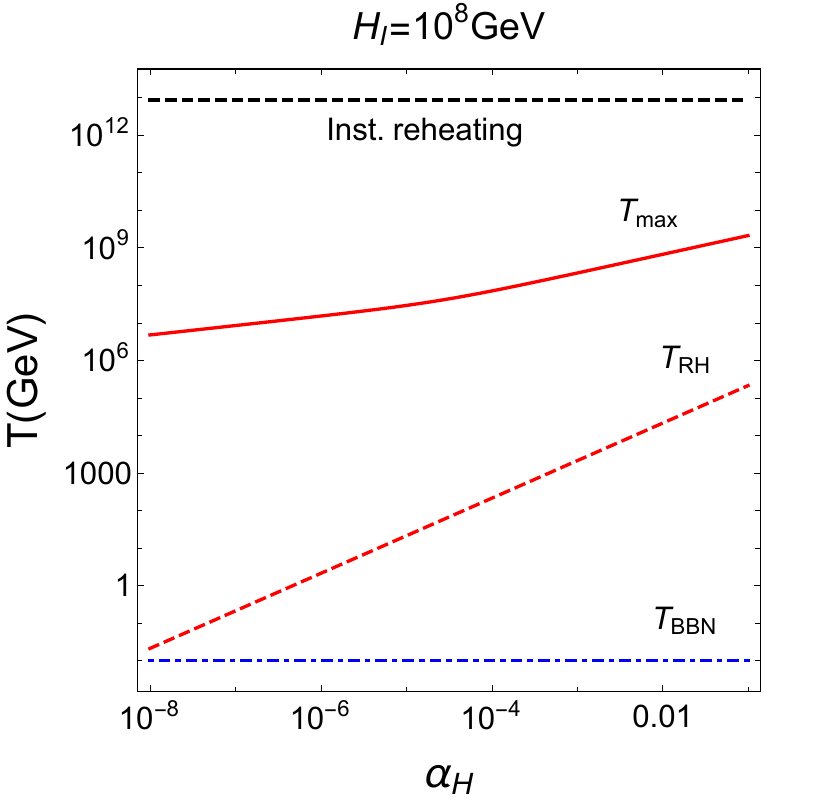} 
\includegraphics[width=0.325\textwidth,clip]{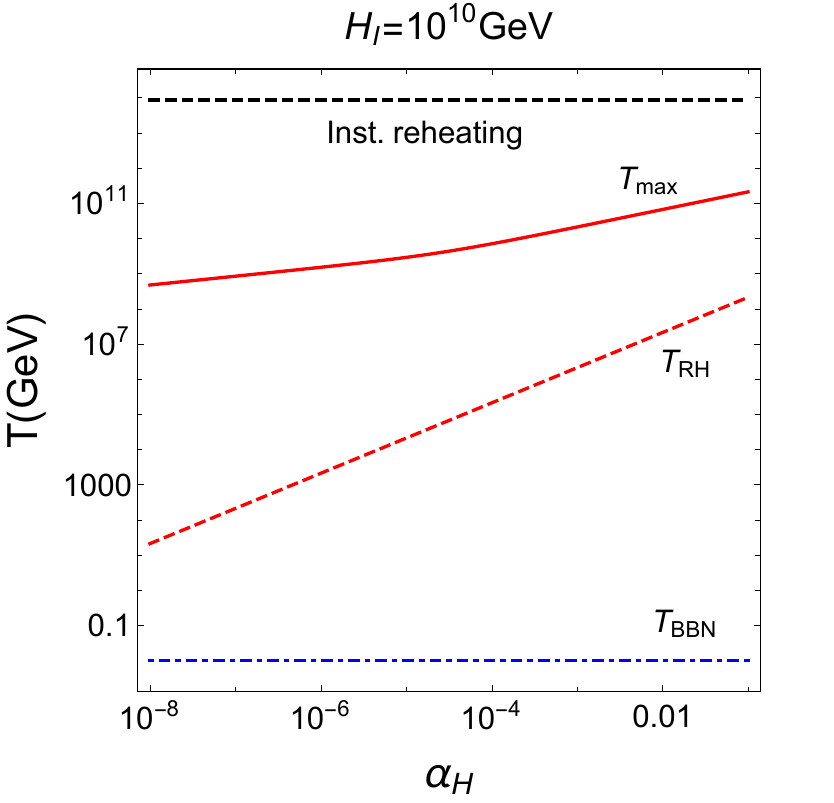} 
\end{center}
\caption{Maximum and reheating temperatures as a function of $\alpha_H=\kappa_1/\lambda_\chi$ in red solid and dashed lines, respectively, for $H_I=10^6, 10^8, 10^{10}\,{\rm GeV}$ from left to right.
We took $m_{\psi,0}=0$ and $m_1=100\,{\rm H}_I$.  Black dashed lines correspond to the instantaneous reheating and blue dot-dashed lines correspond to $T_{\rm RH}=10\,{\rm MeV}$, which is the lower bound from BBN. }
\label{fig:temp}
\end{figure}	

In Fig.~\ref{fig:temp}, we show the maximum and reheating temperatures as a function of $\alpha_H=\kappa_1/\lambda_\chi$ in red solid and dashed lines, respectively. We chose $H_I=10^6, 10^8, 10^{10}\,{\rm GeV}$ from left to right. The lower bound on the reheating temperature for BBN is set to $T_{\rm BBN}=10\,{\rm MeV}$, given in the blue dashed lines, and the black dashed lines correspond to instantaneous reheating, for comparison. Consequently, for a relatively small Hubble scale as in the left-most plot, the waterfall coupling to the Higgs field should be sizable for a successful reheating, for instance, $\alpha_H\gtrsim 5\times 10^{-6}$ for $H_I=10^6\,{\rm GeV}$.
Howe
ver, for a high Hubble scale, as in the middle and right-most plots, small waterfall couplings can be compatible with BBN, for which the preheating effect as shown in Fig.~\ref{fig:comp} becomes important for reheating.

\subsection{Dark matter production}

If the $Z'_2$ symmetry for the waterfall field $\chi_2$ is unbroken in the vacuum, the waterfall field $\chi_2$ can be a dark matter candidate. Dark matter can be produced from preheating, but there is no production from the decay of the waterfall field $\chi_1$ due to the kinematic blocking, namely, $m_{{\tilde \chi}_1}<2m_{\chi_2}$, from eqs.~(\ref{chi1mass}) and (\ref{chi2mass}).

From the result in eq.~(\ref{preheating}), preheating leads to the number density for dark matter $X=\chi_2$ as
\bea
n_X(t_*)=  10^{-3} m^3_1 f(\alpha_X,1.3)/\alpha_X, \label{dmnumber}
\eea
with $\alpha_X={\bar\lambda}_\chi/\lambda_\chi$.
Then, the number density produced from preheating becomes red-shifted at the reheating temperature, as follows,
\bea
n^{\rm pre}_X(T_{\rm RH})=\bigg(\frac{a_*}{a(t_{\rm RH})}\bigg)^3 n_X(t_*)=\bigg(\frac{g_* \pi^2 A T^4_{\rm RH}}{24\Gamma^2_{\chi_1}M^2_P}\bigg)^2 n_X(t_*) \label{dmpre}
\eea
where eq.~(\ref{scaleRH}) at the reheating temperature is used in the second equality. 
As a result, the relic density for dark matter $\chi_2$ at present is given by
\bea
\Omega_{\rm DM} h^2&=& \frac{m_{\chi_2}n_X(t_0)}{\rho_c/h^2}  \nonumber \\
&=&5.9\times 10^6\bigg(\frac{n^{\rm pre}_X(T_{\rm RH})}{T^3_{\rm RH}} \bigg) \bigg(\frac{m_{\chi_2}}{1\,{\rm GeV}} \bigg).
\eea

\begin{figure}[!t]
\begin{center}
\includegraphics[width=0.325\textwidth,clip]{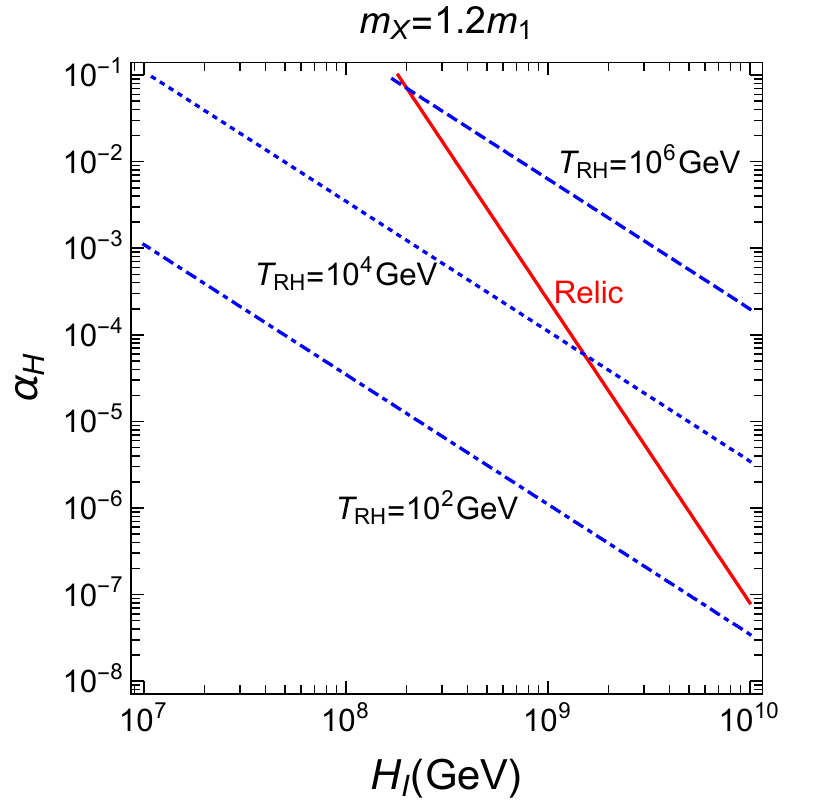}  \includegraphics[width=0.325\textwidth,clip]{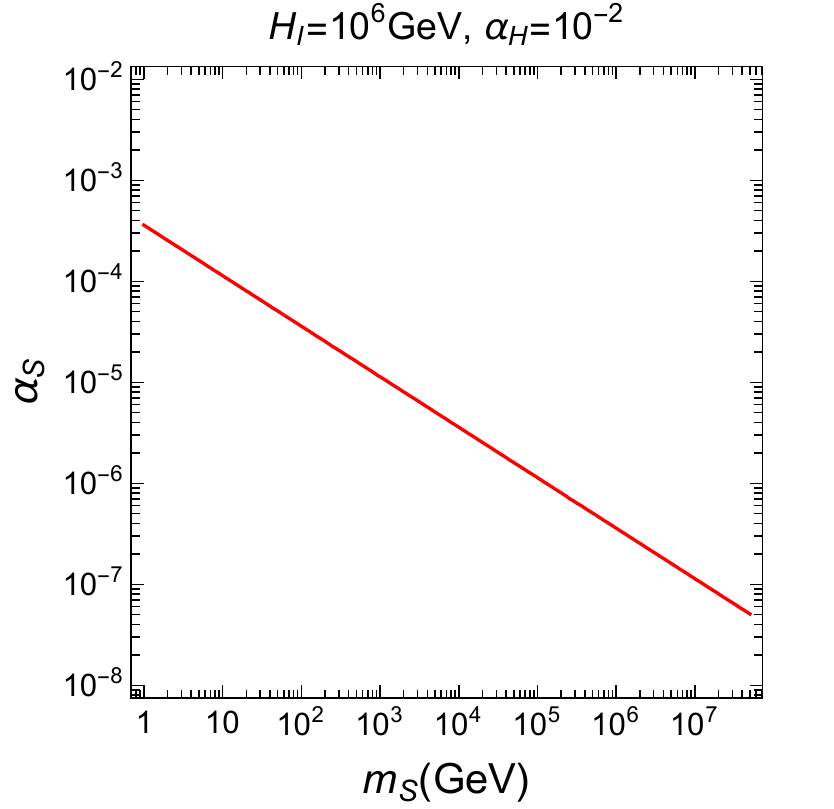}  \includegraphics[width=0.325\textwidth,clip]{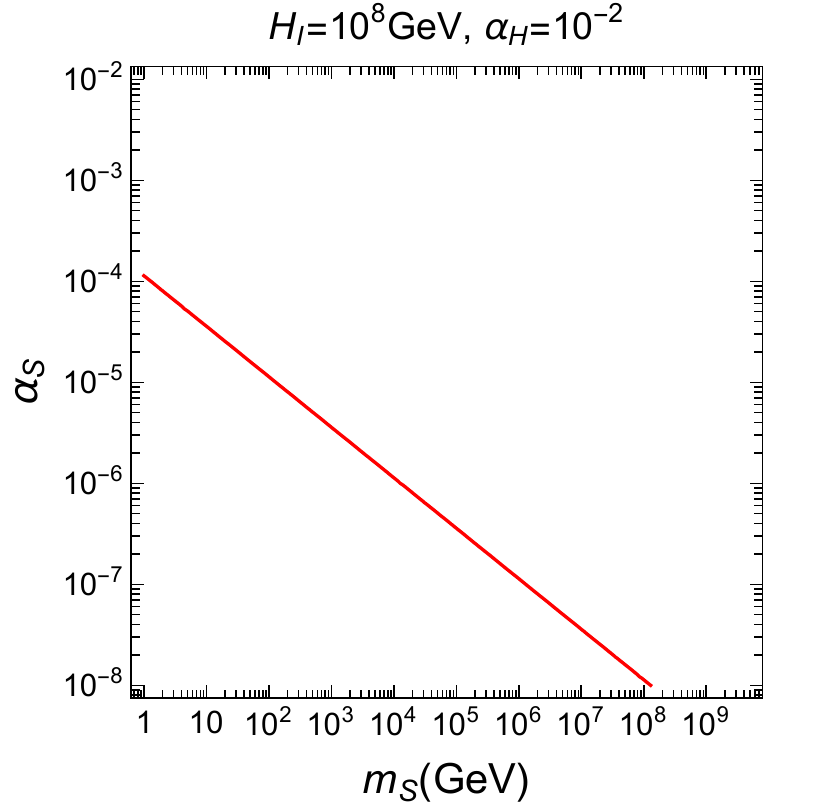} 
\end{center}
\caption{Parameter space satisfying the relic density in red lines. (Left) $H_I$ vs $\alpha_H$ with the relic density from dark matter $\chi_2$. We chose $m_X=1.2m_1$ and drew the contours with reheating temperature, $T_{\rm RH}=10^2, 10^4, 10^6\,{\rm GeV}$, in blue dashed, dotted and dot-dashed lines, respectively. The results are insensitive to $\alpha_X={\bar\lambda}_\chi/\lambda_\chi$ as far as $\alpha_X\lesssim \alpha_H$. (Middle, Right) $m_S$ vs $\alpha_S$ with the relic density from dark matter $S$. We chose $\alpha_H=10^{-2}$ for both plots, but $H_I=10^6, 10^8\,{\rm GeV}$ in the middle and right plots in order.  We took $m_1=100 H_I$ for all the plots. }
\label{fig:dm}
\end{figure}

In the left-most plot of Fig.~\ref{fig:dm}, we show the parameter space for $H_I$ vs $\alpha_H=\kappa_1/\lambda_\chi$, satisfying the relic density for dark matter $\chi_2$ in red line. We took the dark matter mass to $m_X=1.2m_1$ and $m_1=100\,H_I$. The results are insensitive to the dark matter coupling, $\alpha_X$, as far as $\alpha_X\lesssim \alpha_H$, because the numerical formula for the number density of dark matter in eq.~(\ref{dmnumber}) does not depend much on $\alpha_X$.  We show the contours of the reheating temperature, $T_{\rm RH}=10^2, 10^4, 10^6\,{\rm GeV}$, in blue dashed, dotted and dot-dashed lines, respectively. 
In this case, we find that the correct relic density can be achieved for $\alpha_H=0.1-10^{-7}$ and $H_I=2\times 10^8-10^{10}\,{\rm GeV}$. Thus, we need a relatively high reheating temperature for the relic density.

Suppose that there is a real scalar dark matter $S$, which does not participate in the waterfall transition, transforms by $S\to -S$ under the $Z'_2$ symmetry and takes the following effective squared mass,
\bea
m^2_{S,{\rm eff}}=m^2_{S,0} + \lambda_S \chi^2_1 (t)
\eea
where $m^2_{S,0}$ is the squared bare mass and $\lambda_S$ is the coupling between $S$ and the waterfall field $\chi_1$.
Then, $S$ particles can be produced not only from preheating but also from the decay of the waterfall field $\chi_1$. 
The Boltzmann equation for dark matter density after preheating is given by
\bea
{\dot n}_S + 3H n_S =2B_{\chi_1}\Gamma_{\chi_1} \frac{\rho_{\chi_1}}{m_{\chi_1}}
\eea 
where $B_{\chi_1}$ is the decay branching ratio for $\chi_1\to SS$. Then, solving the above Boltzmann equation with eq.~(\ref{RH}), we obtain the number density at the reheating temperature from the decay of the waterfall field $\chi_1$ \cite{Kaneta:2019zgw} as
\bea
n^{\rm decay}_S(T_{\rm RH}) = \frac{g_*(T_{\rm RH})\pi^2 B_{\chi_1}}{18m_{\chi_1}}\, T^4_{\rm RH}.  \label{dmdecay}
\eea
As a result, we obtain the relic density for the scalar dark matter $S$ at present, as follows,
\bea
\Omega_{\rm DM} h^2&=& \frac{m_S n_S(t_0)}{\rho_c/h^2}  \nonumber \\
&=&5.9\times 10^6\bigg(\frac{n_S(T_{\rm RH})}{T^3_{\rm RH}} \bigg) \bigg(\frac{m_S}{1\,{\rm GeV}} \bigg)
\eea
where the total number density for dark matter is given by the sum of both the preheating and the decay contributions,
\bea
n_S(T_{\rm RH})= n^{\rm pre}_S(T_{\rm RH})+ n^{\rm decay}_S(T_{\rm RH}).
\eea
Here,  $ n^{\rm pre}_S(T_{\rm RH})$ is the preheating contribution, obtained from eqs.~(\ref{dmnumber}) and (\ref{dmpre}) with $\alpha_X$ being replaced by $\alpha_S$.

In the middle and right-most plots of Fig.~\ref{fig:dm}, we present the parameter space for $m_S$ vs $\alpha_S$, satisfying the correct relic density for $S$ in the case with $H_I=10^6, 10^8\,{\rm GeV}$, respectively. We chose $\alpha_H=10^{-2}$ and $m_1=100 H_I$ for both plots. As a result, we find that dark matter with masses up to $m_S=1-10^8\,{\rm GeV}$ can be produced with a correct relic density for $10^{-4}\gtrsim \alpha_S\gtrsim 10^{-8}$.

\section{Microscopic realizations}

In this section we show that a Dark QCD model, where the waterfall couples to light dark quarks, can be a microscopic realization of our model.

\bea
{\cal L}_{dQCD}=-m_u u_1 u^c_1-m_u u_2 u^c_2  -y \Phi_1 u^c_1 d-y' \Phi_1 u_1 d^c -iy \Phi_2 u^c_2 d-iy' \Phi^*_2 u_2 d^c +{\rm h.c.}
 \label{micromodel}
\eea

The model has has an underlying $Z_2$ symmetry, such that $\Phi_1\to i\Phi_2$ and $\Phi_2\to-i\Phi_1$. We take the Yukawa couplings to be real.
Then, we can identify the waterfall fields of our model, $\chi_1$ and $\chi_2$, with the real part of the complex scalar fields, $\Phi_1, \Phi_2$, carrying the same dark PQ charges.

For $m_d\ll \Lambda_h\ll m_u$ with $\Lambda_h$ being the dark QCD scale, after integrating out $u_i, u^c_i$ and plugging $\Phi_{1,2}=\frac{1}{\sqrt{2}} \chi_{1,2} \, e^{i\phi/(4f)}$ into eq.~(\ref{micromodel}), we obtain the effective Yukawa couplings for $d,d^c$, as follows,
\bea
{\cal L}_{\rm dQCD, eff} &=&-\frac{yy'}{m_u} (\Phi^2_1-\Phi^2_2)  \, d d^c +{\rm h.c.} \nonumber \\
&=&-\frac{yy'}{2m_u}(\chi^2_1-\chi^2_2)  \, e^{i\phi/(2f)}\, d d^c +{\rm h.c.}. \label{dquarkmass}
\eea
Then, after the dark QCD condensation, integrating out dark mesons and making a shift by $\phi/(2f)\to \phi/(2f)+\pi$, we get the effective waterfall field couplings by
\bea
{\cal L}_{\rm eff}=\frac{1}{2}\mu^2 (\chi^2_1-\chi^2_2) \sin\Big(\frac{\phi}{2f} \Big) \label{effwater}
\eea
with
\bea
\mu^2 =\frac{|y y'|}{m_u}\, \Lambda^3_h. \label{micro}
\eea
Thus, the $Z_2$ symmetry remains unbroken in the effective interactions generated after the dark QCD condensation.
On the other hand, we remark that the inflaton potential in eq.~(\ref{inflaton}) can be also originated from an extra dark QCD condensation at a scale $\Lambda$, respecting the $Z_2$ symmetry with $\phi\to -\phi$. 

The masses of the dark quarks will depend on the VEVs of the waterfall fields, 

\bea
m_{d,{\rm eff}} = \frac{\mu^2v^2_\chi}{\Lambda^3_h}.  \label{dquarkeff}
\eea

Therefore, from $m_{d,{\rm eff}}\ll \Lambda_h \ll m_u$ with eqs.~(\ref{micro}) and (\ref{dquarkeff}), we get $\mu^2 v^2_\chi \ll \Lambda^4_h$ and $\mu^2\ll |yy'| \Lambda^2_h$. 
It follows from $\Lambda^4\ll \mu^2 v^2_\chi\sim 4V_0 \ll \Lambda^4_h$ that we need a hierarchy of the condensation scales by $\Lambda_h\gg \Lambda$.
The waterfall part of the potential arises from
\bea
\Delta V_W=m^2_\chi( |\Phi_1|^2+|\Phi_2|^2)+ \lambda_\chi (|\Phi_1|^4+|\Phi_2|^4)+2{\bar\lambda}_\chi  |\Phi_1|^2 |\Phi_2|^2.
\eea

which is $Z_4$ invariant.

\chapter{Inflation at the pole} \label{HiggsPole} 

\begin{small}
    “We are trying to prove ourselves wrong as quickly as possible, because only in that way can we find progress.”\\
     Richard P. Feynman
        \end{small}
        \vspace{5mm}

In \textbf{section \ref{sec:higgsinflation}}, as well as in \textbf{chapters \ref{chap_UV}} and \textbf{\ref{chap_sugra}}, when we considered Higgs inflation, we relied upon a large non-minimal coupling. 
In this chapter we propose an alternative realization of inflation, in which we exploit conformal symmetry.
We consider an expansion of the inflaton field around the pole of the Einstein frame kinetic term, ensuring perturbativity. 
This model of inflation at the pole belongs to the class of $\alpha$ attractors in supergravity.
We first consider two particular applications for this set up: Higgs inflation and PQ inflation.

The general set up will is given by the Lagrangian,

\bea
\frac{{\cal L}_J}{\sqrt{-g_J}} = -\frac{1}{2}M^2_P\, \Omega(\phi) R(g_J) + |D_\mu \phi|^2 -V_J(\phi) \label{LJhp}
\eea
where the most general form of the non-minimal coupling function and the effective potential are given by
\bea
\Omega(\phi) &=& 1-\frac{1}{3M^2_P}|\phi|^2+ \sum_{n=2}^\infty\frac{b_n}{\Lambda^{2n-4}} \, |\phi|^{2n}, \label{framefunc} \\
V_J(\phi) &=& c_2 |\phi|^2 + c_4|\phi|^4 + \sum_{n=3}^\infty \frac{c_n}{\Lambda^{2n-4}} \, |\phi|^{2n}.  \label{HEFT}
\eea
Here, $\Lambda$ is the cutoff scale, $b_n, c_n$ are dimensionless parameters and $\phi$ is the inflaton field. For Higgs inflation, $\phi=H$, and for PQ inflation, $\phi=\Phi$.
Furthermore, for the case of Higgs inflation, $c_2= \mu$ and $c_4=\lambda$.
We notice that the inflaton field is conformally coupled to gravity but the conformal symmetry is still broken by the masses and the higher order interactions, appearing in both the non-minimal coupling and the scalar potential.

We consider the Lagrangian in eq.~(\ref{LJhp}) and assume for simplicity that all the $b_n$'s in eq.~ (\ref{framefunc}) are zero.
Then, after a Weyl transformation $g_{J,\mu\nu}=g_{E,\mu\nu}/\Omega$ with $\Omega=1-\frac{1}{3M^2_P}|H|^2$,

\bea
\frac{{\cal L}_E}{\sqrt{-g_E}} &=&-\frac{1}{2} M^2_P R(g_E) + \frac{|D_\mu \phi|^2}{\big(1-\frac{1}{3M^2_P}|\phi|^2\big)^2} -V_E(\phi) \nonumber \\
&&-\frac{1}{3M^2_P\big(1-\frac{1}{3M^2_P}|\phi|^2\big)^2}\bigg(|\phi|^2 |D_\mu \phi|^2-\frac{1}{4}\partial_\mu |\phi|^2 \partial^\mu |\phi|^2\bigg). \label{LE}
\eea
\section{Higgs inflation at the pole}
\label{sec:higgsatthepole}

In the case of Higgs inflation, we find the Einstein-frame potential
\bea
V_E(H)= \frac{V_J(H)}{\big(1-\frac{1}{3M^2_P}|H|^2\big)^2}.
\eea

In the unitary gauge, $H^T=(0,h)^T/\sqrt{2}$, the first two terms in the second line of eq.~(\ref{LE}) cancel, allowing us to simplify the Einstein-frame Lagrangian,
\bea
\frac{{\cal L}_E}{\sqrt{-g_E}} =-\frac{1}{2} M^2_P R +\frac{1}{2}\,\frac{(\partial_\mu h)^2}{\big(1-\frac{1}{6M^2_P}h^2\big)^2} - \frac{V_J\big(\frac{1}{\sqrt{2}}h\big)}{\big(1-\frac{1}{6M^2_P}h^2\big)^2}, \label{Linf}
\eea

where $h$ is identified with the Higgs boson.
Still, the kinetic terms in eq. (\ref{Linf}) are not canonical. To find the field redefinition that canonicalizes them we just use that
$\left(\partial_\mu K\left(\rho\right)\right)^2 = \left(\dv{K(\rho)}{\rho}\right)^2 = \left(\dv{k(\rho)}{\rho}\right)^2 \left(\partial_\mu \rho\right)^2$,
\bea
K(\rho) = \frac{1}{\sqrt{6}M_P}\tanh^{-1}\left(\frac{1}{\sqrt{6}M_p}\rho\right).\label{canhigs}
\eea

Using eq.~(\ref{canhigs}) we redefine the Higgs field as
\bea
h=\sqrt{6}M_P \tanh\Big(\frac{\phi}{\sqrt{6}M_P}\Big), \label{can}
\eea

and we obtain the canonically normalized Einstein frame Lagrangian,
\bea
\frac{{\cal L}_E}{\sqrt{-g_E}} =-\frac{1}{2} M^2_P R + \frac{1}{2}(\partial_\mu\phi)^2 - V_E(\phi),
\eea
with 
\bea
V_E(\phi)= \cosh^4\Big(\frac{\phi}{\sqrt{6}M_P}\Big) V_J\bigg(\sqrt{3}\tanh\Big(\frac{\phi}{\sqrt{6}M_P}\Big)\bigg). \label{Einpot}
\eea

Looking at eq. (\ref{Einpot}) we realize that the success of slow-roll inflation depends on the particular shape of the Jordan frame potential. 
For now, we take the minimal form of the effective potential, and we will worry about generalizing it in the next sections.
We consider the case in which only the a single power m contributes to eq.~(\ref{HEFT}), 
\bea
V_J(H) = c_m \Lambda^{4-2m} |H|^{2m} \bigg(1-\frac{1}{3M^2_P}|H|^2\bigg)^2,  \label{poleJ}
\eea 
or, equivalently, we take the Einstein frame potential to be
\bea
V_E(H)=c_m \Lambda^{4-2m} |H|^{2m}. \label{poleE}
\eea

Then, in terms of the canonical Higgs field $\phi$ given in eq.~(\ref{can}), 
\bea
V_E(\phi)&=& \frac{c_m}{2^m}  \Lambda^{4-2m} h^{2m} \nonumber \\
&=& 3^m c_m\Lambda^{4-2m}M^{2m}_P \bigg[  \tanh\Big(\frac{\phi}{\sqrt{6}M_P}\Big)\bigg]^{2m}. \label{inflatonpot}
\eea 
Therefore, the slow-roll conditions are satisfied at large field values, $|\phi|$, or $|h|\sim \sqrt{6} M_P$. Since inflation takes places near the pole of the Higgs kinetic term in eq.~(\ref{Linf}), we dub this possibility the \textit{Higgs pole inflation}.

\subsection{Inflationary predictions}\label{sec:higgspoleinfla}

We take the inflaton potential in eq.~(\ref{inflatonpot}) to be
\bea
V_E(\phi)=V_I \bigg[  \tanh\Big(\frac{\phi}{\sqrt{6}M_P}\Big)\bigg]^{2m},
\eea
with $V_I\equiv 3^m c_m\Lambda^{4-2m}M^{2m}_P$ being the vacuum energy during inflation.
Then, we first obtain the slow-roll parameters (\ref{eq:slowrolleps}) and (\ref{eq:slowrolleta}) as,
\bea
\epsilon &=& \frac{4}{3} m^2 \bigg[  \sinh\Big(\frac{2\phi}{\sqrt{6}M_P}\Big)\bigg]^{-2} ,  \label{ep} \\
\eta &=&-\frac{4m}{3} \bigg[  \cosh\Big(\frac{2\phi}{\sqrt{6}M_P}\Big)-2m \bigg]  \bigg[\sinh\Big(\frac{2\phi}{\sqrt{6}M_P}\Big)\bigg]^{-2}, \label{eta}
\eea

and the number of e-foldings (\ref{efoldnum}) is given by
\bea
N&=&\frac{3}{4m}\, \bigg[ \cosh\Big(\frac{2\phi_*}{\sqrt{6}M_P}\Big)- \cosh\Big(\frac{2\phi_e}{\sqrt{6}M_P}\Big)  \bigg] \label{efold}
\eea
where $\phi_*, \phi_e$ are the values of the Higgs boson at horizon exit and the end of inflation, respectively. Here, we note that $\epsilon=1$ determines $\phi_e$.
As a result, using  eqs.~(\ref{ep}), (\ref{eta}), (\ref{efold}) and $N\simeq \frac{3}{4m}\,  \cosh\Big(\frac{2\phi_*}{\sqrt{6}M_P}\Big)$ for $\phi_*\gg \sqrt{6} M_P$ during inflation, we obtain the slow-roll parameters at horizon exit in terms of the number of e-foldings as
\bea
\epsilon_* &\simeq & \frac{3}{4\big(N^2-\frac{9}{16m^2}\big)},  \\
\eta_* &\simeq& \frac{3-2N}{2\big(N^2-\frac{9}{16m^2}\big)}.
\eea
Thus, we get the spectral index (\ref{eq:planckn}) and the tensor-to-scalar ratio (\ref{eq:planckr}) in terms of the number of e-foldings, 
\bea
n_s&=& 1-\frac{4N+3}{2\big(N^2-\frac{9}{16m^2}\big)}, \label{sindex}\\
r&=&  \frac{12}{N^2-\frac{9}{16m^2}}. \label{ratiohiggspole}
\eea
Therefore, the inflationary predictions are pretty much insensitive to the value of $m$.
For the usual $N=60$ e-folds, eq.~(\ref{sindex}) and eq.~(\ref{ratiohiggspole}) predict $n_s=0.966$ and $r=0.0033$, both of them in perfect agreement with Plack data. Furthermore, from the CMB normalization (\ref{eq:CMBnormalization}) we get limits on the $c_m$ parameters,

\bea
3^m c_m \Big(\frac{\Lambda}{M_P}\Big)^{4-2m}= (3.1\times 10^{-8}) \,r=1.0\times 10^{-10}, \label{CMBhp}
\eea
where we took $r=0.0033$, in the second equality. 

In the case $m=0$, the inflaton potential in Einstein frame becomes just a constant, which we can interpret as the cosmological constant observed today. 
For $m=1$, we recover Starobinsky inflation. In this case, we identify the $c_1 \Lambda^2=\mu^2_H$ in eq.~(\ref{HEFT}), which must be positive.  For $m=1$, the vacuum energy density is given by $V_I=3\mu^2_H M^2_P$, so that the CMB normalization constrains the Higgs mass to $\mu_H= 1.4\times10^{13}\,{\rm GeV}$.
For the SM Higgs, we need to take $m>1$ in order to accommodate electroweak symmetry breaking.
Finally, if we take $m=2$ in eq.~(\ref{inflatonpot}), we identify  $c_2=\lambda_H$ in eq.~(\ref{HEFT}) and the vacuum energy during inflation becomes $V_I=9 c_2M^4_P=9\lambda_H M^4_P$. In this case, the CMB normalization fixes $\lambda_H=1.1\times 10^{-11}$.

\subsection{General Higgs potential and inflationary bounds}
We take the generalized Higgs potential 
\bea
V_E(H) = V_0+ \mu^2_H |H|^2 +\lambda_H |H|^4 + \sum^\infty_{m=3} c_m \Lambda^{4-2m} |H|^{2m}. 
\eea

In terms of the canonical Higgs field~(\ref{can}), and taking the unitary gauge, 
\bea
V_E(\phi)&=& V_0 +3\mu^2_HM^2_P   \tanh^2\Big(\frac{\phi}{\sqrt{6}M_P}\Big)+9\lambda_HM^4_P  \tanh^4\Big(\frac{\phi}{\sqrt{6}M_P}\Big)\nonumber \\
&&+ \sum^\infty_{m=3} 3^m c_m \Lambda^{4-2m} M^{2m}_P   \tanh^{2m}\Big(\frac{\phi}{\sqrt{6}M_P}\Big).
\eea

Setting the vacuum energy $V_0=0$, we can generalize the CMB normalization in eq.~(\ref{CMBhp})  to
\bea
\frac{3\mu^2_H}{M^2_P} +9\lambda_H+ \sum^\infty_{m=3} 3^m c_m \Big(\frac{\Lambda}{M_P}\Big)^{4-2m}=1.0\times 10^{-10}. \label{gCMB}
\eea

Unless there are fine-tuned cancellations, each individual term in the Higgs potential needs to be small.
To realize successful inflation, we need to take $9\lambda_H M^4_P\gtrsim 3|\mu^2_H|M^2_P, \; V_0$ during inflation. 
It sufficient to take $\lambda_H\gtrsim  \frac{|\mu^2_H|}{3M^2_P}=1.1 \times 10^{-15}$ for the correct electroweak symmetry breaking \footnote{It is interesting to notice that the CMB normalization only leads to the bound on the Higgs mass parameter as $|\mu_H|\lesssim 1.4\times 10^{13}\,{\rm GeV}$, although we need to understand the Higgs mass hierarchy.} with $|\mu_H|=88\,{\rm GeV}$; $\lambda_H\gtrsim  \frac{V_0}{9M^2_4}=4.6 \times 10^{-85}$ for the observed dark energy, $V_0=(2.2\times 10^{-3}\,{\rm eV})^4$.
Therefore, the CMB constraint $\lambda_H= 1.1\times 10^{-11}$, obtained in the previous subsection, is consistent with both electroweak symmetry breaking and dark energy. Such a tiny Higgs quartic coupling during inflation is achievable due to the RG running of the Higgs quartic coupling, although it is subject to the precise values of the low energy parameters in the SM such as the top quark Yukawa coupling \cite{Elias-Miro:2011sqh,Degrassi:2012ry} and the threshold or running contributions from new particles \cite{threshold}.

Including higher order terms in the potential with $m\geq 3$ is possible, as far as $3^m c_m\big(\frac{\Lambda}{M_P}\big)^{4-2m}\gtrsim 9\lambda_H, \frac{3|\mu^2_H|}{M^2_P} , \frac{V_0}{M^4_P}$. In this case, from the CMB normalization in eq.~(\ref{CMBhp}), the Higgs quartic term is bounded by $\lambda_H\lesssim 1.1\times 10^{-11}$, whereas the observed values of the other renormalizable parameters, $|\mu^2_H|$ and $V_0$, are small enough. 
In the case in which the power $m$ is dominating the dynamics of inflaton, these higher order terms with $m+k$, $k\geq 1$ will be suppressed for $c_m\gtrsim 3^k \big(\frac{M_P}{\Lambda}\big)^{2k} c_{m+k}$.
\subsection{Reheating} \label{sec:higgspolereheating}

In this section we study the reheating dynamics for a general equation of state. We discuss the perturbative processes responsible for reheating and compute the reheating temperature.
After inflation, the higgs field no longer has super-Planckian values, so we can take the approximations for  $|H|\ll \sqrt{6}M_P$, where $\phi\simeq h$, so the Einstein-frame potential reads

\bea
V_E(\phi)\simeq \frac{c_m}{2^m} \Lambda^{4-2m} \phi^{2m}\equiv \alpha_m \phi^{2m}, \label{rehpot}
\eea

In the general case in which $m>1$, the field will oscillate anharmonically. 

We define the end of inflation when $\ddot a=0$, where $a$ is the cosmological scale factor. 
The inflaton field value at that time is given by \cite{Ellis:2015pla, Garcia:2020wiy}
\begin{align}
   \phi_{\rm end} &\simeq 
   \sqrt{\frac{3}{8}}M_P\ln\left[\frac{1}{2} + \frac{2m}{3}\left(2m + \sqrt{4m^2+3}\right)\right].
   \label{phiend}
\end{align}

The condition $\ddot a=0$ is equivalent to $\dot\phi_{\rm end}^2 = V_E(\phi_{\rm end})$
and thus the inflaton energy density at $\phi_{\rm end}$ is nothing but  $\rho_{\rm end}=\frac{3}{2}V_E(\phi_{\rm end})$.

Applying the virial theorem to $S\equiv \phi {\dot\phi}$,
\bea
\Big\langle\frac{dS}{dt} \Big\rangle=0,
\eea
we obtain
\bea
\Big\langle \frac{1}{2}{\dot\phi}^2\Big\rangle=-\Big\langle \frac{1}{2} \phi {\ddot\phi} \Big\rangle=m\langle V_E(\phi)\rangle.
\eea
Here, we ignored the Hubble friction term and used the equation of motion for the inflaton, ${\ddot\phi}=-V'$.
We also used the inflaton potential during reheating from eq.~(\ref{rehpot}). Then, the averaged energy density and pressure for the inflaton becomes
\bea
\rho_\phi &=&\Big\langle \frac{1}{2}{\dot\phi}^2\Big\rangle+\langle V_E(\phi)\rangle=(m+1) \langle V_E(\phi)\rangle, \label{infdens} \\
p_\phi &=& \Big\langle \frac{1}{2}{\dot\phi}^2\Big\rangle-\langle V_E(\phi)\rangle=(m-1) \langle V_E(\phi)\rangle.
\eea
And the averaged equation of state for the inflaton during reheating is given by
\bea
\langle w_\phi\rangle=\frac{p_\phi}{\rho_\phi} =\frac{m-1}{m+1}.  \label{eos1}
\eea
In general, the universe will not be matter dominated, $w_\phi\neq 0$.

We write the inflaton during reheating as $\phi=\phi_0(t) {\cal P}(t)$, where $\phi_0(t)$ is the amplitude of the inflaton oscillation, that stays constant over a single oscillation, and ${\cal P}(t)$ is the periodic envelope function.  
From eq.~(\ref{infdens}) with the energy conservation, we obtain $\rho_\phi=(m+1) \langle V_E(\phi)\rangle=V_E(\phi_0)$. Then, we get $\langle {\cal P}^{2m}\rangle=\frac{1}{m+1}$. 

From the energy density for the inflaton,
\bea
 \frac{1}{2}{\dot\phi}^2+V_E(\phi)=V_E(\phi_0),
\eea
we obtain the equation for the periodic function ${\cal P}$, as follows,
\bea
{\dot{\cal P}}^2 =\frac{2\rho_\phi}{\phi^2_0} \Big(1-{\cal P}^{2m}\Big)= \frac{m^2_\phi}{m(2m-1)}\,  \Big(1-{\cal P}^{2m}\Big) \label{Peq}
\eea
where we used the effective inflaton mass in the second equality,
\bea
m^2_\phi =V^{\prime\prime}_E(\phi_0) =2\alpha_m  m (2m-1) \phi^{2m-2}_0. \label{inflatonmasshiggspole}
\eea

Thus, from the integral of  eq.~(\ref{Peq}), we get the angular frequency of the inflaton oscillation \cite{Garcia:2020wiy} as
\bea
\omega = m_\phi\sqrt{\frac{\pi m}{2m-1}}\, \frac{\Gamma\big(\frac{1}{2}+\frac{1}{2m}\big)}{\Gamma\big(\frac{1}{2m}\big)}. \label{freq}
\eea
As a result, we can make a Fourier expansion of the periodic function  $\cal P$ by
\bea
{\cal P}(t) =\sum_{n=-\infty}^\infty {\cal P}_n \,e^{-in\omega t}. 
\eea

If we include the Hubble friction and the inflaton decay terms through a $\Gamma_\phi$ factor, the equation of motion becomes

\bea
{\ddot\phi}+ (3H+\Gamma_\phi) {\dot\phi} +V'_E=0
\eea
where $\Gamma_\phi$  is given by
\bea
\Gamma_\phi=\sum_f\Gamma_{\phi\to f{\bar f}} + \sum_{V=W,Z}\Gamma_{\phi\phi\to VV}.
\eea

The above equation can be approximated to the Boltzmann equation for the averaged energy density,
\bea
{\dot\rho}_\phi + 3(1+w_\phi)H \rho_\phi\simeq -\Gamma_\phi (1+w_\phi) \rho_\phi.
\label{Boltzmann_phi}
\eea
while the Boltzmann equation governing the radiation energy density $\rho_R$ is given by
\bea
{\dot\rho}_R + 4 H\rho_R =\Gamma_\phi (1+w_\phi) \rho_\phi. \label{Boltzmann_rad}
\eea

Furthermore, the Yukawa couplings of the Higgs inflaton to the SM fermions, 

\bea
{\cal L}_{\rm int}=-\frac{1}{\sqrt{2}}y_f\phi {\bar f} f
\eea
provide another decay channel \cite{Ichikawa:2008ne,Garcia:2020wiy},
\bea
\Gamma_{\phi\to f{\bar f}}=\frac{1}{8\pi (1+w_\phi)\rho_\phi} \sum_{n=-\infty}^\infty |M^f_n|^2 (E_n\beta^f_n)
\eea
where $E_n=n \omega$ and
\bea
|M^f_n|^2&=&y^2_f \phi^2_0 |{\cal P}_n|^2 E^2_n \beta^2_n, \\
\beta^f_n &=& \sqrt{1-\frac{4m^2_f}{E^2_n}}. 
\eea
Then, averaging over oscillations, we get
\bea
\langle\Gamma_{\phi\to f{\bar f}}\rangle&=&\frac{y^2_f \phi^2_0 \omega^3}{8\pi (1+w_\phi)\rho_\phi}\, \sum_{n=-\infty}^\infty  n^3 |{\cal P}_n|^2 \langle\beta^3_n\rangle \nonumber \\
&=& \frac{y^2_f \omega^3}{8\pi m^2_\phi}\,(m+1)(2m-1) \Sigma_m^f \Bigg\langle\bigg(1-\frac{4m^2_f}{\omega^2 n^2}\bigg)^{3/2}\Bigg\rangle \label{decayratepq}
\eea
where $\Sigma_m^f =\sum_{n=1}^\infty  n^3 |{\cal P}_n|^2$ whose values  are given as a function of the parameter $m$ in Table \ref{table:1}.
Here, the effective mass for the fermion is given by $m_f=\frac{1}{\sqrt{2}} y_f\phi(t)=(m_f/v)\phi(t)$ where $v$ is the VEV of the Higgs field in the true vacuum.

Similarly, we have a contribution due to the gauge interactions,

\bea
{\cal L}_{\rm int}=\frac{1}{4}g^2 \phi^2 W^+_\mu W^{-\mu}+\frac{1}{8} (g^2+g^{\prime 2})\phi^2 Z_\mu Z^\mu
\eea

Taking the unitarity gauge, we obtain 
\bea
\Gamma_{\phi\phi\to VV}=\frac{1}{8\pi (1+w_\phi)\rho_\phi} \sum_{n=1}^\infty |M^V_n|^2 (E_n\beta^V_n)
\eea

with 
\bea
|M^W_n|^2&=&\frac{1}{4}g^4 \phi^4_0 |({\cal P}^2)_n|^2 \bigg(3+\frac{4E^4_n}{m^4_W}-\frac{4E^2_n}{m^2_W}\bigg), \\
|M^Z_n|^2&=&\frac{1}{8}(g^2+g^{\prime 2})^2 \phi^4_0 |({\cal P}^2)_n|^2 \bigg(3+\frac{4E^4_n}{m^4_Z}-\frac{4E^2_n}{m^2_Z}\bigg),
\eea
and $\beta^V_n = \sqrt{1-\frac{m^2_V}{E^2_n}}$ with $V=W, Z$. Here, the effective masses for the gauge bosons are given by $m^2_W=\frac{1}{4} g^2 \phi^2(t)$ and $m^2_Z=\frac{1}{4}(g^2+g^{\prime 2}) \phi^2(t)$, and $({\cal P}^2)_n$ are the Fourier coefficients of the expansion, ${\cal P}^2=\sum_{n=-\infty}^\infty ({\cal P}^2)_n \,e^{-in\omega t}$.
Then, the averaged scattering rate for the inflaton is given by
\bea
\langle\Gamma_{\phi\phi\to WW}\rangle&=&\frac{g^4 \phi^2_0 \omega}{16\pi m^2_\phi}\,(m+1)(2m-1) \Phi_W, \\
\langle\Gamma_{\phi\phi\to ZZ}\rangle&=&\frac{(g^2+g^{\prime 2})^2 \phi^2_0 \omega}{32\pi m^2_\phi}\,(m+1)(2m-1) \Phi_Z ,
\eea
with
\bea
\Phi_V\equiv \Sigma_m^V \bigg\langle\frac{\beta^V_n(3+3(\beta^V_n)^4-2(\beta^V_n)^2) }{(1-(\beta^V_n)^2)^2} \bigg\rangle, \quad V=W, Z
\eea
where $\Sigma_m^V=\sum_{n=1}^\infty n  |({\cal P}^2)_n|^2 $ whose values are given as a function of the parameter $m$ in Table \ref{table:1}.

\begin{table}[htbp!]
    \centering
    \begin{tabular}{|c|c|c|}
    \hline
        $m$  & $\Sigma_m^f$  & $\Sigma_m^V$ \\
        \hline
        1 &  0.250 & 0.125\\
        \hline
        2 &   0.241 &   0.125   \\
        \hline
        3 &    0.244    &    0.124  \\
        \hline
        4 &    0.250       &     0.122 \\
        \hline
        5 &     0.257      &       0.120        \\
        \hline
        6 &  0.264        &    0.119 \\
        \hline 
        7 &    0.270      &    0.117      \\
        \hline
        8  &     0.276      &     0.116      \\
        \hline
        9 &      0.281        &       0.115      \\
        \hline
        10 &      0.287        &       0.114       \\
        \hline
        
    \end{tabular}
    \caption{Sums of the Fourier coefficients appearing in the decay rates of the Higgs inflaton. We chose some values of the equation of state parameter $m$ during reheating.}
    \label{table:1}
\end{table}

We note that the decay and scattering rates of the inflaton scale with the inflaton energy density by $\Gamma_{\phi\to f{\bar f}}=\gamma_\phi \rho^l_\phi$ with $l=\frac{1}{2}-\frac{1}{2m}$ and $\Gamma_{\phi\phi\to VV}={\hat\gamma}_\phi \rho^n_\phi$ with $n=\frac{3}{2m}-\frac{1}{2}$  \cite{Garcia:2020wiy}. For instance, for the quadratic potential with $m=1$, we get $l=0$ and $n=1$, so the scattering rate is smaller than the decay rate for similar inflaton couplings. In this case, the
scattering process cannot achieve the reheating mechanism alone, without any decay of the
inflation background. However, for $m=2$, both the decay and scattering rates scale by $\rho^\frac{1}{2}_\phi$, so they are comparable. Moreover, the $m>2$ case makes the decay rate being of higher power in the inflaton energy density, that is, $l>n$, tending to be suppressed as compared to the scattering rate. 

 From eq.~(\ref{freq}) with eqs.~(\ref{inflatonmasshiggspole}) and eq.~(\ref{rehpot}), we have $\omega^2\sim m^2_\phi\sim c_m \Lambda^{4-2m}\phi^{2m-2}_0$. Then, together with $m_f\sim y_f \phi_0$ and $m_V\sim g \phi_0$, we get $m^2_f/\omega^2\sim (y^2_f/c_m)(\phi_0/\Lambda)^{4-2m}\lesssim 1$ and $m^2_V/\omega^2\sim (g^2/c_m)(\phi_0/\Lambda)^{4-2m}\lesssim 1$. Then, for $c_m\sim 10^{-10}$ from the CMB normalization in eq.~(\ref{eq:CMBnormalization}), we can avoid the kinematic suppression for the inflaton decay or scattering, only if $y_f, g\lesssim \sqrt{c_m}\sim 10^{-5}(\Lambda/\phi_0)^{4-2m}$. 
 
 The effect of the time-dependent effective masses for fermions and bosons can be determined by averaging over the oscillations the effective decay or scattering rates. The phase-space dependence on the effective masses leads to a suppression of the production rates, even if the production is never completely blocked, as for any coupling there will be a time interval around $\phi=0$ during which the production is allowed \cite{Garcia:2020wiy}. The resulting suppression in the production for large couplings to the inflaton can be numerically evaluated. 
 
 In the initial stage of the inflaton oscillation with $\phi_0\sim \Lambda\sim M_P$ just after inflation, as in eq.~(\ref{phiend}),  only light quarks and leptons in the SM can be produced abundantly from the perturbative decays of the Higgs inflaton. However, heavy quarks or gauge bosons can be produced only in the small field values with $\phi_0\ll \Lambda$ (or at the later time) for $m<2$ and in the large field values with $\phi_0\sim\Lambda$ (or at the earlier time) for $m>2$, when the decay or scattering channels of the inflaton are kinematically open. Thus, as will be discussed in the later section, the parametric resonance during preheating could be also relevant for the gauge boson production.

 \subsection*{Reheating temperature}

 For $a_{\rm end} \ll a\ll a_{\rm RH}$ where $a_{\rm RH}$ is the scale factor at the time reheating is complete, we can ignore the inflaton decay rate and integrate  eq.~(\ref{Boltzmann_phi}) without the decay rate to obtain
 \be
 \rho_{\rm \phi}(a)\simeq \rho_{\rm end}\left(\frac{a_{\rm end}}{a}\right)^{\frac{6m}{m+1}}. \label{inflatondensity}
 \ee
 This is due to the general equation of state during reheating, given in eq.~(\ref{eos1}).
 When the reheating process is dominated by the perturbative decays of the inflaton into light fermions in the SM, we obtain the reheating  temperature, depending on the power of the leading inflaton potential, $V(\phi)=\alpha_m \phi^{2m}$, during reheating, as follows  \cite{Garcia:2020wiy},
 \bea
 T_{\rm RH}= \left\{\begin{array}{cc} \bigg(\frac{30}{\pi g_*(T_{\rm RH})}\bigg)^{1/4} \bigg[ \frac{2m}{7-2m}\, \sqrt{3}M_P \gamma_\phi\bigg]^{\frac{m}{2}}, \qquad\qquad 7-2m>0, \vspace{0.3cm}\\  \bigg(\frac{30}{\pi g_*(T_{\rm RH})}\bigg)^{1/4}\bigg[\frac{2m}{2m-7}\,\sqrt{3}M_P\gamma_\phi\,( \rho_{\rm end})^{\frac{2m-7}{6m}}\bigg]^{\frac{3m}{4(m-2)}}, \quad 7-2m<0 \end{array} \right.
 \eea
 where we approximated the decay rate of the inflaton from eq.~(\ref{decayrate}) by $\Gamma_\phi\simeq \Gamma_{\phi\to f{\bar f}} \equiv \gamma_\phi \rho^l_\phi$, with $l=\frac{1}{2}-\frac{1}{2m}$, and
 \bea
 \gamma_\phi\equiv  \sum_f \frac{1}{8} N_c y^2_f\sqrt{2\pi}  m^2 (m+1)\, (\alpha_m)^{\frac{1}{2m}}\,\bigg(\frac{\Gamma\big(\frac{1}{2}+\frac{1}{2m}\big)}{\Gamma\big(\frac{1}{2m}\big)}\bigg)^3 \Sigma^f_m
 \eea
 where $f$ runs over all the SM fermions that can be produced and $N_c$ indicates the number of colors for each fermion.
 Here, $g_*(T_{\rm RH})$ is the effective number of degrees of freedom at the reheating temperature $T_{\rm RH}$, and $\rho_{\rm end}$ is the inflaton energy density at the end of inflation, and we have neglected the average effective masses for the light fermions generated by their couplings to the inflaton during reheating.
 We note that only for $2m<7$ the reheating temperature is independent of $\rho_{\rm end}$.

 We now consider the effects of  time-dependent effective masses of fermions during the decay process of the Higgs inflaton. These effects are included in the calculations of the decay rate (\ref{decayrate}) with the averaged effective kinetic factors. For small effective masses, we can neglect the kinematic factors. However, for large Yukawa couplings to the Higgs inflaton, the production of fermions is suppressed.
In the limit of large effective masses, we can replace the averaged kinetic factor in the decay rate by 
 \bea
 \Gamma_\phi &=& \gamma_\phi\rho_\phi^l\Bigg\langle\bigg(1-\frac{4m^2_f}{\omega^2 n^2}\bigg)^{3/2}\Bigg\rangle  \nonumber \\
 & \simeq&  \gamma_\phi\rho_\phi^l\times\mathcal{R}^{-1/2},
 \eea
 with $\mathcal{R} = \left(\frac{2m_f}{\omega}\right)^2_{\phi\rightarrow \phi_0}$,  where  the approximation in the second line is valid if $\mathcal{R}\gg1$ \cite{Garcia:2020wiy}. We consider only the first Fourier mode of the Higgs inflaton background in the effective masses of fermions as the dominant contribution. The higher order modes also contribute with much smaller Fourier coefficients, making their contributions negligible up to $\sim 10\%$ level. We performed the numerical integration of the Boltzmann equations in eqs.~(\ref{Boltzmann_phi}) and (\ref{Boltzmann_rad}), taking into account the averaged kinetic factor in the limit $\mathcal{R}\gg 1$ for heavy fermions during the integration.  As a result, we determine the reheating temperature from the perturbative decays of the Higgs inflaton into fermions, as given in Table \ref{table:2}. 
 
 \begin{table}[htbp!]
     \centering
     \begin{tabular}{|c|c|}
     \hline
        $m$   &   $T_{\rm RH}$ $[\rm GeV]$    \\
     \hline
        1  &  $5.1\times10^{13}$  \\
     \hline
       2 &        $2.6\times10^{9}$ \\
       \hline
       3 &  260 \\
       \hline
       4 & $9.4\times 10^5$ \\
       \hline
       5 & $2.1\times 10^7$ \\
       \hline
       6 & $1.1\times 10^8$ \\
       \hline
       7 & $2.8\times10^8 $ \\
       \hline
       8 & $4.9\times10^8$ \\
     \hline   
     9 & $8.4\times10^8$ \\
     \hline
     10 & $1.2\times10^9$ 
     \\
     \hline
     \end{tabular}
     \caption{Reheating temperature $T_{\rm RH}$, determined from the decays of the Higgs inflaton into the SM fermions, including the kinematic suppression for the effective fermion masses. We chose some values of the equation of state parameter $m$ during reheating. }
     \label{table:2}
 \end{table}
 
 \subsection{Preheating}
 
 In the limit of sub-Planckian Higgs fields below the pole, we can approximate the Higgs perturbation, $\varphi_k$, with the comoving momentum $k$, as the following modified Klein-Gordon equation, 
 \bea
 {\ddot\varphi}_k + 3H {\dot \varphi}_k + \bigg(\frac{k^2}{a^2}+m^2_\varphi(t) +6\xi_H \Big(\frac{\ddot{a}}{a}+\frac{{\dot a}^2}{a^2}\Big)\bigg) \varphi_k =0 \label{higgspert}
 \eea
 where $m_\varphi(t)$ is  the effective mass  for the Higgs perturbation during reheating, $m^2_\varphi(t)=m^2_\phi(t) {\cal P}^{2m-2}$, with $m^2_\phi(t)$ being given by eq.~(\ref{inflatonmass}). Here, $\xi_H$ is the non-minimal coupling to the SM Higgs, which is chosen to $\xi_H=-\frac{1}{6}$ in our model. Then, we make a change of variables with $z=\omega t$ where $\omega$ is the angular frequency for the inflaton oscillation, given in eq.~(\ref{freq}).
 Then, from eq.~(\ref{higgspert}), we get the equation for the rescaled perturbation, $H_k=\omega^{1/(1-m)} \varphi_k$, as
 \bea
 H^{\prime\prime}_k+   \bigg(\kappa^2+\frac{ m^2 m^2_\varphi(t)}{\omega^2}+\frac{2\xi_H(m+1)(2-m)}{3(1-m)^2} \cdot\bigg(\frac{\omega'}{\omega}\bigg)^2 \bigg) H_k=0 \label{pert2}
 \eea
 where 
 \bea
 \kappa^2\equiv \frac{m^2 k^2}{\omega^2 a^2},
 \eea
 the prime denotes the derivative with respect to $z$ with $dz=\frac{\omega}{m}\, dt$, and we used ${\dot\omega}=\frac{1-m}{m}\, \frac{\omega}{t}$. 
 This is the master equation governing the generalized preheating with an anharmonic inflaton potential. 
The last term in eq.~(\ref{pert2}) is due to the Hubble expansion, so it can be neglected given that particle production is efficient over an oscillation.
 
 In order to derive eq.~(\ref{pert2}), we assumed that the scale factor scales as $a\propto t^{\frac{m+1}{3m}}$ during reheating, so the Hubble parameter becomes $H=\frac{m+1}{3m}\, \frac{1}{t}$. Then, from the averaged Friedmann equation, $\langle H^2 \rangle=\frac{1}{3M^2_P}\, \langle \rho\rangle$, and the averaged inflaton density in eq.~(\ref{infdens}), $\langle \rho\rangle\propto \langle V_E(\phi)\rangle\sim \phi^{2m}_0$, the inflaton condensate scales by $\phi_0(t)\propto t^{-1/m}$, leading to $\omega\propto m_\phi\propto \phi^{m-1}_0\propto t^{1/m-1}$ and ${\dot\omega}=\frac{1-m}{m}\, \frac{\omega}{t}$. 
 
 As a result, we find that the effective mass term in eq.~(\ref{pert2}) becomes
 \bea
 \frac{m^2m^2_\varphi(t)}{\omega^2}= \frac{1}{\pi}\, m(2m-1)\bigg(\frac{\Gamma\big(\frac{1}{2m}\big)}{\Gamma\big(\frac{1}{2}+\frac{1}{2m}\big)}\bigg)^2 {\cal P}^{2m-2}(t),
 \eea
 which does not depend on the  amplitude of the inflaton oscillation, $\phi_0$, but instead it depends on the power of the inflaton potential during reheating.  This case is similar to the scale-invariant model for inflation with a quartic inflaton model during reheating \cite{Greene:1997fu,Choi:2019osi}, namely, the $m=2$ case, but the oscillating part is different if $m\neq 2$.

 Similarly, the perturbation equations for $W, Z$ gauge bosons are obtained from the one for the Higgs perturbation in eq.~(\ref{pert2}), after the effective mass $m^2_\varphi(t)$ is replaced by those for the gauge bosons, $m^2_V(t)$, with $V=W, Z$, and $\xi_H=0$. In this case, the effective mass term for the $W$ boson perturbation becomes
 \bea
 \frac{m^2m^2_W(t)}{\omega^2} = \frac{g^2}{8\pi \alpha_m}\, \bigg(\frac{\Gamma\big(\frac{1}{2m}\big)}{\Gamma\big(\frac{1}{2}+\frac{1}{2m}\big)}\bigg)^2 \phi^{4-2m}_0(t) {\cal P}^2(t),  \label{gaugemass}
 \eea
 which depends on the  amplitude of the inflaton oscillation, $\phi_0$, unlike the case for the Higgs perturbation. 
 For the effective mass term for the $Z$ boson perturbation, we only have to replace $g^2$ with $g^2+g^{\prime 2}$ in the above result.  
 
 As a result, $z=\omega t \propto t^{1/m}$ and from $\phi_0\propto t^{-1/m}\propto z^{-1}$, the effective mass for the gauge boson perturbation depends on the amplitude of the inflaton oscillation because $ \phi^{4-2m}_0(t) \propto z^{2m-4}$.
   Therefore, for $m<2$, kinematic blocking only exists during the early stages of reheating. Preheating becomes important here.
 On the other hand, for $m>2$, the perturbative decay of the inflaton is important from the beginning.
 We also note that the coefficients in the effective mass terms for the gauge boson perturbations in eq.~(\ref{gaugemass}) are proportional to $g^2/\alpha_m$, which is large for $\alpha_m\lesssim 10^{-10}$, as required by the CMB normalization.
  Therefore, preheating could lead to a higher reheating temperature than the one obtained from the perturbative reheating, as in the original Higgs inflation with a large non-minimal coupling \cite{Bezrukov:2008ut,Garcia-Bellido:2008ycs}.
 
 For the early period of the inflaton oscillation, the case with $m>2$ effectively introduces a large coupling to the inflaton as time goes by, so the parametric resonances for the particle production are extended to large momentum modes \cite{Greene:1997fu, Choi:2019osi}.  On the other hand,  for the case with $m<2$, the resonant production of particles can be limited to relatively smaller momentum modes. 
 The detailed discussion on preheating and particle production is beyond the scope of this study.

 \section{Peccei-Quinn Inflation at the pole} \label{PQPole} 

 In this section we perform a similar analysis to the one in \textbf{section \ref{sec:higgsatthepole}}, but for the Peccei-Quinn field.
 For the case with PQ conservation, the discussion for inflation is completely parallel, but if we include PQ violating terms, we end up with a two-field inflation scenario.

 \subsection*{Setup}
We introduce an extra complex scalar field $\Phi$,  charged under the $U(1)_{PQ}$ symmetry responsible for the QCD axion.
The angular component of the PQ field is identified as the axion, while the radial component is regarded as the saxion. 
The QCD anomalies are generated due to an extra vector-like quark as in the KSVZ axion model \cite{Kim:1979if,Shifman:1979if}, providing the axion-gluon coupling necessary for the dynamical solution to the strong CP problem. 

In this case, the non-minimal coupling function and the PQ potential are defined as
\bea
\Omega(\Phi) &=& 1-\frac{1}{3M^2_P}|\Phi|^2,\nonumber \\
V_E(\Phi)&=& V'_0+\frac{\beta_m}{M^{2m-4}_P} |\Phi|^{2m} -m^2_\Phi |\Phi|^2+  \bigg(\sum_{k=0}^{[n/2]}\frac{c_k}{2M^{n-4}_P} \, |\Phi|^{2k}\Phi^{n-2k} +{\rm h.c.}\bigg).  \nonumber
\eea
Here, $V_0$ denotes the cosmological constant, $\beta_m$ is a PQ conserving dimensionless parameter, and the $c_k$'s are dimensionless parameters characterizing the explicit breaking of the PQ symmetry at the Planck scale \cite{Kamionkowski:1992mf, Barr:1992qq}.
We note that $n>4$, and $[n/2]$ is the largest integer smaller than $n/2$. We can always add higher order terms to the PQ violating potential, but they will be subdominant.

If we take $m=2$, the PQ conserving part of the potential is given by

\bea
V_{\rm PQ} = V_0+ \lambda_\Phi\bigg(|\Phi|^2-\frac{f^2_a}{2}\bigg)^2
\eea

where we can redefine the parameters as $\beta_m=\lambda_\Phi$, $m^2_\phi=\lambda_\Phi f^2_a$ and 
\bea
V'_0=V_0 +\frac{1}{4}\lambda_\Phi f^4_a \nonumber
\eea

The Yukawa couplings, including the new vector-like quark Q, are given by

\bea
{\cal L}_{Q,{\rm int}} = - y_Q \Phi {\bar Q}_R Q_L +{\rm h.c.}
\eea

where the quarks $Q_L$ and $Q_R$ carry charge $+1$ and $-1$ respectively, and $\Phi$ has a charge of $-2$ under the $U(1)_{\rm PQ}$ symmetry.  

In order to achieve reheating, we also introduce Higgs-portal interactions between the SM Higgs and the PQ field,
\bea
\Delta V_E = \lambda_{H\Phi} |\Phi|^2 |H|^2. 
\eea






Now we take the PQ field in the polar coordinate representation,
\bea
\Phi=\frac{1}{\sqrt{2}}\,\rho\,e^{i\theta}
\eea

where $\rho$ and $\theta$ are the radial and angular modes of the PQ field, respectively. For $m=2$, the PQ conserving potential, we get the Einstein-frame Lagrangian
\bea
\frac{{\cal L}_E}{\sqrt{-g_E}} =-\frac{1}{2} M^2_P R +\frac{1}{2}\,\frac{(\partial_\mu \rho)^2}{\big(1-\frac{1}{6M^2_P}\rho^2\big)^2}+\frac{1}{2}\frac{\rho^2 (\partial_\mu\theta)^2}{\big(1-\frac{1}{6M^2_P}\rho^2\big)} - V_E(\rho,\theta) \label{Linfpqpole}
\eea

with
\bea
V_E(\rho,\theta)=V_0 + \frac{1}{4} \lambda_\Phi (\rho^2-f^2_a)^2 +\frac{\rho^n}{2^{n/2}M^{n-4}_P} \sum_{k=0}^{[n/2]} |c_k|\, \cos\Big((n-2k)\theta+A_k\Big), \nonumber
\eea
and  $c_k=|c_k| e^{iA_k}$.

In order to canonicalize the kinetic term we change variables to
\bea
\rho=\sqrt{6}M_P \tanh\Big(\frac{\phi}{\sqrt{6}M_P}\Big).
\eea

In the canonical variables, the Einstein frame Lagrangian in \ref{LE} becomes

\bea
\frac{{\cal L}_E}{\sqrt{-g_E}} =-\frac{1}{2} M^2_P R + \frac{1}{2}(\partial_\mu\phi)^2 +3M^2_P\sinh^2\Big(\frac{\phi}{\sqrt{6} M_P}\Big)\,(\partial_\mu\theta)^2- V_E(\phi,\theta), \label{Elag}
\eea

where we divide the potential into the PQ conserving and the PQ violating parts,
\bea
V_E(\phi,\theta)&=& V_{\rm PQ}(\phi) +V_{\rm \cancel{PQ}}(\rho,\theta),  \label{Einpotpqpole}
\eea
with
\bea
V_{\rm PQ}(\phi) &=&V_0 +  \frac{1}{4} \lambda_\Phi \Big(6M^2_P  \tanh^2\Big(\frac{\phi}{\sqrt{6}M_P}\Big)-f^2_a\Big)^2, \nonumber \\
V_{\rm \cancel{PQ}}(\rho,\theta) &=& 3^{n/2}M^4_P \tanh^n\Big(\frac{\phi}{\sqrt{6}M_P}\Big) \sum_{k=0}^{[n/2]} |c_k| \cos\Big((n-2k)\theta+A_k\Big).\nonumber
\eea

In the true vacuum, the PQ field satisfies the condition

\bea
6 \Mpl^2 \tanh^2 \left(\frac{\langle \phi \rangle}{\sqrt{6}\Mpl}\right) \simeq f_a^2 .
\eea

Then, $\langle\rho\rangle\sim\langle\phi\rangle\sim f_a$ for $f_a\ll M_P$. 
The cosmological constant is given by 

\bea
V_0\simeq - M^4_P\bigg(\frac{f_a}{\sqrt{2}M_P}\bigg)^n \,\sum_{k=0}^{[n/2]} |c_k| \cos\Big((n-2k)\langle\theta\rangle+A_k\Big). 
\eea 

In the true vacuum, the kinetic term for the axion recovers its canonical form,

\bea
3M^2_P\sinh^2\big(\frac{\langle\phi\rangle}{\sqrt{6} M_P}\big)\,(\partial_\mu\theta)^2\simeq \frac{1}{2} f^2_a\,(\partial_\mu\theta)^2
\eea

where we can identify the axion of the strong CP problem, $a=f_a\,\theta$.

If instead of taking $m=2$, we consider a general PQ conserving potential, 

\bea
V_{\rm PQ}(\phi) =V'_0+3^m\beta_mM^4_P  \bigg[\tanh\Big(\frac{\phi}{\sqrt{6}M_P}\Big)\bigg]^{2m} - 3m^2_\Phi M^2_P  \tanh^2\Big(\frac{\phi}{\sqrt{6}M_P}\Big), \nonumber
\eea

where for $\phi\ll \sqrt{6} M_P$, the VEV of the PQ field satisfies the condition

\bea2m \beta_m \phi^{2m-1}/(2^m M^{2m-4}_P)=m^2_\Phi\phi.
\eea
That is, $\langle\rho\rangle\sim\langle\phi\rangle\sim f_a=(m^2_\Phi M^{2m-4}_P)^{1/(2m-2)}$.
For instance, for $m=3$ we get $f_a=\sqrt{m_\Phi M_P}$. 

\subsection{Axion Quality Problem}

In the presence of PQ anomalies, eq.~(\ref{axiongluon}) is slightly modified to

\bea
{\cal L}_{\rm gluons}=\frac{g^2_s}{32\pi^2} \Big({\bar\theta}+\xi\frac{a}{f_a}\Big) G^a_{\mu\nu}{\tilde G}^{a\mu\nu}
\eea

where $\xi$ is the PQ anomaly coefficient. In KSVZ models, as is our case, $\xi=1$.

After the QCD phase transition, we get an extra contribution to the axion potential, 
\bea
\Delta V_E =-\Lambda^4_{\rm QCD} \cos\Big({\bar\theta}+\xi\frac{a}{f_a}\Big). \label{QCDpot}
\eea

When the radial mode settles in the minimum of the potential, i.e. $\langle\rho\rangle\simeq f_a$, from eqs.~(\ref{Einpotpqpole}) and (\ref{QCDpot}), the effective potential for the axion after the QCD phase transition is given by
\bea
V_{\rm eff}(a) =V_0-\Lambda^4_{\rm QCD} \cos\Big({\bar\theta}+\xi\frac{a}{f_a}\Big)+M^4_P\bigg(\frac{f_a}{\sqrt{2}M_P}\bigg)^n \,\sum_{k=0}^{[n/2]} |c_k| \cos\Big((n-2k)\frac{a}{f_a}+A_k\Big).  \nonumber
\eea

Furthermore, to resolve the strong CP problem we need to satisfy the EDM bound \cite{Kamionkowski:1992mf,Barr:1992qq},
\bea
|\theta_{\rm eff}|=\bigg|{\bar\theta}+\xi\frac{\langle a\rangle}{f_a}\bigg|<10^{-10}. \label{EDMbound}
\eea

From the minima of the potential, we get
\bea
a_{\rm phys}\equiv a+\frac{f_a}{\xi} {\bar\theta} \simeq \frac{f^{n-1}_a}{2^{n/2}M^{n-4}_P m^2_a}\, \sum_{k=0}^{[n/2]} |c_k| (n-2k)\sin\Big( A_k-\frac{n-2k}{\xi}\,{\bar\theta}\Big) \label{axionVEV}
\eea
where $m^2_a=\frac{\xi^2}{f^2_a}\,\Lambda^4_{\rm QCD}$ is the squared mass for the axion generated by QCD, and we assumed $(n-2k) \frac{a_{\rm phys}}{f_a}\ll 1$ and 
\bea
(n-2k)^2|c_k|f^{n-2}_a/(2^{n/2}M^{n-4}_P)\cos\big( A_k-\frac{n-2k}{\xi}\,{\bar\theta}\big)\lesssim m^2_a
\eea
 for all $k$. 

From eq.~(\ref{axionVEV}) with eq.~(\ref{EDMbound}), we see that we can solve the strong CP problem if
\bea
\frac{\xi f^{n-2}_a}{2^{n/2}M^{n-4}_P m^2_a}\, \sum_{k=0}^{[n/2]} |c_k| (n-2k)\sin\Big( A_k-\frac{n-2k}{\xi}\,{\bar\theta}\Big) < 10^{-10}.
\eea
Unless there are fine tuned cancellations between the different contributions,each term in the PQ violating potential must be constrained by
\bea
\bigg(\frac{f_a}{M_P}\bigg)^{n}\lesssim \frac{2^{n/2}\xi}{(n-2k)|c_k|} \bigg(\frac{\Lambda_{\rm QCD}}{M_P}\bigg)^4.
\eea 
The bound on n depends on the value of $f_a$. For example, for $f_a=10^{12}\,{\rm GeV}$, $\xi=1$ and $|c_k|={\cal O}(1)$, we require $n\gtrsim 12$ in order to solve the Axion Quality Problem.
For $f_a=10^6\,{\rm GeV}$, we would need $n\gtrsim 6$. 

However, we will also need to consider CMB bounds, and as we will see in the next section, they tend to be more constringent.

\subsection{Axion-photon coupling}

If charged fermions are also charged under the PQ symmetry, they will contribute to the axion-photon coupling,

\bea
{\cal L}_{\rm photon} =\frac{1}{4} g_{a\gamma\gamma} \, F_{\mu\nu} {\tilde F}^{\mu\nu}
\eea

with
\bea
g_{a\gamma\gamma} =\frac{\alpha}{2\pi f_a/\xi} \bigg(\frac{E}{N}-1.92\bigg).
\eea
In the minimal case, the only contribution to the axion-photon coupling arises from the extra heavy quark, with $\xi=1$ and $E/N=0$.

\subsection{Inflation with PQ conservation}

We separate the study of inflation in two cases: inflation with PQ conservation and with PQ violation.
Our model differs from previous works because we consider conformal couplings \cite{Fairbairn:2014zta, Nakayama:2015pba, Ballesteros:2016euj,Ballesteros:2016xej}.

Let's first derive the equations of motion for the fields during inflation. 
From eq.~(\ref{Elag}), we obtain the dynamics of the radial component, 
\bea
{\ddot\phi}+3H {\dot \phi} -\sqrt{6}M_P\,\sinh\Big(\frac{\phi}{\sqrt{6}M_P}\Big) \cosh\Big(\frac{\phi}{\sqrt{6}M_P}\Big) \, {\dot\theta}^2=-\frac{\partial V_E}{\partial \phi}, \label{eom-phi}
\eea

Using the slow-roll condition,  ${\ddot\phi}\ll H {\dot\phi}$ and ${\dot\theta}\ll H$, we find the velocity of the radial mode to obey the equation
\bea
{\dot\phi}\simeq -\frac{1}{3H} \frac{\partial V_E}{\partial \phi}=-\sqrt{2\epsilon_\phi}\, M_P H \label{slow-phi}
\eea

where

\bea
\epsilon_\phi=\frac{M^2_P}{2 V^2_E} \big(\frac{\partial V_E}{\partial\phi}\big)^2.
\eea

Furthermore, the Hubble parameter is defined as

\bea
H^2=\frac{1}{3M^2_P}\, \bigg(\frac{1}{2}(\partial_\mu \phi)^2+ 3M^2_P\sinh^2\Big(\frac{\phi}{\sqrt{6} M_P}\Big)\,(\partial_\mu\theta)^2+ V_E\bigg).
\eea
which, during inflation can be approximated to 
\bea 
H^2\simeq \frac{V_E}{3M^2_P}.
\eea 

Similarly, from eq.~(\ref{Elag}) we obtain the equation of motion for the axial component,
\bea
6M^2_P\sinh^2\Big(\frac{\phi}{\sqrt{6}M_P}\Big)\bigg[{\ddot\theta}+ 3H {\dot \theta} + \frac{2}{\sqrt{6} M_P} \coth\Big(\frac{\phi}{\sqrt{6}M_P}\Big)\, {\dot\phi}\,{\dot\theta}\bigg] = -\frac{\partial V_E}{\partial\theta}.  \label{eom-theta}
\eea

During inflation we can take ${\ddot\theta}\ll H {\dot\theta}$ and ${\dot\phi}\ll H$ such that
\bea
{\dot\theta}\simeq  -\frac{1}{3H} \, \frac{\frac{\partial V_E}{\partial\theta}}{6M^2_P\sinh^2\big(\frac{\phi}{\sqrt{6}M_P}\big)}=-\frac{\sqrt{2\epsilon_\theta} \,H}{6\sinh^2\big(\frac{\phi}{\sqrt{6}M_P}\big)} \label{slow-theta}
\eea

where
\bea
\epsilon_\theta=\frac{1}{2V^2_E} \big(\frac{\partial V_E}{\partial\theta}\big)^2.
\eea

Eq.~ (\ref{slow-theta}) is crucial for the post-inflationary dynamics, since $\dot{\theta}(\phi_{\rm{end}})$ is the non-zero velocity required from the kinetic misalignment mechanism.
We see that the velocity of the axial mode in eq.~(\ref{slow-phi}), is suppressed by $\phi\gg \sqrt{6} M_P$. At its heart, this suppression is due to the way we have built our model, i.e, it is due to the presence of the non-canonical kinetic terms.

During slow-roll inflation,  $\rho\sim \sqrt{6} M_P\gg f_a$ is satisfied and we can ignore the bare cosmological constant term in eq.~(\ref{Einpotpqpole}). 
Then, the potential for inflation becomes
\bea
V_E(\phi,\theta)\simeq  9 \lambda_\Phi M^4_P\tanh^4\Big(\frac{\phi}{\sqrt{6}M_P}\Big) + V_n(\theta) \tanh^n\Big(\frac{\phi}{\sqrt{6}M_P}\Big),
\eea
where we defined
\bea
V_n(\theta)\equiv 3^{n/2}M^4_P \sum_{k=0}^{[n/2]} |c_k| \cos\Big((n-2k)\theta+A_k\Big).
\eea

If the PQ conserving term are dominant, we can rewrite the potential in~(\ref{inflatonpot}) as, 
\bea
V_E(\phi)\simeq V_I \bigg[  \tanh\Big(\frac{\phi}{\sqrt{6}M_P}\Big)\bigg]^{2m}, \label{inflation1}
\eea

with $V_I\equiv 3^m \beta_m M^4_P$, which has the same shape as the potential in \textbf{ section \ref{sec:higgspoleinfla}}.
Then, from eq.~(\ref{sindex}), we get $n_s=0.966$ for $N=60$, in perfect agreement with Planck data.
Moreover, we obtain the tensor-to-scalar ratio, $r=0.0033$ for $N=60$, which is again compatible with  the bound from the combined Planck and Keck data, $r<0.036$.
Again, from CMB normalization, we get the bound
$3^m \beta_m= (3.1\times 10^{-8}) \,r=1.0\times 10^{-10}$. 
For instance, for $m=2$, we identify $\beta_m=\lambda_\Phi$, leading to $\lambda_\Phi=1.1\times 10^{-11}$ during inflation.

For a PQ conserving model of inflation, we find the bounds over the PQ breaking parameters to be
\bea
V_n(\theta_i)/M^4_P= 3^{n/2}\sum_{k=0}^{[n/2]} |c_k| \cos\Big((n-2k)\theta_i+A_k\Big) <1.0\times 10^{-10}. \label{PQVbound}
\eea

\subsection{Inflation with PQ violation}

Now we move to case where the inflaton potential is dominated by the PQ violating terms,
\bea
V_E(\phi,\theta)\simeq V_n(\theta) \cdot W_n(\phi) . \label{inflation2}
\eea 

with
\bea
W_n(\phi)\equiv  \bigg[\tanh\Big(\frac{\phi}{\sqrt{6}M_P}\Big)\bigg]^n.
\eea

This case corresponds to a scenario of multi-field inflation \cite{Gong:2011cd}, since as far as the potential remains positive definite during inflation, both fields could drive inflation. 
Generically, the angular mode should be far from the minimum of the potential during inflation, since otherwise the potential would always be negative. 

Even if the PQ violating terms can be relevant at the beginning of inflation, they eventually become subdominant, and the dynamics of the radial field end up determining the evolution at the end of inflation.
In this case, the slow-roll parameters are derived from the PQ violating part of the potential, where
\bea
 \epsilon_\phi\simeq \frac{M^2_P}{2W^2_n} \left(\frac{\partial W_n}{\partial\phi}\right)^2
 \eea
  and 
  \bea
  \epsilon_\theta\simeq \frac{1}{2V^2_n} \left(\frac{\partial V_n}{\partial\theta}\right)^2
  \eea. 
  
We notice that now, while the slow-roll parameter for the radial mode is still small for $\phi\gg \sqrt{6} M_P$, the one for the angular mode is already of order one.
Nonetheless, the velocity of the angular mode is still sufficiently small for $\phi\gg \sqrt{6} M_P$, suppressed by the non-canonical kinetic terms. 
We define the effective slow-roll parameters by taking into account the non-canonical kinetic terms,
\bea
\epsilon=g^{\phi\phi}\epsilon_\phi + g^{\theta\theta} \epsilon_\theta \label{genslow}
\eea 
with $g^{\phi\phi}=1$ and $g^{\theta\theta}=1/\big[6M^2_P\sinh^2\big(\frac{\phi}{\sqrt{6}M_P}\big)\big]$.

When the axion field settles to $\theta=\theta_i$, the CMB sets the following constraints,
\bea
V_n(\theta_i)= 3^{n/2}\sum_{k=0}^{[n/2]} |c_k| \cos\Big((n-2k)\theta_i+A_k\Big) =(3.1\times 10^{-8}) \,r
\eea
where $r=16\epsilon_*$, where we used eq.~(\ref{genslow}) to evaluate  $\epsilon_*$ at horizon exit.

\subsection*{Post-inflationary evolution of the fields}

After inflation, the radial mode becomes sub-Planckian. Then, the equations of motion simplify to
\bea
{\ddot\phi}+3H {\dot \phi} -\phi \, {\dot\theta}^2&\simeq& -\frac{\partial V_E}{\partial \phi},  \label{approx-phi} \\
\phi^2 ({\ddot\theta}+ 3H {\dot \theta}) + 2\phi {\dot\phi}\,{\dot\theta}&\simeq&  -\frac{\partial V_E}{\partial\theta} \label{approx-theta}
\eea

where
\bea
V_E\simeq V_0+\frac{1}{4} \lambda_\Phi (\phi^2-f^2_a)^2 + \frac{1}{2^{n/2}} \bigg(\frac{\phi}{M_P}\bigg)^n M^4_P  \sum_{k=0}^{[n/2]} |c_k| \cos\Big((n-2k)\theta+A_k\Big).
\eea

\begin{figure}[!t]
    \begin{center}
     \includegraphics[width=0.45\textwidth,clip]{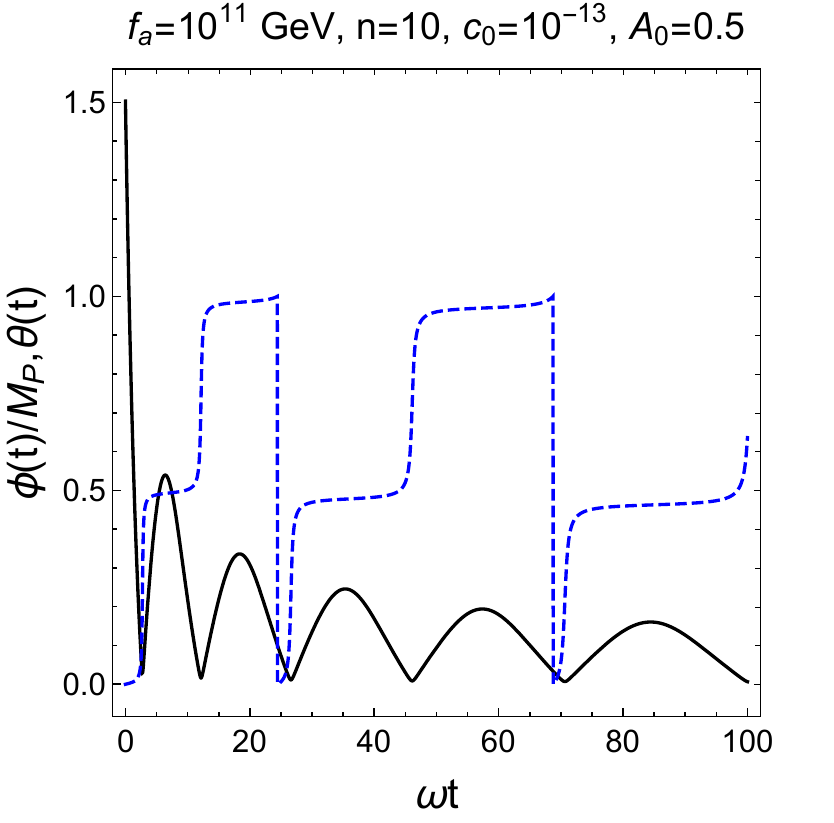}\,\,  \includegraphics[width=0.45\textwidth,clip]{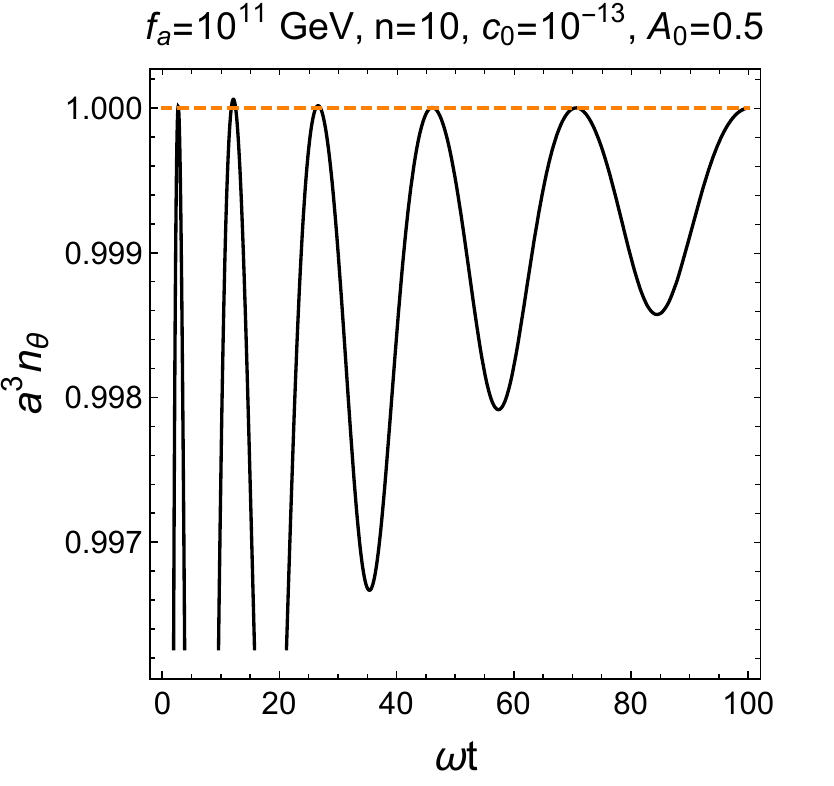}  
     \end{center}
    \caption{ $f_a=10^{11}\,{\rm GeV}$  $n=10$, $c_0=10^{-13}$, $A_0=0.5$, and $c_k=0$ for $k\neq 0$.}
    \label{fig:oscillation}
    \end{figure}

In this work, we consider that the PQ conserving terms are dominant. The radial mode oscillates with an amplitude that decreases gradually until the field settles around the VEV. 
Because the PQ violating terms are subdominant, the PQ Noether charge is approximately conserved after inflation \cite{Co:2019jts}, as seen in the left panel of \ref{fig:oscillation}. For $n_\theta=\phi^2 {\dot\theta}$, eq.~(\ref{approx-theta}) leads to $\frac{d}{dt}(a^3 n_\theta)=0$.

Then, the equation of motion for the radial component eq.~(\ref{approx-phi}) simplifies to
\bea
{\ddot\phi}+3H {\dot \phi} \simeq \frac{C^2}{a^6 \phi^3} -\lambda_\Phi \phi^3
\eea

in the regime in which $f_a\lesssim \phi\ll M_P$.
Here, $C=a_e^3 \phi^2_e{\dot\theta}_e$ is just an integration constant. The angular motion generates a centrifugal forces that becomes less and less relevant as the universe expands, and by the time the radial field settles in the minimum, the potential is dominated by the quartic term. The radial field oscillates then coherently in a quartic potential.

In Fig.~\ref{fig:oscillation}, we show the time evolution of the radial and angular modes on the left, and the total Noether charge on the right, as a function of $\omega t$.
We took $f_a=10^{11}\,{\rm GeV}$, and the parameters in the PQ violating potential to $n=10$, $c_0=10^{-13}$, $A_0=0.5$, and $c_k=0$ for $k\neq 0$. 
From the left plot, we see that the radial component(the inflaton) behaves as a damped harmonic oscillator, while from the right plot we confirm that, even if the Noether charge is oscillating, it approaches a constant value, which is fixed by the initial conditions of inflation.

Finally, a comment is in order. We notice that, had we considered non-negligible PQ violating terms, the Noether charge would evolve in a non-trivial way and we would have to solve a system of equations similar to the one studied on the Higgs-sigma reheating section. 

\subsection{Reheating}
In a parallel discussion with that in \textbf{section \ref{sec:higgspolereheating}}, we define $V_E(\phi)\simeq \alpha_m \phi^{2m}$,
with $\alpha_m=\beta_m M^{4-2m}_P/2^m$.  For $m=2$, we identify $\beta_m=\lambda_\Phi$. 
And let's remind us that the inflation field value at the end of inflation is determined by $\ddot a=0$ \cite{Clery:2023ptm}, such that
\begin{align}
   \phi_{\rm end} &\simeq 
   \sqrt{\frac{3}{8}}M_P\ln\left[\frac{1}{2} + \frac{2m}{3}\left(2m+ \sqrt{4m^2+3}\right)\right]. 
\end{align}

\subsection*{Boltzmann equations for reheating}

In this case, the contributions to the decay rate are:
\bea
\Gamma_\phi=\sum_f\Gamma_{\phi\to f{\bar f}} + \Gamma_{\phi\phi\to HH}+\Gamma_{\phi\phi\to aa}.
\eea

In order to study reheating, we choose a Cartesian basis for the PQ field by $\Phi=\frac{1}{\sqrt{2}}(\phi+ia)$. In this basis, the kinetic term for the PQ field during reheating takes a canonical form

\bea
{\cal L}_{\rm kin}=\frac{1}{2}(\partial_\mu\phi)^2 +\frac{1}{2}(\partial_\mu a)^2
\eea
 and the PQ invariant potential gives rise to the interaction term  ${\cal L}_{\rm int}=-\frac{1}{2}\lambda_\Phi \phi^2 a^2$, where the inflaton condensate is $\phi=\phi_0(t) {\cal P}(t)$. 
The scattering rate of the inflaton condensate \cite{Garcia:2020wiy,Clery:2023ptm} for $\phi\phi\to aa$, is given as follows, 
\bea
\Gamma_{\phi\phi\to aa}=\frac{1}{8\pi (1+w_\phi)\rho_\phi} \sum_{n=1}^\infty |M^a_n|^2 (E_n\beta^{a}_n),
\eea
with 
\bea
|M^a_n|^2 =  4\lambda^2_\Phi \phi^4_0 |({\cal P}^2)_n|^2,
\eea
and $\beta^{a}_n= \sqrt{1-\frac{m^2_{a}}{E^2_n}}$.

We notice that here, $m^2_{a}=\lambda_\Phi \phi^2$ for $\phi\gg f_a$.
From eq.~(\ref{freq}), we get $\omega^2=0.72\lambda_\Phi \phi^2_0$ for $m=2$, so $2\omega> m_a$, and the channel  $\phi\phi\to aa$ is available for reheating.
The Fourier coefficients are given by ${\cal P}^2=\sum_{n=-\infty}^\infty ({\cal P}^2)_n e^{-in\omega t}$. 
In the case in which $m=2$, the first non-zero coefficients are given by $2({\cal P}^2)_2=0.4972, 2({\cal P}^2)_4=0.04289$ and $2({\cal P}^2)_6=0.002778$.
 Then, the averaged scattering rate is
\bea
\langle \Gamma_{\phi\phi\to aa}\rangle =\frac{\lambda^2_\Phi \phi^2_0 \omega}{2\pi m^2_\phi}\,(m+1)(2m-1) \Sigma^a_m \bigg\langle \bigg(1-\frac{m^2_a}{\omega^2n^2}\bigg)^{1/2}\bigg\rangle \label{scatteringrate1}
\eea
with $\Sigma^a_m=\sum^\infty_{n=1}n |({\cal P}^2)_n|^2$,

and the Higgs-portal interactions lead to
\bea
\Gamma_{\phi\phi\to HH}=\frac{1}{8\pi (1+w_\phi)\rho_\phi} \sum_{n=1}^\infty |M^H_n|^2 (E_n\beta^H_n),
\eea
with 
\bea
|M^H_n|^2 = 4\lambda_{H\Phi}^2 \phi^4_0|({\cal P}^2)_n|^2
\eea
and $\beta^H_n = \sqrt{1-\frac{m^2_H}{E^2_n}}$. 
The effective Higgs masses are  $m^2_H=m^2_{H,0}+\frac{1}{2}\lambda_{H\Phi} \phi^2(t)$ where $m^2_{H,0}$ denotes the bare Higgs mass parameter.  Thus, the corresponding averaged scattering rate for the inflaton is given by
\bea
\langle \Gamma_{\phi\phi\to HH}\rangle =\frac{\lambda^2_{H\Phi}\phi^2_0 \omega}{\pi m^2_\phi}\,(m+1)(2m-1) \Sigma^H_m \bigg\langle \bigg(1-\frac{m^2_H}{\omega^2n^2}\bigg)^{1/2}\bigg\rangle \label{scatteringrate2}
\eea
with $\Sigma^H_m=\sum^\infty_{n=1}n |({\cal P}^2)_n|^2$. 

The quartic coupling needs to remain small during inflation and enforces $|\lambda_{H\Phi}|\lesssim 10^{-5}$. 

Furthermore, in order to solve the strong CP problem, we introduced the Yukawa couplings to extra heavy quarks $Q$, as in KSVZ axion models.
 Moreover, the PQ inflaton can be responsible for the generation of the masses of right-handed neutrinos, $N_i(i=1,2,3)$.
We obtain the decay rate
\bea
\Gamma_{\phi\to f{\bar f}}=\frac{1}{8\pi (1+w_\phi)\rho_\phi} \sum_{n=1}^\infty |M^f_n|^2 (E_n\beta^f_n)
\eea
where $E_n=n \omega$,
\bea
|M^f_n|^2=N_c y^2_f \phi^2_0 |{\cal P}_n|^2 E^2_n (\beta^f_n)^2,
\eea
with $N_c$ being the number of colors for the extra fermion $f$, $\beta^f_n = \sqrt{1-\frac{4m^2_f}{E^2_n}}$, and ${\cal P}_n$ the Fourier coefficients of the expansion, ${\cal P}=\sum_{n=-\infty}^\infty {\cal P}_n e^{-in\omega t}$. 
For $m=2$, the first non-zero coefficients are  $2{\cal P}_1=0.9550, 2{\cal P}_3=0.04305$ and $2{\cal P}_5=0.001859$.
Then, averaging over oscillations, we get
\bea
\langle\Gamma_{\phi\to f{\bar f}}\rangle&=&\frac{N_c y^2_f \phi^2_0 \omega^3}{8\pi (1+w_\phi)\rho_\phi}\, \sum_{n=1}^\infty  n^3 |{\cal P}_n|^2 \langle\beta^3_n\rangle \nonumber \\
&=& \frac{N_c y^2_f \omega^3}{8\pi m^2_\phi}\,(m+1)(2m-1) \Sigma^f_m \Bigg\langle\bigg(1-\frac{4m^2_f}{\omega^2 n^2}\bigg)^{3/2}\Bigg\rangle \label{decayrate}
\eea
with $\Sigma^f_m =\sum_{n=1}^\infty  n^3 |{\cal P}_n|^2$.\\
If the extra fermions get their masses only from the Yukawa couplings to the PQ inflaton, they would be given by

\bea
m_f=\frac{1}{\sqrt{2}} y_f\phi(t)=(m_{f,0}/f_a)\phi(t)
\eea
 where $f_a$ is the VEV of the PQ field and $m_{f,0}$ is the fermion mass in the true vacuum.
Two more comments are in order. Firstly, there is no decay into Standard Model fermions, since we would need to break the Electroweak symmetry during reheating.
Secondly, the Yukawa couplings to the PQ field must satisfy $y_f\lesssim 10^{-3}$, so that they don't contribute radiatively to the quartic coupling $\lambda_\Phi$.

The Yukawa couplings introduced for the QCD anomalies lead to
\bea
\Gamma_{\phi\phi\to f{\bar f}}=\frac{1}{8\pi (1+w_\phi)\rho_\phi} \sum_{n=1}^\infty |{\hat M}^f_n|^2 (E_n {\hat \beta}^f_n)
\eea
where 
\bea
 |{\hat M}^f_n|^2=\frac{32N_c y^4_f m^2_f \phi^4_0 ({\hat \beta}^f_n)^2}{E^2_n}
\eea
and ${\hat \beta}^f_n=\sqrt{1-\frac{m^2_f}{E^2_n}}$. Thus, the averaged scattering rate is given by
\bea
\langle \Gamma_{\phi\phi\to f{\bar f}}\rangle &=& \frac{4N_c y^4_f  \phi^4_0 }{\pi (1+w_\phi)\rho_\phi \omega}\, \sum_{n=1}^\infty  n^{-1}|({\cal P}^2)_n|^2\langle m^2_f ({\hat \beta}^f_n)^3 \rangle \nonumber \\
&=&\frac{4N_c y^4_f \phi^2_0 }{\pi m^2_\phi \omega}\, (m+1)(2m-1)\,{\hat\Sigma}^f_m\bigg\langle  m^2_f \bigg(1-\frac{m^2_f}{n^2\omega^2}\bigg)^{3/2}\bigg\rangle,
\eea
with $ {\hat\Sigma}^f_m=\sum_{n=1}^\infty n^{-1}|({\cal P}^2)_n|^2$.

The decay and scattering rates of the inflaton scale with the inflaton energy density as $\Gamma_{\phi \to f{\bar f}}=\gamma_1 \rho^l_\phi$ with $l=\frac{1}{2}-\frac{1}{2m}$ and $\Gamma_{\phi\phi\to HH}=\gamma_2 \rho^n_\phi$ with $n=\frac{3}{2m}-\frac{1}{2}$, respectively \cite{Garcia:2020wiy,Clery:2023ptm}. 
That means that for $m=2$, $l=n=\frac{1}{4}$, and both the decay and the scattering rates are comparable.
However, for $m>2$, $l>n$, and the decay rate gets suppressed.

\subsection*{Reheating temperature}

For a general PQ invariant potential, the scaling of the inflaton energy density is given by
\bea
\rho_{\rm \phi}(a)\simeq \rho_{\rm end}\left(\frac{a_{\rm end}}{a}\right)^{\frac{6m}{m+1}}, 
\eea
and the reheating temperature \cite{Clery:2023ptm} by
\bea
T_{\rm RH}= \left\{\begin{array}{cc} \bigg(\frac{30}{\pi g_*(T_{\rm RH})}\bigg)^{1/4} \bigg[ \frac{2m}{4+m-6mk}\,(\sqrt{3}M^{2(1-2k)}_P\gamma_\phi)\bigg]^{\frac{1}{2(1-2k)}}, \qquad 4+m-6m k>0, \vspace{0.3cm}\\  \bigg(\frac{30}{\pi g_*(T_{\rm RH})}\bigg)^{1/4}\bigg[\frac{2m}{6mk-4-m}\,(\sqrt{3}M^{2(1-2k)}_P\gamma_\phi)\,( \rho_{\rm end})^{\frac{6mk-m-4}{6m}}\bigg]^{\frac{3m}{4(m-2)}}, \quad 4+m-6mk<0. \end{array} \right.
\eea
Where we parametrized the decay or scattering rates as $\Gamma_\phi=\gamma_\phi \rho^k_\phi/M^{4k-1}_P$.

Taking the case with $m=2$, we find that 
\bea
\gamma_\phi|_{\rm decay} &\simeq &\frac{3\sqrt{\pi}N_c}{2}\, y^2_f \lambda^{1/4}_\Phi\,\bigg(\frac{\Gamma\big(\frac{3}{4}\big)}{\Gamma\big(\frac{1}{4}\big)}\bigg)^3\,(0.5\Sigma^f_2 {\cal R}^{-1/2}_f), \\
\gamma_\phi|_{\rm scattering} &\simeq & \frac{6}{\sqrt{\pi}}\bigg(\frac{\Gamma\big(\frac{3}{4}\big)}{\Gamma\big(\frac{1}{4}\big)}\bigg) {\rm max}\bigg[\frac{\lambda^2_{H\Phi}}{\lambda^{3/4}_\Phi}\, \,\Sigma^H_2,  \frac{4N_cy^4_f}{\lambda^{3/4}_\Phi}\, \frac{m^2_f}{\omega^2}\,({\hat\Sigma}^f_2 {\hat {\cal R}}_f^{-1/2}) \bigg].
\eea
Here, $\Sigma^f_2=0.2406$, $\Sigma^H_2=0.1255$, ${\hat\Sigma}^f_2=0.2282$, and we approximated the averaged phase space factor for $2m_f\gg w$ by ${\cal R}_f\equiv 4m^2_f/w^2$ and ${\hat {\cal R}}_f\equiv m^2_f/w^2$ \cite{Garcia:2020wiy}.
Thus, we can determine the reheating temperature by the inflaton decay into a pair of extra heavy quarks or the inflaton scattering into a pair of the SM Higgs bosons, respectively, as follows,
\begin{align}
T_{\rm RH}|_{\rm decay} &\simeq  2.9\times 10^4\,{\rm GeV} \bigg(\frac{100}{g_*(T_{\rm reh})}\bigg)^{1/4}\bigg(\frac{y_f}{10^{-4}}\bigg)\bigg(\frac{\lambda_\Phi}{10^{-11}}\bigg)^{1/4}, \label{decayhp}  \\
T_{\rm RH}|_{\rm scattering} &\simeq 6.0\times 10^{11}\,{\rm GeV}  \bigg(\frac{100}{g_*(T_{\rm reh})}\bigg)^{1/4} \bigg(\frac{{\rm max}[\lambda_{H\Phi},\sqrt{4N_c}y^2_f m_f/\omega]}{10^{-7}}\bigg)^2 \bigg(\frac{10^{-11}}{\lambda_\Phi}\bigg)^{3/4}. \label{scattering}
\end{align}
Therefore, the inflaton scattering through the Higgs-portal is the most efficient contribution for reheating. 
\subsection{Dark radiation from axions}

The produced amount of axions during inflation can become a source of Dark Radiation, so we need to make sure that we are not in tension with the bounds on the effective number of neutrinos,  $\Delta N_{\rm eff}$.
From eqs.~(\ref{scatteringrate1}) and (\ref{scatteringrate2}), we obtain
\bea
\frac{\Gamma_{\phi\phi\to aa}}{\Gamma_{\phi\phi\to HH}}\simeq \frac{\lambda^2_\Phi}{2\lambda^2_{H\Phi}}.
\eea

If the produced axions remain out of equilibrium after reheating, the correction to the effective number of neutrino species is given by\cite{Salvio:2013iaa},
\bea
\Delta N_{\rm eff}= 0.02678\, \bigg(\frac{Y_a}{Y^{\rm eq}_a}\bigg) \bigg(\frac{106.75}{g_{*s}(T_{\rm reh})}\bigg)^{4/3}
\eea
where $Y_a$ is the axion abundance produced from the inflaton scattering and $Y^{\rm eq}_a$ is the abundance at equilibrium, 
 \bea
 Y^{\rm eq}_a=\frac{45 \zeta(3)}{2\pi^4 g_{*s}(T_{\rm reh})}.
 \eea
In the SM, $N^{\rm SM}_{\rm eff}=3.0440$  \cite{Bennett:2019ewm,Bennett:2020zkv,Akita:2020szl}. If $Y_a\gtrsim 10 Y^{\rm eq}_a$, the excess in the effective number of neutrinos is in tension with the current bounds, $N_{\rm eff}=2.99\pm 0.17$ \cite{Planck:2018vyg}.

As it was discussed in the reheating section, the scattering channel $\phi\phi\to HH$ is dominant for $\lambda_{H\Phi}\gtrsim \sqrt{4N_c}y^2_f m_f/\omega$. Then, the number of axions produced during reheating is given by
\bea
n_a\simeq {\rm BR}(\phi\phi\to aa)\,\frac{\rho_\phi}{\omega} \label{na}.
\eea
To derive eq.~(\ref{na}), we used that ${\rm BR}(\phi\phi\to aa)=\Gamma_{\phi\phi\to aa}/(\Gamma_{\phi\phi\to aa}+\Gamma_{\phi\phi\to HH})\simeq \lambda^2_\Phi/(2\lambda^2_{H\Phi})$ for $\lambda_{H\Phi}\gtrsim \frac{1}{\sqrt{2}}\lambda_\Phi$, and we approximated the inflaton to the first Fourier mode with $E=\omega$, since higher Fourier modes are suppressed.
At the end of reheating, $\rho_\phi=\rho_R$,  $Y_a= \frac{n_a}{s} $ is given by
\bea
Y_a=  {\rm BR}(\phi\phi\to aa)\,\frac{\rho_\phi}{\omega s} = {\rm BR}(\phi\phi\to aa)\,\frac{\rho_R}{\omega s} ={\rm BR}(\phi\phi\to aa)\, \frac{T_{\rm reh}}{4\omega}.
\eea
Then, using $\rho_R/s=3T_{\rm reh}/4$ and $\omega\simeq  0.85 \lambda^{1/2}_\Phi \phi_0$  for $m=2$, we get
\bea
\frac{Y_a}{Y^{\rm eq}_a} =\frac{3 \lambda^{3/2}_\Phi g_{*s}(T_{\rm reh})T_{\rm reh}}{2.2\lambda^2_{H\Phi}  \phi_0}, \label{axionprod}
\eea
and from $\phi_0\simeq 1.5M_P$ and $\lambda_\Phi=1.1\times 10^{-11}$, we obtain $Y_a\gtrsim Y^{\rm eq}_a$, provided that 
\bea
T_{\rm reh}\gtrsim 6.8\times 10^{10}\,{\rm GeV} \,\bigg(\frac{\lambda_{H\Phi}}{10^{-11}}\bigg)^2.
\eea

However, there is another possibility. If the reheating temperature is sufficiently large, the axions can thermalize with the SM plasma \cite{Masso:2002np,Salvio:2013iaa,Graf:2010tv}, 
\bea
T_{\rm reh}\gtrsim 1.7\times 10^9\,{\rm GeV}\bigg(\frac{f_a}{10^{11}\,{\rm GeV}}\bigg)^{2.246}\equiv T_{a, {\rm eq}}. \label{axionthermal}
\eea
Then, for $T_{\rm reh}>T_{a, {\rm eq}}$, we can compute the contribution of the axions to the effective number of neutrino species just from the abundance in thermal equilibrium $Y^{\rm eq}_a$, as follows,
\bea
\Delta N_{\rm eff}=\frac{4}{7}\bigg(\frac{T_{a,0}}{T_{\nu,0}}\bigg)^4=\frac{4}{7}\bigg(\frac{11}{4}\bigg)^{4/3} \bigg(\frac{g_{*s}(T_0)}{g_{*s}(T_{a, {\rm eq}})}\bigg)^{4/3} \label{Neff}
\eea
where $T_{\nu,0}, T_{a,0}$ are the neutrino and axion temperatures, respectively, at present, and $g_{*s}(T_0)=3.91$. Thus, we get $\Delta N_{\rm eff}=0.02678$ for  $g_{*s}(T_{a, {\rm eq}})=106.75$; $\Delta N_{\rm eff}=0.02229$ for $g_{*s}(T_{a, {\rm eq}})=122.5$ (adding one charge-neutral heavy quark and three right-handed neutrinos to the SM); $\Delta N_{\rm eff}=0.02363$ for $g_{*s}(T_{a, {\rm eq}})=117.25$ (adding one charge-neutral heavy quark to the SM). 
Hence, this excess can be tested in the future CMB experiments.

In the case in which $T_{\rm reh}<T_{a, {\rm eq}}$, the axions could have never been in thermal equilibrium. 
Using eqs.~(\ref{axionprod}) and (\ref{axionthermal}), we find 
\bea
\frac{Y_a}{Y^{\rm eq}_a}  = 0.025 \bigg(\frac{T_{\rm reh}}{T_{a, {\rm eq}}}\bigg)\bigg(\frac{10^{-11}}{\lambda_{H\Phi}}\bigg)^2\bigg(\frac{f_a}{10^{11}\,{\rm GeV}}\bigg)^{2.246}, \label{axionnonthermal}
\eea
which has an extra suppression factor.

Preheating may also affect axion production, since the effective mass of the axion during reheating is field-dependent as $m^2_a=\lambda_\Phi \phi^2$.
 Then, the evolution of the axion becomes non-adiabatic and axions can be produced through parametric resonance, contributing to Dark Radiation.
Making use the results in Ref.~\cite{Clery:2023ptm}, the  axion perturbation, $A_k=\omega^{1/(1-m)}a_k$, evolves as
\bea
A^{\prime\prime}_k +\bigg(\kappa^2 +\frac{m^2m^2_a}{\omega^2}\bigg)A_k=0
\eea
with 
\bea
\kappa^2\equiv \frac{m^2k^2}{\omega^2 a^2}.  
\eea
We used the prime for derivatives with respect to $z$ with $dz=\frac{\omega}{m}\, dt$, and the effective mass term for the axion perturbation becomes $\frac{m^2m^2_a}{\omega^2}=1.39 m^2{\cal P}^2(t)$.
Then, for $m=2$, the axion perturbations falling in the range $2.78<\kappa^2<3.21$ can grow and lead to axionic dark radiation \cite{Choi:2019osi,Greene:1997fu}.

\section{Axion dark matter from the kinetic misalignment}

In this section, we determine the axion dark matter abundance in the case of inflation with PQ conservation.

\subsection{Evolution of the axion velocity}

The initial velocity for the axion is obtained from eq.~(\ref{slow-theta}), 
\bea
{\dot\theta}_{\rm end}\simeq -\frac{\sqrt{2\epsilon_{\theta,{\rm end}}} H_{\rm end}}{6\sinh^2\big(\frac{\phi_{\rm end}}{\sqrt{6}M_P}\big)},
\eea
and the PQ charge, 
\bea
n_{\theta,{\rm end}}=6M^2_P \sinh^2\Big(\frac{\phi_{\rm end}}{\sqrt{6}M_P}\Big) |{\dot\theta}_{\rm end}|\simeq M^2_P \sqrt{2\epsilon_{\theta,{\rm end}}} \, H_{\rm end}.
\eea
The parameters, $\epsilon_{\theta,{\rm end}}, H_{\rm end}, \phi_{\rm end}$ are evaluated at the end of inflation, given by 
\bea
\epsilon_{\theta,{\rm end}} = \frac{1}{2 V^2_E} \Big(\frac{\partial V_E}{\partial \theta}\Big)^2 \lesssim 1
\eea
where the PQ violating terms are taken to be subdominant with respect to the PQ conserving ones. 

The total Noether charge for the PQ symmetry is conserved approximately after inflation, so we can take $a^3 n_\theta=a^3 \phi^2 {\dot\theta}\simeq$ const. 
 Then, during reheating, the inflaton condensate is radiation-like, so the radial mode scales with the scale factor by $\phi\propto a^{-1}$.
Then, the axion velocity decreases by ${\dot\theta}\propto a^{-1}$. 
However, if reheating takes a significant amount of time, as the temperature drops, the quadratic term in the PQ potential becomes dominant, leading to an early period of matter domination.
 After reheating, the inflaton settles down to the minimum of the potential, $\phi=f_a$, so the axion velocity scales by ${\dot\theta}\propto a^{-3}$. 
 Therefore, as the kinetic energy density for the axion  is given by $\rho_\theta=\frac{1}{2}\phi^2 {\dot\theta}^2$, its scaling with the scale factor changes from $a^{-4}$ to $a^{-6}$, and we enter a kination era.

As a result, the PQ Noether charge density from the axion rotation red-shifts at the end of reheating by
\bea
n_\theta(T_{\rm RH})=n_{\theta,{\rm end}}\,\bigg(\frac{a_{\rm end}}{a_{\rm RH}}\bigg)^3
\eea
where $a_{\rm end}, a_{\rm RH}$ are the values of the scale factor at the end of inflation and the reheating completion, respectively.

The critical reheating temperature is defined as,
\bea
T^c_{\rm RH}&\equiv& \bigg(\frac{90\lambda_\Phi}{8\pi^2 g_*}\bigg)^{1/4} 2f_a \nonumber \\
&=& \bigg(\frac{100}{g_*}\bigg)^{1/4} \bigg(\frac{f_a}{10^{11}\,{\rm GeV}}\bigg)\, (1.2\times 10^8\,{\rm GeV}).
\eea

If $\phi(a_{\rm RH})>3f_a$, the reheating temperature is large $T_{\rm RH}>T^c_{\rm RH}$, and using 
\bea
\frac{a_{\rm end}}{a_{\rm RH}}=\bigg(\frac{\rho_{\rm RH}}{\rho_{\rm end}}\bigg)^{1/4}, 
\eea
with $\rho_{\rm RH}=\frac{\pi^2}{30} g_*(T_{\rm RH}) T^4_{\rm RH}$ and $\rho_{\rm end}=\frac{3}{2} V_E(\phi_{\rm end})$, we obtain
the PQ Noether charge density at the reheating temperature,
\bea
n_\theta(T_{\rm RH}) = n_{\theta,{\rm end}}\,\bigg(\frac{\pi^2 g_*(T_{\rm RH})T^4_{\rm RH}}{45 V_E(\phi_{\rm end})}\bigg)^{3/4}. \label{density1}
\eea
Here, $g_*(T_{\rm RH}), g_*(T_*)$ are the number of the effective entropy degrees of freedom at the reheating temperature and the onset of the axion oscillation, respectively.
 Thus, the  PQ Noether charge density at $T=T_{\rm RH}$ is independent of the reheating temperature, and so the axion abundance is insensitive to the reheating temperature. 

 In the case in which reheating is delayed, $T_{\rm RH}<T^c_{\rm RH}$, the energy density of the inflation scales during reheating as 
\bea
\rho_\phi =\rho_{\rm end} \bigg(\frac{a_{\rm end}}{a_c}\bigg)^4 \bigg(\frac{a_c}{a_{\rm RH}}\bigg)^3
\eea
where $a_c$ is the scalar factor when $\phi(a_c)=3f_a$ such that the inflation becomes matter-like for $a>a_c$.  Then, using 
\bea
\frac{a_c}{a_{\rm RH}}=\bigg(\frac{\rho_{\rm RH}}{\rho_{\phi,c}}\bigg)^{1/3}=\bigg(\frac{T_{\rm RH}}{T^c_{\rm RH}}\bigg)^{4/3}, \label{MD}
\eea
we get the PQ Noether charge density at the end of reheating  for $T_{\rm RH}<T^c_{\rm RH}$ as 
\bea
n_\theta(T_{\rm RH})&=&n_{\theta,{\rm end}}\,\bigg(\frac{a_{\rm end}}{a_c}\bigg)^3\bigg(\frac{a_c}{a_{\rm RH}}\bigg)^3 \nonumber \\
&=&n_{\theta,{\rm end}}\ \bigg(\frac{\pi^2 g_*(T_{\rm RH}) T^4_{\rm RH}}{45 V_E(\phi_{\rm end})}\bigg)^{3/4} \bigg(\frac{T_{\rm RH}}{T^c_{\rm RH}}\bigg). \label{density2}
\eea
Here, in eqs.~(\ref{MD}) and (\ref{density2}), we took $\rho_\phi(a_{\rm RH})=\rho_{\rm RH}$ at reheating completion, $a=a_{\rm RH}$, and $\rho_{\phi,c}$ is the energy density of the inflaton at $a=a_c$, which is rewritten as $\rho_{\phi,c}=\rho_{\rm end}\big(\frac{a_{\rm end}}{a_c}\big)^4\equiv\frac{\pi^2}{30} g_*(T_{\rm RH}) (T^c_{\rm RH})^4$.

During the matter domination era, the Hubble rate is smaller, so there's an extra expansion of the universe that can be seen in the 
the factor $\frac{T_{\rm RH}}{T^c_{\rm RH}}$ on the Noether density.

\subsection{Dark matter abundance from axions}

After the QCD phase transition, the QCD instantons contribute to the axion potential, so the axion is confined to one of the local minima when the kinetic energy of the axion is comparable to the potential of the axion, namely, $\frac{1}{2} f^2_a {\dot\theta}^2(T_*)=2m^2_a(T_*) f^2_a $.
Thus, we need ${\dot\theta}(T_*)=2m_a(T_*)$ and $m_a(T_*) \geq 3 H(T_*)$ for the axion oscillation \cite{Co:2019jts}. 
Therefore, we obtain the condition for the axion kinetic misalignment as 

\bea
{\dot\theta}(T_*)\geq 6H(T_{\rm osc})
\eea.

The non-zero velocity of the axion actually delays the onset of oscillations, $T_*\leq T_{\rm osc}$.
 Here, $T_{\rm osc}$ is the temperature of the standard misalignment, determined by $m_a(T_{\rm osc})=3H(T_{\rm osc})$, which is given by

\bea
T_{\rm osc} =\bigg(\frac{10}{\pi^2 g_*}\bigg)^{1/12}\Big(m_a(0) M_P \Lambda^4_{\rm QCD}\Big)^{1/6}
\eea
where $m_a(0)$ is the axion mass at zero temperature whose precise value is given \cite{Gorghetto:2018ocs} by
\bea
m_a(0)= 5.691(51)\times 10^{-3}\,{\rm eV} \bigg(\frac{10^9\,{\rm GeV}}{f_a}\bigg).
\eea
For instance, for $f_a=10^9\,{\rm GeV}$, $\Lambda_{\rm QCD}=150\,{\rm MeV}$ and $g_*=75.75$, we find $T_{\rm osc}=3.05\,{\rm GeV}$.

The axion relic abundance from the kinetic misalignment is given by
\bea
\Omega_a h^2 = 0.12 \bigg(\frac{10^9\,{\rm GeV}}{f_a}\bigg) \bigg(\frac{Y_\theta}{40}\bigg)  \label{relic}
\eea
where $Y_\theta$ is the abundance for the axion  given by $Y_\theta=\frac{n_\theta(T_{\rm RH})}{s(T_{\rm RH})}$ with $n_\theta(T_{\rm RH})$  and  $s(T_{\rm RH})$ being the Noether charge density and the entropy density at reheating, respectively. 
 Here, we used $\frac{\rho_a}{s}=Cm_a(0) Y_\theta$ with $C\simeq 2$ to convert the PQ charge abundance to the relic density \cite{Co:2019jts}.
 Here, we note that there is an uncertainty of order one in  $C$ due to the particle production from the axion fragmentation \cite{Eroncel:2022vjg}. So, we would need to rely on the lattice simulations for a more precise calculation of the axion kinetic misalignment, which is beyond the scope of our work.

In the usual misalignment mechanism, we note that the axion abundance is determined by $Y_{a,{\rm mis}}=\frac{n_a}{s}$ at the onset of the axion oscillation at $T=T_{\rm osc}$ \cite{Bae:2008ue,Wantz:2009it,GrillidiCortona:2015jxo,Borsanyi:2016ksw}, namely, 
\bea
Y_{a,{\rm mis}}=0.11 \bigg(\frac{f_a}{10^9\,{\rm GeV}}\bigg)^{13/6}.
\eea
Then, the dominance with the axion kinetic misalignment, namely, $Y_\theta>Y_{a,{\rm mis}}$,  requires $f_a<1.5\times 10^{11}\,{\rm GeV}$ \cite{Co:2019jts}.

\begin{figure}[!t]
\begin{center}
 \includegraphics[width=0.45\textwidth,clip]{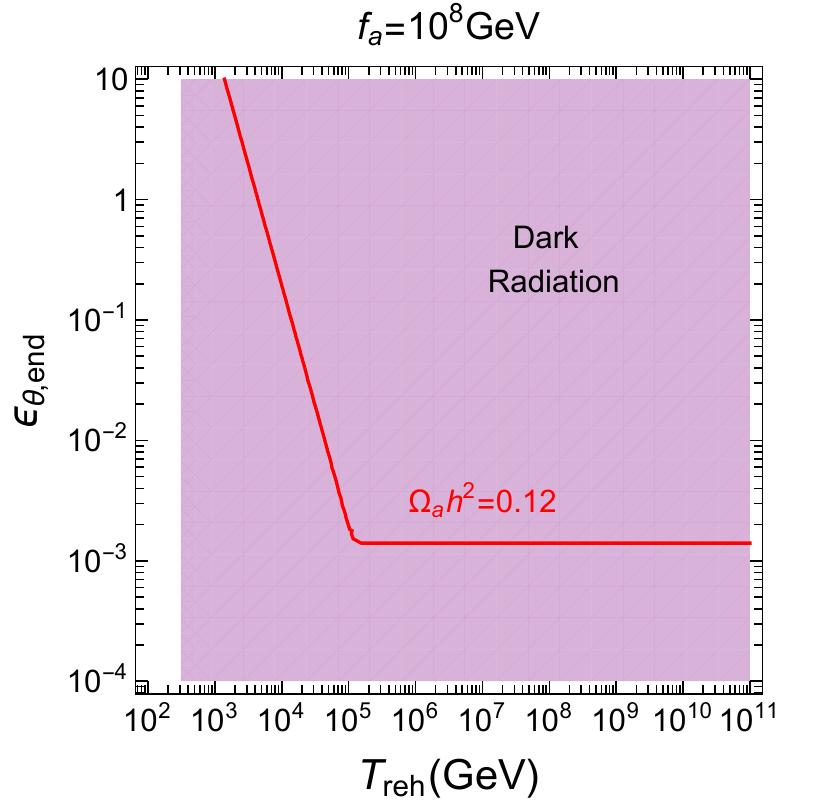}\,\,  \includegraphics[width=0.45\textwidth,clip]{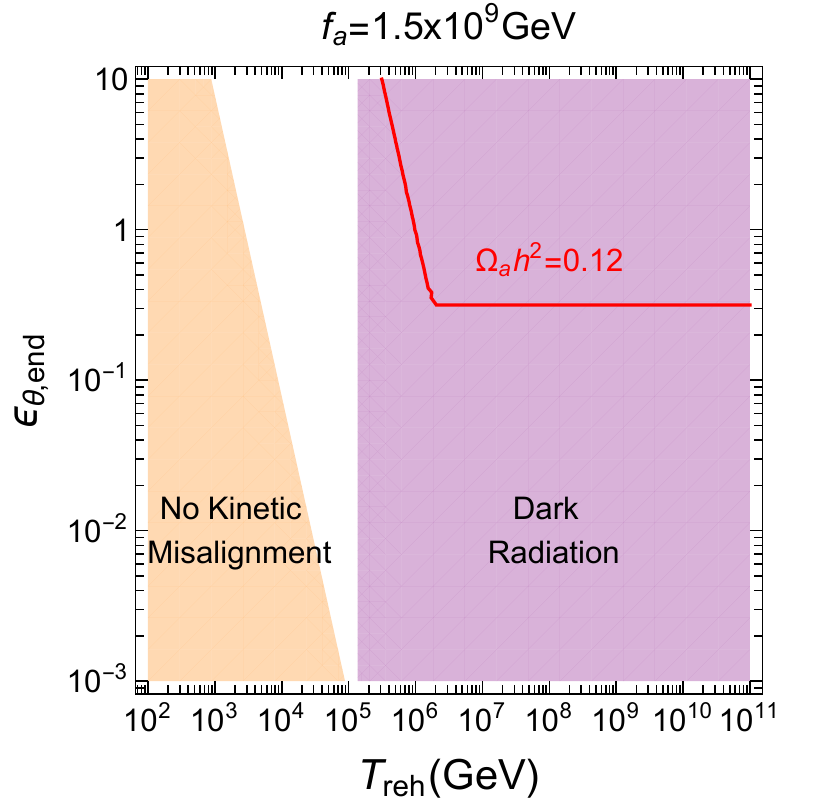}  
 \end{center}
\caption{Parameter space in the reheating temperature $T_{\rm reh}$ vs $\epsilon_{\theta,{\rm end}}$ for axion dark matter with kinetic misalignment. The correct relic density is obtained by the axion kinetic misalignment along the red line. The kinetic misalignment is sub-dominant in orange region, and the axions produced from the inflaton scattering were in thermal equilibrium, becoming dark radiation at a detectable level in purple regions. We took $f_a=10^8\,{\rm GeV}, 1,5\times 10^{9}\,{\rm GeV}$ on the left and right plots, respectively. }
\label{fig:relic1}
\end{figure}

In Fig.~\ref{fig:relic1}, we depict the parameter space for the reheating temperature and the slow-roll parameter for the axion, $\epsilon_{\theta,{\rm end}}$, satisfying the correct relic density by the axion kinetic misalignment in red lines. We show that the axions produced from the saxion scattering  becomes thermalized after reheating and provides a detectable dark radiation in purple regions, whereas the kinetic misalignment is sub-dominant in the orange region. The correct relic density is achievable even for a relatively small axion decay constant such as $f_a=10^8\,{\rm GeV}$ and $1.5\times 10^{9}\,{\rm GeV}$ in the left and right plots, respectively. Here, for the axion relic density, we used eq.~(\ref{relic}) with eqs.~(\ref{density1}) or (\ref{density2}), depending on whether $T_{\rm RH}>T^c_{\rm RH}$ or not, and took $V_E(\phi_{\rm end})=0.089 V_I$ with the inflaton potential energy $V_I$ being constrained by the CMB normalization in eq.~(\ref{eq:CMBnormalization}) and $\phi_{\rm end}=1.5M_P$. 

It is worthwhile to remark that the reheating temperature corresponding to the correct relic density is achieved from the inflaton decay with the Yukawa coupling $y_f$ in eq.~(\ref{decayhp}), and/or  the inflaton scattering with the Higgs-portal coupling $\lambda_{H\Phi}$ in eq.~(\ref{scattering}).  A low reheating temperature below $T_{\rm RH}\sim 10^4\,{\rm GeV}$ can be obtained from the inflaton decay with $y_f\lesssim 10^{-4}$, but a high reheating temperature  up to $T_{\rm RH}\sim 10^{11}\,{\rm GeV}$ is possible due to the inflaton scattering with $\lambda_{H\Phi}\lesssim 10^{-7}$, being consistent with a small running quartic coupling for the PQ field. We also note that eq.~(\ref{Neff}) is sufficient for computing the dark radiation abundance in the parameter space explaining the correct relic density. This is because axions are thermalized with the SM plasma, namely, $T_{\rm RH}\gtrsim 310 \,{\rm GeV}, 1.4\times 10^5\,{\rm GeV}$ is satisfied for $f_a=10^8\,{\rm GeV}, 1.5\times 10^9\,{\rm GeV}$, respectively, and the dark radiation becomes independent of the initial abundance produced from the inflaton scattering given in eq.~(\ref{axionnonthermal}).

\begin{figure}[!t]
\begin{center}
 \includegraphics[width=0.45\textwidth,clip]{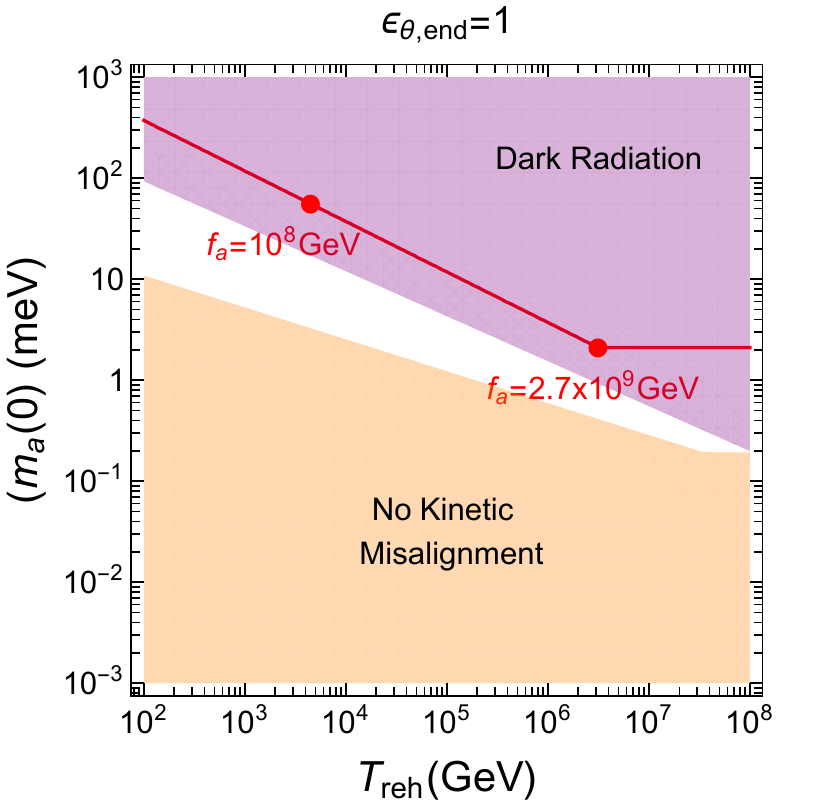}\,\,  \includegraphics[width=0.45\textwidth,clip]{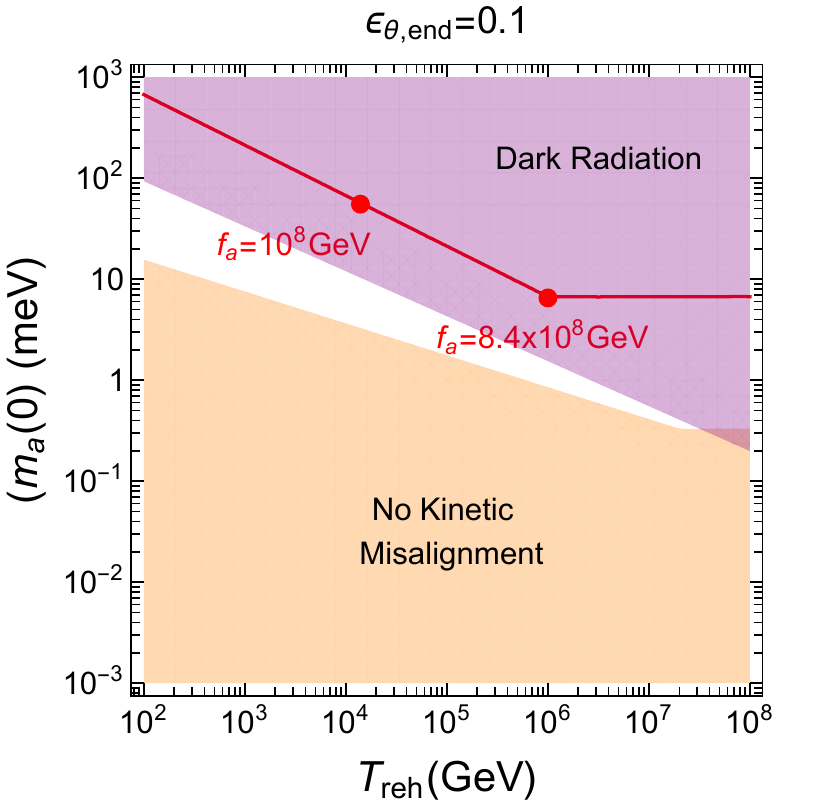}  
 \end{center}
\caption{Parameter space in the reheating temperature $T_{\rm reh}$ vs the axion mass $m_a(0)$  for axion dark matter with kinetic misalignment. The correct relic density is obtained by the axion kinetic misalignment along the red line. The kinetic misalignment is sub-dominant in orange regions and the axions produced from the inflaton scattering were in thermal equilibrium, becoming dark radiation at a detectable level in purple regions. We took $\epsilon_{\theta,{\rm end}}=1, 0.1$ on the left and right plots, respectively. }
\label{fig:relic2}
\end{figure}

In Fig.~\ref{fig:relic2}, we show the correlation between the reheating temperature and the axion mass at zero temperature, accounting for the correct relic density from the axion kinetic misalignment in red lines as in Fig.~\ref{fig:relic1}. 
We chose the slow-roll parameter of the axion at the end of inflation to $\epsilon_{\theta,{\rm end}}=1, 0.1$ in the left and right plots, respectively. We show the corresponding axion decay constant with $f_a\geq 10^8\,{\rm GeV}$ satisfying the correct relic density for dark matter and the reheating temperature needs to be above $T_{\rm reh}=4\times10^3\,{\rm GeV}$ for $\epsilon_{\theta,{\rm end}}=1$, and above $T_{\rm reh}=10^4\,{\rm GeV}$  for $\epsilon_{\theta,{\rm end}}=0.1$, respectively. Here, the lower end of $f_a$ is taken from the bounds from Supernova 1987A \cite{Raffelt:2006cw, Chang:2018rso}, which is $f_a>(1-4)\times 10^8\,{\rm GeV}$. As in Fig.~\ref{fig:relic1}, the region with a detectable dark radiation  is shown in purple regions and  the kinetic misalignment is sub-dominant in orange regions.

\chapter{Conclusions}\label{chap:summary}

\begin{small}
    ``Running isn't just about crossing the finish line. It's about embracing the journey with all your heart.”
    \\ Eliud Kipchoge.
    \end{small} 
    \vspace{5mm}

In the 1980s, inflation was proposed as an elegant solution to the Big Bang puzzles. Ever since, measurements of the CMB anisotropies have become more and more precise, allowing us 
to probe our theories wrong. At the same time, our searches for new particles in colliders have returned empty-handed, pressing towards Dark Matter theories beyond WIMPs and casting doubts upon Supersymmetry.
In this thesis, we explored different inflationary models and linked them to several of the open questions in Particle Physics, such as the nature of Dark Matter or the possible signatures from supersymmetry.

In \textbf{Chapters \ref{chap_UV} and \ref{chap_sugra}}, we proposed possible UV completions for Higgs inflation, which we dubbed \textit{Higgs-sigma models}.
We saw that, in order to achieve successful inflation, a particular linear class of sigma models coming from the $R^2$
term is preferred. We discussed the implications of the Higgs-sigma interactions for the running of the Higgs quartic coupling in the early Universe and the vacuum stability problem of the Standard Model.
We then provided an in depth study of these UV completions: their inflationary predictions and the subsequent perturbative reheating, including the possibility of Dark Matter production.
We provided a supergravity embedding of the Higgs-sigma models, and showed that slow-roll inflation can be realized from the mixture of the SM
Higgs and the real part of T (identified with the $\sigma$ field).  In order to ensure stability of the inflationary potential,
we introduced an extra quartic coupling for S in the frame function. As low-energy remnants of the Higgs-$R^2$ supergravity, we have suggested the possibilities
for SUSY breaking, in the presence of either modified higher curvature terms
or an extra singlet chiral superfield of O’Raifeartaigh type. After SUSY breaking, the non-minimal
coupling to the Higgs fields and the couplings between the dual superfields and the Higgs
give rise to naturally small contributions to the µ term.

In \textbf{Chapter \ref{chapter_hybrid}}, motivated by the naturalness problem of inflation, we presented a model for natural inflation with twin waterfall fields where
a $Z_2$ symmetry in the waterfall sector protects the inflaton potential against quantum corrections.
Inflation ends here  due to a tachyonics instability developed by one of the waterfall fields. 
In this scenario we do not need a trans-Planckian axion decay constant neither
fine-tuning on the initial conditions for inflation.
 Furthermore, in the presence of an extra $Z_2$ symmetry, the lightest waterfall field can provide the correct relic abundance for Dark Matter.

Finally, in \textbf{chapter \ref{HiggsPole}} we explored the possibility of realizing inflation at the pole of the kinetic term in the Einstein frame. Here, perturbativity is ensured by construction and we can realize Higgs inflation without the need for a large-non minimal coupling.
The downside of it is that we require a extremely small quartic coupling during inflation, that calls for a non-trivial extension of the SM. We considered two possible extensions: first with an extra scalar field and  then in supergravity.
Finally, we applied the pole structure to the Peccei Quinn field. In this case, the angular component of the PQ field can be identified with the QCD axion, which acquires a non-zero velocity and allows for the production of sufficient Dark Matter through the kinetic misalignment mechanism.

All in all, progress in the field keeps ongoing. 
Whatever nature has in store for inflation, Dark Matter or the patterns underlying the structure of the Standard Model, we still have a long way to go.
In the years to come, we can only keep a keen eye on our Dark Matter direct and indirect searches, the improving CMB measurements and the whole new window recently opened from gravitational wave signals.
It remains to be seen if we can provide a self-consistent picture of Early Universe Cosmology compatible with our more close to home low energy collider data.

\end{doublespace}

\lhead{} \chead{\leftmark} \rhead{}
\doublespacing
\bibliographystyle{unsrt}
\bibliography{thesis.bib}

\begin{thebibliography}{100}

\bibitem{WMAP:2012fli}
C.~L. Bennett et~al.
\newblock {Nine-Year Wilkinson Microwave Anisotropy Probe (WMAP) Observations: Final Maps and Results}.
\newblock {\em Astrophys. J. Suppl.}, 208:20, 2013.

\bibitem{ParticleDataGroup:2018ovx}
M.~Tanabashi et~al.
\newblock {Review of Particle Physics}.
\newblock {\em Phys. Rev. D}, 98(3):030001, 2018.

\bibitem{Co:2019wyp}
Raymond~T. Co and Keisuke Harigaya.
\newblock {Axiogenesis}.
\newblock {\em Phys. Rev. Lett.}, 124(11):111602, 2020.

\bibitem{Co:2020jtv}
Raymond~T. Co, Nicolas Fernandez, Akshay Ghalsasi, Lawrence~J. Hall, and Keisuke Harigaya.
\newblock {Lepto-Axiogenesis}.
\newblock {\em JHEP}, 03:017, 2021.

\bibitem{Kawamura:2021xpu}
Junichiro Kawamura and Stuart Raby.
\newblock {Lepto-axiogenesis in minimal SUSY KSVZ model}.
\newblock {\em JHEP}, 04:116, 2022.

\bibitem{Gouttenoire:2021jhk}
Yann Gouttenoire, Geraldine Servant, and Peera Simakachorn.
\newblock {Kination cosmology from scalar fields and gravitational-wave signatures}.
\newblock 11 2021.

\bibitem{Eroncel:2022vjg}
Cem Er\"oncel, Ryosuke Sato, Geraldine Servant, and Philip S\o{}rensen.
\newblock {ALP dark matter from kinetic fragmentation: opening up the parameter window}.
\newblock {\em JCAP}, 10:053, 2022.

\bibitem{Fairbairn:2014zta}
Malcolm Fairbairn, Robert Hogan, and David J.~E. Marsh.
\newblock {Unifying inflation and dark matter with the Peccei-Quinn field: observable axions and observable tensors}.
\newblock {\em Phys. Rev. D}, 91(2):023509, 2015.

\bibitem{Nakayama:2015pba}
Kazunori Nakayama and Masahiro Takimoto.
\newblock {Higgs inflation and suppression of axion isocurvature perturbation}.
\newblock {\em Phys. Lett. B}, 748:108--112, 2015.

\bibitem{Ballesteros:2016euj}
Guillermo Ballesteros, Javier Redondo, Andreas Ringwald, and Carlos Tamarit.
\newblock {Unifying inflation with the axion, dark matter, baryogenesis and the seesaw mechanism}.
\newblock {\em Phys. Rev. Lett.}, 118(7):071802, 2017.

\bibitem{Ballesteros:2016xej}
Guillermo Ballesteros, Javier Redondo, Andreas Ringwald, and Carlos Tamarit.
\newblock {Standard Model\textemdash{}axion\textemdash{}seesaw\textemdash{}Higgs portal inflation. Five problems of particle physics and cosmology solved in one stroke}.
\newblock {\em JCAP}, 08:001, 2017.

\bibitem{Hui:2018cvg}
Howard Hui et~al.
\newblock {BICEP Array: a multi-frequency degree-scale CMB polarimeter}.
\newblock {\em Proc. SPIE Int. Soc. Opt. Eng.}, 10708:1070807, 2018.

\bibitem{Hensley:2021ydb}
Brandon~S. Hensley et~al.
\newblock {The Simons Observatory: Galactic Science Goals and Forecasts}.
\newblock 11 2021.

\bibitem{Abazajian:2019eic}
Kevork Abazajian et~al.
\newblock {CMB-S4 Science Case, Reference Design, and Project Plan}.
\newblock 7 2019.

\bibitem{Hazumi:2019lys}
M.~Hazumi et~al.
\newblock {LiteBIRD: A Satellite for the Studies of B-Mode Polarization and Inflation from Cosmic Background Radiation Detection}.
\newblock {\em J. Low Temp. Phys.}, 194(5-6):443--452, 2019.

\bibitem{NASAPICO:2019thw}
Shaul Hanany et~al.
\newblock {PICO: Probe of Inflation and Cosmic Origins}.
\newblock 3 2019.

\bibitem{Guth:1997wk}
Alan~H. Guth.
\newblock {\em {The inflationary universe: The quest for a new theory of cosmic origins}}.
\newblock 1997.

\bibitem{Linde:1982zj}
Andrei~D. Linde.
\newblock {Coleman-Weinberg Theory and a New Inflationary Universe Scenario}.
\newblock {\em Phys. Lett. B}, 114:431--435, 1982.

\bibitem{Albrecht:1982wi}
Andreas Albrecht and Paul~J. Steinhardt.
\newblock {Cosmology for Grand Unified Theories with Radiatively Induced Symmetry Breaking}.
\newblock {\em Phys. Rev. Lett.}, 48:1220--1223, 1982.

\bibitem{Linde:1983gd}
Andrei~D. Linde.
\newblock {Chaotic Inflation}.
\newblock {\em Phys. Lett. B}, 129:177--181, 1983.

\bibitem{Guth:2000hz}
Alan~H. Guth.
\newblock {Inflationary models and connections to particle physics}.
\newblock In {\em {Pritzker Symposium and Workshop on the Status of Inflationary Cosmology}}, 2 2000.

\bibitem{Baumann:2009ds}
Daniel Baumann.
\newblock {Inflation}.
\newblock In {\em {Theoretical Advanced Study Institute in Elementary Particle Physics}: {Physics of the Large and the Small}}, pages 523--686, 2011.

\bibitem{Riotto:2018pcx}
A.~Riotto.
\newblock {Inflation and the Theory of Cosmological Perturbations}.
\newblock 2018.

\bibitem{Planck:2019evm}
Y.~Akrami et~al.
\newblock {Planck 2018 results. VII. Isotropy and Statistics of the CMB}.
\newblock {\em Astron. Astrophys.}, 641:A7, 2020.

\bibitem{Linde:1993cn}
Andrei~D. Linde.
\newblock {Hybrid inflation}.
\newblock {\em Phys. Rev. D}, 49:748--754, 1994.

\bibitem{Zeldovich:1972zz}
Ya.~B. Zeldovich.
\newblock {A Hypothesis, unifying the structure and the entropy of the universe}.
\newblock {\em Mon. Not. Roy. Astron. Soc.}, 160:1P--3P, 1972.

\bibitem{Harrison:1969fb}
Edward~R. Harrison.
\newblock {Fluctuations at the threshold of classical cosmology}.
\newblock {\em Phys. Rev. D}, 1:2726--2730, 1970.

\bibitem{Bezrukov:2007ep}
Fedor~L. Bezrukov and Mikhail Shaposhnikov.
\newblock {The Standard Model Higgs boson as the inflaton}.
\newblock {\em Phys. Lett. B}, 659:703--706, 2008.

\bibitem{Callan:1970ze}
Curtis~G. Callan, Jr., Sidney~R. Coleman, and Roman Jackiw.
\newblock {A New improved energy - momentum tensor}.
\newblock {\em Annals Phys.}, 59:42--73, 1970.

\bibitem{Birrell:1982ix}
N.~D. Birrell and P.~C.~W. Davies.
\newblock {\em {Quantum Fields in Curved Space}}.
\newblock Cambridge Monographs on Mathematical Physics. Cambridge Univ. Press, Cambridge, UK, 2 1984.

\bibitem{Burgess:2009ea}
C.~P. Burgess, Hyun~Min Lee, and Michael Trott.
\newblock {Power-counting and the Validity of the Classical Approximation During Inflation}.
\newblock {\em JHEP}, 09:103, 2009.

\bibitem{Burgess:2010zq}
C.~P. Burgess, Hyun~Min Lee, and Michael Trott.
\newblock {Comment on Higgs Inflation and Naturalness}.
\newblock {\em JHEP}, 07:007, 2010.

\bibitem{Barbon:2009ya}
J.~L.~F. Barbon and J.~R. Espinosa.
\newblock {On the Naturalness of Higgs Inflation}.
\newblock {\em Phys. Rev. D}, 79:081302, 2009.

\bibitem{Hertzberg:2010dc}
Mark~P. Hertzberg.
\newblock {On Inflation with Non-minimal Coupling}.
\newblock {\em JHEP}, 11:023, 2010.

\bibitem{Bezrukov:2014ina}
Fedor Bezrukov and Mikhail Shaposhnikov.
\newblock {Why should we care about the top quark Yukawa coupling?}
\newblock {\em J. Exp. Theor. Phys.}, 120:335--343, 2015.

\bibitem{Butenschoen:2017ays}
Mathias Butenschoen, Bahman Dehnadi, Andr\'e~H. Hoang, Vicent Mateu, Moritz Preisser, and Iain~W. Stewart.
\newblock {Top quark mass calibration for Monte-Carlo event generators}.
\newblock {\em PoS}, Hadron2017:189, 2018.

\bibitem{Espinosa:2016nld}
Jose~R. Espinosa, Mathias Garny, Thomas Konstandin, and Antonio Riotto.
\newblock {Gauge-Independent Scales Related to the Standard Model Vacuum Instability}.
\newblock {\em Phys. Rev. D}, 95(5):056004, 2017.

\bibitem{Rubio:2018ogq}
Javier Rubio.
\newblock {Higgs inflation}.
\newblock {\em Front. Astron. Space Sci.}, 5:50, 2019.

\bibitem{Giudice:2010ka}
Gian~F. Giudice and Hyun~Min Lee.
\newblock {Unitarizing Higgs Inflation}.
\newblock {\em Phys. Lett. B}, 694:294--300, 2011.

\bibitem{Burgess:2003zw}
C.~P. Burgess, James~M. Cline, and R.~Holman.
\newblock {Effective field theories and inflation}.
\newblock {\em JCAP}, 10:004, 2003.

\bibitem{Ema:2017rqn}
Yohei Ema.
\newblock {Higgs Scalaron Mixed Inflation}.
\newblock {\em Phys. Lett. B}, 770:403--411, 2017.

\bibitem{Gorbunov:2018llf}
Dmitry Gorbunov and Anna Tokareva.
\newblock {Scalaron the healer: removing the strong-coupling in the Higgs- and Higgs-dilaton inflations}.
\newblock {\em Phys. Lett. B}, 788:37--41, 2019.

\bibitem{He:2018mgb}
Minxi He, Ryusuke Jinno, Kohei Kamada, Seong~Chan Park, Alexei~A. Starobinsky, and Jun'ichi Yokoyama.
\newblock {On the violent preheating in the mixed Higgs-$R^2$ inflationary model}.
\newblock {\em Phys. Lett. B}, 791:36--42, 2019.

\bibitem{Cheong:2020rao}
Dhong~Yeon Cheong, Hyun~Min Lee, and Seong~Chan Park.
\newblock {Beyond the Starobinsky model for inflation}.
\newblock {\em Phys. Lett. B}, 805:135453, 2020.

\bibitem{Ema:2020zvg}
Yohei Ema, Kyohei Mukaida, and Jorinde van~de Vis.
\newblock {Higgs inflation as nonlinear sigma model and scalaron as its $\sigma$-meson}.
\newblock {\em JHEP}, 11:011, 2020.

\bibitem{Ema:2020evi}
Yohei Ema, Kyohei Mukaida, and Jorinde van~de Vis.
\newblock {Renormalization group equations of Higgs-R$^{2}$ inflation}.
\newblock {\em JHEP}, 02:109, 2021.

\bibitem{Linde:1981mu}
Andrei~D. Linde.
\newblock {A New Inflationary Universe Scenario: A Possible Solution of the Horizon, Flatness, Homogeneity, Isotropy and Primordial Monopole Problems}.
\newblock {\em Phys. Lett. B}, 108:389--393, 1982.

\bibitem{Adams:1990pn}
Fred~C. Adams, Katherine Freese, and Alan~H. Guth.
\newblock {Constraints on the scalar field potential in inflationary models}.
\newblock {\em Phys. Rev. D}, 43:965--976, 1991.

\bibitem{Freese:1990rb}
Katherine Freese, Joshua~A. Frieman, and Angela~V. Olinto.
\newblock {Natural inflation with pseudo - Nambu-Goldstone bosons}.
\newblock {\em Phys. Rev. Lett.}, 65:3233--3236, 1990.

\bibitem{Giudice:2003jh}
G.~F. Giudice, A.~Notari, M.~Raidal, A.~Riotto, and A.~Strumia.
\newblock {Towards a complete theory of thermal leptogenesis in the SM and MSSM}.
\newblock {\em Nucl. Phys. B}, 685:89--149, 2004.

\bibitem{Peccei:1977hh}
R.~D. Peccei and Helen~R. Quinn.
\newblock {CP Conservation in the Presence of Instantons}.
\newblock {\em Phys. Rev. Lett.}, 38:1440--1443, 1977.

\bibitem{Wilczek:1977pj}
Frank Wilczek.
\newblock {Problem of Strong $P$ and $T$ Invariance in the Presence of Instantons}.
\newblock {\em Phys. Rev. Lett.}, 40:279--282, 1978.

\bibitem{Weinberg:1977ma}
Steven Weinberg.
\newblock {A New Light Boson?}
\newblock {\em Phys. Rev. Lett.}, 40:223--226, 1978.

\bibitem{Preskill:1982cy}
John Preskill, Mark~B. Wise, and Frank Wilczek.
\newblock {Cosmology of the Invisible Axion}.
\newblock {\em Phys. Lett. B}, 120:127--132, 1983.

\bibitem{Abbott:1982af}
L.~F. Abbott and P.~Sikivie.
\newblock {A Cosmological Bound on the Invisible Axion}.
\newblock {\em Phys. Lett. B}, 120:133--136, 1983.

\bibitem{Dine:1982ah}
Michael Dine and Willy Fischler.
\newblock {The Not So Harmless Axion}.
\newblock {\em Phys. Lett. B}, 120:137--141, 1983.

\bibitem{Co:2019jts}
Raymond~T. Co, Lawrence~J. Hall, and Keisuke Harigaya.
\newblock {Axion Kinetic Misalignment Mechanism}.
\newblock {\em Phys. Rev. Lett.}, 124(25):251802, 2020.

\bibitem{Elias-Miro:2012eoi}
Joan Elias-Miro, Jose~R. Espinosa, Gian~F. Giudice, Hyun~Min Lee, and Alessandro Strumia.
\newblock {Stabilization of the Electroweak Vacuum by a Scalar Threshold Effect}.
\newblock {\em JHEP}, 06:031, 2012.

\bibitem{Starobinsky:1980te}
Alexei~A. Starobinsky.
\newblock {A New Type of Isotropic Cosmological Models Without Singularity}.
\newblock {\em Phys. Lett. B}, 91:99--102, 1980.

\bibitem{Fan:2019udt}
JiJi Fan, Matthew Reece, and Yi~Wang.
\newblock {An Inflationary Probe of Cosmic Higgs Switching}.
\newblock {\em JHEP}, 05:042, 2020.

\bibitem{He:2020qcb}
Minxi He.
\newblock {Perturbative Reheating in the Mixed Higgs-$R^2$ Model}.
\newblock {\em JCAP}, 05:021, 2021.

\bibitem{Ichikawa:2008ne}
Kazuhide Ichikawa, Teruaki Suyama, Tomo Takahashi, and Masahide Yamaguchi.
\newblock {Primordial Curvature Fluctuation and Its Non-Gaussianity in Models with Modulated Reheating}.
\newblock {\em Phys. Rev. D}, 78:063545, 2008.

\bibitem{Nurmi:2015ema}
Sami Nurmi, Tommi Tenkanen, and Kimmo Tuominen.
\newblock {Inflationary Imprints on Dark Matter}.
\newblock {\em JCAP}, 11:001, 2015.

\bibitem{Kainulainen:2016vzv}
Kimmo Kainulainen, Sami Nurmi, Tommi Tenkanen, Kimmo Tuominen, and Ville Vaskonen.
\newblock {Isocurvature Constraints on Portal Couplings}.
\newblock {\em JCAP}, 06:022, 2016.

\bibitem{Battefeld:2008bu}
Diana Battefeld and Shinsuke Kawai.
\newblock {Preheating after N-flation}.
\newblock {\em Phys. Rev. D}, 77:123507, 2008.

\bibitem{Choi:2008et}
Ki-Young Choi, Jinn-Ouk Gong, and Donghui Jeong.
\newblock {Evolution of the curvature perturbation during and after multi-field inflation}.
\newblock {\em JCAP}, 02:032, 2009.

\bibitem{Battefeld:2009xw}
Diana Battefeld, Thorsten Battefeld, and John~T. Giblin.
\newblock {On the Suppression of Parametric Resonance and the Viability of Tachyonic Preheating after Multi-Field Inflation}.
\newblock {\em Phys. Rev. D}, 79:123510, 2009.

\bibitem{Braden:2010wd}
Jonathan Braden, Lev Kofman, and Neil Barnaby.
\newblock {Reheating the Universe After Multi-Field Inflation}.
\newblock {\em JCAP}, 07:016, 2010.

\bibitem{Meyers:2013gua}
Joel Meyers and Ewan R.~M. Tarrant.
\newblock {Perturbative Reheating After Multiple-Field Inflation: The Impact on Primordial Observables}.
\newblock {\em Phys. Rev. D}, 89(6):063535, 2014.

\bibitem{Elliston:2014zea}
Joseph Elliston, Stefano Orani, and David~J. Mulryne.
\newblock {General analytic predictions of two-field inflation and perturbative reheating}.
\newblock {\em Phys. Rev. D}, 89(10):103532, 2014.

\bibitem{Hotinli:2017vhx}
Selim~C. Hotinli, Jonathan Frazer, Andrew~H. Jaffe, Joel Meyers, Layne~C. Price, and Ewan R.~M. Tarrant.
\newblock {Effect of reheating on predictions following multiple-field inflation}.
\newblock {\em Phys. Rev. D}, 97(2):023511, 2018.

\bibitem{Leung:2012ve}
Godfrey Leung, Ewan R.~M. Tarrant, Christian~T. Byrnes, and Edmund~J. Copeland.
\newblock {Reheating, Multifield Inflation and the Fate of the Primordial Observables}.
\newblock {\em JCAP}, 09:008, 2012.

\bibitem{Huston:2013kgl}
Ian Huston and Adam~J. Christopherson.
\newblock {Isocurvature Perturbations and Reheating in Multi-Field Inflation}.
\newblock 2 2013.

\bibitem{Leung:2013rza}
Godfrey Leung, Ewan R.~M. Tarrant, Christian~T. Byrnes, and Edmund~J. Copeland.
\newblock {Influence of Reheating on the Trispectrum and its Scale Dependence}.
\newblock {\em JCAP}, 08:006, 2013.

\bibitem{Watanabe:2015eia}
Yuki Watanabe and Jonathan White.
\newblock {Multifield formulation of gravitational particle production after inflation}.
\newblock {\em Phys. Rev. D}, 92:023504, 2015.

\bibitem{DeCross:2015uza}
Matthew~P. DeCross, David~I. Kaiser, Anirudh Prabhu, C.~Prescod-Weinstein, and Evangelos~I. Sfakianakis.
\newblock {Preheating after Multifield Inflation with Nonminimal Couplings, I: Covariant Formalism and Attractor Behavior}.
\newblock {\em Phys. Rev. D}, 97(2):023526, 2018.

\bibitem{DeCross:2016cbs}
Matthew~P. DeCross, David~I. Kaiser, Anirudh Prabhu, Chanda Prescod-Weinstein, and Evangelos~I. Sfakianakis.
\newblock {Preheating after multifield inflation with nonminimal couplings, III: Dynamical spacetime results}.
\newblock {\em Phys. Rev. D}, 97(2):023528, 2018.

\bibitem{DeCross:2016fdz}
Matthew~P. DeCross, David~I. Kaiser, Anirudh Prabhu, Chanda Prescod-Weinstein, and Evangelos~I. Sfakianakis.
\newblock {Preheating after multifield inflation with nonminimal couplings, II: Resonance Structure}.
\newblock {\em Phys. Rev. D}, 97(2):023527, 2018.

\bibitem{Schimmrigk:2017jwa}
Rolf Schimmrigk.
\newblock {Multifield Reheating after Modular $j$-Inflation}.
\newblock {\em Phys. Lett. B}, 782:193--197, 2018.

\bibitem{Gonzalez:2018jax}
Pablo Gonz\'alez, Gonzalo~A. Palma, and Nelson Videla.
\newblock {Covariant evolution of perturbations during reheating in two-field inflation}.
\newblock {\em JCAP}, 12:001, 2018.

\bibitem{Martin:2021frd}
J\'er\^ome Martin and Lucas Pinol.
\newblock {Opening the reheating box in multifield inflation}.
\newblock {\em JCAP}, 12(12):022, 2021.

\bibitem{Chung:1998rq}
Daniel J.~H. Chung, Edward~W. Kolb, and Antonio Riotto.
\newblock {Production of massive particles during reheating}.
\newblock {\em Phys. Rev. D}, 60:063504, 1999.

\bibitem{Giudice:2000ex}
Gian~Francesco Giudice, Edward~W. Kolb, and Antonio Riotto.
\newblock {Largest temperature of the radiation era and its cosmological implications}.
\newblock {\em Phys. Rev. D}, 64:023508, 2001.

\bibitem{Ellis:2015pla}
John Ellis, Marcos A.~G. Garcia, Dimitri~V. Nanopoulos, and Keith~A. Olive.
\newblock {Calculations of Inflaton Decays and Reheating: with Applications to No-Scale Inflation Models}.
\newblock {\em JCAP}, 07:050, 2015.

\bibitem{Ellis:2015jpg}
John Ellis, Marcos A.~G. Garcia, Dimitri~V. Nanopoulos, Keith~A. Olive, and Marco Peloso.
\newblock {Post-Inflationary Gravitino Production Revisited}.
\newblock {\em JCAP}, 03:008, 2016.

\bibitem{Garcia:2017tuj}
Marcos A.~G. Garcia, Yann Mambrini, Keith~A. Olive, and Marco Peloso.
\newblock {Enhancement of the Dark Matter Abundance Before Reheating: Applications to Gravitino Dark Matter}.
\newblock {\em Phys. Rev. D}, 96(10):103510, 2017.

\bibitem{BICEP:2021xfz}
P.~A.~R. Ade et~al.
\newblock {Improved Constraints on Primordial Gravitational Waves using Planck, WMAP, and BICEP/Keck Observations through the 2018 Observing Season}.
\newblock {\em Phys. Rev. Lett.}, 127(15):151301, 2021.

\bibitem{Chowdhury:2018tzw}
Debtosh Chowdhury, Emilian Dudas, Ma\'\i{}ra Dutra, and Yann Mambrini.
\newblock {Moduli Portal Dark Matter}.
\newblock {\em Phys. Rev. D}, 99(9):095028, 2019.

\bibitem{Kaneta:2019zgw}
Kunio Kaneta, Yann Mambrini, and Keith~A. Olive.
\newblock {Radiative production of nonthermal dark matter}.
\newblock {\em Phys. Rev. D}, 99(6):063508, 2019.

\bibitem{Anastasopoulos:2020gbu}
Pascal Anastasopoulos, Kunio Kaneta, Yann Mambrini, and Mathias Pierre.
\newblock {Energy-momentum portal to dark matter and emergent gravity}.
\newblock {\em Phys. Rev. D}, 102(5):055019, 2020.

\bibitem{Brax:2020gqg}
Philippe Brax, Kunio Kaneta, Yann Mambrini, and Mathias Pierre.
\newblock {Disformal dark matter}.
\newblock {\em Phys. Rev. D}, 103(1):015028, 2021.

\bibitem{Kaneta:2021pyx}
Kunio Kaneta, Pyungwon Ko, and Wan-Il Park.
\newblock {Conformal portal to dark matter}.
\newblock {\em Phys. Rev. D}, 104(7):075018, 2021.

\bibitem{Dudas:2017rpa}
Emilian Dudas, Yann Mambrini, and Keith Olive.
\newblock {Case for an EeV Gravitino}.
\newblock {\em Phys. Rev. Lett.}, 119(5):051801, 2017.

\bibitem{Garcia:2020eof}
Marcos A.~G. Garcia, Kunio Kaneta, Yann Mambrini, and Keith~A. Olive.
\newblock {Reheating and Post-inflationary Production of Dark Matter}.
\newblock {\em Phys. Rev. D}, 101(12):123507, 2020.

\bibitem{Garcia:2020wiy}
Marcos A.~G. Garcia, Kunio Kaneta, Yann Mambrini, and Keith~A. Olive.
\newblock {Inflaton Oscillations and Post-Inflationary Reheating}.
\newblock {\em JCAP}, 04:012, 2021.

\bibitem{Mambrini:2021zpp}
Yann Mambrini and Keith~A. Olive.
\newblock {Gravitational Production of Dark Matter during Reheating}.
\newblock {\em Phys. Rev. D}, 103(11):115009, 2021.

\bibitem{Clery:2021bwz}
Simon Clery, Yann Mambrini, Keith~A. Olive, and Sarunas Verner.
\newblock {Gravitational portals in the early Universe}.
\newblock {\em Phys. Rev. D}, 105(7):075005, 2022.

\bibitem{Hall:2009bx}
Lawrence~J. Hall, Karsten Jedamzik, John March-Russell, and Stephen~M. West.
\newblock {Freeze-In Production of FIMP Dark Matter}.
\newblock {\em JHEP}, 03:080, 2010.

\bibitem{Edsjo:1997bg}
Joakim Edsjo and Paolo Gondolo.
\newblock {Neutralino relic density including coannihilations}.
\newblock {\em Phys. Rev. D}, 56:1879--1894, 1997.

\bibitem{Watanabe:2010vy}
Yuki Watanabe.
\newblock {Rate of gravitational inflaton decay via gauge trace anomaly}.
\newblock {\em Phys. Rev. D}, 83:043511, 2011.

\bibitem{Choi:2019osi}
Soo-Min Choi, Yoo-Jin Kang, Hyun~Min Lee, and Kimiko Yamashita.
\newblock {Unitary inflaton as decaying dark matter}.
\newblock {\em JHEP}, 05:060, 2019.

\bibitem{Garny:2015sjg}
Mathias Garny, McCullen Sandora, and Martin~S. Sloth.
\newblock {Planckian Interacting Massive Particles as Dark Matter}.
\newblock {\em Phys. Rev. Lett.}, 116(10):101302, 2016.

\bibitem{Garny:2017kha}
Mathias Garny, Andrea Palessandro, McCullen Sandora, and Martin~S. Sloth.
\newblock {Theory and Phenomenology of Planckian Interacting Massive Particles as Dark Matter}.
\newblock {\em JCAP}, 02:027, 2018.

\bibitem{Tang:2017hvq}
Yong Tang and Yue-Liang Wu.
\newblock {On Thermal Gravitational Contribution to Particle Production and Dark Matter}.
\newblock {\em Phys. Lett. B}, 774:676--681, 2017.

\bibitem{Bernal:2018qlk}
Nicol\'as Bernal, Ma\'\i{}ra Dutra, Yann Mambrini, Keith Olive, Marco Peloso, and Mathias Pierre.
\newblock {Spin-2 Portal Dark Matter}.
\newblock {\em Phys. Rev. D}, 97(11):115020, 2018.

\bibitem{Barman:2021ugy}
Basabendu Barman and Nicol\'as Bernal.
\newblock {Gravitational SIMPs}.
\newblock {\em JCAP}, 06:011, 2021.

\bibitem{Haque:2022kez}
Md~Riajul Haque and Debaprasad Maity.
\newblock {Gravitational reheating}.
\newblock {\em Phys. Rev. D}, 107(4):043531, 2023.

\bibitem{Haque:2021mab}
Md~Riajul Haque and Debaprasad Maity.
\newblock {Gravitational dark matter: Free streaming and phase space distribution}.
\newblock {\em Phys. Rev. D}, 106(2):023506, 2022.

\bibitem{Bringmann:2016din}
Torsten Bringmann, Felix Kahlhoefer, Kai Schmidt-Hoberg, and Parampreet Walia.
\newblock {Strong constraints on self-interacting dark matter with light mediators}.
\newblock {\em Phys. Rev. Lett.}, 118(14):141802, 2017.

\bibitem{Yamaguchi:2011kg}
Masahide Yamaguchi.
\newblock {Supergravity based inflation models: a review}.
\newblock {\em Class. Quant. Grav.}, 28:103001, 2011.

\bibitem{Kaku:1978nz}
M.~Kaku, P.~K. Townsend, and P.~van Nieuwenhuizen.
\newblock {Properties of Conformal Supergravity}.
\newblock {\em Phys. Rev. D}, 17:3179, 1978.

\bibitem{Kaku:1978ea}
M.~Kaku and P.~K. Townsend.
\newblock {POINCARE SUPERGRAVITY AS BROKEN SUPERCONFORMAL GRAVITY}.
\newblock {\em Phys. Lett. B}, 76:54--58, 1978.

\bibitem{Townsend:1979ki}
P.~K. Townsend and P.~van Nieuwenhuizen.
\newblock {Simplifications of Conformal Supergravity}.
\newblock {\em Phys. Rev. D}, 19:3166, 1979.

\bibitem{Kugo:1982cu}
Taichiro Kugo and Shozo Uehara.
\newblock {Conformal and Poincare Tensor Calculi in $N=1$ Supergravity}.
\newblock {\em Nucl. Phys. B}, 226:49--92, 1983.

\bibitem{Kugo:1983mv}
T.~Kugo and S.~Uehara.
\newblock {$N=1$ Superconformal Tensor Calculus: Multiplets With External Lorentz Indices and Spinor Derivative Operators}.
\newblock {\em Prog. Theor. Phys.}, 73:235, 1985.

\bibitem{Cecotti:1987sa}
S.~Cecotti.
\newblock {HIGHER DERIVATIVE SUPERGRAVITY IS EQUIVALENT TO STANDARD SUPERGRAVITY COUPLED TO MATTER. 1.}
\newblock {\em Phys. Lett. B}, 190:86--92, 1987.

\bibitem{Einhorn:2009bh}
Martin~B. Einhorn and D.~R.~Timothy Jones.
\newblock {Inflation with Non-minimal Gravitational Couplings in Supergravity}.
\newblock {\em JHEP}, 03:026, 2010.

\bibitem{Ferrara:2010yw}
Sergio Ferrara, Renata Kallosh, Andrei Linde, Alessio Marrani, and Antoine Van~Proeyen.
\newblock {Jordan Frame Supergravity and Inflation in NMSSM}.
\newblock {\em Phys. Rev. D}, 82:045003, 2010.

\bibitem{Ferrara:2010in}
Sergio Ferrara, Renata Kallosh, Andrei Linde, Alessio Marrani, and Antoine Van~Proeyen.
\newblock {Superconformal Symmetry, NMSSM, and Inflation}.
\newblock {\em Phys. Rev. D}, 83:025008, 2011.

\bibitem{Lee:2021dgi}
Hyun~Min Lee and Adriana~G. Menkara.
\newblock {Cosmology of linear Higgs-sigma models with conformal invariance}.
\newblock {\em JHEP}, 09:018, 2021.

\bibitem{Kallosh:2013lkr}
Renata Kallosh and Andrei Linde.
\newblock {Superconformal generalizations of the Starobinsky model}.
\newblock {\em JCAP}, 06:028, 2013.

\bibitem{Lee:2010hj}
Hyun~Min Lee.
\newblock {Chaotic inflation in Jordan frame supergravity}.
\newblock {\em JCAP}, 08:003, 2010.

\bibitem{Kallosh:2010ug}
Renata Kallosh and Andrei Linde.
\newblock {New models of chaotic inflation in supergravity}.
\newblock {\em JCAP}, 11:011, 2010.

\bibitem{Kallosh:2010xz}
Renata Kallosh, Andrei Linde, and Tomas Rube.
\newblock {General inflaton potentials in supergravity}.
\newblock {\em Phys. Rev. D}, 83:043507, 2011.

\bibitem{Kallosh:2011qk}
Renata Kallosh, Andrei Linde, Keith~A. Olive, and Tomas Rube.
\newblock {Chaotic inflation and supersymmetry breaking}.
\newblock {\em Phys. Rev. D}, 84:083519, 2011.

\bibitem{Dalianis:2014aya}
I.~Dalianis, F.~Farakos, A.~Kehagias, A.~Riotto, and R.~von Unge.
\newblock {Supersymmetry Breaking and Inflation from Higher Curvature Supergravity}.
\newblock {\em JHEP}, 01:043, 2015.

\bibitem{Intriligator:2007py}
Kenneth~A. Intriligator, Nathan Seiberg, and David Shih.
\newblock {Supersymmetry breaking, R-symmetry breaking and metastable vacua}.
\newblock {\em JHEP}, 07:017, 2007.

\bibitem{Shih:2007av}
David Shih.
\newblock {Spontaneous R-symmetry breaking in O'Raifeartaigh models}.
\newblock {\em JHEP}, 02:091, 2008.

\bibitem{Kitano:2006wz}
Ryuichiro Kitano.
\newblock {Gravitational Gauge Mediation}.
\newblock {\em Phys. Lett. B}, 641:203--207, 2006.

\bibitem{Giudice:1988yz}
G.~F. Giudice and A.~Masiero.
\newblock {A Natural Solution to the mu Problem in Supergravity Theories}.
\newblock {\em Phys. Lett. B}, 206:480--484, 1988.

\bibitem{Randall:1998uk}
Lisa Randall and Raman Sundrum.
\newblock {Out of this world supersymmetry breaking}.
\newblock {\em Nucl. Phys. B}, 557:79--118, 1999.

\bibitem{Giudice:1998xp}
Gian~F. Giudice, Markus~A. Luty, Hitoshi Murayama, and Riccardo Rattazzi.
\newblock {Gaugino mass without singlets}.
\newblock {\em JHEP}, 12:027, 1998.

\bibitem{Falkowski:2005zv}
Adam Falkowski, Hyun~Min Lee, and Christoph Ludeling.
\newblock {Gravity mediated supersymmetry breaking in six dimensions}.
\newblock {\em JHEP}, 10:090, 2005.

\bibitem{Garcia-Bellido:2001dqy}
Juan Garcia-Bellido and Ester Ruiz~Morales.
\newblock {Particle production from symmetry breaking after inflation}.
\newblock {\em Phys. Lett. B}, 536:193--202, 2002.

\bibitem{Choi:2016eif}
Soo-Min Choi and Hyun~Min Lee.
\newblock {Inflection point inflation and reheating}.
\newblock {\em Eur. Phys. J. C}, 76(6):303, 2016.

\bibitem{Aoki:2022dzd}
Shuntaro Aoki, Hyun~Min Lee, Adriana~G. Menkara, and Kimiko Yamashita.
\newblock {Reheating and dark matter freeze-in in the Higgs-R$^{2}$ inflation model}.
\newblock {\em JHEP}, 05:121, 2022.

\bibitem{Elias-Miro:2011sqh}
Joan Elias-Miro, Jose~R. Espinosa, Gian~F. Giudice, Gino Isidori, Antonio Riotto, and Alessandro Strumia.
\newblock {Higgs mass implications on the stability of the electroweak vacuum}.
\newblock {\em Phys. Lett. B}, 709:222--228, 2012.

\bibitem{Degrassi:2012ry}
Giuseppe Degrassi, Stefano Di~Vita, Joan Elias-Miro, Jose~R. Espinosa, Gian~F. Giudice, Gino Isidori, and Alessandro Strumia.
\newblock {Higgs mass and vacuum stability in the Standard Model at NNLO}.
\newblock {\em JHEP}, 08:098, 2012.

\bibitem{threshold}
Joan Elias-Miro, Jose~R. Espinosa, Gian~F. Giudice, Hyun~Min Lee, and Alessandro Strumia.
\newblock {Stabilization of the Electroweak Vacuum by a Scalar Threshold Effect}.
\newblock {\em JHEP}, 06:031, 2012.

\bibitem{Greene:1997fu}
Patrick~B. Greene, Lev Kofman, Andrei~D. Linde, and Alexei~A. Starobinsky.
\newblock {Structure of resonance in preheating after inflation}.
\newblock {\em Phys. Rev. D}, 56:6175--6192, 1997.

\bibitem{Bezrukov:2008ut}
F.~Bezrukov, D.~Gorbunov, and M.~Shaposhnikov.
\newblock {On initial conditions for the Hot Big Bang}.
\newblock {\em JCAP}, 06:029, 2009.

\bibitem{Garcia-Bellido:2008ycs}
Juan Garcia-Bellido, Daniel~G. Figueroa, and Javier Rubio.
\newblock {Preheating in the Standard Model with the Higgs-Inflaton coupled to gravity}.
\newblock {\em Phys. Rev. D}, 79:063531, 2009.

\bibitem{Kim:1979if}
Jihn~E. Kim.
\newblock {Weak Interaction Singlet and Strong CP Invariance}.
\newblock {\em Phys. Rev. Lett.}, 43:103, 1979.

\bibitem{Shifman:1979if}
Mikhail~A. Shifman, A.~I. Vainshtein, and Valentin~I. Zakharov.
\newblock {Can Confinement Ensure Natural CP Invariance of Strong Interactions?}
\newblock {\em Nucl. Phys. B}, 166:493--506, 1980.

\bibitem{Kamionkowski:1992mf}
Marc Kamionkowski and John March-Russell.
\newblock {Planck scale physics and the Peccei-Quinn mechanism}.
\newblock {\em Phys. Lett. B}, 282:137--141, 1992.

\bibitem{Barr:1992qq}
Stephen~M. Barr and D.~Seckel.
\newblock {Planck scale corrections to axion models}.
\newblock {\em Phys. Rev. D}, 46:539--549, 1992.

\bibitem{Gong:2011cd}
Jinn-Ouk Gong and Hyun~Min Lee.
\newblock {Large non-Gaussianity in non-minimal inflation}.
\newblock {\em JCAP}, 11:040, 2011.

\bibitem{Clery:2023ptm}
Simon Cl\'ery, Hyun~Min Lee, and Adriana~G. Menkara.
\newblock {Higgs inflation at the pole}.
\newblock {\em JHEP}, 10:144, 2023.

\bibitem{Salvio:2013iaa}
Alberto Salvio, Alessandro Strumia, and Wei Xue.
\newblock {Thermal axion production}.
\newblock {\em JCAP}, 01:011, 2014.

\bibitem{Bennett:2019ewm}
Jack~J. Bennett, Gilles Buldgen, Marco Drewes, and Yvonne Y.~Y. Wong.
\newblock {Towards a precision calculation of the effective number of neutrinos $N_{\rm eff}$ in the Standard Model I: the QED equation of state}.
\newblock {\em JCAP}, 03:003, 2020.
\newblock [Addendum: JCAP 03, A01 (2021)].

\bibitem{Bennett:2020zkv}
Jack~J. Bennett, Gilles Buldgen, Pablo~F. De~Salas, Marco Drewes, Stefano Gariazzo, Sergio Pastor, and Yvonne Y.~Y. Wong.
\newblock {Towards a precision calculation of $N_{\rm eff}$ in the Standard Model II: Neutrino decoupling in the presence of flavour oscillations and finite-temperature QED}.
\newblock {\em JCAP}, 04:073, 2021.

\bibitem{Akita:2020szl}
Kensuke Akita and Masahide Yamaguchi.
\newblock {A precision calculation of relic neutrino decoupling}.
\newblock {\em JCAP}, 08:012, 2020.

\bibitem{Planck:2018vyg}
N.~Aghanim et~al.
\newblock {Planck 2018 results. VI. Cosmological parameters}.
\newblock {\em Astron. Astrophys.}, 641:A6, 2020.
\newblock [Erratum: Astron.Astrophys. 652, C4 (2021)].

\bibitem{Masso:2002np}
Eduard Masso, Francesc Rota, and Gabriel Zsembinszki.
\newblock {On axion thermalization in the early universe}.
\newblock {\em Phys. Rev. D}, 66:023004, 2002.

\bibitem{Graf:2010tv}
Peter Graf and Frank~Daniel Steffen.
\newblock {Thermal axion production in the primordial quark-gluon plasma}.
\newblock {\em Phys. Rev. D}, 83:075011, 2011.

\bibitem{Gorghetto:2018ocs}
Marco Gorghetto and Giovanni Villadoro.
\newblock {Topological Susceptibility and QCD Axion Mass: QED and NNLO corrections}.
\newblock {\em JHEP}, 03:033, 2019.

\bibitem{Bae:2008ue}
Kyu~Jung Bae, Ji-Haeng Huh, and Jihn~E. Kim.
\newblock {Update of axion CDM energy}.
\newblock {\em JCAP}, 09:005, 2008.

\bibitem{Wantz:2009it}
Olivier Wantz and E.~P.~S. Shellard.
\newblock {Axion Cosmology Revisited}.
\newblock {\em Phys. Rev. D}, 82:123508, 2010.

\bibitem{GrillidiCortona:2015jxo}
Giovanni Grilli~di Cortona, Edward Hardy, Javier Pardo~Vega, and Giovanni Villadoro.
\newblock {The QCD axion, precisely}.
\newblock {\em JHEP}, 01:034, 2016.

\bibitem{Borsanyi:2016ksw}
Sz. Borsanyi et~al.
\newblock {Calculation of the axion mass based on high-temperature lattice quantum chromodynamics}.
\newblock {\em Nature}, 539(7627):69--71, 2016.

\bibitem{Raffelt:2006cw}
Georg~G. Raffelt.
\newblock {Astrophysical axion bounds}.
\newblock {\em Lect. Notes Phys.}, 741:51--71, 2008.

\bibitem{Chang:2018rso}
Jae~Hyeok Chang, Rouven Essig, and Samuel~D. McDermott.
\newblock {Supernova 1987A Constraints on Sub-GeV Dark Sectors, Millicharged Particles, the QCD Axion, and an Axion-like Particle}.
\newblock {\em JHEP}, 09:051, 2018.

\bibitem{Whitt:1984pd}
Brian Whitt.
\newblock {Fourth Order Gravity as General Relativity Plus Matter}.
\newblock {\em Phys. Lett. B}, 145:176--178, 1984.

\end{thebibliography}

\section*{List publications}

\begin{enumerate}
\item Shuntaro Aoki, Hyun Min Lee, and Adriana G. Menkara. Inflation and supersymmetry breaking in Higgs-$R^2$ supergravity. JHEP, 10:178, 2021.
\item  Shuntaro Aoki, Hyun Min Lee, Adriana G. Menkara, and Kimiko Yamashita. Reheating and dark matter freeze-in in the Higgs-R2
inflation model. JHEP, 05:121, 2022.
\item Simon Clery, Hyun Min Lee, and Adriana G. Menkara. Higgs inflation at the pole.  ́
JHEP, 10:144, 2023.
\item Yoo-Jin Kang, Hyun Min Lee, Adriana G. Menkara, and Jiseon Song. Flux-mediated
Dark Matter. JHEP, 06:013, 2021.
\item Seong-Sik Kim, Hyun Min Lee, Adriana G. Menkara, and Kimiko Yamashita.
SU(2)D lepton portals for the muon g-2, W-boson mass, and dark matter. Phys. Rev.
D, 106(1):015008, 2022.
\item Seong-Sik Kim, Hyun Min Lee, Adriana Guerrero Menkara, and Kimiko Yamashita.
The SU(2)D lepton portals for muon g - 2, W boson mass and dark matter. SciPost
Phys. Proc., 12:045, 2023.
\item Hyun Min Lee and Adriana G. Menkara. Cosmology of linear Higgs-sigma models
with conformal invariance. JHEP, 09:018, 2021.
\item Hyun Min Lee and Adriana G. Menkara. Pseudo-Nambu-Goldstone inflation with
twin waterfalls. Phys. Lett. B, 834:137483, 2022.
\item Hyun Min Lee and Adriana G. Menkara. Graceful exit from inflation and reheating
with twin waterfall scalar fields. Phys. Rev. D, 107(11):115019, 2023.
\item Hyun Min Lee, Adriana G. Menkara, Myeong-Jung Seong, and Jun-Ho Song. Peccei-
Quinn Inflation at the Pole and Axion Kinetic Misalignment. 10 2023.
\end{enumerate}
     

\newpage

\appendix
\chapter*{\LARGE Appendix I\linebreak[2] \huge{\spaceskip=0.3em plus 0.5em minus 0.5em{Equations of motion with kinetic mixing}}}
\addcontentsline{toc}{chapter}{Appendix I. Equations of motion with kinetic mixing }

In this appendix, we derive the general equations of motion in a FRW background.
The most general Lagrangian will also contain kinetic mixing between the fields,

\begin{equation}\label{apa:lj}
    \mathcal{L} = \sqrt{- g}\left( -\frac{1}{2} G_{ab} \partial_\mu \phi^a \partial^\mu \phi^ b - V \right),
\end{equation}

where $G_{ab}$ denotes the metric in field space. Then, the variation of the Lagrangian in eq.~(\ref{apa:lj}) reads

\begin{equation}
    \delta \mathcal{L} = \sqrt{-g} \left( - G_{ab} \partial_\mu \delta \phi^a \partial ^\mu \phi^b - \frac{1}{2} G_{ab,c} \delta \phi ^c \partial_\mu \phi^a \partial^\mu \phi ^b - V_a \delta \phi^a\right).
\end{equation}

Integrating by parts,

\begin{equation} \label{eqn:vari}
    \delta \mathcal{L} = \text{t.d.} + \left(\partial_\mu \left( \sqrt{-g} G_{ab} \partial^\mu \phi^b\right) + \sqrt{-g}\left(- \frac{1}{2} G_{bc,a}\partial_\mu\phi^b \partial^\mu \phi^c -V_a \right)
    \right)\delta \phi^a.
\end{equation}

Now, using that the covariant derivative of the metric vanishes, we get

\begin{equation}
    \sqrt{-g} \nabla_\mu \left(G_{ab}\partial^\mu \phi^b\right) = \partial_\mu \left( \sqrt{-g} G_{ab} \partial^\mu \phi^b\right),
\end{equation}

where $\nabla_\mu$ is the covariant derivative. We can rewrite eq.~(\ref{eqn:vari}) as 

\begin{equation}
    G_{ab} \nabla_\mu \partial^\mu \phi^b + G_{ab,c} \partial_\mu \phi^c \partial^\mu \phi^b - \frac{1}{2} G_{bc,a}\partial_\mu\phi^b \partial^\mu \phi^c - V_a = 0,
\end{equation}

so that

\begin{equation}
    \nabla_\mu \partial^\mu \phi^a + G^{ab}\left(G_{bc,d} \partial_\mu \phi^c \partial^\mu \phi ^d - \frac{1}{2} G_{cd,b}\partial_\mu \phi^c \partial^\mu \phi^d\right) - G^{ab}V_b = 0.
\end{equation}

$G_{ab}$ has associated to it the field-space Christoffel symbols,

\begin{equation}\label{Chrissymb}
    \Gamma^a_{cd} = G^{ab}\left(G_{bc,d} \partial_\mu \phi^c \partial^\mu \phi ^d - \frac{1}{2} G_{cd,b}\partial_\mu \phi^c \partial^\mu \phi^d\right).
\end{equation}

Using (\ref{Chrissymb}) we finally write our equation of motion in a general case

\begin{equation}\label{apa:eom}
    \nabla_\mu \partial^\mu \phi^a + \Gamma^a_{cd} \partial_\mu \phi^c \partial^\mu \phi^d - G^{ab} V_b = 0.
\end{equation}

\section*{FRW background}
Until here the discussion has been completely general. Notice that, if there is no kinetic mixing, then $G_{ab}$ is just the a diagonal matrix, not necessarily proportional to the identity.
For a FRW background, given by the metric element,

\begin{equation}
    ds^2 = -dt^2 + a(t)^2 d\vec{x}^2.
\end{equation}

The non-zero Christoffel symbols are

\bea
    \Gamma^{t}_{xx} &=& \Gamma^{t}_{yy}= \Gamma^{t}_{zz}=a(t) a'(t),\\
    \Gamma^{x}_{xt} &=& \Gamma^{y}_{yt}= \Gamma^{z}_{zt} = \frac{a'(t)}{a(t)},
\eea

so that the equation of motion in (\ref{apa:eom}) becomes
\bea
    \nabla_\mu \partial^\mu \phi^a &=& \partial_\mu \partial^\mu \phi^a + \Gamma^{\mu}_{\mu\nu} \partial^\nu \phi^a \nonumber\\
    &=& - \ddot{\phi}^a + \Gamma^i_{i0} (-1) \dot{\phi}^a \nonumber \\
    &=& - \ddot{\phi}^a - 3H \dot{\phi}^a,
\eea

and 

\begin{equation} \label{apa:finaleom}
    \ddot{\phi}^a + 3H\dot{\phi}^a - \Gamma^a_{bc} \dot{\phi}^b \dot{\phi}^c + G^{ab}V_b = 0.
\end{equation}

Eq.~(\ref{apa:finaleom}) was derived for the scalar fields, but ot top of these propagating degrees of freedom, we also have the 10 components of the metric.
Taking the variation with respect to the metric we would have derived Einstein's equations instead.

\chapter*{\LARGE Appendix II \linebreak[2] \huge{\spaceskip=0.3em plus 0.5em minus 0.5em{$f(R)$ theories of gravity}}}
\addcontentsline{toc}{chapter}{Appendix II. $f(R)$ theories of gravity } \label{appendix:eom}

The exponential expansion predicted by inflation is incompatible with General Relativity. Modifying Einstein gravity by including a non-linear function $f(R)$ can lead to cosmological expansion, making these models suitable to explain inflation.
It can be shown that an $f(R)$ theory is indeed equivalent to the addition of new scalar degrees of freedom \cite{Whitt:1984pd}.
Let's start from the action
\begin{equation}
    S = \frac{1}{2}\int \mathrm{d}^4 x \sqrt{-g} f\left(R\right)\label{eq:fr}.
\end{equation}

The energy momentum tensor associated to (\ref{eq:fr}) reads 

\begin{equation}
    T_{\mu\nu} = f'\left(R\right) R_{\mu \nu} - \frac{1}{2}f\left(R\right)g_{\mu \nu} - \nabla_\mu \nabla_\nu f'\left(R\right) + g_{\mu\nu} \Box f'\left(R\right) \label{emt},
\end{equation}

where we have made use of the short hand notation  $f'\left(R\right) \equiv \dv{f}{R}$.

We take now the trace of (\ref{emt}),

\begin{equation}
    3 \Box f' + f' R - 2f = T \label{trace}.
\end{equation}

This is the equation of motion of a self-interacting scalar field, if we interpret the terms $f' R - 2f$ as the self-interaction.
To make the connection clearer, we can use eq.~(\ref{trace}) to obtain an algebraic relation $R = R\left(f'\right)$,

\begin{align}
    \phi &\equiv f',\\
    V\left(\phi\right) &\equiv R\left(\phi\right) f' - f\left(\phi\right).
\end{align}

The Einstein equations then become

\begin{equation}
    G_{\mu\nu} = \frac{T_{\mu \nu}}{\phi} - \frac{V\left(\phi\right)}{2\phi}g_{\mu \nu} + \frac{1}{\phi} \left(\nabla_\mu \nabla_\nu - g_{\mu\nu} \Box \phi\right).
\end{equation}

We can show that the Lagrangian

\begin{equation}
    S\left[g_{\mu\nu, \phi}\right] = \frac{1}{2}\int \mathrm{d}^4 \sqrt{-g} \left( \phi R - V\left(\phi\right)\right) \label{Brans-Dicke-lag}
\end{equation}

raises the same equations of motion. Indeed, the Lagrangian in (\ref{Brans-Dicke-lag}) represents a Brans-Dicke scalar theory in which $\omega \rightarrow 0$, so that there is no derivative term for the $\phi$ field.
We can conclude that the Lagrangians (\ref{eq:fr}) and (\ref{Brans-Dicke-lag}) are indeed equivalent. 

\section*{Conformal transformation}

Performing a conformal transformation takes us from the Jordan frame to the Einstein frame. Since re-scaling the metric doesn't alter the physics, both frames are equivalent (at least at classical level).
In the Jordan frame, we have an explicit coupling between gravity (through the Ricci scalar) and the matter field. 
In this frame, there is no manifest scalar degree of freedom in the metric $g_{\mu\nu}$.
Once we perform a conformal transformation, we obtain the familiar Einstein-Hilbert term plus some terms in which the scalar field is totally decoupled from gravity.
In the Einstein frame we have an extra scalar field, arising from the transformation of the Ricci scalar, which contained non-canonical but nonetheless dynamical degrees of freedom.
In this appendix, we will explicitly derive the details of the conformal transformation.
The re-scaling of the metric is as follows,

\begin{align} \label{eq:conformaltransformetric}
    g_{\mu \nu} &= \Omega^2 \bar{g_{\mu \nu}},\\
    g^{\mu \nu } &= \Omega^{-2}\bar{g}^{\mu\nu},\\
    \sqrt{g} &= \Omega^D \sqrt{\bar{g}}.
\end{align}

\section*{Christoffel symbols}

\begin{align}
    \Gamma^{\alpha}_{\mu\nu} =& \frac{1}{2} g^{\alpha \beta}\left( g_{\beta\mu, \nu} + g_{\beta\nu, \mu} - g_{\mu\nu, \beta}\right) \nonumber\\
     =&\bar{\Gamma}^{\alpha}_{\mu \nu} + \Omega^{-1}\bar{g}^{\alpha \beta} \left(\bar{g}_{\beta\mu}\partial_\nu \Omega + \bar{g}_{\beta\nu}\partial_\mu \Omega - \bar{g}_{\mu\nu} \partial_\beta \Omega\right) \nonumber\\
     =&\bar{\Gamma}^{\alpha}_{\mu \nu} + \delta^\alpha_\mu \partial_\nu \ln \Omega + \delta^\alpha_\nu \partial_\mu \ln \Omega - \bar{g}^{\alpha \beta} \bar{g}_{\mu \nu}\partial_\beta \ln \Omega \nonumber \\
     \equiv &\bar{\Gamma}^{\alpha}_{\mu \nu} + \Delta ^{\alpha}_{\mu \nu}.
\end{align}

We see that the Christoffel symbols in the new frame get a correction $ \Delta ^{\alpha}_{\mu \nu}$.

\section*{Riemann Tensor}

Once the Christoffel symbols are defined, we can compute the Riemann tensor

\begin{align}
    R^{\alpha}_{\beta \mu \nu} =& \Gamma^{\alpha}_{\beta \nu, \mu} - \Gamma^{\alpha}_{\beta \mu, \nu} + \Gamma^{\alpha}_{\sigma \mu} \Gamma^{\sigma}_{\beta \nu} -  \Gamma^{\alpha}_{\sigma \nu} \Gamma^{\sigma}_{\beta \mu} \nonumber \\
    =& \bar{\Gamma}^{\alpha}_{\beta \nu, \mu} - \bar{\Gamma}^{\alpha}_{\beta \mu, \nu}   + \Delta^{\alpha}_{\beta \nu, \mu} - \Delta^{\alpha}_{\beta \mu, \nu} \nonumber \\
     +& \left( \bar{\Gamma}^{\alpha}_{\sigma \mu} + \Delta^\alpha_{\sigma \mu}\right)\left( \bar{\Gamma}^{\sigma}_{\beta \nu} + \Delta^\sigma_{\beta \nu}\right) - \left( \bar{\Gamma}^{\alpha}_{\sigma \nu} + \Delta^\alpha_{\sigma \nu}\right)\left( \bar{\Gamma}^{\sigma}_{\beta \mu} + \Delta^\sigma_{\beta \mu}\right) \nonumber\\
    =&\bar{\Gamma}^{\alpha}_{\beta \nu, \mu} - \bar{\Gamma}^{\alpha}_{\beta \mu, \nu} + \bar{\Gamma}^{\alpha}_{\sigma \mu} \bar{\Gamma}^{\sigma}_{\beta \nu} -  \bar{\Gamma}^{\alpha}_{\sigma \nu} \bar{\Gamma}^{\sigma}_{\beta \mu} \nonumber \\
    +& \Delta^{\alpha}_{\beta \nu, \mu} - \Delta^{\alpha}_{\beta \mu, \nu} + \bar{\Gamma}^{\alpha}_{\sigma \mu}\Delta^\sigma_{\beta \nu} + \Delta^{\alpha}_{\sigma \mu} \bar{\Gamma}^{\sigma}_{\beta \nu} - \bar{\Gamma}^\alpha_{\sigma \nu} \Delta^{\sigma}_{\beta \mu} - \Delta^{\alpha}_{\sigma \nu}\bar{\Gamma}^{\sigma}_{\beta \mu}\nonumber\\
    +& \Delta^\alpha_{\sigma \mu} \Delta^\sigma_{\beta \nu} - \Delta^{\alpha}_{\sigma\nu}\Delta^{\sigma}_{\beta \mu}
\end{align}

Now, we can group the terms as follows,

\begin{equation}
    \Delta^\alpha_{\beta \nu ; \mu} = \Delta^{\alpha}_{\beta \nu , \mu} + \Delta^\sigma_{\beta \nu}\bar{\Gamma}^\alpha_{\sigma \mu} - \Delta^{\alpha}_{\sigma\mu}\bar{\Gamma}^\sigma_{\beta \mu} - \Delta^{\alpha}_{\beta \sigma}\bar{\Gamma}^{\sigma}_{\nu \mu}.
\end{equation}

We notice that 
\begin{equation}
    \Delta ^{\alpha}_{\beta \nu; \mu} = \Delta ^{\alpha}_{\beta \nu, \mu} + \Delta^{\sigma}_{\beta\nu}\bar{\Gamma}^{\alpha}_{\sigma \mu} - \Delta^{\alpha}_{\sigma \nu}\bar{\Gamma}^{\sigma}_{\beta \mu} - \Delta^{\alpha}_{\beta\sigma} \bar{\Gamma}^{\sigma}_{\nu \mu},
\end{equation}

so we can rewrite the Riemann tensor as
\begin{align}
    R^{\alpha}_{\beta \mu \nu} &= \bar{R}^{\alpha}_{\beta \mu \nu} + \left(\Delta^{\alpha}_{\beta\nu; \mu} + \Delta^{\alpha}_{\beta\sigma}\bar{\Gamma}^{\sigma}_{\nu\mu}\right) - \left( \Delta^{\alpha}_{\beta\mu; \nu} + \Delta^{\alpha}_{\beta \sigma}\bar{\Gamma}^{\sigma}_{\mu\nu}\right) + \Delta^{\alpha}_{\sigma\mu} \Delta^{\sigma}_{\beta \nu} - \Delta^{\alpha}_{\sigma \nu}\Delta^{\sigma}_{\beta\mu}\nonumber\\
     &= \bar{R}^{\alpha}_{\beta \mu \nu} + \Delta^{\alpha}_{\beta\nu; \mu} - \Delta^{\alpha}_{\beta\mu; \nu} + \Delta^{\alpha}_{\sigma\mu} \Delta^{\sigma}_{\beta \nu} - \Delta^{\alpha}_{\sigma \nu}\Delta^{\sigma}_{\beta\mu}.
\end{align}

\section*{Ricci tensor}

\begin{equation}
    R_{\beta \nu} = \bar{R}_{\beta \nu} + \Delta^{\alpha}_{\beta \nu ; \alpha} - \Delta^\alpha_{\beta \alpha ; \nu} + \Delta^{\alpha}_{\sigma\alpha}\Delta^\sigma_{\beta\nu} - \Delta^{\alpha}_{\sigma\nu}\Delta^\sigma_{\beta \alpha}.
\end{equation}

Assuming $\bar{g}_{\alpha \beta; \mu}=0$, and using that the conformal transformation does not affect the coordinates, $\bar{\nabla}_\mu = \partial_\mu$, we get

\begin{align}
    \Delta^{\alpha}_{\beta\nu ;\alpha} &= 2 \partial_\beta \partial_\nu \ln \Omega - \bar{g}_{\beta\nu}\partial^2_\mu \ln \Omega, \\
    - \Delta^{\alpha}_{\beta \alpha; \nu} &= - D \partial_\nu \partial_\beta \ln \Omega.
\end{align}

Then, we find the Ricci tensor,

\begin{align}
    R_{\beta \nu} &= \bar{R}_{\beta \nu} + \left(2 -D\right)\partial_\beta \partial_\nu \ln \Omega + \left(D-2\right)\left(\partial_\beta \ln \Omega\right) \left(\partial_\nu \ln \Omega\right)\nonumber \\
     &- \bar{g}_{\beta \nu}\left( \left(D-2\right)\left(\partial_\mu \ln \Omega\right)^2 + \partial_\mu^2\ln{\Omega}\right).
\end{align}

\section*{Ricci scalar}
From
\begin{align}
    R \Omega^{2}&= \Omega^{2} g^{\beta \nu}R_{\beta \nu} \nonumber\\
    &= \bar{R} - \bar{g}^{\beta \nu} \left(D-2\right)\partial_{\beta}\partial_\nu \ln \Omega + \bar{g}^{\beta \nu} \left(D-2\right)\left(\partial_\beta \ln \Omega\right)\left(\partial_\nu \ln \Omega\right) \nonumber\\
    &- \bar{g}^{\beta \nu} \bar{g}_{\beta \nu} \left( \left(D-2\right)\left( \partial_\mu \ln \Omega\right)^2 + \partial_\mu^2 \ln \Omega
    \right),
\end{align}

we get
\begin{equation}
    R = \Omega^{-2} \left( \bar{R} - 2\left(D-1\right)\partial_\mu^2\ln\Omega - \left(D-2\right)\left(D-1\right)\left(\partial_\mu \ln \Omega\right)^2
    \right).
\end{equation}

We can further simplify the expression for the Ricci scalar if we use 

\begin{align}
    \partial_\mu^2 \ln \Omega &= - \Omega^{-2}\left(\partial_\mu \Omega\right)^2 + \Omega^{-1}\partial_\mu^2 \Omega,\\
    \left(\partial_\mu \ln \Omega\right)^2 &= \Omega^{-2}\left(\partial_\mu\Omega\right)^2.
\end{align}
Then,
\begin{equation} \label{eq:ricciconformaltrans}
    R = \Omega^{-2} \left( \bar{R} + \Omega^{-2}\left(\partial_\mu \Omega\right)^2\left[2\left(D-1\right)-\left(D-2\right)\left(D-1\right)
        \right] - 2 \left(D-1\right)\Omega^{-1}\partial_\mu^2\Omega\right).
\end{equation}

\section*{Application example} \label{conformaltranspole}

Now we apply the result from the previous sections to the particular case of the Higgs-Pole inflation, studied in \textbf{Chapter \ref{HiggsPole}}.
In this case, the Jordan frame Lagrangian is given by
\bea
\frac{{\cal L}_J}{\sqrt{-g_J}} = -\frac{1}{2}M^2_P\, \Omega(H) R(g_J) + |D_\mu H|^2 -V_J(H),
\eea
with 
\bea
\Omega(H)^2 &=& 1-\frac{1}{3M^2_P}|H|^2 \\
V_J(H) &=& \mu^2_H |H|^2 +\lambda_H |H|^4 + \sum_{n=3}^\infty \frac{c_n}{\Lambda^{2n-4}} \, |H|^{2n}. 
\eea

We use eq.~(\ref{eq:conformaltransformetric}) and eq.~(\ref{eq:ricciconformaltrans}),

\bea
R_J &=& \Omega^2 R_E + \frac{6}{\Omega}\square \Omega\\
\frac{{\cal L}_E}{\sqrt{-g_E}\Omega^{-4}} &=& -\frac{1}{2} \Omega^2 \left(\Omega^2 R_E + \frac{6}{\Omega}\square \Omega\right) + \Omega^2 g^{\mu \nu}_E |D_\mu H||D_\nu H| -V_J.
\eea

Hence, in the Einstein frame,

\bea
\label{appendixeinstein}
\frac{{\cal L}_E}{\sqrt{-g_E}} = -\frac{1}{2}R_E - \frac{3}{\Omega^3}\square \Omega +  \frac{g^{\mu \nu}_E}{\Omega^2} |D_\mu H||D_\nu H| - V_E,
\eea

where we defined 

\bea
V_E=\frac{V_J}{\Omega^4}.
\eea

To find the form that we used in \textbf{Chapter \ref{HiggsPole}}, we rewrite first the second term in eq. (\ref{appendixeinstein}) as

\bea
- \frac{3}{\Omega^3} \square \Omega &=& \frac{3}{4\Omega^4} g^{\mu\nu}_E \partial_\mu{\Omega^2} \partial_\nu{\Omega^2}\nonumber\\
&=& \frac{1}{12}\frac{\partial_\mu |H|^2 \partial^\mu |H|^2}{\Omega^4}
\eea

From the third term in the equation we get

\bea
\frac{g_E^{\mu\nu}}{\Omega^2}|D_\mu H|^2 |D_\nu H|^2 &=& \frac{1}{\Omega^4}\Omega^2 |D_\mu H|^2\nonumber\\
&=& \left(1 -\frac{1}{3M_p^2}|H|^2\right)\frac{|D_\mu H|^2}{\Omega^4}\nonumber\\
&=&\frac{|D_\mu H|^2}{\Omega^4} -\frac{1}{3 M_p^2}\frac{|H|^2 |D_\mu H|^2}{\Omega^4}.
\eea

The total Lagrangian is then given by

\bea
\frac{\mathcal{L}}{\sqrt{-g_E}} &=& -\frac{1}{2}R_E - \frac{1}{3M_p^2 \left( 1 -\frac{1}{3M_p^2|H|^2}\right)^2}\left(|H|^2 |D_\mu H|^2 - \frac{1}{4} \partial_\mu |H|^2 \partial^\mu|H|^2\right) \nonumber\\
 &+& \frac{|D_\mu H|^2}{\left( 1 -\frac{1}{3M_p^2|H|^2}\right)^2} - V_E.
\eea

    \begin{kabstract}

우주 급팽창 이론은 우주의 평탄성 문제와 지평선 문제를 해결할 뿐 아니라 우주 구조 형성에 필요한 씨앗도 제공하기에 현대 우주론의 기둥으로 자리잡고 있습니다. 무수히 많은 우주 급팽창 모형을 구별하기 위해서는 급팽창 너머의 재가열 시기를 살펴볼 필요가 있습니다. 본 학위연구에서는 이러한 초기우주에 대한 지식과 입자물리학의 지식을 연결짓는 한 걸음을 내딛으려 합니다. 이를 위해 여기에서는 크게 다음의 세 질문을 살펴보려 합니다. i) 인플라톤의 본질은 무엇인가, ii) 우주 급팽창과 재가열을 통해 어떻게 입자물리의 수수께끼를 밝혀낼 수 있는가, 그리고 iii) 자연스럽게 우주 급팽창을 설명하는 모형을 어떻게 구축할 수 있는가.

우선 힉스 급팽창 모형을 살펴보았습니다. 힉스 급팽창 모형은 유니테리 성질을 충족시키지 못하는 문제를 가지고 있는 것으로 알려져 있습니다. 우리는 이 유니테리 성질이 플랑크 에너지 아래에서는 만족될 수 있는, 고차 곡률항에서 비롯된 UV 완전한 모형들을 제안하였습니다. 이에 따라 새롭게 도입되는 자유도가 진공의 안정성에 미치는 영향과, 우주 재가열 시기 및 그 이후에 암흑물질을 생성할 가능성도 논하였습니다. 또한 높은 에너지의 초기우주에서는 양자중력이 중요해지는만큼 이러한 모형을 초중력 이론에 포함시키는 방법을 살펴보았습니다. 초대칭성으로 추가되는 장들에 의한 타키온 불안정성을 고치기 위해 고차 곡률 항을 포함시킬 필요성과 초대칭성 깨짐이 미칠 현상론적 효과도 논하였습니다.

다음으로 쌍둥이 폭포를 갖는 자연스러운 급팽창 모형을 고려하였습니다. 여기서는 $Z_2$ 대칭성이 복사 보정에 대한 퍼텐셜의 안정성을 책임지는데, 별도의 $Z_2'$ 대칭성이 있으면 쌍둥이 폭포 중 하나를 암흑물질장으로 간주할 수 있고 우주 재가열 중에 충분한 암흑물질을 생성하게 됩니다.

마지막으로 우리는 $\alpha$-끌개 유형의 우주 급팽창을 위한 모형을 제안하였습니다. 여기서는 모형의 건설에서부터 이론의 섭동성이 보장됩니다. 우리는 등각 결합을 취해 아인슈타인 프레임에서의 운동에너지 항의 극점에서 전개했습니다. 이 극점에서의 우주 급팽창 구조를 통해 큰 비최소 결합 없이도 성공적인 힉스 급팽창을 얻을 수 있었습니다. 또한 이 모형을 강한 CP 문제 해결을 위한 Peccei-Quinn 장에 적용하였고, 액시온의 운동학적 오정렬 메커니즘을 통해 관측된 암흑물질 잔존 밀도를 재현하였습니다.

\keywords{키워드 : 우주 급팽창, 암흑 물질, 재가열}
    \end{kabstract}

    \begin{acknowledgement}
    \thispagestyle{empty}
    \pagenumbering{gobble}
5년전 저는 지구 반대편을 떠나 처음으로 이곳에 왔습니다. 그리고 지난 5년을 돌아보면 이 모험을 함께한 많은 분들이 계셨습니다. 
먼저 저에게 정말 많은 것을 가르쳐 주신 이현민 교수님께 진심으로 감사를 전합니다. 교수님의 끊임없는 열정 덕분에 저희는 세미나, 워크샵 및 토론으로 가득한 활기찬 그룹을 유지할 수 있었습니다.
    또한 제 작업물을 검토 및 심사하시고 개선이 필요한 부분을 찾아내어 수정하는데 도움을 주신 제 논문심사 교수님들께 깊은 감사를 표합니다.
    그리고 제가 한국에 적응하는데 가장 먼저 도와주고 어려운 순간이 있을 때마다 항상 함께 해준 저의 첫 한국인 친구 강유진에게 감사를 전하고 싶습니다. 
또한 저의 연구 동료들인 최수민, 심성보, 김종국에게 유용한 토론을 함께 할 수 있는 협력자들이 되어주어 감사합니다. 
    여러분 덕분에 연구실 분위기가 매우 편안하고 친근했고 연구를 하는 시간을 즐길 수 있었습니다.
    I am profoundly grateful to Shuntaro and Kimiko, who answered relentlessly my infinite amount of questions, and to Seongsik (notice how I very gracefully included you in the Japanese team), to whom I shall buy a few energy drinks in return for all his help along these years.
    I learned a lot from the fun projects and discussions with Simon, MJ, Junho and Jiseon.
    I am also very thankful for my labmate and flatmate Rojalin, even though you made me cry on a chili-pepper basis.
    Now Carlo, as promised, I am officially dis-acknowledging you, because while I appreciated the physics discussions, you are also a terrible singer.

 진행하기도 했던 우주론 스터디를 포함하여 이성묵, 민의, 김희주, 박중현, 정동연, 박예지, 이윤하, 조정훈과 함께한 토론에 깊은 감사를 표합니다. 또한 길을 함께한 박사과정 학생들과의 Bulgaa, Wanda, 유성문, 조원섭, 김태훈 대화에도 감사합니다.
    한국어 선생님과 동기들께도 감사의 말씀을 전합니다. 특히 정예림, 박소영님의 무한한 인내심에 감사드립니다.
    아마도 한국어 선생님들과 동기들 덕분에 지금 한국어로 쓰고 있을지도 모릅니다. 
    더불어 CAUON에서 함께 뛰어준 모든 이들에게 곽경원, 류상원, 한수진, 이주현, 김여훈, 김주영, 김재준, 신경민, Lucas, 조아성, 유지민, 임예슬, 이영학님 등, 매번 저와 함께 뛰며 저의 첫 번째 마라톤 도전을 응원해 주셔서 정말 감사합니다. 
    저는 이 박사과정 동안 스트레스를 받을때마다 CAUON 함께 달리기가 저에게 마음의 평화를 가져다 준다는 것을 깨닫게 되었습니다. 

    Still in Korea, but outside of the physics department and the Heukseok runbase, I want to thank Geet, Hyun, Cath, 의청, Shirley and Jerikah, for being my home far away from home.
    Without you, this PhD would have been a very steep mountain to climb.

\vspace{0.5cm}

I acknowledge the economical support I received from KPS, that allowed me to spend six incredible months at CERN.
I deeply thank everyone at THEP for the insights and discussions shared at R1, during the coffee breaks or in Cosmo Ice-Creams.
If I were to name everyone, then these already long acknowledgments would never end, but let me thank DongWoo Kang and Gauthier Durieux, for their help as I got into the Gegenbauer world; and Geraldine Servant and Guillermo Ballesteros, for their quick feedback as soon as I stepped a foot into axion territory.
Above all, I am extremely grateful for the great deal I could learn in just six months (and that I keep learning) from my incredible collaborators, Matthew and Ennio. 
\vspace{0.5cm}

As many of my peers, I started this PhD right before covid, which means that the first half of my PhD could feel a bit isolating sometimes... or at least it did until I went to Les Houches Summer School in 2021.
I'm extremely indebted for the fantastic lectures, but also for the wonderful PhD students I met there, and that I've kept meeting around the world since: Guillermo, Maud, Charles, J, Arman, Nudzeim, Raquel, Quim, Dom, Federica, Pueh, Alessandro, Anna...

A mention goes here to both Tracy Slatyer and Bibhushan Shakya, firstly for the big deal I learnt from you about Dark Matter, but also (and most importantly) for the vegan-Jeju research work.

So far, most of my talks have been online (from the cosiness of my home, perks of doing a PhD during a pandemic). However, when attending online events, a lot is lost, and so, I'm extremely thankful for those chances at an offline seminar or a student talk.
I'd like to thank Yann Mabrini, for the fantastic astroparticles symposiums that lead to a lot of inspiring discussions.
In particular, I thank Marcos Garcia, Kunio Kaneta and Daniel Figueroa for their insights.
I also want to thank David Weir, for both the insights on gravitational waves shared at CAU and the seminar invitation at Helsinki University.

During the last 5 years, I had the opportunity to attend countless seminars, workshops and conferences within Korea.
I’d specially like to thank Seodong Shin for inviting me to JBNU and Seokhoon Yun for the opportunity to talk at IBS,
as well as the professors and postdocs from whom I learnt a lot through discussions and feedback: Meshkat Rajaee, Pouya Bakhti, Wan-Il Park, Minxi He, Seong Chan Park and Deog Ki Hong.

\vspace{0.5cm}
Back at home, I always felt supported by my crazy, brave, longstanding friends. Mirna, Pria and Angelica,
I really have no words to express what it meant to have you here in Korea for my defense, 
and to have been able to share with you the place I've been calling home. Las quiero fleje.
Bini, your endless support, in both my physics and my sportive endeavors, certainly got me through some of the most difficult days of this adventure. 
And Julia, thanks for being always one phone call away: Hvala što me držiš normalnom.

Sometimes I wonder what my life-path would have been if it weren't for my incredible highschool teachers: Juanjo, who is, in first instance, responsible for my love for physics,
and Guillermo, from whom I inherited a not so healthy passion for learning languages. You may not be here today to celebrate this milestone with me, but what I learnt from you I'll always carry. 

I have the best parents one could ask for. They have always supported each one of my decisions, even that day when I just came home and stated ``I'm leaving to South Korea to get a PhD", like if that was the only reasonable and logical thing to do. 
My mother, who immersed herself in the Korean culture with me, and my father, the first one to hug me at the airport every time I went back home, didn't matter if at 5am on a Monday morning or at 11pm on a Christmas Eve.
You always say how you are proud of me, but truth is, it is me the one who is proud to call you my parents.

Finalment, el meu més profund agraïment és i sempre serà per a Iván, el meu col·laborador en aquest enorme projecte que hem començat junts.
Perdona'm per anar-me'n a viure a l'altre costat del món... i gràcies per demostrar-me que som més forts que 5 anys a 9900km i 14 hores de vol. 

\vspace{5mm}

Seoul, Korea,

May 2024.

    \end{acknowledgement}
\end{document}